\documentclass[aps,prx,reprint,twocolumn,superscriptaddress,nofootinbib,floatfix]{revtex4-2}
 
\usepackage{float}
\usepackage{verbatim}
\usepackage[utf8]{inputenc}
\usepackage[T1]{fontenc}
\usepackage[english]{babel}
\usepackage{dsfont}
\usepackage[normalem]{ulem}
\usepackage{amsmath}
\usepackage{amssymb}
\usepackage{mathrsfs}
\usepackage{amsthm}
\usepackage{mathtools}
\usepackage{wasysym}
\usepackage{breqn}

\usepackage{bbm}
\usepackage{nicefrac}
\usepackage{soul}
\usepackage{stackrel}
\usepackage{braket}
\usepackage{enumitem}
\usepackage{graphicx}
\usepackage{tikz}
\usepackage{mdframed}
\usepackage{booktabs}
\usepackage{siunitx}
\usepackage{fancyhdr}

\usepackage{calc}
\usepackage{accents}

\usepackage{multirow}
\usepackage{array}
\newcolumntype{P}[1]{>{\centering\arraybackslash}p{#1}}

\usepackage[hidelinks]{hyperref}
\usepackage{framed}

\def\diracspacing{0.7pt}

\newcommand{\ketbra}[2]{| \hspace{\diracspacing} #1 \rangle \langle #2 \hspace{\diracspacing} |} % ketbra with different vectors

\addto\captionsenglish{}
\newcommand{\norm}[1]{\left\|#1\right\|}
\newcommand{\abs}[1]{\left|#1\right|}
\DeclareMathOperator{\Tr}{Tr}

\newcommand{\pia}{\Pi^\alpha_{\mathcal{Z}}}

\newcommand{\one}{\mathds{1}}
\newcommand{\Erf}{\operatorname{Erf}}

\newcommand{\proj}[1]{\ketbra{#1}{#1}}
\usepackage{cases}

\newtheorem{theorem}{Theorem}

\newtheorem{remark}{Remark}
\newtheorem{lemma}[theorem]{Lemma}

\addto\captionsenglish{}
\addto\captionsenglish{}

\usepackage{algorithm}

\makeatletter
\newenvironment{protocol}
{% \begin{protocol}
		\renewcommand{\ALG@name}{Protocol}% Update algorithm name
		\refstepcounter{algorithm}% New algorithm
		\hrule height.8pt depth0pt \kern2pt% \@fs@pre for \@fs@ruled
		\renewcommand{\caption}[2][\relax]{% Make a new \caption
			{\raggedright\textbf{\fname@algorithm~\thealgorithm} ##2\par}%
			\ifx\relax##1\relax % #1 is \relax
			\addcontentsline{loa}{algorithm}{\protect\numberline{\thealgorithm}##2}%
			\else % #1 is not \relax
			\addcontentsline{loa}{algorithm}{\protect\numberline{\thealgorithm}##1}%
			\fi
			\kern2pt\hrule\kern2pt
		}
	}{% \end{protocol}
	\kern2pt\hrule\relax% \@fs@post for \@fs@ruled
}
\makeatother

\usepackage{natbib}
\makeatletter
\let\cat@comma@active\@empty
\makeatother

\begin{document}
\title{Quantum Key Distribution with Basis-Dependent Detection Probability}

\author{Federico Grasselli}
\email[Corresponding author: ]{federico.grasselli@hhu.de}
\affiliation{Institut f\"ur Theoretische Physik III, Heinrich-Heine-Universit\"at D\"usseldorf, Universit\"atsstra\ss{}e 1, 40225 D\"usseldorf, Germany}
\affiliation{Leonardo Innovation Labs -- Quantum Technologies, Via Tiburtina km 12,400, 00131 Rome, Italy}

\author{Giovanni Chesi}
\affiliation{QUIT Group, Physics Department, Univ.~Pavia, INFN Sez.~Pavia, Via Bassi 6, 27100 Pavia, Italy}

\author{Nathan Walk}
\affiliation{Dahlem Center for Complex Quantum Systems, Freie Universit\"at Berlin, 14195 Berlin, Germany}

\author{Hermann Kampermann}
\affiliation{Institut f\"ur Theoretische Physik III, Heinrich-Heine-Universit\"at D\"usseldorf, Universit\"atsstra\ss{}e 1, 40225 D\"usseldorf, Germany}

\author{Adam Widomski}
\author{Maciej Ogrodnik}
\author{Micha\l{}  Karpi\'nski}
\affiliation{Faculty of Physics, University of Warsaw, Pasteura 5, 02-093 Warszawa, Poland}

\author{Chiara Macchiavello}
\affiliation{QUIT Group, Physics Department, Univ.~Pavia, INFN Sez.~Pavia, Via Bassi 6, 27100 Pavia, Italy}

\author{Dagmar Bru\ss{}}

\author{Nikolai Wyderka}
\email[Corresponding author: ]{wyderka@hhu.de}
\affiliation{Institut f\"ur Theoretische Physik III, Heinrich-Heine-Universit\"at D\"usseldorf, Universit\"atsstra\ss{}e 1, 40225 D\"usseldorf, Germany}

\begin{abstract}

\noindent Quantum Key Distribution (QKD) is a promising technology for secure communication. Nevertheless, QKD is still treated with caution in certain contexts due to potential gaps between theoretical models and actual QKD implementations. A common assumption in security proofs is that the detection probability at the receiver, for a given input state, is independent of the measurement basis, which might not always be verified and could lead to security loopholes. This paper presents a security proof for QKD protocols that does not rely on the above assumption and is thus applicable in scenarios with detection probability mismatches, even when induced by the adversary. We demonstrate, through simulations, that our proof can extract positive key rates for setups vulnerable to large detection probability mismatches. This is achieved by monitoring whether an adversary is actively exploiting such vulnerabilities, instead of considering the worst-case scenario as in previous proofs. Our work highlights the importance of accounting for basis-dependent detection probabilities and provides a concrete solution for improving the security of practical QKD systems.
\end{abstract}

\maketitle

\section{Introduction}

Quantum Key Distribution (QKD) stands as one of the most studied, developed and commercialized quantum technologies of the past decades \cite{QKDmarkets}. With QKD, two users can, in principle, establish an information-theoretically secure key when linked by an insecure quantum channel and an authenticated classical channel \cite{pir2019advances}, thereby providing a solution for long-term secure communication.

However, QKD is still facing some challenges that hinder its widespread adoption and standardization \cite{positionBSI}. From a security point of view, the main challenge is the implementation security of QKD protocols \cite{ETSI,Zapatero2023implementationQKD}, which arises due to a disagreement between the theoretical models used by the security proofs and the actual implementation of QKD protocols. As a matter of fact, discrepancies between theory and experiment have been exploited to conduct successful quantum hacking attacks on QKD setups \cite{quantumhacking}.

A crucial quantity in the security of prepare-and-measure QKD schemes with two measurement bases is the phase error rate \cite{ShorPreskill}. In simple terms, this is the error rate in a basis complementary to the key basis -- usually called the test basis -- which characterizes the pulses detected in key generation rounds. Standard QKD security proofs assume that the detection probability of a given state is independent of the measurement basis \cite{Tomamichel2017,finite-key-decoyBB84-security,tamaki2014loss}, such that the phase error rate reduces to the (observed) bit error rate of the test measurements. However, this assumption is typically not verified by QKD experimental setups, potentially opening a security loophole that can be exploited by an eavesdropper.

A basis-dependent detection probability --or efficiency, we will use the two terms interchangeably-- could originate from an asymmetry in the nominal efficiency or dark count probability of the detectors used in the two measurement bases \cite{trushechkin-mismatch1,trushechkin-mismatch2}. The resulting mismatch in the detection probability of the two bases is typically independent of the optical mode incident upon the detector and can be easily characterized. On the other hand, a mode-dependent detection efficiency mismatch could originate from an asymmetric coupling of the two measurement bases to the incoming mode. A notable example are tailored light pulses prepared by the adversary to control which basis clicks in time-frequency QKD setups \cite{Nunn2013,BourassaLo}. Regardless of the origin, a basis-dependent detection probability introduces a vulnerability in the QKD protocol that must be treated with hardware countermeasures or security proof fixes. Indeed, popular attacks on QKD systems, such as detector blinding attacks \cite{detector-blinding}, time-shift attacks \cite{time-shift-attack}, and spatial-mode attacks on free-space QKD systems \cite{satelliteQKDmismatch}, can be traced back to basis-dependent detection probabilities.

In this work, we address the security of prepare-and-measure QKD protocols with basis-dependent detection probabilities. In particular, we derive an analytical security proof where the assumption about the independence of the detection probability of a state on the measurement basis is dropped. This implies that the phase error rate is no longer identified with the bit error rate in the test basis. Our proof directly applies to all QKD setups where one of the two measurements is a time-of-arrival measurement, e.g., time-bin QKD \cite{Zbinden-time-bin2,Zbinden-time-bin3,Islam-SciAdv,Islam-QST} and time-frequency QKD \cite{time-freq-QKD-dispersiveoptics,time-freq-QKD-dispersiveoptics2,Nunn2013,Roediger,time-freq-analysis,time-freq-QKD,chang2023large}, but could also be extended to other setups. Our proof is obtained in the asymptotic regime and under collective attacks.

A key role in our solution is played by a high-speed tunable beam splitter (TBS) that is used by the receiver to redirect the signal to the two measurement bases. Our proof processes the rich measurement statistics enabled by the TBS with advanced techniques, including the detector decoy technique \cite{detector-decoy}, to quantify the mismatch in the detection probability of the two bases and reduce the key rate accordingly.

We apply our proof to an experimental time-encoded QKD setup which is prone to efficiency mismatches \cite{TalboteffectWarsaw, ExpWarsawArxiv} and show that it generates positive key rates for honest implementations. We then design a sophisticated attack that induces significant asymmetries in the detection efficiency of the two bases by preparing tailored pulses. We show that our proof reduces the key rate proportionately to the extent of the attack, while the standard BB84 key rate would return overly optimistic key rates due to its inability to detect the attack. Moreover, we argue that previous security proofs applicable to such scenarios would return pessimistic rates since they consider the worst-case scenario, i.e.\ the scenario of the attack, even when the eavesdropper is not present.

The paper is organized as follows. In Sec.~\ref{sec:loopholes} we discuss security loopholes and countermeasures linked to basis-dependent detection probabilities and describe an attack exploiting asymmetric nominal efficiencies. In Sec.~\ref{sec:protocol} we illustrate a generic prepare-and-measure QKD protocol, which may feature basis-dependent detection probabilities. In Sec.~\ref{sec:security} we highlight the main points of the protocol's security proof, while we defer the fully-detailed proof to Appendix~\ref{app:security-proof}. In Sec.~\ref{sec:simulations}, we simulate the key rate of a time-encoded QKD scheme with our security proof both in an honest implementation and with attack-induced detection efficiency asymmetries. We provide further simulation details in Appendix~\ref{app:simulations}. We discuss the results of the simulations in Sec.~\ref{sec:discussion} and conclude in Sec.~\ref{sec:conclusion}. In Appendix~\ref{app:notation} we summarize the notation adopted in the manuscript, while in Appendix~\ref{app:phase-error-rate-formula} we report the formula for the phase error rate resulting from our security proof. Finally, in Appendix~\ref{app:decoy} we derive some of the bounds required by our proof with the decoy-state method.

\section{Security loopholes and countermeasures} \label{sec:loopholes}

Several proposed and implemented QKD schemes can present asymmetries in the detection probability of the two measurement bases, for certain input states.  If their security proof fails to account for this fact, an eavesdropper could exploit the security loophole to invalidate the security claim on the established keys. In this paper, we adopt the common nomenclature of QKD for which Alice is the sender of the pulses and Bob the receiver, while Eve is the eavesdropper.

A mode-independent detection probability mismatch is easier to characterize and treat since the magnitude of the mismatch is independent of the light mode measured by Bob. It can be caused, e.g., by detectors with unequal nominal efficiencies and/or unequal dark count probabilities. Nevertheless, here we provide an example where ignoring an asymmetry in the nominal detection efficiency of two detectors opens the door for a successful attack by Eve. To our knowledge, this type of attack has not been listed in the recent report by the German cyber security agency \emph{Bundesamt für Sicherheit in der Informationstechnik} (BSI) about implementation attacks on QKD \cite{BSIonQKDattacks}.

Consider the BB84 protocol where Bob has an active basis choice and, for simplicity, Alice has a deterministic single-photon source (the attack would also work with decoy BB84). Suppose that Eve performs an intercept-and-resend attack where she intercepts and measures in the $Z$ or $X$ basis, each  with probability $p$, while she lets the signal go undisturbed with probability $1-2p$. Since Eve's attack generates noise in Bob's measurements, the key rate is reduced accordingly. In particular, the fraction of secret key bits in the sifted rounds is given by the well-known BB84 asymptotic key rate:
\begin{align}
    r_{\mathrm{BB84}}= 1-h(Q_X) - h(Q_Z) \label{keyrateBB84-ideal}.
\end{align}
Now, if both the $Z$-basis and $X$-basis detectors have unit efficiency and if the quantum channel is ideal, except for Eve's attack, the quantum bit error rate (QBER) in the two bases reads: $Q_X = Q_Z = p/2$, where the factor $1/2$ accounts for the fact that, when Eve's and Bob's bases differ, Bob's outcome is random. Now, suppose that the $X$-basis detector has efficiency $\eta_X$, with $\eta_X <1$, while the $Z$-basis detector is still ideal. In this case, Eve performs the same attack as before, but this time she re-sends a pulse with $n$ photons (one photon) when she measures in the $X$ basis ($Z$ basis). This has the effect of spoiling the statistics of Bob's $X$-basis outcomes, by flooding them with events where there is no error since Eve's and Bob's bases agree, thereby decreasing  the QBER in the $X$ basis. Indeed, the error probability in the $X$ basis is unchanged -- the errors are still caused by the one photon sent by Eve when she chooses the wrong basis, i.e., the $Z$ basis -- but the detection  probability increases due to more photons arriving when Eve's and Bob's bases coincide. Overall, this decreases the QBER in the $X$ basis, which now reads:
\begin{align}
    Q_X = \frac{p}{2} \left[1 + p \frac{1-\eta_X}{\eta_X} \left[1- (1-\eta_X)^{n-1} \right]\right]^{-1} < \frac{p}{2}.
\end{align}
On the other hand, $Q_Z$ remains unchanged by Eve's new attack, if we assume that multi-click events in the $Z$ basis are assigned randomly to one of the outcomes. As a consequence, the fraction of secret key bits in the sifted rounds, as per Eq.~\eqref{keyrateBB84-ideal}, increases compared to the case with ideal detectors, but Eve's knowledge about the key remains the same (indeed, note that the rounds discarded due to a no-click event in the $X$ basis are announced publicly, hence known to Eve). This implies that the key rate provided in the case of inefficient detectors is an overestimation of the actual secure key rate. The security of the key is thus compromised.

It is worth mentioning that security fixes in the case of detectors with mismatching nominal efficiencies have already been laid out, e.g., in \cite{trushechkin-mismatch1,trushechkin-mismatch2}, for the BB84 protocol with an active basis choice. 

Mismatching detection probabilities can also be a result of the different way in which the two measurement bases are constructed. For instance, Ref.~\cite{Zbinden-time-bin2} features a two-dimensional time-bin encoding where the key basis measurement consists of a simple time-of-arrival measurement with a single-photon detector. Conversely, the test basis measurement is realized by an unbalanced Michelson interferometer with a delay in one of its arms, thereby spreading the whole signal over the range of three time bins, followed by a detector in only one of its output ports. This fact, combined with the fact that the outer time bins overlap with the bins of the neighboring rounds, implies that about $50\%$ of potential detections are unobserved in the test basis compared to the key basis. This violates the basis-independent detection probability assumption of the security proofs \cite{finite-key-decoyBB84-security,tamaki2014loss} adopted by Ref.~\cite{Zbinden-time-bin2}. Similarly, in the BB84 protocols implemented in \cite{Zbinden-time-bin3,Zahidy2024}, the test basis measurement setup includes, among other optical elements,  an interferometer where only one output is detected, such that the overall efficiency is reduced compared to the key basis measurements. By generalizing the setup of Ref.~\cite{Zbinden-time-bin2}, the authors of Ref.~\cite{Islam-SciAdv} utilize a cascade of interferometers to detect the relative phases between pulses of a four-dimensional time-bin QKD protocol. Despite presenting a detector at each output port, they discard the detections in all but the central time bin, thus reducing the detection efficiency of the test basis by $75\%$ under nominal conditions. In such setups \cite{Zbinden-time-bin2,Islam-SciAdv}, an analogous attack to the one described above, where Eve adds photons when measuring in the more lossy basis, would successfully spoil the security of the established key. Therefore, measures to avoid such attacks or modified security proofs are required to restore implementation security.

Mode-dependent detection probability mismatches depend on the coupling between the incoming light mode and each of the two measurement bases. Hence, they might not emerge when the QKD experiment is operating nominally. Crucially, however, the assumption about a basis-independent detection probability must be verified not only by the experimental signals, but also by any possible signal prepared by an adversary. For instance, prepare-and-measure protocols exploiting the complementarity of time and frequency (energy) degrees of freedom \cite{time-freq-QKD-dispersiveoptics,time-freq-QKD-dispersiveoptics2,Nunn2013,Roediger,time-freq-analysis,time-freq-QKD,chang2023large} are particularly prone to mode-dependent detection efficiency mismatches. In these setups the key basis is typically a time-of-arrival measurement, while the test basis consists of a frequency measurement. This is done either directly with spectrometers or indirectly through group delay dispersion. In the latter case, a dispersive medium spreads out a signal with varying delays depending on its frequency components, such that from the arrival time of a dispersed signal one can infer a frequency value. Regardless of the implemented frequency measurement, the finite detection windows in time and frequency represent a vulnerability that can be exploited by an eavesdropper. Indeed, as pointed out in \cite{Nunn2013,BourassaLo}, Eve can intercept the signals and measure either their time or frequency, after which she prepares a very narrow pulse in time (if she measured in time) or in frequency (if she measured the frequency) and sends it to Bob. The detection probability of such signals at Bob strongly depends on the measurement basis. In fact, in the opposite basis to Eve's, the Fourier transform of a narrow pulse spreads out much more than the finite detection window of Bob's apparatus, thereby decreasing the detection probability significantly. This ensures that the majority of the detection events are those where Bob's and Eve's choice of basis matches, hence invalidating security (unless no-detection events are kept).

Current security proofs that could account for mode-dependent detection probabilities are limited in scope and, more importantly, would return overly pessimistic key rates \cite{detection-efficiency-mismatch-Lo,detection-efficiency-mismatch-Ma,detection-efficiency-mismatch-Lydersen,detection-mismatch-Luetkenhaus}. In particular, the analytical proofs in \cite{detection-efficiency-mismatch-Lo,detection-efficiency-mismatch-Ma} restrict Bob's input to be a single-photon signal, while the numerical security proof in \cite{detection-mismatch-Luetkenhaus} allows for multi-photon states received by Bob but assumes a deterministic single-photon source used by Alice.
The main limitation of such proofs, however, is that they require a prior characterization of the mode-dependent detection efficiency of both measurement bases, for any possible incoming mode. Then, the worst-case efficiency mismatch between the two bases, among all potential input modes, determines the secure key rates of Refs.~\cite{detection-efficiency-mismatch-Lo,detection-efficiency-mismatch-Ma,detection-efficiency-mismatch-Lydersen,detection-mismatch-Luetkenhaus}. Hence, the resulting key rates are overly pessimistic in all the practical scenarios where there is no eavesdropper tailoring the signals to those modes with the largest efficiency mismatch.

On the contrary, our proof can handle both mode-independent and mode-dependent detection probability mismatches, it does not confine Alice's output or Bob's input to the single-photon subspace, and is applicable to time-encoded QKD protocols like the ones discussed in this section. Moreover, it can generate significantly higher key rates than previous proofs for honest implementations of protocols which are particularly sensitive to mode-dependent detection efficiency mismatches, e.g., time-frequency QKD (see Sec.~\ref{sec:discussion}).

\section{QKD protocol} \label{sec:protocol}

We consider a prepare-and-measure QKD protocol with two measurement bases and $d$ outcomes per measurement. Bob measures the states sent by Alice with a measurement apparatus that is partially characterized. Namely, we assume that the mode-independent detection efficiencies (e.g.\ nominal detector efficiency, coupling loss) and the dark count probabilities of both measurement bases are known. Nevertheless, an imperfect characterization of such quantities would not affect security: any deviation between the characterized values and the setup's actual behavior will be treated by our proof as an attack attempt. Importantly, opposed to previous proofs \cite{detection-efficiency-mismatch-Lo,detection-efficiency-mismatch-Ma,detection-efficiency-mismatch-Lydersen,detection-mismatch-Luetkenhaus}, we do not require a careful characterization of (possibly adversarial) mode-dependent efficiency mismatches.

\subsection{Alice's states}

Alice prepares states from two sets, named the $Z$-basis and $X$-basis states, respectively. The $Z$-basis states are primarily used for key generation while the $X$-basis states are used for testing. 
In both bases, Alice prepares phase-randomized coherent states (PRCSs), as required by the decoy-state method \cite{Decoy2,Decoy3,decoy-bounds}, with three different intensities: $\mu_1$, $\mu_2$, and $\mu_3$. The three intensities satisfy: $\mu_1 >\mu_2 + \mu_3$ and $\mu_2>\mu_3 \geq 0$.

The $Z$-basis states form the set $\{\rho_{Z_j}\}_{j=0}^{d-1}$, where:
\begin{align}
    \rho_{Z_j}(\mu_i) = {\textstyle\sum_{n=0}^\infty} \Pr(n|\mu_i) \ketbra{n_{Z_j}}{n_{Z_j}},  \label{stateKG}
\end{align}
is the state corresponding to the $j$-th symbol of the $Z$ basis, with intensity $\mu_i$ ($i=1,2,3$). In \eqref{stateKG} we defined the Fock state of $n$ photons in the mode of the $j$-th symbol:
\begin{align}
    \ket{n_{Z_j}} := \frac{1}{\sqrt{n!}} \left(a^\dag_{Z_j}\right)^n \ket{vac} \label{n-psij},
\end{align}
where $a^\dag_{Z_j}$ is the creation operator, as well as the Poissonian distribution of photons typical of PRCSs:
\begin{align}
    \Pr(n|\mu_i) = e^{-\mu_i}\frac{\mu_i^n}{n!}. \label{Pr(n|mu)}
\end{align}
The $X$-basis states are given by $\{\rho_{X_k}\}_{k=0}^{d-1}$, where:
\begin{align}
    \rho_{X_k}(\mu_i) = {\textstyle\sum_{n=0}^\infty} \Pr(n|\mu_i) \ketbra{n_{X_k}}{n_{X_k}}, \label{stateTest}
\end{align}
is the state corresponding to the $k$-th symbol of the $X$ basis, where $\ket{n_{X_k}}=(a^\dag_{X_k})^n\ket{vac}/\sqrt{n!}$ is the Fock state of $n$ photons in the mode of the $k$-th symbol and $a^\dag_{X_k}$ is the creation operator. The only requirement on the states prepared by Alice is that their single-photon components do not reveal Alice's choice of basis to a potential eavesdropper. In other words, the average state prepared by Alice in the $Z$ and $X$ basis must coincide, when restricted to the single-photon subspace \cite{finite-key-decoyBB84-security,Tomamichel2017}:
\begin{align}
S:= {\textstyle\sum_{j=0}^{d-1}} \ketbra{1_{Z_j}}{1_{Z_j}} = {\textstyle\sum_{k=0}^{d-1}} \ketbra{1_{X_k}}{1_{X_k}} .\label{equal-sum-twobases}
\end{align}
This condition, satisfied by many QKD implementations (e.g.\ polarization, time-bin, etc), can be achieved if, e.g., the two bases are linked by a discrete Fourier transform:
\begin{align}
    a^\dag_{X_k} &= \frac{1}{\sqrt{d}} {\textstyle\sum_{j=0}^{d-1}} e^{\frac{-2\pi i}{d} k j} a^\dag_{Z_j}. \label{Fourier-transform}
\end{align}

\subsection{Bob's measurement} \label{sec:Bob-apparatus}

\begin{figure}[htb]
    \centering
    \includegraphics[width=0.6\linewidth,keepaspectratio]{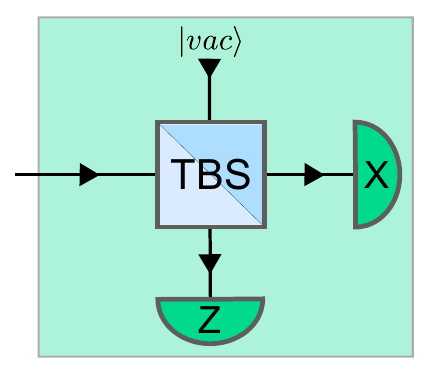}
    \caption{Bob's measurement apparatus. The incoming signal (entering Bob's lab from the left) is split into two modes by a tunable beam splitter (TBS). The transmitted mode is measured by an $X$-basis detector and the reflected mode by a $Z$-basis detector. The clicks in the $Z$ detector are used to generate key bits while the joint detection statistics from both detectors are used for testing.}
    \label{fig:Bobs-setup}
\end{figure}

Bob's measurement apparatus is schematized in Fig.~\ref{fig:Bobs-setup}. For concreteness, here we describe the scenario where Bob's $Z$-basis detector measures the time of arrival of Alice's pulses. Nevertheless, the protocol and its proof are general and can be applied to other scenarios where Alice and Bob employ different photonic degrees of freedom.

The tunable beam splitter (TBS) splits the signal received by Bob into a transmitted and a reflected mode. The reflected mode is measured by Bob's $Z$-basis detector, whose outcomes form Bob's raw key. The $Z$-basis detector is characterized by a mode-independent, or nominal, detector efficiency $\eta_Z$, which includes the insertion loss of the TBS. We indicate with $\mathcal{Z}$ the total time window in which the detector may click in a single measurement round. The time window $\mathcal{Z}$ is partitioned into discrete time-bins $\mathcal{Z}_j$, with $\mathcal{Z}=\cup_{j=0}^{d-1} \mathcal{Z}_j$, that match those used by Alice to encode the key symbols. When a multi-click event occurs, i.e., a measurement with multiple clicks in different intervals $\mathcal{Z}_j$, Bob must map it to a single measurement outcome $j\in\{0,\dots,d-1\}$ according to an arbitrary (potentially probabilistic) mapping; however, Bob cannot map it to the no-detection outcome, $\emptyset$.  The $Z$-basis detector could also be affected by dark counts. With $p^Z_d$ we indicate the probability that the $Z$-basis detector clicks in the interval $\mathcal{Z}_j$, for every $j$, given that no photon arrived in that interval.

The mode transmitted by the TBS is collected by Bob's $X$-basis detector, which can implement any measurement allowing Bob to discern the test symbols sent by Alice. The $X$-basis detector is characterized by a mode-independent, or nominal, detector efficiency $\eta_X$, which includes the TBS insertion loss and is assumed to be smaller than the efficiency of the key generation measurement, $\eta_X\leq \eta_Z$, otherwise the two measurements can be exchanged. Let $\mathcal{X}_k$ be the physical mode corresponding to the $k$-th outcome and let $\mathcal{X}=\cup_{k=0}^{d-1}\mathcal{X}_k$ be the set of all modes that can cause a click in the $X$-basis detector. Similarly to the $Z$-basis measurement, we assume that Bob maps each multi-click event to one of the $d$ measurement outcomes: $k\in\{0,\dots,d-1\}$, excluding the no-detection outcome $\emptyset$, with an arbitrary mapping. Finally, we define $p_d^X$ to be the dark count probability in mode $\mathcal{X}_k$, for each $k$.

We remark that we  allow the effective detection efficiency of the two measurement bases to arbitrarily differ due, e.g., to tailored light modes that couple differently to the measurement apparatus of the two bases.

\begin{figure}[t]
    \centering
    \includegraphics[width=0.8\linewidth,keepaspectratio]{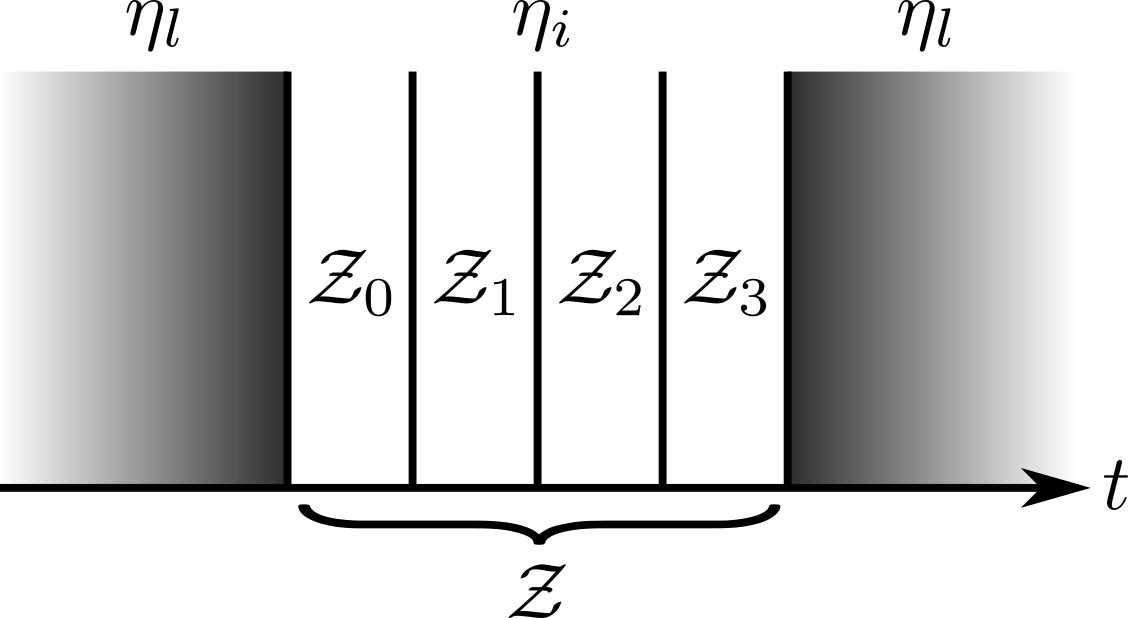}
    \caption{Schematic representation of one measurement round. Only during the time interval $\mathcal{Z} = \bigcup_{j=0}^{d-1}\mathcal{Z}_j$ does the $Z$-basis detector register clicks and assign the corresponding outcome. The tunable beam splitter switches between transmittance $\eta_l$ outside of the detection window $\mathcal{Z}$ and $\eta_i$ inside of $\mathcal{Z}$. This TBS setting is labeled by $(\eta_i, \eta_l)$. The statistics generated by different choices for $\eta_i$ and $\eta_l$ are used in the security proof to bound Eve's knowledge of the key.}
    \label{fig:detection_round}
\end{figure}

We now turn to describe the capabilities of the TBS. The TBS can be tuned at high speed such that its transmittance can be set to different values within one measurement round. In particular, with $(\eta_i,\eta_l)$ we indicate a setting where the transmittance of the TBS, in one round, is set to $\eta_i$ within the duration of the time interval $\mathcal{Z}$ and to $\eta_l$ outside of that interval. Note that the duration of each measurement round is typically longer than the time interval $\mathcal{Z}$ reserved for measuring in the key basis (due, e.g., to a longer detection time required in the test basis). A graphical representation of a detection round and the corresponding TBS settings is given in Fig.~\ref{fig:detection_round}. In our proof, we assume that the TBS cannot be influenced by eavesdroppers and, in particular, that its transmittance settings are mode-independent.

Due to the TBS finite extinction ratio, we assume that the TBS transmittances are always contained in the range: $[\eta_\downarrow,\eta_\uparrow]\subset [0,1]$, where $\eta_\uparrow$ ($\eta_\downarrow$) represents the maximal (minimal) transmittance allowed by the TBS. The maximal and minimal transmittances need to satisfy the following two constraints, required by our security proof:
\begin{align}
    \eta_\uparrow &> \frac{\eta_\downarrow}{1-\eta_\downarrow} \label{constraint1} \\
    \frac{\eta_X}{\eta_Z} &> (\eta_\downarrow)^{-1} \left(1-\sqrt{1-\frac{\eta_\downarrow}{\eta_\uparrow}}\right) \label{constraint2}.
\end{align}
We argue that the two constraints are not particularly stringent. For example, consider that, typically, the tuning of the TBS is symmetric such that $\eta_\downarrow=1-\eta_\uparrow$. Then, the condition in \eqref{constraint1} becomes: $\eta_\uparrow> (\sqrt{5}-1)/2 \approx 0.62$. This inequality is expected to be satisfied by TBSs that function as basis selectors. Similarly, the condition in \eqref{constraint2}, for $\eta_\downarrow=1-\eta_\uparrow$, can be satisfied, e.g., with $\eta_X/\eta_Z \geq 0.6$, and $\eta_\uparrow>0.87$, which should be achievable values. In particular, we observe that a high-speed TBS, as required by our proof and satisfying the constraints \eqref{constraint1} and \eqref{constraint2}, could be experimentally implemented by, e.g., a dual-output Mach-Zehnder modulator \cite{TBS1,TBS2}. 

As a final remark, if Bob measures another photonic degree of freedom than the arrival time in key generation rounds, then $\mathcal{Z}$ would represent the set of modes that are detected by the $Z$-basis detector. Notably, the framework and the proof we provide are still applicable, as far as the TBS can be dynamically tuned along the new degree of freedom measured in the $Z$ basis.

In the following, we describe the generic QKD protocol whose security is proven in Sec.~\ref{sec:security}.\\

\begin{protocol} \label{prot} \caption{QKD protocol}
\begin{enumerate}[wide, labelwidth=!, labelindent=0pt]
\item The following describes one measurement round of the protocol and it must be iterated for a sufficient amount of rounds.
\begin{enumerate}[label*=\arabic*.]
    
    \item 
    Alice prepares a $Z$-basis state (event $T=Z$) with probability $p_Z$ and an $X$-basis state (event $T=X$) with probability $1-p_Z$. When preparing a $Z$-basis ($X$-basis) state, Alice draws the symbol $j\in\{0,\dots,d-1\}$ ($k\in\{0,\dots,d-1\}$) uniformly at random and records the outcome in the random variable $Z_A$ ($X_A$). She then chooses an intensity $\mu_i \in\mathcal{S}:=\{\mu_1, \mu_2, \mu_3\}$ with probabilities $p_{\mu_1}$, $p_{\mu_2}$ and $p_{\mu_3}=1-p_{\mu_1}-p_{\mu_2}$ and records it in a random variable $I_A$. Based on her choices, Alice prepares the phase-randomized coherent state $\rho_{Z_j}(\mu_i)$ ($\rho_{X_k}(\mu_i)$) given in Eq.~\eqref{stateKG} (Eq.~\eqref{stateTest}) and sends it to Bob via an insecure quantum channel. The states prepared by Alice in the two bases satisfy \eqref{equal-sum-twobases}.
    
    \item  In each round, Bob chooses one of seven different TBS settings, namely: $(\eta_i,\eta_\uparrow)$ for $i=1,2,3$, $(\eta_i,\eta_\downarrow)$ for $i=1,2,3$, and $(\eta_2,\eta_2)$. The transmittances are chosen as: $\eta_1=\eta_\uparrow$,  $\eta_3=\eta_\downarrow$ and $\eta_2$ satisfies: $\eta_\downarrow<\eta_2<\eta_\uparrow$ (a close-to-optimal choice for maximizing the key rate is: $\eta_2 = (1/4)(\sqrt{\eta_\downarrow} + \sqrt{\eta_\uparrow})^2$). In particular, Bob chooses the setting $(\eta_\downarrow,\eta_\downarrow)$ with probability $p_Z$ or one of the other six settings with probability $(1-p_Z)/6$ each.\\
    If the $Z$ ($X$) detector clicks and returns outcome $j$ ($k$), Bob sets $Z_B=j$ ($X_B=k$). Otherwise, if the $Z$ ($X$) detector does not click, Bob sets $Z_B=\emptyset$ ($X_B=\emptyset$).
\end{enumerate}

    \item Public announcements: For each round, the parties announce the following information over an authenticated public channel. Alice announces the type of round she performed ($T$) and her intensity choice ($I_A$), while Bob announces the TBS setting and whether the $Z$ detector clicked ($Z_B\neq\emptyset$) or not ($Z_B=\emptyset$). The parties label the rounds where: $T=Z$, $Z_B\neq \emptyset$, and Bob selects the TBS setting $(\eta_\downarrow,\eta_\downarrow)$ as ``key generation rounds'', while the other rounds are labeled as ``test rounds''. For the test rounds, Bob additionally announces $X_B$ over the public channel.
    
    \item Gains estimation: From the information announced, the parties estimate the $Z$-basis gains:
    \begin{align}
        G^Z_{\mu_j,(\eta_l,\eta_l)} &= \Pr(Z_B \neq \emptyset|T=Z,I_A=\mu_j,(\eta_l,\eta_l)) \label{gainKG} ,
    \end{align}
    for each $\mu_j\in\mathcal{S}$ and $\eta_l \in\{\eta_\uparrow,\eta_2,\eta_\downarrow\}$. They also estimate the $X$-basis gains:
    \begin{align}
    G^{X,\checkmark}_{\mu_j,(\eta_i,\eta_l)} &= \Pr(X_B \neq \emptyset,Z_B \neq \emptyset|T=X,I_A=\mu_j,(\eta_i,\eta_l)) \label{gaintest-Zclick} \\
        G^{X,\emptyset}_{\mu_j,(\eta_i,\eta_l)} &= \Pr(X_B \neq \emptyset, Z_B= \emptyset|T=X,I_A=\mu_j,(\eta_i,\eta_l)) \label{gaintest-Znoclick},
    \end{align}
    for $\mu_j\in\mathcal{S}$ and for $\eta_i\in\{\eta_\uparrow,\eta_2,\eta_\downarrow\}$ and $\eta_l \in\{\eta_\uparrow,\eta_\downarrow\}$.
    
    \item Errors estimation: Bob reveals $Z_B$ for a subset of the key generation rounds in order for Alice to compute the quantum bit error rate (QBER) of the key generation rounds, for each intensity chosen by Alice:
    \begin{align}
        Q_{Z,\mu_j} = \Pr(Z_A \neq Z_B |T=Z, I_A=\mu_j, (\eta_\downarrow,\eta_\downarrow),Z_B\neq \emptyset) \label{QBER-Z}.
    \end{align}
    Alice also computes the QBERs of the test rounds with a detection in the $X$-basis detector ($X$-basis QBERs):
    \begin{align}
        &Q_{X,\mu_j,(\eta_i,\eta_\uparrow),\checkmark}= \nonumber\\
        & \Pr(X_A \neq X_B |T=X, I_A=\mu_j ,(\eta_i,\eta_\uparrow),X_B \neq \emptyset,Z_B \neq \emptyset) \label{QBER-X-Zclick}\\
        &Q_{X,\mu_j,(\eta_i,\eta_\uparrow),\emptyset} = \nonumber\\
        &\Pr(X_A \neq X_B |T=X, I_A=\mu_j ,(\eta_i,\eta_\uparrow) ,X_B \neq \emptyset, Z_B=\emptyset) \label{QBER-X-Znoclick},
    \end{align}
    for $\eta_i\in\{\eta_\uparrow,\eta_2,\eta_\downarrow\}$ and $\mu_j \in \mathcal{S}$.
    
    \item Classical post-processing: The parties, by performing error correction and privacy amplification, extract a shared secret key from the variables $Z_A$ (for Alice) and $Z_B$ (for Bob) relative to the key generation rounds where Bob did not reveal $Z_B$. The fraction of shared secret key bits established per protocol round, in the asymptotic limit of infinitely many rounds, is given by:
    \begin{align}
        r_\infty = &p_Z^2 {\textstyle\sum_{j=1}^3} p_{\mu_j} \left\lbrace e^{-\mu_j} \underline{Y^Z_{0,(\eta_\downarrow,\eta_\downarrow)}} \log_2 d  \right.\nonumber\\
        &\left. + e^{-\mu_j} \mu_j \underline{Y^Z_{1,(\eta_\downarrow,\eta_\downarrow)}} \left[\log_2 \left( \frac{1}{c} \right) - u(\,\overline{\tilde{e}_{X,1}}\,)\right] \right.\nonumber\\
        &\left.- G^Z_{\mu_j,(\eta_\downarrow,\eta_\downarrow)} u(Q_{Z,\mu_j}) \right\rbrace      \label{protocol-rate},
    \end{align}
    and is called the asymptotic key rate of the protocol. Below we define the quantities appearing in the key rate expression.
\end{enumerate}
\end{protocol}\vspace{2ex}

We define the compatibility coefficient $c$ of Alice's states in the one-photon subspace \cite{Tomamichel2017}:
\begin{align}
    c= \max_{0 \leq j,k \leq d-1} \abs{\bra{1_{Z_j}} S^{-1} \ket{1_{X_k}}}^2 \,,  \label{compatibility}
\end{align}
where $S^{-1}$ is the generalized inverse of $S$ (the inverse on its support), with $S$ given in \eqref{equal-sum-twobases}. The key rate is maximal when the compatibility coefficient is minimal (i.e., $c=1/d$), and this occurs if Alice's states, in the one-photon subspace, form two mutually unbiased sets (proof in  Remark~\ref{rmk:compatibility} in Appendix~\ref{app:security-proof}):
\begin{align}
    \abs{\braket{1_{Z_j}|1_{X_{k}}}}^2 =\frac{1}{d} \quad\forall\, j,k. \label{mutually-unbiased-states}
\end{align}
In turn, the condition in \eqref{mutually-unbiased-states} can be obtained, e.g., by combining the discrete Fourier transform \eqref{Fourier-transform} with the orthogonality of the states in one basis ($\braket{1_{Z_j}|1_{Z_{j'}}}=\delta_{j,j'}$).

Next, the function $u(x)$ appearing in \eqref{protocol-rate} is defined as:
\begin{align}
    u(x)= \left\lbrace \begin{array}{ll}
       h(x) + x \log_2 (d-1)  & \mbox{if } x\in\left( 0,1-\frac{1}{d}\right)  \\[1ex]
        \log_2 d &  x \in \left[1-\frac{1}{d},1\right)
    \end{array}\right. \label{u(x)},
\end{align}
with $h(x)=-x \log_2 x -(1-x)\log_2 (1-x)$ the binary entropy.

One of the main results of the paper is the derivation of the upper bound on the phase error rate of the protocol, $\overline{\tilde{e}_{X,1}}$. Due to its cumbersome expression, we report it in Appendix~\ref{app:phase-error-rate-formula} as a function of (bounds on) the yields and test-round bit error rates.

Finally, $\underline{Y^Z_{n,(\eta_l,\eta_l)}}$ denotes a statistical lower bound on the $n$-photon yield in the $Z$-basis, i.e., the probability of a $Z$-detector click given that Alice sent $n$ photons and Bob chose the TBS setting $(\eta_l,\eta_l)$. The explicit expressions of $\underline{Y^Z_{1,(\eta_\downarrow,\eta_\downarrow)}}$, $\underline{Y^Z_{0,(\eta_\downarrow,\eta_\downarrow)}}$, and of all the other bounds on the yields and bit error rates that appear in $\overline{\tilde{e}_{X,1}}$ are due to the decoy-state method and are reported in Appendix~\ref{app:decoy}.

For a complete overview of the notation and quantities defined in this paper, we refer the reader to Appendix~\ref{app:notation}.

\section{Security proof} \label{sec:security}

We analytically prove the asymptotic security of Protocol~1 under collective attacks by the eavesdropper, thereby deriving the key rate expression in \eqref{protocol-rate}. We remark that  security against coherent attacks cannot be directly inferred by invoking de Finetti-type results (e.g.\ post-selection technique \cite{postselection-tech,postselection-tech-Lutken}). Indeed, such results require a reduction to finite dimensions, which contrasts with the generality of our approach (Bob can receive states in an infinite number of different optical modes).

Unlike standard QKD security proofs \cite{tamaki2014loss,finite-key-decoyBB84-security,Tomamichel2017}, our proof does not rely on the assumption that the detection probability of the key generation measurement and of test measurement coincide, for every input state. In fact, we allow Eve to add photons to Alice's single-photon pulses or prepare states in continuous degrees of freedom (e.g.\ time, frequency) \cite{Niu-Furrer-Shapiro,Walk2016}, such that the detection probability of Eve's states depends on the measurement basis. We remark that such attacks might not manifest themselves in asymmetric gains observed in an experiment, as Eve might be able to conceal her attack by acting symmetrically on both bases. Existing proofs \cite{detection-efficiency-mismatch-Lo,detection-efficiency-mismatch-Ma,detection-efficiency-mismatch-Lydersen,detection-mismatch-Luetkenhaus} accounting for detection efficiency mismatches require, at the very least, that mode-dependent mismatches are fully characterized and known a priori.

The security proof presented here can avoid such a characterization by actively estimating the relevant quantities from the rich statistics generated with the TBS. More specifically, it employs the well-known decoy-state method \cite{Decoy2,Decoy3,decoy-bounds} to focus on the rounds where Alice sends exactly one photon. In parallel, it uses the TBS to apply the detector decoy technique \cite{detector-decoy} and estimate the photon number distribution of Bob's input states. Finally, it can estimate eventual mode-dependent mismatches in the detection efficiency of the two measurement bases, thus not requiring their prior characterization.

In the following, we provide a simplified version of the security proof highlighting the steps that set it apart from previous proofs. The proof in full detail is reported in Appendix~\ref{app:security-proof}.

\begin{proof}
According to the description of Protocol~1 in Sec.~\ref{sec:protocol}, the parties extract the shared secret key from the rounds labeled as key generation rounds, where: Alice sends a $Z$-basis state, Bob selects the TBS setting $(\eta_\downarrow,\eta_\downarrow)$, and the $Z$-basis detector clicks. We label the intersection of these three events with $\Omega_Z$.

The asymptotic secret key rate that can be extracted from the events $\Omega_Z$, under collective attacks, is lower bounded by the Devetak-Winter (DW) rate \cite{DW}:
\begin{align}
    r \geq \Pr(\Omega_Z) \left[H(Z_A|I_A E)_{\rho|\Omega_Z} - H(Z_A|I_A Z_B)_{\rho|\Omega_Z} \right], \label{simplifiedproof-DWrate}
\end{align}
where $H(Z_A|I_A E)$ and $H(Z_A| I_A Z_B)$ are von Neumann entropies computed on the state shared by Alice, Bob and Eve in a generic round, conditioned on the event $\Omega_Z$. This state is explicitly derived in Appendix~\ref{app:proof1state}. Note that we can express:
\begin{align}
    \Pr(\Omega_Z) = p_Z^2 \Pr(Z_B \neq \emptyset|T=Z,(\eta_\downarrow,\eta_\downarrow)). \label{Pr(Omega_Z)}
\end{align}
The goal of the proof is to show that the key rate given in \eqref{protocol-rate} is a lower bound of \eqref{simplifiedproof-DWrate}.

The second entropy in \eqref{simplifiedproof-DWrate} quantifies Bob's uncertainty about Alice's key bit $Z_A$ and represents the information leakage of an optimal error correction scheme. In Appendix~\ref{app:proof2leakage} we show how to bound this term with standard arguments in terms of the QBERs of the key generation rounds, \eqref{QBER-Z}, as follows:
\begin{align}
    H(Z_A|I_A Z_B)_{\rho|\Omega_Z} &\leq {\textstyle\sum_{j=1}^3} \frac{p_{\mu_j} G^Z_{\mu_j,(\eta_\downarrow,\eta_\downarrow)} u(Q_{Z,\mu_j})}{\Pr(Z_B \neq \emptyset|T=Z,(\eta_\downarrow,\eta_\downarrow))}  \label{simplifiedproof-Bobs-entropy-bound},
\end{align}
where the function $u(x)$ is defined in \eqref{u(x)} and the gain $G^Z_{\mu_j,(\eta_\downarrow,\eta_\downarrow)}$ is obtained as explained in step~3 of the protocol.

The remainder of the proof is devoted to deriving a lower bound on the first entropy in \eqref{simplifiedproof-DWrate}, in terms of observed statistics.

The first step consists in discarding the contributions to $H(Z_A|I_A E)_{\rho|\Omega_Z}$ relative to the rounds where Alice sends more than one photon to Bob. This is because such rounds are vulnerable to photon-number-splitting attacks \cite{Decoy1} and hence cannot lead to shared secret bits. Thus, we show in Appendix~\ref{app:proof3Amultiphoton} that the following inequality holds:
\begin{align}
    &H(Z_A|I_A E)_{\rho|\Omega_Z} \geq  {\textstyle\sum_{j=1}^3} \frac{p_{\mu_j}}{\Pr(Z_B\neq\emptyset|T=Z,(\eta_\downarrow,\eta_\downarrow))} \nonumber\\
    &\left[ e^{-\mu_j} Y^Z_{0,(\eta_\downarrow,\eta_\downarrow)} \log_2 d +  e^{-\mu_j} \mu_j Y^Z_{1,(\eta_\downarrow,\eta_\downarrow)} H(Z_A | E)_{\rho|1,\Omega_Z} \right], \label{simplifiedproof-Alice-entropy}
\end{align}
where $Y^Z_{n,(\eta_l,\eta_l)}$ is the $n$-photon $Z$-basis yield (cf.\ Appendix~\ref{app:notation}), i.e., the probability that the $Z$-basis detector clicks, given that Alice sent $n$ photons encoded in a~$Z$-basis state and Bob selected the TBS setting $(\eta_l,\eta_l)$, while $H(Z_A | E)_{\rho|1,\Omega_Z}$ is the entropy conditioned on a key generation round where Alice sent exactly one photon. Note that the conditional entropy is maximal when Alice sends the vacuum: $H(Z_A | E)_{\rho|0,\Omega_Z}=\log_2 d$, since Eve cannot be correlated to Alice's symbol when Alice sends the vacuum.

The next step, which is detailed in Appendix~\ref{app:proof3Buncertainty}, employs the uncertainty relation for von Neumann entropies \cite{entropicUncert,UncertRelSmooth,Furrer:2012um,Furrer:2014ig} to derive a lower bound on the entropy $H(Z_A | E)_{\rho|1,\Omega_Z}$. In particular, thanks to the condition on Alice's states in the one-photon subspace \eqref{equal-sum-twobases}, we can equivalently describe a key generation round in the entanglement-based picture. That is, there exists an entangled state $\ket{\Psi_\tau}_{AB}$, given by the purification of the state $\tau_B=S/d$ with $S$ in \eqref{equal-sum-twobases}, and two fictitious measurements for Alice, called Alice's $Z$-basis and $X$-basis measurement \cite{Tomamichel2017}. Then, Bob's conditional state, when Alice performs the $Z$-basis ($X$-basis) measurement on system $A$ of $\ket{\Psi_\tau}_{AB}$ and obtains outcome $Z_A=j$ ($X_A=k$), coincides with the state $\ket{1_{Z_j}}$ ($\ket{1_{X_k}}$) sent to Bob in Protocol~1 when Alice draws the symbol $Z_A=j$ ($X_A=k$) and sends one photon (recall that Alice prepares probabilistic mixtures of Fock states with a Poissonian distribution, \eqref{stateKG}).

Now, we recall that the entropy $H(Z_A | E)_{\rho|1,\Omega_Z}$ is computed on the state of a key generation round, i.e., the state received by Bob and post-selected on a detection of the $Z$-basis detector with TBS setting $(\eta_\downarrow,\eta_\downarrow)$. Hence, we update the state $\ket{\Psi_\tau}_{AB}$ to the following entangled state:
\begin{align}
    \ket{\Psi} &= \left(\one_{AE} \otimes \frac{\sqrt{Z_\checkmark}}{\sqrt{Y^Z_{1,(\eta_\downarrow,\eta_\downarrow)}}}\right)(\one_A \otimes U_{BE}) \ket{\Psi_\tau}_{AB} \otimes\ket{0}_E,
\end{align}
where $U_{BE}$ is a unitary applied by Eve, jointly on Bob's system and Eve's system $E$, describing her collective attack, while $Z_\checkmark$ is the POVM element corresponding to a detection of a key generation round, i.e., a detection in the $Z$-basis detector with TBS setting $(\eta_\downarrow,\eta_\downarrow)$. The denominator in the last expression ensures normalization. In this way, by applying the uncertainty relation on $\ket{\Psi}$ with Alice's two fictitious measurements, we obtain:
\begin{align}
    H(Z_A | E)_{\rho|1,\Omega_Z} \geq \log_2 \frac{1}{c} - H(X_A|B)_\sigma \label{simplifiedproof-uncer-rel},
\end{align}
where $c$ is given in \eqref{compatibility} and where the entropy on the left-hand side is computed on the state $\ket{\Psi}$ after Alice's $Z$-basis measurement, which results in the entropy that needs to be bounded in \eqref{simplifiedproof-Alice-entropy}. The entropy on the right-hand side, instead, is computed on the state $\ket{\Psi}$ after Alice's $X$-basis measurement, which reads:
\begin{align}
    \sigma_{X_A B}  &=\frac{1}{ Y^Z_{1,(\eta_\downarrow,\eta_\downarrow)} d} \sum_{k=0}^{d-1}  \ketbra{k}{k}_{X_A} \otimes \sqrt{Z_\checkmark}  \sigma_k \sqrt{Z_\checkmark} \label{simplifiedproof-sigmaXB2},
\end{align}
where $\sigma_k$ is the state received by Bob when Alice sends the one-photon $X$-basis state relative to symbol $X_A=k$:
\begin{align}
    \sigma_k &= \Tr_E \left[ U_{BE} \proj{1_{X_k}}_B \otimes \proj{0}_E U_{BE}^\dag \right] \label{simplifiedproof-sigmak}.
\end{align}
We remark that the state in \eqref{simplifiedproof-sigmaXB2} is properly normalized thanks to \eqref{equal-sum-twobases}.

We can now focus on deriving an upper bound on the entropy $H(X_A|B)_\sigma$ appearing in \eqref{simplifiedproof-uncer-rel}. Since the entropy $H(X_A|B)_\sigma$ is conditioned on a quantum system, system $B$, we can apply a measurement map on $B$ such that the resulting conditional Shannon entropy can be estimated with measurement statistics. To this aim, we first define Bob's test measurement by the POVM: $\{X_0,X_1,\dots,X_{d-1},X_\emptyset\}$, with $X_k$ the POVM element describing the detection of outcome $X_B=k$ in the $X$-basis detector, with TBS setting $(\eta_\uparrow,\eta_\uparrow)$. We can also define $X_\checkmark=\sum_{k=0}^{d-1} X_k$ to be the POVM element corresponding to a detection and $X_\emptyset=\one-X_\checkmark$ corresponding to a no detection. Then, we consider the following measurement map:
\begin{align}
    \mathcal{E}^X_{B \to  \tilde{X}_B} (\cdot) =  \sum_{k'=0}^{d-1}  \Tr\left[ \tilde{X}_{k'} \,\cdot\, \right] \proj{k'}_{\tilde{X}_B}   ,  \label{simplifiedproof-measurementmap}
\end{align}
where $\{\tilde{X}_{k'}\}_{k'=0}^{d-1}$ is a POVM defined from Bob's test measurement as follows:
\begin{align}
    \tilde{X}_k &= (\sqrt{X_\checkmark})^{-1} X_k (\sqrt{X_\checkmark})^{-1}  \oplus \frac{\one^{\perp}_{X_\checkmark}}{d},  \label{simplifiedproof-Xprimek}
\end{align}
with $(\sqrt{X_\checkmark})^{-1}$ the inverse of $\sqrt{X_\checkmark}$ over its support and $\one^{\perp}_{X_\checkmark}$ the projector on the complement of the  support of $X_\checkmark$. Then, by applying the measurement map \eqref{simplifiedproof-measurementmap} on system $B$ of \eqref{simplifiedproof-sigmaXB2}, the following inequality holds \cite{NielsenChuang}:
\begin{align}
    H(X_A|B)_{\sigma} &\leq H(X_A|\tilde{X}_B)_{\mathcal{E}^X(\sigma)}  \nonumber\\
    &\leq u(\tilde{e}_{X,1}) \label{simplifiedproof-dataprocessing},
\end{align}
where in the second inequality we used Fano's inequality and defined the phase error rate:
\begin{align}
    \tilde{e}_{X,1} &= \sum_{k=0}^{d-1} \sum_{k'\neq k}
    \frac{\Tr\left[ \tilde{X}_{k'} \sqrt{Z_\checkmark}  \sigma_k \sqrt{Z_\checkmark} \right]}{Y^Z_{1,(\eta_\downarrow,\eta_\downarrow)} d}\label{simplifiedproof-phase-error-rate}.
\end{align}

We can now lower-bound the Devetak-Winter rate in \eqref{simplifiedproof-DWrate} by plugging in \eqref{simplifiedproof-Bobs-entropy-bound} and \eqref{simplifiedproof-Alice-entropy}, where we further bound the latter using \eqref{simplifiedproof-uncer-rel} and subsequently \eqref{simplifiedproof-dataprocessing}. This leads to the following expression: 
\begin{align}
        r \geq &p_Z^2 {\textstyle\sum_{j=1}^3} p_{\mu_j} \left\lbrace e^{-\mu_j} \underline{Y^Z_{0,(\eta_\downarrow,\eta_\downarrow)}} \log_2 d  \right.\nonumber\\
        &\left. + e^{-\mu_j} \mu_j \underline{Y^Z_{1,(\eta_\downarrow,\eta_\downarrow)}}
        \left[\log_2 \left( \frac{1}{c} \right) - u(\,\tilde{e}_{X,1}\,)\right] \right.\nonumber\\
        &\left.- G^Z_{\mu_j,(\eta_\downarrow,\eta_\downarrow)} u(Q_{Z,\mu_j}) \right\rbrace, \label{intermediate-rate-proof}
\end{align}
where we used \eqref{Pr(Omega_Z)} to replace $\Pr(\Omega_Z)$ and replaced the $Z$-basis yields with their respective lower bounds, obtained with the decoy-state method (see \eqref{YZ0-low} and \eqref{YZ1-low}).

The remainder of the proof is devoted to deriving a meaningful upper bound on the phase error rate, Eq.~\eqref{simplifiedproof-phase-error-rate}, in terms of the statistics collected in the test rounds. Indeed, the fact that $u(x)$ given in \eqref{u(x)} is a monotonically non-decreasing function, implies that the upper bound can be used to replace $\tilde{e}_{X,1}$ in \eqref{intermediate-rate-proof} and obtain another lower bound on the DW rate. 

In standard QKD security proofs \cite{finite-key-decoyBB84-security,Tomamichel2017} it is assumed that the detection probability of Bob's key generation measurement is equal to the detection probability of Bob's test measurement, for every input state. Mathematically, this amounts to an equality between the POVM elements corresponding to a detection in the two measurements:
\begin{align}
    Z_\checkmark = X_\checkmark \label{common-assumption}.
\end{align}
By combining the assumption in \eqref{common-assumption} with \eqref{simplifiedproof-Xprimek}, the phase error rate reduces to the one-photon bit error rate of Bob's test measurement, as we derive explicitly in Appendix~\ref{app:proof3Creduction}:
\begin{align}
    \tilde{e}_{X,1} &= \sum_{k=0}^{d-1} \sum_{k'\neq k}
    \frac{\Tr\left[ X_{k'} \sigma_k \right]}{Y^X_{1,(\eta_\uparrow,\eta_\uparrow)} d}\label{simplifiedproof-reduced-phase-error-rate},
\end{align}
which can be readily estimated with the decoy-state method. The challenge in the more general setting addressed by this paper is to estimate the phase error rate when \eqref{common-assumption} is not assumed to hold.

\subsection{Phase error rate estimation}

Here we briefly sketch the main ideas behind the estimation of the phase error rate. The full argument can be found in Appendix~\ref{sec:phase-error-rate-bound}, starting with a detailed overview of the main steps in Appendix~\ref{app:PER0}.

The first step in estimating the phase error rate in \eqref{simplifiedproof-phase-error-rate} consists in reducing the calculation to the subspace containing at most one photon in the set of modes $\mathcal{Z}$ detected by the $Z$-basis detector. We call this the ($\leq 1$)-subspace. The details of this step are deferred to the appendices from Appendix~\ref{app:PER1} to Appendix~\ref{app:PER4} and in Appendix~\ref{app:PER9}. Indeed, we expect most of the state $\bar{\sigma}$, 
\begin{align}
    \bar{\sigma}={\sum_{k=0}^{d-1}} \frac{\sigma_k}{d} ,
\end{align}
which is the average state received by Bob, when Alice sends one photon, to lie in the ($\leq 1$)-subspace in an honest implementation of the protocol\footnote{In an adversarial implementation, we cannot discard the possibility that Eve adds photons to Alice's pulse before it reaches Bob.}. The reduction to the ($\leq 1$)-subspace leads us to the following upper bound, reported in Eq.~\eqref{simplifiedproof-phase-error-rate-uppbound2}:
\begin{align}
    \tilde{e}_{X,1} Y^Z_{1,(\eta_\downarrow,\eta_\downarrow)} &\leq \frac{{\textstyle\sum_{k=0}^{d-1}} \Tr\left[\frac{\sigma_k}{d}  M^{\leq 1}_{\mathcal{Z}} {\textstyle\sum_{k'\neq k}} X_{k'} M^{\leq 1}_{\mathcal{Z}} \right]}{p^{\mathcal{X}}_d + \eta_r \eta_\uparrow (1-p^{\mathcal{X}}_d)} \nonumber\\
    &\quad+ \frac{\eta_r \eta_\uparrow(1-p^{\mathcal{X}}_d)}{p^{\mathcal{X}}_d + \eta_r \eta_\uparrow (1-p^{\mathcal{X}}_d)} \Tr\left[\Bar{\sigma}  M^{\leq 1}_{\mathcal{Z}} \Pi^0_{\mathcal{X}} M^{\leq 1}_{\mathcal{Z}} \right] \nonumber\\
    &\quad + \overline{\Delta_2} + \overline{w}^{>1}_{\mathcal{Z}} \label{simplifiedproof-phase-error-rate-uppbound2-synopsis}.
\end{align}
In the last expression, $\overline{w}^{>1}_{\mathcal{Z}}$ denotes an upper bound on the weight of the state $\bar{\sigma}$ outside of the $(\leq 1$)-subspace,
\begin{align}
    {\textstyle\sum_{\alpha=2}^\infty} \Tr[\bar{\sigma} \pia] \leq \overline{w}^{>1}_{\mathcal{Z}} \,,
\end{align}
with $\pia$ being the projector onto the subspace containing $\alpha$ photons in the set of modes $\mathcal{Z}$. $\overline{\Delta_2}$ is a function of both $\overline{w}^{>1}_{\mathcal{Z}}$ and ${Y^Z_{1,(\eta_\downarrow,\eta_\downarrow)}}$ and is derived in \eqref{Delta2bar}. The operator $M^{\leq 1}_{\mathcal{Z}}$ is defined as:
\begin{align}
    M^{\leq 1}_{\mathcal{Z}}=\sqrt{p_{\checkmark|0}} \Pi^0_{\mathcal{Z}} + \sqrt{p_{\checkmark|1}} \Pi^1_{\mathcal{Z}},
\end{align}
where the probability $p_{\checkmark|\alpha}$ is defined in Eq.~\eqref{Pr(oneclick|alpha)}. Similarly to $\pia$ for the $Z$-basis detector, we introduced the projector $\Pi^\beta_{\mathcal{X}}$ that projects on the subspace with $\beta$ photons in the set of modes $\mathcal{X}$.  Finally, $\eta_r=\eta_X/\eta_Z$ is the ratio between the detector efficiencies of the two bases (by assumption $\eta_r \leq 1$) and $p^{\mathcal{X}}_d$ is the total probability of a dark count in the $X$-basis detector.

The remainder of the proof derives upper bounds on the first two terms on the right-hand side of Eq.~\eqref{simplifiedproof-phase-error-rate-uppbound2-synopsis} by using the detection statistics of the different TBS settings and by solving systems of linear equations. For the first term, the solution of a linear system yields the variables $ \overline{\mathbbm{E}}^0_{\eta_\uparrow}$, $\overline{\mathbbm{E}}^{0,1}_{\eta_\uparrow}$, $\overline{\mathbbm{E}}^{1t}_{\eta_\uparrow}$ and $\overline{\mathbbm{E}}^{1r}_{\eta_\uparrow}$ and allows us to derive the following bound (see Appendices~\ref{app:PER5} to \ref{app:PER9}):
\begin{align}
    &{\textstyle\sum_{k=0}^{d-1}} \Tr\left[\frac{\sigma_k}{d}  M^{\leq 1}_{\mathcal{Z}} {\textstyle\sum_{k'\neq k}} X_{k'} M^{\leq 1}_{\mathcal{Z}} \right] \leq 
    p_{\checkmark|0} \overline{\mathbbm{E}}^0_{\eta_\uparrow} \nonumber\\
    &+  \sqrt{p_{\checkmark|0} p_{\checkmark|1}} \, \sqrt{\eta_\uparrow} \overline{\mathbbm{E}}^{0,1}_{\eta_\uparrow} + p_{\checkmark|1} \left( \eta_\uparrow \overline{\mathbbm{E}}^{1t}_{\eta_\uparrow} + (1-\eta_\uparrow) \overline{\mathbbm{E}}^{1r}_{\eta_\uparrow}\right). \label{boundonfirstterm}
\end{align}
The second term in \eqref{simplifiedproof-phase-error-rate-uppbound2-synopsis}, in turn, is upper bounded via 
\begin{align}
   \Tr\left[\bar{\sigma}  M^{\leq 1}_{\mathcal{Z}}  \Pi^0_{\mathcal{X}} M^{\leq 1}_{\mathcal{Z}} \right] &\leq p_{\checkmark|0} \overline{\left[w^0_{\mathcal{X}}\right]^0_{\mathcal{Z}}} + p_{\checkmark|1}  \overline{\left[w^0_{\mathcal{X}}\right]^1_{\mathcal{Z}}} \nonumber\\
   &+2\sqrt{p_{\checkmark|0}p_{\checkmark|1}}  \left(\,\overline{\left[w^0_{\mathcal{X}}\right]^1_{\mathcal{Z}}}\,\right)^{\frac{1}{2}}, \label{boundonsecondterm}
\end{align}
where the quantities $\overline{\left[w^0_{\mathcal{X}}\right]^0_{\mathcal{Z}}}$ and $\overline{\left[w^0_{\mathcal{X}}\right]^1_{\mathcal{Z}}}$ are derived with the detector decoy technique. The full derivation is reported in Appendices~\ref{app:PER10} and \ref{app:PER11}.

By inserting the bounds \eqref{boundonfirstterm} and \eqref{boundonsecondterm} into \eqref{simplifiedproof-phase-error-rate-uppbound2-synopsis}, we arrive at the final upper bound on the phase error rate, $\tilde{e}_{X,1} \leq \overline{\tilde{e}_{X,1}}$, where:
\begin{widetext}
\begin{align}
    \overline{\tilde{e}_{X,1}} &= \frac{1}{\underline{Y^Z_{1,(\eta_\downarrow,\eta_\downarrow)}}[p^{\mathcal{X}}_d + \eta_r \eta_\uparrow (1-p^{\mathcal{X}}_d)]}  \left[p_{\checkmark|0} \overline{\mathbbm{E}}^0_{\eta_\uparrow} +  \sqrt{p_{\checkmark|0} p_{\checkmark|1}} \, \sqrt{\eta_\uparrow} \overline{\mathbbm{E}}^{0,1}_{\eta_\uparrow} + p_{\checkmark|1} \left( \eta_\uparrow \overline{\mathbbm{E}}^{1t}_{\eta_\uparrow} + (1-\eta_\uparrow) \overline{\mathbbm{E}}^{1r}_{\eta_\uparrow}\right)\right] \nonumber\\
    &+ \frac{\eta_r \eta_\uparrow(1-p^{\mathcal{X}}_d)}{\underline{Y^Z_{1,(\eta_\downarrow,\eta_\downarrow)}}[p^{\mathcal{X}}_d + \eta_r \eta_\uparrow (1-p^{\mathcal{X}}_d)]}   \left[ p_{\checkmark|0} \overline{\left[w^0_{\mathcal{X}}\right]^0_{\mathcal{Z}}} + 2\sqrt{p_{\checkmark|0}p_{\checkmark|1}} \left(\,\overline{\left[w^0_{\mathcal{X}}\right]^1_{\mathcal{Z}}}\,\right)^{\frac{1}{2}} + p_{\checkmark|1} \overline{\left[w^0_{\mathcal{X}}\right]^1_{\mathcal{Z}}} \right] + \frac{1}{\underline{Y^Z_{1,(\eta_\downarrow,\eta_\downarrow)}}} \left( \overline{\Delta_2} + \overline{w}^{>1}_{\mathcal{Z}}\right).\label{phase-error-rate-bound-synopsis}
\end{align}
\end{widetext}
Each quantity in the last expression is either known or directly expressed in terms of observed statistics via the relations given in Appendix~\ref{app:phase-error-rate-formula}. In Appendix~\ref{app:decoy}, the formulas for the bounds on the yields are derived using the decoy-state method.

By employing the bound \eqref{phase-error-rate-bound-synopsis} in the key rate expression \eqref{intermediate-rate-proof}, we recover the asymptotic key rate of the protocol, \eqref{protocol-rate}, thus demonstrating that it is a lower bound of the DW rate in \eqref{simplifiedproof-DWrate}. This concludes the proof.
\end{proof}

\section{Simulations}  \label{sec:simulations}

\begin{figure}[t]
    \centering
    \includegraphics[width=1.0\linewidth,keepaspectratio]{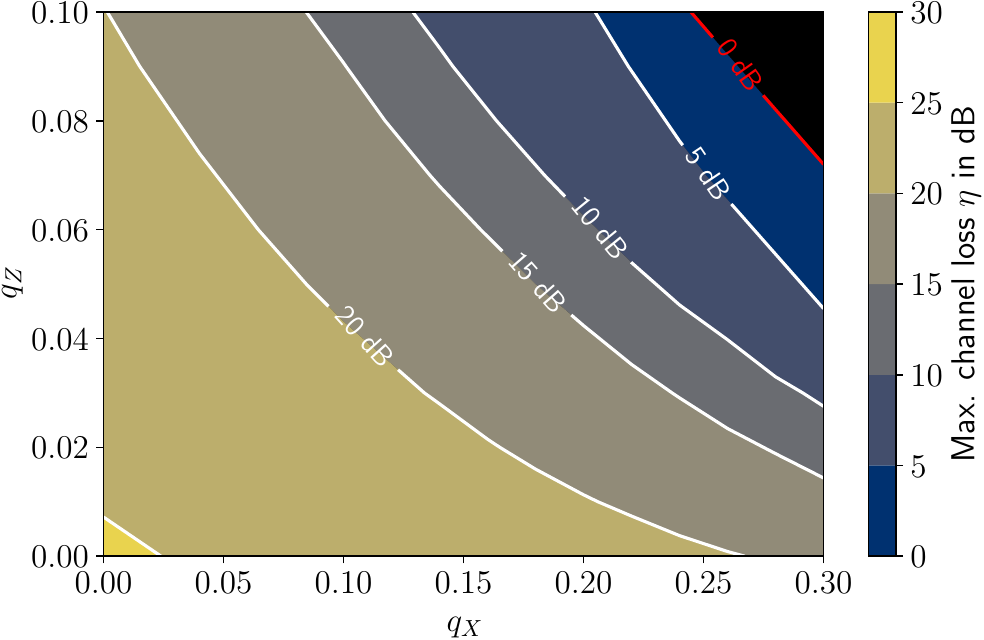}\\[1ex]
\caption{Maximal tolerable channel loss for a positive key rate of a four-dimensional time-bin QKD protocol with nominal detection efficiency mismatch $\eta_X / \eta_Z = 10^{-1/10}$, for varying channel-intrinsic QBERs $q_Z$ (key generation basis) and $q_X$ (test basis). The key rate is the one obtained from our proof,  Eq.~\eqref{protocol-rate}, and is optimized over the pulse intensity $\mu_1$. Furthermore, we fix: the decoy intensities $\mu_2 = 2\mu_3 =  2\cdot 10^{-6}$, the detectors' mode-independent efficiencies $\eta_Z = 0.9\cdot 10^{-1/10}$, $\eta_X = \eta_Z\cdot 10^{-1/10}$, the TBS parameters $\eta_\uparrow = 1-\eta_\downarrow = 0.9$, $\eta_2 = 0.4$ and the dark count probabilities $p^{\mathcal{X}}_d = 1.2 \, p^{\mathcal{Z}}_d=1.2 \cdot 10^{-4}$. }
    \label{fig:keyrate_maxloss}
\end{figure}

In this section we analyze the performance of our security proof on an experimental four-dimensional time-bin QKD protocol (Protocol~1), both in the case of an honest implementation, Sec.~\ref{sec:honest-implementation}, and under an attack where Eve can partially control the detection probabilities of the two bases, Sec.~\ref{sec:adversarial-implementation}. We benchmark the key rate resulting from our proof \eqref{protocol-rate} with the asymptotic key rate of the standard decoy-BB84 protocol, with four outcomes per basis, which implicitly assumes basis-independent detection probabilities.

\subsection{High-dimensional time-bin QKD} \label{sec:honest-implementation}
We first consider an honest implementation of the protocol, without any attack in the quantum channel.

In the experimental setup of \cite{TalboteffectWarsaw, ExpWarsawArxiv}, Alice prepares $d=4$ mutually orthogonal time-bins for key generation, while the testing symbols are encoded in the relative phases between such pulses. In particular, if $a^\dag_{Z_j}$ is the creation operator of one photon in the $j$-th time bin mode, the creation operators of the testing modes are given by the discrete Fourier transform of the time bin modes \eqref{Fourier-transform}, such that the condition in \eqref{equal-sum-twobases} is satisfied and the compatibility coefficient of Alice's states is $c=1/4$. Alice prepares the pulses in both bases as phase-randomized coherent states, given by \eqref{stateKG} for the $Z$ basis and \eqref{stateTest} for the $X$ basis, with intensities $\mu_1$, $\mu_2=2 \mu_3$ and $\mu_3=10^{-6}$ (note that weak decoy intensities are optimal). In the asymptotic limit, the optimal number of test rounds is negligible and hence we set $p_Z=1$ and $p_{\mu_1}=1$.

Bob's measurement apparatus is described in Sec.~\ref{sec:Bob-apparatus}, where the $Z$-basis measurement is realized with a single detector which detects the arrival time of the pulse, while the $X$-basis measurement is implemented by a dispersive medium followed by a time-of-arrival detection. From the arrival time of the dispersed signal, it is possible to infer the relative phases in the superposition of time bins sent by Alice in the test rounds. More details on Bob's $X$-basis measurement can be found in Ref.~\cite{TalboteffectWarsaw, ExpWarsawArxiv}. We set the largest and smallest transmittance allowed by the TBS to $\eta_\uparrow=0.9$ and $\eta_\downarrow=0.1$, respectively, such that \eqref{constraint1} is satisfied. We also fix $\eta_2 = (1/4)(\sqrt{\eta_\downarrow} + \sqrt{\eta_\uparrow})^2=0.4$. We fix the insertion loss of the TBS and of the dispersive medium to $1$~dB each, while the efficiency of the two detectors is $0.9$. Thus, we have: $\eta_Z=0.9 \cdot 10^{-1/10}$ and $\eta_X=\eta_Z \cdot 10^{-1/10}$, such that \eqref{constraint2} is satisfied. Finally, the dark count probability of the $Z$ ($X$) basis detector in the whole detection window is: $p^{\mathcal{Z}}_d=10^{-4}$ ($p^{\mathcal{X}}_d=1.2 \cdot 10^{-4}$ due to a longer detection window).

We adopt a simple channel model where Alice's pulses go through a bosonic channel with transmittance $\eta$, such that the pulses arriving at Bob are completely contained in the detection window $\mathcal{Z}$ used for key generation measurements (cf.~Fig.~\ref{fig:detection_round}). Moreover, we choose the duration of the detection window of the $X$-basis detector longer than in the $Z$ basis and long enough such that a negligible fraction of the dispersed signal (i.e., the signal exiting the dispersive medium) falls outside of the detection window.  We avoid modeling noise sources and instead assume that the signals arriving at Bob carry an intrinsic QBER $q_Z$ ($q_X$) in the $Z$ ($X$) basis. Therefore,  the QBERs observed in protocol, namely $Q_{Z,\mu_j}$, $Q_{X,\mu_j,(\eta_i,\eta_\uparrow),\checkmark}$, and $Q_{X,\mu_j,(\eta_i,\eta_\uparrow),\emptyset}$, are the result of the intrinsic QBERs and of the dark counts in the detectors. We report the explicit formulas used for the gains and QBERs in Appendix~\ref{app:simulations-honest}.

Using this channel model, in Fig.~\ref{fig:keyrate_maxloss} we investigate the adaptability of our proof to different noise scenarios by computing the maximal tolerable channel loss, such that a positive secret key rate is obtained with our formula \eqref{protocol-rate}, for varying intrinsic QBERs $q_X$ and $q_Z$. We observe that our proof can tolerate up to $30$~dB of loss and is resilient to noise, especially in the test basis. For example, for $q_Z=2\%$ and $20$~dB loss, we can extract a positive key rate for test-basis QBERs up to $q_X \approx 16\%$.

\begin{figure}[t]
    \centering
    \includegraphics[width=1.0\linewidth,keepaspectratio]{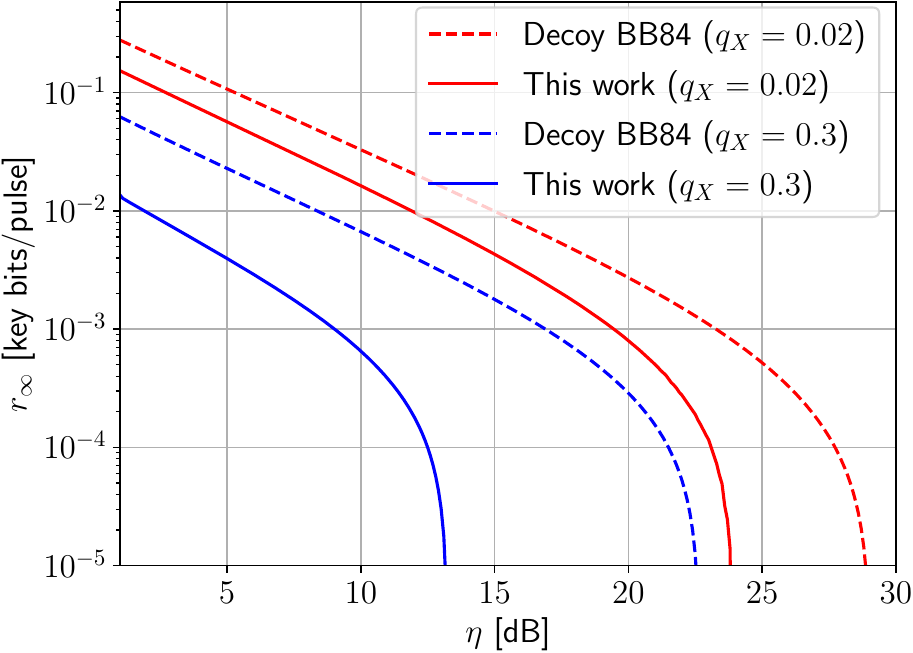}\\[1ex]
    \caption{Comparison of the secret key rate from our proof (Eq.~\eqref{protocol-rate}, solid) with the decoy-BB84 key rate (Eq.~\eqref{protocol-rate-withassumption}, dashed), as a function of the channel loss $\eta$, for an honest implementation of a four-dimensional time-bin QKD protocol with nominal detection efficiency mismatch $\eta_X / \eta_Z = 10^{-1/10}$. We set $q_Z = 0.02$ and fix different values of the intrinsic test-basis QBER, $q_X$. All key rates are optimized over the intensity $\mu_1$. We further fix: the decoy intensities $\mu_2 = 2\mu_3 =  2\cdot 10^{-6}$, the detectors' mode-independent efficiencies $\eta_Z = 0.9\cdot 10^{-1/10}$, $\eta_X = \eta_Z\cdot 10^{-1/10}$, the TBS parameters $\eta_\uparrow = 1-\eta_\downarrow = 0.9$, $\eta_2 = 0.4$ and the dark count probabilities $p^{\mathcal{Z}}_d = 10^{-4}$ and $p^{\mathcal{X}}_d=1.2 \cdot 10^{-4}$. In this honest implementation, the detection mismatch is mode-independent and is only caused by different dark count rates and different nominal detection efficiencies for the two measurement bases. Even in this case, the decoy-BB84 key rate would not be applicable as it does not contemplate detection efficiency mismatches.}
    \label{fig:keyrate_eta}
\end{figure}
By virtue of the considered honest implementation, different detection probabilities for the two bases are only caused by mode-independent asymmetries in the efficiencies and dark count probabilities of the two detectors ($\eta_Z \neq \eta_X$, $p^{\mathcal{Z}}_d \neq p^{\mathcal{X}}_d$). Still, they could be exploited by an attacker as discussed in Sec.~\ref{sec:loopholes}. The key rate we provide in \eqref{protocol-rate} allows for asymmetric detection probabilities and is secure even when the asymmetries are exploited or caused by an adversary. Conversely, the standard decoy-BB84 key rate for $d$-dimensional encoding \cite{Zbinden-time-bin3},
\begin{align}
        &r_{\rm decBB84} = p_Z^2 {\textstyle\sum_{j=1}^3} p_{\mu_j} \left\lbrace e^{-\mu_j} \underline{Y^Z_{0}} \log_2 d  \right.\nonumber\\
        &\quad\left. + e^{-\mu_j} \mu_j \underline{Y^Z_{1}} \left[\log_2 d - u(\,\overline{e_{X,1}}\,)\right] - G^Z_{\mu_j} u(Q_{Z,\mu_j}) \right\rbrace      \label{protocol-rate-withassumption},
\end{align}
is derived under the assumption that the detection probability in the two bases coincides for every input state, i.e., that \eqref{common-assumption} holds. This effectively results in substituting the phase error rate upper bound in our rate, $\overline{\tilde{e}_{X,1}}$, with an upper bound on the bit error rate in the test basis, $\overline{e_{X,1}}$. However, the considered experimental setup \cite{ExpWarsawArxiv} violates the assumption in \eqref{common-assumption} due to mode-independent (and potentially mode-dependent) asymmetries, implying that the decoy-BB84 key rate \eqref{protocol-rate-withassumption} cannot be applied to distill secure keys.

Nevertheless, in Fig.~\ref{fig:keyrate_eta} we benchmark the performance of the key rate from our proof, Eq.~\eqref{protocol-rate}, with the key rate in \eqref{protocol-rate-withassumption}, where $d=4$, for the honest implementation described above. Note that the TBS is not required by the security proof underlying Eq.~\eqref{protocol-rate-withassumption}. Hence, when computing the key rate via \eqref{protocol-rate-withassumption}, we only employ the TBS settings $(\eta_\uparrow,\eta_\uparrow)$ and $(\eta_\downarrow,\eta_\downarrow)$ to select the $X$ and $Z$ basis, respectively. The explicit expressions of the quantities appearing in \eqref{protocol-rate-withassumption} are reported in Appendix~\ref{app:simulations-honest}. We observe that our key rate presents a gap to the decoy-BB84 key rate and that the gap widens for larger values of the QBER $q_X$.

\subsection{Attack-induced efficiency mismatch} \label{sec:adversarial-implementation}
We now consider an adversarial implementation of the four-dimensional time-bin QKD protocol (Protocol~1). Eve's attack could exploit one of the two vulnerabilities of the protocol, namely, the nominal detection efficiency mismatch between the two bases ($\eta_Z>\eta_X$), or a detection probability asymmetry induced by tailored light modes. An attack based on the first vulnerability is already illustrated in Sec.~\ref{sec:loopholes} and it involves adding photons to the one-photon signals sent from Alice. This would cause an increase of the gain of the lossier basis and could be noticed by Alice and Bob when compared to other runs where Eve is not present. Nevertheless, the standard decoy-BB84 key rate does not account for  asymmetric nominal efficiencies, hence the attack would be successful.

In this section, we consider the second vulnerability and devise an intercept-resend attack where the eavesdropper can actively control which basis clicks with tailored light pulses. In particular, in each round, with probability $w_Z$ ($w_X$), Eve replaces the quantum channel with an apparatus that intercepts Alice's pulse and measures it in the  ($X$) basis, such that the measurement can perfectly distinguish Alice's states when Alice's and Eve's bases coincide. Then, Eve prepares a highly-localized one-photon pulse in the time (frequency) domain, with width parametrized by $s$ ($\sigma$), which corresponds to the outcome she observed and sends it through a quantum channel with transmittance $\xi_Z$ ($\xi_X$) to Bob. If Bob measures the pulse with the same basis as Eve, his outcome is almost perfectly correlated with Eve's. Otherwise, when Bob measures in the opposite basis, the outcome is approximately uniformly random and the detection probability is reduced. For example, if Eve intercepts in the $Z$ basis and Bob measures in the $X$ basis, her time-localized pulse would be stretched by the dispersive medium over a large time interval such that only a fraction of the outgoing pulse is contained in the detection window of the $X$-basis detector, thereby decreasing the detection probability in the $X$ basis. Moreover, the portion of the stretched signal within the detection window of the $X$ basis is nearly constant in amplitude, such that every outcome is approximately equally likely. When Eve does not perform the attack, we adopt the channel model from Sec.~\ref{sec:honest-implementation} where Alice and Bob are linked by a channel with transmittance $\eta$ and intrinsic QBERs given by $q_Z=q_X=0.05$ (these values are chosen for comparison purposes and do not reflect currently experimentally achievable values). In Appendix~\ref{app:simulations-adv} we detail Eve's attack and the corresponding statistics observed by Alice and Bob.

In order to investigate the performance of our proof under the described attack, we assume that Alice is equipped with a deterministic single-photon source (note that this assumption is also made, e.g., in \cite{detection-mismatch-Luetkenhaus}). Therefore, the decoy-state method becomes superfluous and the key rate of Protocol~1 in \eqref{protocol-rate} simplifies to the following expression (for $c=1/d$):
\begin{align}
        r_{\textrm{no decoy}} = &p_Z^2 Y^Z_{1,(\eta_\downarrow,\eta_\downarrow)} \left[\log_2 d - u(\,\overline{\tilde{e}_{X,1}}\,) -  u(e_{Z,1,(\eta_\downarrow,\eta_\downarrow)}) \right]     \label{protocol-rate-nodecoy},
    \end{align}
where the one-photon yield, $Y^Z_{1,(\eta_\downarrow,\eta_\downarrow)}$, and the one-photon error rate in the $Z$ basis, $e_{Z,1,(\eta_\downarrow,\eta_\downarrow)}$, are directly-observed quantities, replacing the corresponding gains and QBERs. By removing the decoy-state method, the upper bound on the phase error rate, $\overline{\tilde{e}_{X,1}}$, reduces to a simpler expression, as discussed in Appendix \ref{app:simulations-adv}. Similarly, when assuming a deterministic single-photon source, the decoy-BB84 protocol reduces to the original BB84 protocol, with key rate given by:
\begin{align}
        r_{\rm BB84} = & p_Z^2  Y^Z_{1,(\eta_\downarrow,\eta_\downarrow)} \nonumber\\
        &\left[\log_2 d - u(e_{X,1,(\eta_\uparrow,\eta_\uparrow)}) -  u(e_{Z,1,(\eta_\downarrow,\eta_\downarrow)}) \right]      \label{BB84-rate},
\end{align}
where the phase error rate bound is replaced by the observed one-photon error rate in the $X$ basis, $e_{X,1,(\eta_\uparrow,\eta_\uparrow)}$.

\begin{figure}[ht]
    \centering
    \includegraphics[width=1.0\linewidth,keepaspectratio]{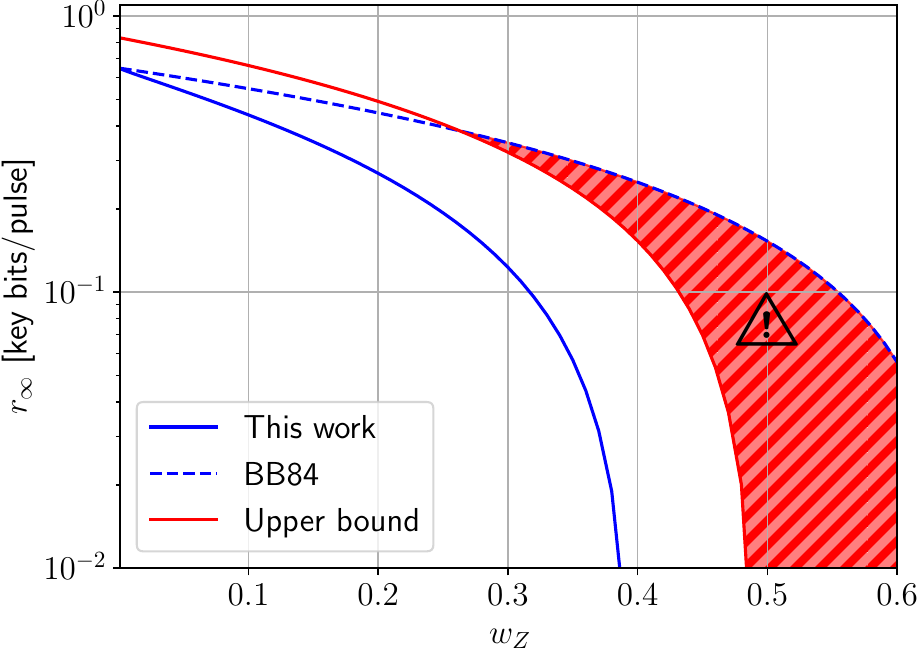}\\[1ex]
    \caption{Comparison of the secret key rate derived in this work (Eq.~\eqref{protocol-rate-nodecoy}, solid blue) with the BB84 key rate (Eq.~\eqref{BB84-rate}, dashed blue) and with an upper bound on the achievable key rate (Eq.~\eqref{keyrate-upperbound}, solid red), as a function of the probability that Eve intercepts and measures in the $Z$ basis (the key basis). All key rates are obtained under the assumption of a deterministic single-photon source (no decoys). With this attack, Eve can partially steer which basis clicks in each round to match the basis she measured in. The result is that the BB84 key rate, which does not account for attack-induced detection efficiency mismatches, overestimates the fraction of secret bits and surpasses the upper bound on the secret key rate, while the key rate from our proof remains below it. The plot parameters are: number of symbols $d=4$, intrinsic QBERs in the channel without Eve $q_Z =q_X = 0.05$, detector' mode-independent efficiencies $\eta_Z = 9/10 \cdot 10^{-1/10}$, $\eta_X = 10^{-1/10} \eta_Z$, TBS settings $\eta_\uparrow = 1-\eta_\downarrow = 0.9$, $\eta_2 = 0.4$, dark count probabilities $p^{\mathcal{Z}}_d = 10^{-4}$, $p^{\mathcal{X}}_d = 1.2 \cdot 10^{-4}$, and channel loss of $1$~dB, $\eta=10^{-1/10}$. We partition the detection window of $\Delta_t = 1.132\,\text{ns}$ in the $Z$ basis into four time bins of width $\Delta_j = 0.283\,\text{ns}$. Likewise, $\Delta_f = 4\cdot \Delta_k = 4\cdot 0.340\,\text{ns}$ in the $X$ basis. The group delay dispersion (GDD) coefficient of the dispersive medium is fixed to $\Phi_2 = 0.01275$~ns$^2$. We refer to  Appendix~\ref{app:adver-implementation} to see how these parameters enter the key rates.}
    \label{fig:keyrate_attack}
\end{figure}

In Fig.~\ref{fig:keyrate_attack}, we plot the key rates in \eqref{protocol-rate-nodecoy} and \eqref{BB84-rate} as a function of the probability of attacking the key basis, $w_Z$. For each value of $w_Z$, we optimize the parameters of Eve's attack ($w_X,s,\sigma,\xi_Z,\xi_X$) such that the statistics observed by Alice and Bob in the BB84 protocol remain approximately equal to those of the scenario without attack. The intention is that of concealing Eve's attack when the parties run the standard BB84 protocol. Additionally, we plot an upper bound on the secret key rate that Alice and Bob can extract (the full derivation is found in Sec.~\ref{app:simulations-adv}). This is obtained by subtracting from Alice's raw key the amount of bits known to Eve, which come from the rounds where Eve correctly guesses Alice's basis and from public error correction information.

From Fig.~\ref{fig:keyrate_attack}, we observe that Eve's attack cannot be completely concealed from the BB84 protocol, as its key rate decreases for increasing $w_Z$. However, the BB84 key rate surpasses the upper bound on the extractable key rate starting from about $w_Z \approx 0.27$, while our key rate remains well below it. This shows that the BB84 protocol cannot fully grasp the extent of the information gained by Eve, returning an overly optimistic and, crucially, an insecure key rate.

\section{Discussion}   \label{sec:discussion}

In Fig.~\ref{fig:keyrate_eta}, we highlighted a gap between our key rate \eqref{protocol-rate} and the decoy-BB84 key rate \eqref{protocol-rate-withassumption}, for the honest implementation of a four-dimensional time-bin QKD protocol. The gap is related to differing nominal efficiencies for the two measurement bases, i.e. $\eta_r=\eta_X/\eta_Z<1$, which could be exploited by an attacker. Intuitively, however, the key rate should not penalized by asymmetric nominal efficiencies, unless an attacker is actively exploiting the asymmetry as illustrated in Sec.~\ref{sec:loopholes}. The attack would entail adding photons to the signal resent to Bob, which would result in an increase of the weight of the state received by Bob in the subspace with more than one photon in the $Z$-basis detection interval, $\overline{w}^{>1}_{\mathcal{Z}}$. We remark that this parameter is otherwise null in an honest implementation without flaws, since it refers to the subset of one-photon signals sent by Alice.

In Appendix~\ref{app:simulations-weightestimation}, we show that our key rate is indeed insensitive to $\eta_r$ -- and matching the decoy-BB84 key rate -- for honest implementations with asymmetric nominal efficiencies, provided that we impose $\overline{w}^{>1}_{\mathcal{Z}}=0$. In other words, when imposing that $\overline{w}^{>1}_{\mathcal{Z}}$ matches its true value for honest implementations, our key rate confirms the intuition that asymmetric nominal efficiencies do not influence the resulting key rate. We emphasize that this conclusion could not have been drawn from the decoy-BB84 rate since it does not apply to scenarios with $\eta_r<1$. Conversely, when using the estimation for $\overline{w}^{>1}_{\mathcal{Z}}$ derived in our proof (cf.~Eq.~\eqref{protocol-yalpha>1}), we observe that our key rate decreases as the asymmetry in the detection efficiency of the two bases increases ($\eta_r$ decreases), generating the gap to the decoy-BB84 key rate. Therefore, the gap in Fig.~\ref{fig:keyrate_eta} can be attributed to a non-tight estimation of the parameter $\overline{w}^{>1}_{\mathcal{Z}}$ with the bound \eqref{protocol-yalpha>1} and could be improved by deriving tighter bounds.

A strong indication on how to improve the bound on the parameter $\overline{w}^{>1}_{\mathcal{Z}}$ comes from Fig.~\ref{fig:keyrate_attack}. In the figure, we observe that our key rate matches the BB84 key rate for $w_Z=0$, i.e., in an honest implementation where Alice uses a deterministic single-photon source. In such an implementation, the yields and one-photon error rates are directly-observed quantities, rather than being estimated with the decoy-state method. One of consequences is that the formula \eqref{protocol-yalpha>1} for $\overline{w}^{>1}_{\mathcal{Z}}$ returns the true value, zero (see Appendix~\ref{app:simulations-adv}), allowing our key rate to match the BB84 key rate. This suggests that increasing the number of decoy intensities (we assumed two decoy intensities in Protocol~1) may improve the estimation of $\overline{w}^{>1}_{\mathcal{Z}}$ and reduce the gap to the decoy-BB84 key rate for honest implementations.

We already remarked that the standard BB84 protocol is not applicable to scenarios with asymmetric detection efficiencies, regardless of them being nominal or induced by the adversary. However, for the honest implementation of Fig.~\ref{fig:keyrate_eta}, the decoy-BB84 key rate  represents de facto the largest amount of key that can be securely extracted, by matching our key rate when setting $\overline{w}^{>1}_{\mathcal{Z}}=0$. This is not anymore the case for the attack implemented in Fig.~\ref{fig:keyrate_attack}. As a matter of fact, the BB84 rate overestimates the fraction of secret key bits that can be extracted, by surpassing the upper bound on the secret key rate. Remarkably, when Eve attacks in the key basis $50\%$ of the time ($w_Z=0.5$), the BB84 key rate certifies more than $10\%$ of secret bits per pulse when in reality no secure key can be extracted. This is caused by the nature of the attack we designed, where Eve can increase the fraction of rounds where she learns the key bits (increasing $w_Z$) without proportionately increasing her footprint: the $X$-basis error rate remains approximately constant. This is achieved by Eve preparing highly time-localized pulses when attacking in the $Z$ basis ($s \approx 0.017 \Delta_j$), such that their detection probability in the $X$ basis is about $4.6$ times less likely than in the same basis as Eve's. From the optimization, we observe that also attacking in the $X$ basis, i.e.\ $w_X>0$, benefits the concealing of Eve's attack. In this case, Bob's $Z$ basis clicks with a probability reduced by a factor of about $2.3$ compared to the $X$ basis. The result is that the vast majority of Bob's $X$-basis detections are events where either Eve attacks in the same basis or does not attack, thereby keeping the $X$-basis error rate low.

Crucially, the augmented statistics enabled by the TBS allow our proof to detect Eve's actions and reduce the key rate accordingly. Indeed, our key rate never surpasses the key rate upper bound in Fig.~\ref{fig:keyrate_attack} and, curiously, presents a similar scaling with respect to $w_Z$. In other words, our security solution is capable of reducing the extractable key rate in proportion to the extent of adversarial attacks inducing asymmetric detection efficiencies in a given protocol run. This feature sets our proof apart from previous security proofs dealing with mode-dependent detection efficiencies \cite{detection-efficiency-mismatch-Lo,detection-efficiency-mismatch-Ma,detection-efficiency-mismatch-Lydersen,detection-mismatch-Luetkenhaus}.

To be more specific, we focus on the proofs that, like ours, do not restrict Eve's actions and allow her to, e.g., add photons to Alice's pulses and induce detection efficiency asymmetries with tailored light modes \cite{detection-efficiency-mismatch-Lydersen,detection-mismatch-Luetkenhaus}. Since such proofs are not capable of discerning whether Eve is exploiting or not the vulnerabilities of the measurement apparatus, they require the prior characterization of the largest detection probability ratio that Eve could induce with specific light modes between two detectors (and potentially two bases); we label this quantity $\kappa$. The key rates provided in \cite{detection-efficiency-mismatch-Lydersen,detection-mismatch-Luetkenhaus} present a penalty depending on the value of $\kappa$, regardless of whether Eve is present or not. From the simulations reported in \cite{detection-efficiency-mismatch-Lydersen,detection-mismatch-Luetkenhaus}, we deduce that for a BB84 protocol with a deterministic single-photon source, $3$~dB loss and QBERs around $5\%$, no key can be extracted for $\kappa \geq 2$, even if Eve is not present. By employing the same parameters ($\eta=10^{-3/10}$, $q_Z=q_X=0.05$ and $d=2$) in the adversarial implementation studied in Sec.~\ref{sec:adversarial-implementation}, our proof certifies $0.137$ secret key bits per pulse when Eve is not present ($w_Z=0$), which is also the rate obtained by naively applying the BB84 protocol. Only when Eve attacks the channel ($w_Z>0$) and generates asymmetries in the detection probabilities of the two bases up to $\kappa = 4.6$, our key rate decreases. We conclude that our proof is able to withstand larger mode-dependent asymmetries ($\kappa$) in the detection probabilities of vulnerable QKD setups, compared to previous proposals. And, crucially, it delivers positive key rates when such vulnerabilities are not actively exploited by an eavesdropper.

\section{Conclusion} \label{sec:conclusion}

We derived an analytical security proof for prepare-and-measure QKD protocols affected by basis-dependent detection probabilities. The proof can handle both nominal (mode-independent) and attack-induced (mode-dependent) detection efficiency mismatches. Compared to previous proposals, our proof requires no prior characterization of the efficiency mismatch generated by different light modes and it makes no assumption on the dimension of the states received by Bob. Moreover, the proof incorporates the decoy-state method and is capable of delivering positive key rates for QKD setups prone to large efficiency mismatches, like the time-bin QKD protocol simulated in Sec.~\ref{sec:simulations}. Our proof achieves this by actively monitoring the presence of attack-induced efficiency mismatches through a tunable beam splitter. As a result, it does not penalize the key rate unless an eavesdropper is actually attempting to control which basis clicks, whereas previous proofs always apply a penalty for the mere \textit{possibility} of controlling the basis that clicks. In the most extreme case, where a QKD setup  allows Eve to fully control which basis clicks, previous proofs cannot generate secret keys. Conversely, our proof would return positive key rates unless an adversary is actively exploiting the setup's flaws, in which case our key rate detects the attack and drops to zero. The merit for this remarkable feature is in part attributed to the tunable beam splitter, which we assume cannot be controlled by the adversary. Relaxing this assumption is an interesting direction for future work.

Further directions worth  pursuing include extending our results to finite-key scenarios, in order to apply our proof to real-world setups. Moreover, proving security against coherent attacks is desirable. Due to apparent difficulties in reducing our general model to finite dimensions, a possible avenue is represented by an adaptation of the approach with entropic uncertainty relations for smooth entropies \cite{Tomamichel2017}. Nonetheless, we believe that the techniques introduced in this work can help address other security vulnerabilities of quantum cryptographic protocols.

\textit{Note added}\quad While preparing the manuscript, we became aware of a related work recently appeared on the preprint server arXiv \cite{Tupkary2024}. This paper provides a security proof with finite-size effects for the decoy BB84 with mode-dependent detection efficiencies and active basis choice, going beyond the results in \cite{detection-mismatch-Luetkenhaus}. However, similarly to \cite{detection-mismatch-Luetkenhaus}, the proof in \cite{Tupkary2024} requires an a priori characterization of the detectors. Indeed, one of the required input parameters is the maximum relative difference in detection efficiencies that the eavesdropper can trigger. Conversely, in our proof no such parameter is required as it is estimated in real time during the execution of the QKD protocol. 

\acknowledgments

FG contributed to this work exclusively on behalf of Heinrich-Heine-Universit\"at D\"usseldorf (previous affiliation). FG acknowledges funding from the Deutsche Forschungsgemeinschaft (DFG, German Research Foundation) through the Individual Grant BR2159/6-1. HK, DB and NWy acknowledge support by the QuantERA project QuICHE via the German Ministry of Education and Research (BMBF grant no.~16KIS1119K). DB and HK acknowledge support by the German Ministry of Education and Research through the project QuNET+ProQuake (BMBF grant no.~16KISQ137) and QuKuK (BMBF grant no.~16KIS1619). NWa acknowledges funding from the BMBF (QPIC-1, Pho-Quant).  MK and MO acknowledge project QuICHE, supported by the National Science Centre, Poland (project no.~2019/32/Z/ST2/00018) under QuantERA, which has received funding from the European Union's Horizon 2020 research and innovation programme under grant agreement no.~731473. MK, MO and AW acknowledge support by the Excellence Initiative---Research University of the University of Warsaw. GC and CM acknowledge the EU H2020 QuantERA ERA-NET Cofund in Quantum Technologies project QuICHE and support from the PNRR MUR Project PE0000023-NQSTI. The authors would like to thank Daniel Gauthier, Brian Smith, and Devashish Tupkary for insightful discussions.

\bibliography{bibliography}

\clearpage
\onecolumngrid

\appendix

\pagestyle{fancy}
\fancyhf{}
\fancyhead[R]{\thepage}
\fancyhead[L]{\leftmark}
\fancyhead[C]{\rightmark}

\section{Notation} \label{app:notation}

\begin{table}[h!t]
\renewcommand\arraystretch{1.3}
\setlength{\tabcolsep}{5pt}
\centering
\caption{Notation adopted in the paper. Note that ``$W$'' is a placeholder for the appropriate yield or error rate. The symbols $\checkmark$ and $\emptyset$ stand for a detection ($Z_B \neq \emptyset$) and a no detection ($Z_B =\emptyset$) in Bob's $Z$-basis detector, respectively.}
\label{tab:notation}
\begin{tabular}[t]{>{\centering}p{0.1\linewidth}>{\centering}p{0.26\linewidth}p{0.6\linewidth}}
\toprule
\textbf{Symbol} & \textbf{Name} & \textbf{Definition} \\
\midrule
$\mu_1,\mu_2,\mu_3$ & / & the three intensities used by Alice to prepare her states\\
 / & TBS & tunable beam splitter\\
$(\eta_i,\eta_l)$ & TBS setting & the transmittance of the TBS is set to $\eta_i$ in the interval $\mathcal{Z}$ and to $\eta_l$ otherwise\\
$\eta_\uparrow$ ($\eta_\downarrow$) & / & maximal (minimal) transmittance achievable by the TBS\\
$\eta_X$ & / &  detection efficiency of the $X$-basis detector, including the TBS insertion loss \\
$\eta_Z$ & / & detection efficiency of the $Z$-basis detector, including the TBS insertion loss  \\
$\eta_r$ &  / & $\eta_r=\eta_X/\eta_Z$ \\
$\mathcal{X}$ & $X$-basis detection interval & the set of all modes detected by the $X$-basis detector: $\mathcal{X}=\cup_{k=0}^{d-1} \mathcal{X}_k$ \\
$\mathcal{Z}$ & $Z$-basis detection interval & the set of all modes detected by the $Z$-basis detector: $\mathcal{Z}=\cup_{j=0}^{d-1} \mathcal{Z}_j$ \\
$p^X_d$ &  /  & probability of a dark count in mode $\mathcal{X}_k$ of the $X$-basis detector, for all $k$ \\
$p^Z_d$ &  /  & probability of a dark count in mode $\mathcal{Z}_j$ of the $Z$-basis detector, for all $j$ \\
$p^{\mathcal{X}}_d$ ($p^{\mathcal{Z}}_d$) & / & total probability of a dark count in the $X$-basis ($Z$-basis) detector, if no photon is localized in $\mathcal{X}$ ($\mathcal{Z}$): $p^{\mathcal{X}}_d=1-(1-p^Z_d)^d$ ($p^{\mathcal{Z}}_d=1-(1-p^Z_d)^d$) \\
$\overline{W}$ ($\underline{W}$) & / &  upper (lower) bound on $W$: $\overline{W} \geq W$ ($\underline{W} \leq W$) \\
/ & ($\leq 1$)-subspace &  the subspace with at most one photon localized in $\mathcal{Z}$ \\
$[W]^{\leq 1}_{\mathcal{Z}}$ & / & $W$ is restricted to the ($\leq 1$)-subspace\\
$\tilde{e}_{X,1}$ & phase error rate & the crucial parameter to be estimated in the security proof \\
$T$ & / & random variable indicating whether Alice prepares a $Z$-basis ($X$-basis) state: $T=Z$ ($T=X$) \\
$I_A$ & / & Alice's intensity choice in one round \\
$N_A$ & / & number of photons sent by Alice in one round\\
$Z_A$ ($X_A$) & / & random variable storing Alice's symbol encoded in a $Z$-basis ($X$-basis) state \\
$Z_B$ ($X_B$) & / & random variable storing the outcome of Bob's $Z$-basis ($X$-basis) detector: $Z_B, X_B \in \{0,1,\dots,d-1,\emptyset\}$\\
$G^Z_{\mu_j,(\eta_l,\eta_l)}$ & $Z$-basis gain & $G^Z_{\mu_j,(\eta_l,\eta_l)} = \Pr(Z_B \neq \emptyset|T=Z,I_A=\mu_j,(\eta_l,\eta_l))$ \\
$G^{X,\checkmark}_{\mu_j,(\eta_i,\eta_l)}$ & $X$-basis gain & $G^{X,\checkmark}_{\mu_j,(\eta_i,\eta_l)} = \Pr(X_B \neq \emptyset,Z_B \neq \emptyset|T=X,I_A=\mu_j,(\eta_i,\eta_l))$ \\
$G^{X,\emptyset}_{\mu_j,(\eta_i,\eta_l)}$ & $X$-basis gain & $G^{X,\emptyset}_{\mu_j,(\eta_i,\eta_l)} = \Pr(X_B \neq \emptyset,Z_B = \emptyset|T=X,I_A=\mu_j,(\eta_i,\eta_l))$ \\
$Q_{Z,\mu_j}$ & QBER of the key generation rounds & $ Q_{Z,\mu_j} = \Pr(Z_A \neq Z_B |T=Z, I_A=\mu_j, (\eta_\downarrow,\eta_\downarrow),Z_B\neq \emptyset)$ \\
$Q_{X,\mu_j,(\eta_i,\eta_\uparrow),\checkmark}$ & $X$-basis QBER & $Q_{X,\mu_j,(\eta_i,\eta_\uparrow),\checkmark}=  \Pr(X_A \neq X_B |T=X, I_A=\mu_j ,(\eta_i,\eta_\uparrow),X_B \neq \emptyset,Z_B \neq \emptyset)$ \\
$Q_{X,\mu_j,(\eta_i,\eta_\uparrow),\emptyset}$ & $X$-basis QBER & $Q_{X,\mu_j,(\eta_i,\eta_\uparrow),\emptyset}=  \Pr(X_A \neq X_B |T=X, I_A=\mu_j ,(\eta_i,\eta_\uparrow),X_B \neq \emptyset,Z_B = \emptyset)$ \\
$Y^Z_{n,(\eta_l,\eta_l)}$ & $n$-photon $Z$-basis yield & $Y^Z_{n,(\eta_l,\eta_l)}=\Pr(Z_B \neq \emptyset|T=Z,N_A=n,(\eta_l,\eta_l))$ \\
$Y^{X,\checkmark}_{n,(\eta_i,\eta_l)}$ & $n$-photon $X$-basis yield \quad[$n=1$: test-round yield] & $Y^{X,\checkmark}_{n,(\eta_i,\eta_l)}=\Pr(X_B \neq \emptyset,Z_B\neq \emptyset|T=X,N_A=n,(\eta_i,\eta_l))$ \\
$Y^{X,\emptyset}_{n,(\eta_i,\eta_l)}$ & $n$-photon $X$-basis yield \quad[$n=1$: test-round yield] & $Y^{X,\emptyset}_{n,(\eta_i,\eta_l)}=\Pr(X_B \neq \emptyset,Z_B = \emptyset|T=X,N_A=n,(\eta_i,\eta_l))$ \\
$e_{X,n,(\eta_i,\eta_l),\checkmark}$ & $n$-photon $X$-basis bit error rate [$n=1$: test-round bit error rate] & $e_{X,n,(\eta_i,\eta_l),\checkmark}=\Pr(X_A \neq X_B |T=X, N_A=n,(\eta_i,\eta_l), X_B \neq \emptyset, Z_B \neq \emptyset)$ \\
$e_{X,n,(\eta_i,\eta_l),\emptyset}$ & $n$-photon $X$-basis bit error rate [$n=1$: test-round bit error rate] & $e_{X,n,(\eta_i,\eta_l),\emptyset}=\Pr(X_A \neq X_B |T=X, N_A=n,(\eta_i,\eta_l), X_B \neq \emptyset, Z_B = \emptyset)$ \\
$\Phi_2$ & GDD coefficient & group delay dispersion coefficient of the setup simulated in Fig.~\ref{fig:keyrate_attack} \\
$\Delta_j$ ($\Delta_k$) & time-bin width & The widths of the time bins in Bob's time-of-arrival measurement for the $Z$ ($X$) basis, used to generate Fig.~\ref{fig:keyrate_attack} \\
\bottomrule
\end{tabular}
\end{table}

\section{Phase error rate bound} \label{app:phase-error-rate-formula}

In this Appendix we report, in a concise manner, the formulas that compose the upper bound on the phase error rate ($\overline{\tilde{e}_{X,1}}$), which appears in the key rate of the protocol \eqref{protocol-rate}. For the notation and symbols used in the formulas below, we refer to Appendix~\ref{app:notation}.\\

The upper bound on the phase error rate is defined as follows:
\begin{align}
    \overline{\tilde{e}_{X,1}} &= \frac{1}{\underline{Y^Z_{1,(\eta_\downarrow,\eta_\downarrow)}}[p^{\mathcal{X}}_d + \eta_r \eta_\uparrow (1-p^{\mathcal{X}}_d)]}  \left[p_{\checkmark|0} \overline{\mathbbm{E}}^0_{\eta_\uparrow} +  \sqrt{p_{\checkmark|0} p_{\checkmark|1}} \, \sqrt{\eta_\uparrow} \overline{\mathbbm{E}}^{0,1}_{\eta_\uparrow} + p_{\checkmark|1} \left( \eta_\uparrow \overline{\mathbbm{E}}^{1t}_{\eta_\uparrow} + (1-\eta_\uparrow) \overline{\mathbbm{E}}^{1r}_{\eta_\uparrow}\right)\right] \nonumber\\
    &+ \frac{\eta_r \eta_\uparrow(1-p^{\mathcal{X}}_d)}{\underline{Y^Z_{1,(\eta_\downarrow,\eta_\downarrow)}}[p^{\mathcal{X}}_d + \eta_r \eta_\uparrow (1-p^{\mathcal{X}}_d)]}   \left[ p_{\checkmark|0} \overline{\left[w^0_{\mathcal{X}}\right]^0_{\mathcal{Z}}} + 2\sqrt{p_{\checkmark|0}p_{\checkmark|1}} \left(\,\overline{\left[w^0_{\mathcal{X}}\right]^1_{\mathcal{Z}}}\,\right)^{\frac{1}{2}} + p_{\checkmark|1} \overline{\left[w^0_{\mathcal{X}}\right]^1_{\mathcal{Z}}} \right] + \frac{\overline{\Delta_2} + \overline{w}^{>1}_{\mathcal{Z}}}{\underline{Y^Z_{1,(\eta_\downarrow,\eta_\downarrow)}}}  \label{protocol-phase-error-rate}.
\end{align}
The expression for $\overline{\tilde{e}_{X,1}}$ contains several quantities, which we now detail. We start by defining  $p_{\checkmark|0}= p^{\mathcal{Z}}_d$ and $p_{\checkmark|1}= 1-(1-p^{\mathcal{Z}}_d) \eta_\downarrow$, which are the probabilities that the $Z$-basis detector clicks, with TBS setting $(\eta_\downarrow,\eta_\downarrow)$, given that the state received by Bob contains zero photons, respectively one photon, localized in $\mathcal{Z}$. Here, we recall that:
\begin{align}
    p^{\mathcal{Z}}_d &= 1-(1-p^Z_d)^d \label{prob-dk-tot-Z} \\
    p^{\mathcal{X}}_d &= 1-(1-p^X_d)^d \label{prob-dk-tot-X}
\end{align}
is the total probability of a dark count in the $Z$-basis (resp. $X$-basis) detector. We also define $\overline{\Delta_2}$ to be:
\begin{align}
    \overline{\Delta_2} =  \sqrt{\left(\overline{Y^Z_{1,(\eta_\downarrow,\eta_\downarrow)}} - \min\left\lbrace \overline{w}^{>1}_{\mathcal{Z}} , \frac{\overline{Y^Z_{1,(\eta_\downarrow,\eta_\downarrow)}}}{2}\right\rbrace \right)  \min\left\lbrace \overline{w}^{>1}_{\mathcal{Z}} , \frac{\overline{Y^Z_{1,(\eta_\downarrow,\eta_\downarrow)}}}{2}\right\rbrace} .
\end{align}

We introduce an upper bound on the weight of the average state received by Bob (when Alice sends one photon) in the subspace with two or more photons localized in $\mathcal{Z}$:
\begin{align}
    \overline{w}^{>1}_{\mathcal{Z}}=\ \min\left\lbrace 1, \frac{\overline{(\eta_\uparrow - \eta_\downarrow)Y^Z_{1,(\eta_2,\eta_2)} - (1-\eta_2)(Y^Z_{1,(\eta_\downarrow,\eta_\downarrow)} - Y^Z_{1,(\eta_\uparrow,\eta_\uparrow)}) -(\eta_\uparrow -\eta_\downarrow) p^{\mathcal{Z}}_d}}{(1-p^{\mathcal{Z}}_d)(\eta_\uparrow - \eta_\downarrow)(1-\eta_2)(1+\eta_2 -\eta_\uparrow - \eta_\downarrow)}  \right\rbrace \label{protocol-yalpha>1}.
\end{align}
Similarly, we report the following upper and lower bounds on the weight of the average state received by Bob, in the subspace with zero photons (resp. one photon) localized in $\mathcal{Z}$:
\begin{align}
    \overline{w}^0_{\mathcal{Z}} &=  \min\left\lbrace 1,\frac{\overline{\eta_\downarrow Y^Z_{1,(\eta_\uparrow,\eta_\uparrow)} -\eta_\uparrow (1-\eta_\uparrow)Y^Z_{1,(\eta_\downarrow,\eta_\downarrow)}} +\eta_\uparrow (1-\eta_\uparrow) -\eta_\downarrow + \eta_\uparrow^2 \eta_\downarrow (1-p^{\mathcal{Z}}_d)}{(1-p^{\mathcal{Z}}_d)(1-\eta_\uparrow) \left[\eta_\uparrow -\eta_\downarrow (1 + \eta_\uparrow)\right]}\right\rbrace
    \label{protocol-y0-upperbound} \\
    \underline{w}^0_{\mathcal{Z}} &=  \max\left\lbrace 0,  \frac{-\left(\overline{-\eta_\downarrow (1-\eta_\downarrow) Y^Z_{1,(\eta_\uparrow,\eta_\uparrow)} + \eta_\uparrow Y^Z_{1,(\eta_\downarrow,\eta_\downarrow)}}\right) +\eta_\uparrow -\eta_\downarrow(1-\eta_\downarrow) - \eta_\downarrow^2 \eta_\uparrow (1-p^{\mathcal{Z}}_d) }{(1-p^{\mathcal{Z}}_d)(1-\eta_\downarrow)\left[\eta_\uparrow (1+\eta_\downarrow) -\eta_\downarrow \right]} \right\rbrace \label{protocol-y0-lowerbound}
\end{align}
and
\begin{align}
     \overline{w}^1_{\mathcal{Z}} &= \min\left\lbrace 1, \frac{\overline{-(1-\eta_\downarrow^2) Y^Z_{1,(\eta_\uparrow,\eta_\uparrow)} +Y^Z_{1,(\eta_\downarrow,\eta_\downarrow)}} -\eta_\downarrow^2 p^{\mathcal{Z}}_d}{(1-p^{\mathcal{Z}}_d)(1-\eta_\downarrow)\left[\eta_\uparrow (1+ \eta_\downarrow) - \eta_\downarrow\right]}\right\rbrace  \label{protocol-y1-upperbound} \\
     \underline{w}^1_{\mathcal{Z}} &=\max\left\lbrace 0, \frac{-\left(\overline{Y^Z_{1,(\eta_\uparrow,\eta_\uparrow)}-(1-\eta_\uparrow^2) Y^Z_{1,(\eta_\downarrow,\eta_\downarrow)}}\right) +\eta_\uparrow^2 p^{\mathcal{Z}}_d}{(1-p^{\mathcal{Z}}_d)(1-\eta_\uparrow)\left[\eta_\uparrow -\eta_\downarrow (1 + \eta_\uparrow) \right]}\right\rbrace, \label{protocol-y1-lowerbound}
\end{align}
respectively. These quantities appear in the upper bounds on the weight of the average state received by Bob in the subspace with zero photons in $\mathcal{X}$ and zero photons (resp. one photon) in $\mathcal{Z}$:
\begin{align}
    \overline{\left[w^0_{\mathcal{X}}\right]^0_{\mathcal{Z}}} &= \min\left\lbrace \overline{w}^0_{\mathcal{Z}}, \frac{(1-\eta_\uparrow \eta_r) \overline{\mathbbm{Y}}^0_{\eta_\downarrow} - \eta_\downarrow \eta_r (1-\eta_\downarrow \eta_r) \underline{\mathbbm{Y}}^0_{\eta_\uparrow}  -\underline{w}^0_{\mathcal{Z}} p^{\mathcal{X}}_d (1-\eta_\uparrow \eta_r)(1-\eta_\downarrow \eta_r)^2 }{\eta_\downarrow \eta_r (1-p^{\mathcal{X}}_d)(2 \eta_\uparrow \eta_r -1 -\eta_r^2 \eta_\uparrow \eta_\downarrow)} + \frac{\overline{w}^0_{\mathcal{Z}}}{1-p^{\mathcal{X}}_d} \right\rbrace \label{protocol-x0-upperbound000} 
\end{align}
and
\begin{align}
    \overline{\left[w^0_{\mathcal{X}}\right]^1_{\mathcal{Z}}} &= \min\left\lbrace \overline{w}^1_{\mathcal{Z}}, \frac{(1-\eta_\uparrow \eta_r) \left[\eta_\downarrow \overline{\mathbbm{Y}}^{1t}_{\eta_\downarrow} + (1-\eta_\downarrow)\overline{\mathbbm{Y}}^{1r}_{\eta_\downarrow} \right] - \eta_\downarrow \eta_r(1-\eta_\downarrow \eta_r) \left[\eta_\uparrow \underline{\mathbbm{Y}}^{1t}_{\eta_\uparrow} + (1-\eta_\uparrow)\underline{\mathbbm{Y}}^{1r}_{\eta_\uparrow} \right]}{\eta_\downarrow \eta_r (1-p^{\mathcal{X}}_d)(2 \eta_\uparrow \eta_r -1 -\eta_r^2 \eta_\uparrow \eta_\downarrow)} \right. \nonumber\\
    &\left.\quad\quad\quad\quad- \frac{\underline{w}^1_{\mathcal{Z}} p^{\mathcal{X}}_d (1-\eta_\uparrow \eta_r)(1-\eta_\downarrow \eta_r)^2 }{\eta_\downarrow \eta_r (1-p^{\mathcal{X}}_d)(2 \eta_\uparrow \eta_r -1 -\eta_r^2 \eta_\uparrow \eta_\downarrow)} + \frac{\overline{w}^1_{\mathcal{Z}}}{1-p^{\mathcal{X}}_d} \right\rbrace \label{protocol-x0-upperbound101} ,
\end{align}
respectively. The bounds in \eqref{protocol-x0-upperbound000} and \eqref{protocol-x0-upperbound101} contain bounds on the quantities: $\mathbbm{Y}^0_{\eta_l}$, $\mathbbm{Y}^{1t}_{\eta_l}$, and $\mathbbm{Y}^{1r}_{\eta_l}$. Such quantities are linked to the test-round yields as follows:
\begin{align}
    \overline{\mathbbm{Y}}^0_{\eta_l} &:= \min \left\lbrace \overline{w}^0_{\mathcal{Z}}, \frac{\sqrt{\eta_2 \eta_\downarrow}}{\left(\sqrt{\eta_\uparrow}-\sqrt{\eta_2}\right)\left(\sqrt{\eta_\uparrow}-\sqrt{\eta_\downarrow}\right)\left(1-p^{\mathcal{Z}}_d\right)}\overline{\left[Y^{X,\emptyset}_{1,(\eta_\uparrow,\eta_l)}\right]^{\leq 1}_{\mathcal{Z}}}  - \frac{\sqrt{\eta_\uparrow \eta_\downarrow} }{\left(\sqrt{\eta_\uparrow}-\sqrt{\eta_2}\right)
   \left(\sqrt{\eta_2}-\sqrt{\eta_\downarrow}\right) \left(1-p^{\mathcal{Z}}_d\right)} \underline{\left[Y^{X,\emptyset}_{1,(\eta_2,\eta_l)} \right]^{\leq 1}_{\mathcal{Z}}} \right. \nonumber\\
   &\left.\quad+\frac{\sqrt{\eta_\uparrow \eta_2}}{\left(\sqrt{\eta_\uparrow}-\sqrt{\eta_\downarrow}\right) \left(\sqrt{\eta_2}-\sqrt{\eta_\downarrow}\right) 
   \left(1-p^{\mathcal{Z}}_d\right)} \overline{\left[Y^{X,\emptyset}_{1,(\eta_\downarrow,\eta_l)}  \right]^{\leq 1}_{\mathcal{Z}}} \right\rbrace \label{protocol-Yetak-upp} \\
   \underline{\mathbbm{Y}}^0_{\eta_l} &:=\max \left\lbrace 0, \frac{\sqrt{\eta_2 \eta_\downarrow}}{\left(\sqrt{\eta_\uparrow}-\sqrt{\eta_2}\right)\left(\sqrt{\eta_\uparrow}-\sqrt{\eta_\downarrow}\right)\left(1-p^{\mathcal{Z}}_d\right)}\underline{\left[Y^{X,\emptyset}_{1,(\eta_\uparrow,\eta_l)}\right]^{\leq 1}_{\mathcal{Z}}}  - \frac{\sqrt{\eta_\uparrow \eta_\downarrow} }{\left(\sqrt{\eta_\uparrow}-\sqrt{\eta_2}\right)
   \left(\sqrt{\eta_2}-\sqrt{\eta_\downarrow}\right) \left(1-p^{\mathcal{Z}}_d\right)} \overline{\left[Y^{X,\emptyset}_{1,(\eta_2,\eta_l)} \right]^{\leq 1}_{\mathcal{Z}}} \right.  \nonumber\\
   &\left.\quad+\frac{\sqrt{\eta_\uparrow \eta_2}}{\left(\sqrt{\eta_\uparrow}-\sqrt{\eta_\downarrow}\right) \left(\sqrt{\eta_2}-\sqrt{\eta_\downarrow}\right) 
   \left(1-p^{\mathcal{Z}}_d\right)} \underline{\left[Y^{X,\emptyset}_{1,(\eta_\downarrow,\eta_l)}  \right]^{\leq 1}_{\mathcal{Z}}} \right\rbrace \label{protocol-Yetak-low}
\end{align}
and
\begin{align}
    \overline{\mathbbm{Y}}^{1t}_{\eta_l} &:= \min\left\lbrace \overline{w}^1_{\mathcal{Z}}, \frac{\overline{\left[Y^{X,\emptyset}_{1,(\eta_\uparrow,\eta_l)} \right]^{\leq 1}_{\mathcal{Z}}}}{\left(\sqrt{\eta_\uparrow}-\sqrt{\eta_2}\right)\left(\sqrt{\eta_\uparrow}-\sqrt{\eta_\downarrow}\right)\left(1-p^{\mathcal{Z}}_d\right)}  - \frac{\underline{\left[Y^{X,\emptyset}_{1,(\eta_2,\eta_l)}  \right]^{\leq 1}_{\mathcal{Z}}} }{\left(\sqrt{\eta_\uparrow}-\sqrt{\eta_2}\right)
   \left(\sqrt{\eta_2}-\sqrt{\eta_\downarrow}\right) \left(1-p^{\mathcal{Z}}_d\right)} \right. \nonumber\\
   &\left.\quad+\frac{\overline{\left[Y^{X,\emptyset}_{1,(\eta_\downarrow,\eta_l)} \right]^{\leq 1}_{\mathcal{Z}}}}{\left(\sqrt{\eta_\uparrow}-\sqrt{\eta_\downarrow}\right) \left(\sqrt{\eta_2}-\sqrt{\eta_\downarrow}\right) 
   \left(1-p^{\mathcal{Z}}_d\right)}  \right\rbrace \label{protocol-Yetak-1t-upp} \\
   \overline{\mathbbm{Y}}^{1r}_{\eta_l} &:= \min\left\lbrace \overline{w}^1_{\mathcal{Z}}, \frac{1}{1-\eta_\downarrow} \overline{\left[Y^{X,\checkmark}_{1,(\eta_\downarrow,\eta_l)} \right]^{\leq 1}_{\mathcal{Z}}} - 
   \frac{p^{\mathcal{Z}}_d}{(1-\eta_\downarrow)(1-p^{\mathcal{Z}}_d)} \underline{\left[Y^{X,\emptyset}_{1,(\eta_\downarrow,\eta_l)}  \right]^{\leq 1}_{\mathcal{Z}}} \right\rbrace  \label{protocol-Yetak-1r-upp} \\
   \underline{\mathbbm{Y}}^{1t}_{\eta_l} &:= \max\left\lbrace 0, \frac{\underline{\left[Y^{X,\emptyset}_{1,(\eta_\uparrow,\eta_l)} \right]^{\leq 1}_{\mathcal{Z}}}}{\left(\sqrt{\eta_\uparrow}-\sqrt{\eta_2}\right)\left(\sqrt{\eta_\uparrow}-\sqrt{\eta_\downarrow}\right)\left(1-p^{\mathcal{Z}}_d\right)}  - \frac{\overline{\left[Y^{X,\emptyset}_{1,(\eta_2,\eta_l)}  \right]^{\leq 1}_{\mathcal{Z}}}}{\left(\sqrt{\eta_\uparrow}-\sqrt{\eta_2}\right)
   \left(\sqrt{\eta_2}-\sqrt{\eta_\downarrow}\right) \left(1-p^{\mathcal{Z}}_d\right)} \right. \nonumber\\
   &\left.\quad+\frac{\underline{\left[Y^{X,\emptyset}_{1,(\eta_\downarrow,\eta_l)} \right]^{\leq 1}_{\mathcal{Z}}}}{\left(\sqrt{\eta_\uparrow}-\sqrt{\eta_\downarrow}\right) \left(\sqrt{\eta_2}-\sqrt{\eta_\downarrow}\right) 
   \left(1-p^{\mathcal{Z}}_d\right)}  \right\rbrace \label{protocol-Yetak-1t-low} \\
   \underline{\mathbbm{Y}}^{1r}_{\eta_l} &:=\max\left\lbrace 0, \frac{1}{1-\eta_\downarrow} \underline{\left[Y^{X,\checkmark}_{1,(\eta_\downarrow,\eta_l)} \right]^{\leq 1}_{\mathcal{Z}}} - 
   \frac{p^{\mathcal{Z}}_d}{(1-\eta_\downarrow)(1-p^{\mathcal{Z}}_d)} \overline{\left[Y^{X,\emptyset}_{1,(\eta_\downarrow,\eta_l)}  \right]^{\leq 1}_{\mathcal{Z}}} \right\rbrace  \label{protocol-Yetak-1r-low}.
\end{align}
In particular, the bounds on the test-round yields restricted to the ($\leq 1$)-subspace read:
\begin{align}
    \underline{\left[Y^{X,\checkmark}_{1,(\eta_i,\eta_l)} \right]^{\leq 1}_{\mathcal{Z}}} &= \max \left\lbrace 0, \underline{Y^{X,\checkmark}_{1,(\eta_i,\eta_l)}}  -   \sqrt{\overline{w}^{>1}_{\mathcal{Z}}} - \overline{w}^{>1}_{\mathcal{Z}} \right\rbrace \label{protocol-l-Y,click} \\
    \overline{\left[Y^{X,\checkmark}_{1,(\eta_i,\eta_l)} \right]^{\leq 1}_{\mathcal{Z}}} &= \min \left\lbrace 1, \overline{Y^{X,\checkmark}_{1,(\eta_i,\eta_l)}}  +   \sqrt{\overline{w}^{>1}_{\mathcal{Z}}} \right\rbrace  \label{protocol-u-Y,click}\\
    \underline{\left[Y^{X,\emptyset}_{1,(\eta_i,\eta_l)}\right]^{\leq 1}_{\mathcal{Z}}} &= \max \left\lbrace 0, \underline{Y^{X,\emptyset}_{1,(\eta_i,\eta_l)}}  -   \sqrt{\overline{w}^{>1}_{\mathcal{Z}}} - \overline{w}^{>1}_{\mathcal{Z}} \right\rbrace  \label{protocol-l-Y,noclick}\\ 
    \overline{\left[Y^{X,\emptyset}_{1,(\eta_i,\eta_l)}\right]^{\leq 1}_{\mathcal{Z}}} &= \min \left\lbrace 1, \overline{Y^{X,\emptyset}_{1,(\eta_i,\eta_l)}}  +   \sqrt{\overline{w}^{>1}_{\mathcal{Z}}} \right\rbrace. \label{protocol-u-Y,noclick}
\end{align}

Now, we report the expressions of the remaining quantities in \eqref{protocol-phase-error-rate} that have not yet been defined, namely: $\overline{\mathbbm{E}}^{0}_{\eta_\uparrow},\overline{\mathbbm{E}}^{0,1}_{\eta_\uparrow},\overline{\mathbbm{E}}^{1t}_{\eta_\uparrow}$ and $\overline{\mathbbm{E}}^{1r}_{\eta_\uparrow}$. These are related to the products of the test-round yields and bit error rates as follows:
\begin{align}
    \overline{\mathbbm{E}}^{0}_{\eta_\uparrow} &:=\min\left\lbrace \overline{w}^0_{\mathcal{Z}}, \frac{\sqrt{\eta_2 \eta_\downarrow}\overline{\left[Y^{X,\emptyset}_{1,(\eta_\uparrow,\eta_\uparrow)} e_{X,1,(\eta_\uparrow,\eta_\uparrow),\emptyset}\right]^{\leq 1}_{\mathcal{Z}}}}{\left(\sqrt{\eta_\uparrow}-\sqrt{\eta_2}\right)\left(\sqrt{\eta_\uparrow}-\sqrt{\eta_\downarrow}\right)\left(1-p^{\mathcal{Z}}_d\right)}  - \frac{\sqrt{\eta_\uparrow \eta_\downarrow} \underline{\left[Y^{X,\emptyset}_{1,(\eta_2,\eta_\uparrow)} e_{X,1,(\eta_2,\eta_\uparrow),\emptyset} \right]^{\leq 1}_{\mathcal{Z}}}}{\left(\sqrt{\eta_\uparrow}-\sqrt{\eta_2}\right)
   \left(\sqrt{\eta_2}-\sqrt{\eta_\downarrow}\right) \left(1-p^{\mathcal{Z}}_d\right)} \right.  \nonumber\\
   &\left.\quad+\frac{\sqrt{\eta_\uparrow \eta_2}}{\left(\sqrt{\eta_\uparrow}-\sqrt{\eta_\downarrow}\right) \left(\sqrt{\eta_2}-\sqrt{\eta_\downarrow}\right) 
   \left(1-p^{\mathcal{Z}}_d\right)} \overline{\left[Y^{X,\emptyset}_{1,(\eta_\downarrow,\eta_\uparrow)} e_{X,1,(\eta_\downarrow,\eta_\uparrow),\emptyset} \right]^{\leq 1}_{\mathcal{Z}}} \right\rbrace   \label{protocol-Eetak-0-upp}\\
   \overline{\mathbbm{E}}^{0,1}_{\eta_\uparrow} &:= \min\left\lbrace \frac{\overline{\left[Y^{X,\emptyset}_{1,(\eta_\uparrow,\eta_\uparrow)} e_{X,1,(\eta_\uparrow,\eta_\uparrow),\emptyset} \right]^{\leq 1}_{\mathcal{Z}}}}{(1-p^{\mathcal{Z}}_d)\sqrt{\eta_\uparrow}}, -\frac{(\sqrt{\eta_2} +  \sqrt{\eta_\downarrow})\underline{\left[Y^{X,\emptyset}_{1,(\eta_\uparrow,\eta_\uparrow)} e_{X,1,(\eta_\uparrow,\eta_\uparrow),\emptyset} \right]^{\leq 1}_{\mathcal{Z}}}}{\left(\sqrt{\eta_\uparrow}-\sqrt{\eta_2}\right)\left(\sqrt{\eta_\uparrow}-\sqrt{\eta_\downarrow}\right)\left(1-p^{\mathcal{Z}}_d\right)}  \right. \nonumber\\
   &\left.\quad+ \frac{(\sqrt{\eta_\uparrow} +  \sqrt{\eta_\downarrow}) \overline{\left[Y^{X,\emptyset}_{1,(\eta_2,\eta_\uparrow)} e_{X,1,(\eta_2,\eta_\uparrow),\emptyset} \right]^{\leq 1}_{\mathcal{Z}}}}{\left(\sqrt{\eta_\uparrow}-\sqrt{\eta_2}\right)\left(\sqrt{\eta_2}-\sqrt{\eta_\downarrow}\right)\left(1-p^{\mathcal{Z}}_d\right)} - \frac{(\sqrt{\eta_\uparrow} +  \sqrt{\eta_2}) \underline{\left[Y^{X,\emptyset}_{1,(\eta_\downarrow,\eta_\uparrow)} e_{X,1,(\eta_\downarrow,\eta_\uparrow),\emptyset}\right]^{\leq 1}_{\mathcal{Z}}}}{\left(\sqrt{\eta_\uparrow}-\sqrt{\eta_\downarrow}\right)\left(\sqrt{\eta_2}-\sqrt{\eta_\downarrow}\right)\left(1-p^{\mathcal{Z}}_d\right)}  \right\rbrace \label{protocol-Eetak-01-upp}\\
   \overline{\mathbbm{E}}^{1t}_{\eta_\uparrow} &:= \min\left\lbrace \overline{w}^1_{\mathcal{Z}}, \frac{\overline{\left[Y^{X,\emptyset}_{1,(\eta_\uparrow,\eta_\uparrow)} e_{X,1,(\eta_\uparrow,\eta_\uparrow),\emptyset} \right]^{\leq 1}_{\mathcal{Z}}}}{\left(\sqrt{\eta_\uparrow}-\sqrt{\eta_2}\right)\left(\sqrt{\eta_\uparrow}-\sqrt{\eta_\downarrow}\right)\left(1-p^{\mathcal{Z}}_d\right)}  - \frac{\underline{\left[Y^{X,\emptyset}_{1,(\eta_2,\eta_\uparrow)} e_{X,1,(\eta_2,\eta_\uparrow),\emptyset} \right]^{\leq 1}_{\mathcal{Z}}}}{\left(\sqrt{\eta_\uparrow}-\sqrt{\eta_2}\right)
   \left(\sqrt{\eta_2}-\sqrt{\eta_\downarrow}\right) \left(1-p^{\mathcal{Z}}_d\right)} \right.  \nonumber\\
   &\left. \quad+\frac{1}{\left(\sqrt{\eta_\uparrow}-\sqrt{\eta_\downarrow}\right) \left(\sqrt{\eta_2}-\sqrt{\eta_\downarrow}\right) 
   \left(1-p^{\mathcal{Z}}_d\right)} \overline{\left[Y^{X,\emptyset}_{1,(\eta_\downarrow,\eta_\uparrow)} e_{X,1,(\eta_\downarrow,\eta_\uparrow),\emptyset}\right]^{\leq 1}_{\mathcal{Z}}} \right\rbrace \label{protocol-Eetak-1t-upp} \\
   \overline{\mathbbm{E}}^{1r}_{\eta_\uparrow} &:= \min\left\lbrace \overline{w}^1_{\mathcal{Z}}, \frac{\overline{\left[Y^{X,\checkmark}_{1,(\eta_\downarrow,\eta_\uparrow)} e_{X,1,(\eta_\downarrow,\eta_\uparrow),\checkmark} \right]^{\leq 1}_{\mathcal{Z}}}}{1-\eta_\downarrow} - 
   \frac{p^{\mathcal{Z}}_d}{(1-\eta_\downarrow)(1-p^{\mathcal{Z}}_d)} \underline{\left[Y^{X,\emptyset}_{1,(\eta_\downarrow,\eta_\uparrow)} e_{X,1,(\eta_\downarrow,\eta_\uparrow),\emptyset} \right]^{\leq 1}_{\mathcal{Z}}} \right\rbrace  . \label{protocol-Eetak-1r-upp}
\end{align}
where the bounds on the products of yields and error rates, restricted to the ($\leq 1$)-subspace, are given by:
\begin{align}
    \underline{\left[Y^{X,\checkmark}_{1,(\eta_i,\eta_l)} e_{X,1,(\eta_i,\eta_l),\checkmark}\right]^{\leq 1}_{\mathcal{Z}}} &= \max \left\lbrace 0, \underline{Y^{X,\checkmark}_{1,(\eta_i,\eta_l)} e_{X,1,(\eta_i,\eta_l),\checkmark}} -  \sqrt{\overline{w}^{>1}_{\mathcal{Z}}} - \overline{w}^{>1}_{\mathcal{Z}} \right\rbrace \label{protocol-l-e,click} \\
    \overline{\left[Y^{X,\checkmark}_{1,(\eta_i,\eta_l)} e_{X,1,(\eta_i,\eta_l),\checkmark}\right]^{\leq 1}_{\mathcal{Z}}} &=  \min \left\lbrace 1, \overline{Y^{X,\checkmark}_{1,(\eta_i,\eta_l)} e_{X,1,(\eta_i,\eta_l),\checkmark}} +   \sqrt{\overline{w}^{>1}_{\mathcal{Z}}} \right\rbrace \label{protocol-u-e,click} \\
    \underline{\left[Y^{X,\emptyset}_{1,(\eta_i,\eta_l)} e_{X,1,(\eta_i,\eta_l),\emptyset}\right]^{\leq 1}_{\mathcal{Z}}} &= \max \left\lbrace 0, \underline{Y^{X,\emptyset}_{1,(\eta_i,\eta_l)} e_{X,1,(\eta_i,\eta_l),\emptyset}}  -  \sqrt{\overline{w}^{>1}_{\mathcal{Z}}} - \overline{w}^{>1}_{\mathcal{Z}} \right\rbrace \label{protocol-l-e,noclick} \\
    \overline{\left[Y^{X,\emptyset}_{1,(\eta_i,\eta_l)} e_{X,1,(\eta_i,\eta_l),\emptyset}\right]^{\leq 1}_{\mathcal{Z}}} &=  \min \left\lbrace 1, \overline{Y^{X,\emptyset}_{1,(\eta_i,\eta_l)} e_{X,1,(\eta_i,\eta_l),\emptyset}}  +   \sqrt{\overline{w}^{>1}_{\mathcal{Z}}} \right\rbrace \label{protocol-u-e,noclick}.
\end{align}

Finally, the upper and lower bounds on the yields and on the product of yields and bit error rates appearing in this Appendix are obtained with the decoy-state method and are reported in Appendix~\ref{app:decoy}.

\section{Security proof} \label{app:security-proof}

In this Appendix we prove the security of Protocol~1.\\

The Devetak-Winter (DW) bound \cite{DW} provides a lower bound on the asymptotic secret key rate of a QKD protocol subjected to collective attacks. According to the protocol's description, the shared secret key is extracted from the key generation rounds, i.e.\ the rounds where: $T=Z$, the TBS is set to $(\eta_\downarrow,\eta_\downarrow)$ and $Z_B \neq \emptyset$. To keep the notation concise, we define the following intersections of events:
\begin{align}
    \Gamma_Z &:= [T=Z] \cap (\eta_\downarrow,\eta_\downarrow) \label{GammaZ} \\
    \Omega_Z &:= \Gamma_Z \cap [Z_B \neq \emptyset] \label{OmegaZ}.
\end{align}
Then, the DW bound of our protocol reads \cite{QKDsoftware}:
\begin{align}
    r \geq \Pr(\Omega_Z) \left[H(Z_A|I_A E)_{\rho|\Omega_Z} - H(Z_A|I_A Z_B)_{\rho|\Omega_Z} \right], \label{DWrate}
\end{align}
where $H(Z_A|I_A E)$ ($H(Z_A| I_A Z_B)$) is the von Neumann entropy of Alice's key outcome $Z_A$ conditioned on Eve's (Bob's) total side information, while the subscript in the entropies indicates that they are computed on the state $\rho$ shared by Alice, Bob and Eve, conditioned on the event $\Omega_Z$. Since Alice (Bob) chooses the setting $T=Z$ ($(\eta_\downarrow,\eta_\downarrow)$) with probability $p_Z$, we have that:
\begin{align}
    \Pr(\Omega_Z) &= p_Z^2 \Pr(Z_B \neq \emptyset|\Gamma_Z). \label{prob-OmegaZ}
\end{align}

The remainder of the proof consists in deriving appropriate bounds on the two entropies in \eqref{DWrate} to show that $r_\infty$, given in \eqref{protocol-rate}, is a lower bound on the DW rate \eqref{DWrate}, hence proving the protocol's security.\\

\subsection{Computing the post-selected state} \label{app:proof1state}
The first step is to obtain a mathematical expression for the quantum state $\rho_{Z_A I_A Z_B E | \Omega_Z}$ on which the two entropies in \eqref{DWrate} are computed. 

To start with, we express the state of Alice's classical registers ($Z_A,I_A$) and of Bob's ($B$) and Eve's ($E$) quantum systems in a $Z$-basis round as follows:
\begin{align}
    &\rho_{Z_A I_A N_A BE} = \sum_{j=0}^{d-1} \sum_{\mu_i\in\mathcal{S}} \frac{p_{\mu_i}}{d} \proj{j}_{Z_A} \otimes\proj{\mu_i}_{I_A} \otimes \sum_{n=0}^\infty \Pr(n|\mu_i) \proj{n}_{N_A} \otimes U_{BE} \proj{n_{Z_j}}_B \otimes \proj{0}_E U_{BE}^\dag. \label{rhoZINBE}
\end{align}
Some comments are due. First, we added a classical register $N_A$, inaccessible to Alice, that contains the exact number of photons sent by Alice. Moreover, we modeled Eve's collective attack as a unitary acting on systems $BE$, which is not restrictive since we make no assumption on the dimension of Eve's quantum register $E$. Finally, the factor $1/d$ is due to the uniform probability with which Alice selects the symbol $Z_A$.

Now, we mathematically describe Bob's measurement in terms of the following POVM:
\begin{align}
    \left\lbrace M^{(\eta_i,\eta_l)}_{j,k},M^{(\eta_i,\eta_l)}_{j,\emptyset},M^{(\eta_i,\eta_l)}_{\emptyset,k},M^{(\eta_i,\eta_l)}_{\emptyset,\emptyset}\right\rbrace ,  \label{Bobs-POVM}
\end{align}
where the POVM element $M^{(\eta_i,\eta_l)}_{j,k}$ represents the outcome $Z_B=j,X_B=k$ when Bob selects $(\eta_i,\eta_l)$ as the TBS setting. Similarly, $M^{(\eta_i,\eta_l)}_{j,\emptyset}$ represents the outcome $Z_B=j$ and no click in the $X$ detector ($X_B=\emptyset$) with TBS setting $(\eta_i,\eta_l)$. And so on for the others. From \eqref{Bobs-POVM}, we can define a POVM that focuses on the outcomes of the $Z$ detector under the TBS setting $(\eta_\downarrow,\eta_\downarrow)$ and can be considered the effective POVM performed by Bob in the key generation rounds:
\begin{align}
    \left\lbrace Z_0, Z_1, \dots, Z_{d-1}, Z_{\emptyset} \right\rbrace, \label{Bobs-Z-measurement}
\end{align}
with:
\begin{align}
    Z_j &= M^{(\eta_\downarrow,\eta_\downarrow)}_{j,\emptyset} + \sum_{k=0}^{d-1} M^{(\eta_\downarrow,\eta_\downarrow)}_{j,k} \\
    Z_\emptyset &= M^{(\eta_\downarrow,\eta_\downarrow)}_{\emptyset,\emptyset} + \sum_{k=0}^{d-1} M^{(\eta_\downarrow,\eta_\downarrow)}_{\emptyset,k}. 
\end{align}
For this, we call such a reduced measurement the ``key generation measurement'' of Bob. From \eqref{Bobs-Z-measurement}, in turn, we define the key generation measurement map, $\mathcal{E}^Z_{B\to \Tilde{B} Z_B}$. This map acts between the input system $B$ and two classical output systems, $Z_B$ and $\Tilde{B}$, where $\tilde{B}$ is a classical register that contains a coarse-grained information about the measurement outcome, namely whether the detector clicked ($\Tilde{B}=\checkmark$) or not  ($\Tilde{B}=\emptyset$). We thus have:
\begin{align}
    \mathcal{E}^Z_{B\to \Tilde{B} Z_B}(\cdot)=&\sum_{j=0}^{d-1} \Tr_B[Z_j \,\,\cdot\,\,] \proj{\checkmark}_{\Tilde{B}} \otimes  \ketbra{j}{j}_{Z_B} + \Tr_B[Z_\emptyset \,\,\cdot\,\,] \ketbra{\emptyset}{\emptyset}_{\Tilde{B}} \otimes \ketbra{\emptyset}{\emptyset}_{Z_B}.  \label{BobKGmeasurement-12}
\end{align}
Now, we recast Bob's key generation measurement map as the concatenation of two quantum maps, by following the formalism developed in \cite{Tomamichel2017,BourassaLo}: $\mathcal{E}^Z_{B \to \Tilde{B} Z_B}= \mathcal{E}^Z_{\Tilde{B} B \to \Tilde{B} Z_B} \circ\mathcal{E}^Z_{B \to \tilde{B} B}$. The first map that is applied can be viewed as a coarse-grained measurement:
\begin{align}
    \mathcal{E}^Z_{B \to \tilde{B} B}(\cdot) = &\proj{\emptyset}_{\Tilde{B}} \otimes \sqrt{Z_\emptyset} \cdot \sqrt{Z_\emptyset}+ \proj{\checkmark}_{\Tilde{B}} \otimes \sqrt{Z_\checkmark} \cdot \sqrt{Z_\checkmark} , \label{BobKGmeasurement-1}
\end{align}
where
\begin{align}
    Z_\checkmark &=\textstyle{\sum_{j=0}^{d-1}} Z_j.  \label{Zclick}
\end{align}
The second map replaces the coarse-grained outcomes of the first map with the actual outcomes of Bob's measurement and traces out the quantum system $B$. Let $\one_{Z_\checkmark}$ be the projector on the support of the operator $Z_\checkmark$. Then, the second map reads:
\begin{align}
    &\mathcal{E}^Z_{\Tilde{B} B \to \Tilde{B} Z_B}(\cdot) =  \sum_{j=0}^{d-1} \ketbra{j}{j}_{Z_B} \otimes \Tr_B \left[\left(\proj{\checkmark}_{\Tilde{B}}\otimes \tilde{Z}_{j}\right) \,\cdot\, \left(\proj{\checkmark}_{\Tilde{B}}\otimes \one_B\right)\right] + \ketbra{\emptyset}{\emptyset}_{Z_B} \otimes  \proj{\emptyset}_{\Tilde{B}} \, \Tr_B[\,\cdot\,] \, \proj{\emptyset}_{\Tilde{B}} , \label{BobKGmeasurement-2}
\end{align}
where:
\begin{align}
    \tilde{Z}_j &= (\sqrt{Z_\checkmark})^{-1} Z_j(\sqrt{Z_\checkmark})^{-1}  \oplus \frac{\one^{\perp}_{Z_\checkmark}}{d},  \label{Zprimej}
\end{align}
with $(\sqrt{Z_\checkmark})^{-1}$ the inverse of $\sqrt{Z_\checkmark}$ over its support: $\sqrt{Z_\checkmark} (\sqrt{Z_\checkmark})^{-1} = (\sqrt{Z_\checkmark})^{-1} \sqrt{Z_\checkmark} = \one_{Z_\checkmark}$ and with $\one^{\perp}_{Z_\checkmark}$ the projector on the complement of the support of $Z_\checkmark$. With the definition in \eqref{Zprimej}, both maps in \eqref{BobKGmeasurement-1} and \eqref{BobKGmeasurement-2} are completely positive and trace preserving (CPTP) and their concatenation returns the original map \eqref{BobKGmeasurement-12}.

In a similar manner, from \eqref{Bobs-POVM} we can define a POVM that focuses on the outcomes of the $X$ detector, which would play the role of Bob's complementary measurement in a traditional QKD security proof:
\begin{align}
    \{X_0,X_1,\dots,X_{d-1}, X_\emptyset\},  \label{Bobs-X-measurement}
\end{align}
with:
\begin{align}
    X_k &= M^{(\eta_\uparrow,\eta_\uparrow)}_{\emptyset,k} + \sum_{j=0}^{d-1} M^{(\eta_\uparrow,\eta_\uparrow)}_{j,k} \\
    X_\emptyset &= M^{(\eta_\uparrow,\eta_\uparrow)}_{\emptyset,\emptyset} + \sum_{j=0}^{d-1} M^{(\eta_\uparrow,\eta_\uparrow)}_{j,\emptyset}. 
\end{align}
For this reason, we call this POVM the ``test measurement'' of Bob. We emphasize that the information gathered by the parties in the rounds labeled as test rounds exceeds what can be obtained by the above POVM, where the $Z$ detector outcomes are ignored and the TBS setting is fixed. Nevertheless, we are free to choose Bob's complementary measurement when defining the phase error rate and our choice will be a POVM connected to \eqref{Bobs-X-measurement}. Then, the goal of the security proof will be to show how the resulting phase error rate can be estimated from the full statistics observed in the test rounds. 

We can thus define Bob's test measurement map as the quantum map implementing the POVM in \eqref{Bobs-X-measurement}:
\begin{align}
    \mathcal{E}^X_{B\to \Tilde{B} X_B}(\cdot)=&\sum_{k=0}^{d-1} \Tr_B[X_k \,\,\cdot\,\,] \proj{\checkmark}_{\Tilde{B}} \otimes  \ketbra{k}{k}_{X_B} + \Tr_B[X_\emptyset \,\,\cdot\,\,] \proj{\emptyset}_{\Tilde{B}} \otimes \proj{\emptyset}_{X_B}, \label{Bobtestmeasurement-12}
\end{align}
and decompose it as the concatenation of two CPTP maps:
\begin{align}
    \mathcal{E}^X_{B \to \Tilde{B} X_B}= \mathcal{E}^X_{\Tilde{B} B \to \Tilde{B} X_B} \circ\mathcal{E}^X_{B \to  \tilde{B} B}, \label{Bobtestmeasurement-12-decomposed}
\end{align}
where the first map reads:
\begin{align}
    \mathcal{E}^X_{B \to \tilde{B} B}(\cdot) = &\proj{\emptyset}_{\Tilde{B}} \otimes \sqrt{X_\emptyset} \cdot \sqrt{X_\emptyset} + \proj{\checkmark}_{\Tilde{B}} \otimes \sqrt{X_\checkmark} \cdot \sqrt{X_\checkmark}, \label{Bobtestmeasurement-1}
\end{align}
with
\begin{align}
    X_\checkmark &= \textstyle{\sum_{k=0}^{d-1}} X_k  \label{Xclick},
\end{align}
while the second map reads:
\begin{align}
    &\mathcal{E}^X_{\Tilde{B} B \to \Tilde{B} X_B}(\cdot) =  \sum_{k=0}^{d-1} \ketbra{k}{k}_{X_B} \otimes \Tr_B \left[\left(\proj{\checkmark}_{\Tilde{B}}\otimes \tilde{X}_{k}\right) \,\cdot\, \left(\proj{\checkmark}_{\Tilde{B}}\otimes \one_B\right)\right] + \ketbra{\emptyset}{\emptyset}_{X_B} \otimes  \proj{\emptyset}_{\Tilde{B}} \, \Tr_B[\,\cdot\,] \, \proj{\emptyset}_{\Tilde{B}} , \label{Bobtestmeasurement-2}
\end{align}
where:
\begin{align}
    \tilde{X}_k &= (\sqrt{X_\checkmark})^{-1} X_k (\sqrt{X_\checkmark})^{-1}  \oplus \frac{\one^{\perp}_{X_\checkmark}}{d},  \label{Xprimek}
\end{align}
with $(\sqrt{X_\checkmark})^{-1}$ the inverse of $\sqrt{X_\checkmark}$ over its support and $\one^{\perp}_{X_\checkmark}$ the projector on the complement of the support of $X_\checkmark$.

At this point, we can express the state of Alice's and Bob's classical registers $Z_A,Z_B$ and Eve's quantum system $E$, in a $Z$-basis round where Bob selected the TBS setting $(\eta_\downarrow,\eta_\downarrow)$, as follows:
\begin{align}
    &\rho_{Z_A I_A N_A \tilde{B} Z_B E|\Gamma_Z} = \mathcal{E}^Z_{B \to \tilde{B}Z_B} \left(\rho_{Z_A I_A N_A B E}\right), \label{finalstate-KG}
\end{align}
where $\rho_{Z_A I_A N_A B E}$ is given in \eqref{rhoZINBE} and where we used the definition of $\Gamma_Z$ in \eqref{GammaZ}. In order to obtain the state on which the entropies in \eqref{DWrate} are computed, we must post-select the state in \eqref{finalstate-KG} on a detection in the $Z$ detector ($Z_B \neq \emptyset$). For this, we perform the projective measurement $\{\proj{\emptyset}_{\Tilde{B}},\proj{\checkmark}_{\Tilde{B}}\}$ on system $\Tilde{B}$ and select the state conditioned on the outcome $\Tilde{B}=\checkmark$ through the following non-linear map:
\begin{align}
    \mathcal{P} (\cdot )= \frac{\proj{\checkmark}_{\Tilde{B}} \,\,\cdot\,\, \proj{\checkmark}_{\Tilde{B}}}{\Tr[\,\,\cdot\,\, \proj{\checkmark}_{\Tilde{B}}]} , \label{postsel-map}
\end{align}
such that the state of a key generation round (i.e., conditioned on the event $\Omega_Z$) is given by \cite{QKDsoftware}:
\begin{align}
    \rho_{Z_A I_A N_A \tilde{B} Z_B E|\Omega_Z} &=\mathcal{P} \left( \rho_{Z_A I_A N_A \tilde{B} Z_B E |\Gamma_Z}\right) \nonumber\\
    &=\frac{\braket{\checkmark|\rho_{Z_A I_A N_A  \tilde{B} Z_B E|\Gamma_Z}|\checkmark}_{\Tilde{B}}\otimes\proj{\checkmark}_{\Tilde{B}}}{\Pr(Z_B \neq \emptyset|\Gamma_Z)} \label{finalstate-KGpost},
\end{align}
where $\Pr(Z_B \neq \emptyset|\Gamma_Z)=\Tr[\rho_{Z_A I_A N_A \tilde{B} Z_B E|\Gamma_Z} \proj{\checkmark}_{\Tilde{B}}]$. The state in \eqref{finalstate-KGpost} is the one needed to compute the entropies in \eqref{DWrate}.\\

\subsection{Computing the leakage} \label{app:proof2leakage}
We first focus on the second entropy in \eqref{DWrate}, which quantifies the optimal leakage due to error correction in the asymptotic regime and is computed on a reduced state of \eqref{finalstate-KGpost}, namely:
\begin{align}
    \rho_{Z_A I_A Z_B |\Omega_Z} 
    &=\frac{\braket{\checkmark|\Tr_{N_A E}[\rho_{Z_A I_A N_A  \tilde{B} Z_B E|\Gamma_Z}]|\checkmark}_{\Tilde{B}}}{\Pr(Z_B \neq \emptyset|\Gamma_Z)} \label{finalstate-KGpost-traced}.
\end{align}
By explicitly calculating the above state through \eqref{finalstate-KG}, we obtain the following expression:
\begin{align}
    &\rho_{Z_A I_A Z_B |\Omega_Z} =\sum_{\mu_i \in\mathcal{S}} \frac{p_{\mu_i}}{\Pr(Z_B \neq \emptyset|\Gamma_Z)} \ketbra{\mu_i}{\mu_i}_{I_A} \otimes\sum_{j,j'=0}^{d-1} \frac{\Tr_B[Z_{j'} \zeta_{B|\mu_i,j}]}{d}  \ketbra{j}{j}_{Z_A} \otimes \ketbra{j'}{j'}_{Z_B} \label{finalstate-KGpost-traced2},
\end{align}
where $\zeta_{B|\mu_i,j}$ is the state received by Bob when Alice prepared the WCP with intensity $\mu_i$ encoding the symbol $Z_A=j$:
\begin{align}
    \zeta_{B|\mu_i,j}:= \sum_{n=0}^\infty \Pr(n|\mu_i)\Tr_E[U_{BE}(\ketbra{n_{Z_j}}{n_{Z_j}}\otimes\ketbra{0}{0})U_{BE}^\dag].
\end{align}
Hence, we recognize that $\Tr_B[Z_{j'} \zeta_{B|\mu_i,j}]=\Pr(Z_B=j'|\Gamma_Z,I_A=\mu_i,Z_A=j)$ is the probability that Bob obtains outcome $Z_B=j'$, given that Alice sent the symbol $Z_A=j$ encoded in a WCP of intensity $\mu_i$ and Bob chose the setting $(\eta_\downarrow,\eta_\downarrow)$. From this quantity we recover the $Z$-basis gain:
\begin{align}
    G^Z_{\mu_i,(\eta_\downarrow,\eta_\downarrow)} &= \Pr(Z_B \neq \emptyset|\Gamma_Z, I_A = \mu_i) \nonumber\\
    &= \sum_{j,j'=0}^{d-1} \frac{\Tr_B[Z_{j'} \zeta_{B|\mu_i,j}]}{d},
\end{align}
and recast the state in \eqref{finalstate-KGpost-traced2} as:
\begin{align}
    &\rho_{Z_A I_A Z_B |\Omega_Z} =\sum_{\mu_i \in\mathcal{S}} \Pr(\mu_i|\Omega_Z) \ketbra{\mu_i}{\mu_i}_{I_A} \otimes\sum_{j,j'=0}^{d-1} \frac{\Tr_B[Z_{j'} \zeta_{B|\mu_i,j}]}{d G^Z_{\mu_i,(\eta_\downarrow,\eta_\downarrow)}}  \ketbra{j}{j}_{Z_A} \otimes \ketbra{j'}{j'}_{Z_B} \label{finalstate-KGpost-traced3},
\end{align}
where we used Bayes' rule and the definition of $\Omega_Z$ in \eqref{OmegaZ} to define:
\begin{align}
    \Pr(\mu_i|\Omega_Z) = \frac{p_{\mu_i}G^Z_{\mu_i,(\eta_\downarrow,\eta_\downarrow)}}{\Pr(Z_B \neq \emptyset|\Gamma_Z)}. \label{bayes}
\end{align}
Then, we observe that the state in \eqref{finalstate-KGpost-traced3} is a classical-quantum (cq) state of the form $\sum_x p_x \ketbra{x}{x} \otimes \rho_x$ with $\rho_x$ normalized states, where the role of $x$ is played by $\mu_i$. Therefore, we can express its conditional entropy as:
\begin{align}
    H(Z_A|I_A Z_B)_{\rho|\Omega_Z} &= \sum_{\mu_i \in\mathcal{S}} \Pr(\mu_i|\Omega_Z) H(Z_A|Z_B)_{\rho|\Omega_Z,\mu_i} \nonumber\\
    &\leq  \sum_{\mu_i \in\mathcal{S}} \Pr(\mu_i|\Omega_Z) [h(Q_{Z,\mu_i}) + Q_{Z,\mu_i} \log_2(d-1)] \nonumber\\
    &\leq \sum_{\mu_i \in\mathcal{S}} \frac{p_{\mu_i}G^Z_{\mu_i,(\eta_\downarrow,\eta_\downarrow)}}{\Pr(Z_B \neq \emptyset|\Gamma_Z)} u(Q_{Z,\mu_i}) \label{Bobs-entropy-bound},
\end{align}
where we used Fano's inequality with $Q_{Z,\mu_i}$ being the QBER of the key generation rounds where Alice used the intensity $\mu_i$, \eqref{QBER-Z}. In the last inequality we used \eqref{bayes} and the definition of $u(x)$ in \eqref{u(x)}.\\ 

\subsection{Lower bounding Eve's uncertainty}\label{app:proof3evesuncertainty}
\subsubsection{Discarding multi-photon contributions}\label{app:proof3Amultiphoton}
We now turn our attention to the first conditional entropy in \eqref{DWrate}, representing Eve's uncertainty about Alice's key bit $Z_A$ in a key generation round. The goal is to obtain a lower bound that only depends on the statistics observed in the protocol. By the strong subadditivity, we can lower bound the conditional entropy as follows:
\begin{align}
    H(Z_A|I_A E)_{\rho|\Omega_Z} \geq H(Z_A|I_A N_A E)_{\rho|\Omega_Z}, \label{step2}
\end{align}
where the right-hand side describes Eve's uncertainty about $Z_A$ if Eve knew the exact photon number $N_A$ of the signal sent by Alice. We now observe that the reduced state on which the conditional entropy is computed, which by definition \eqref{finalstate-KGpost} reads:
\begin{align}
    &\rho_{Z_A  I_A N_A E|\Omega_Z} = \Tr_{\tilde{B}Z_B} \left[ \mathcal{P} \circ \mathcal{E}^Z_{\Tilde{B} B \to \Tilde{B} Z_B} \circ\mathcal{E}^Z_{B \to \tilde{B} B} \left(\rho_{Z_A I_A N_A BE}\right)\right],
\end{align}
is equivalent to:
\begin{align}
    \rho_{Z_A I_A N_A E|\Omega_Z} &= \Tr_{B\tilde{B}} \left[\mathcal{P} \circ \mathcal{E}^Z_{B \to \tilde{B} B} \left(\rho_{Z_A I_A N_A BE}\right)\right]. \label{observation}
\end{align}
In other words, the second part of Bob's key generation measurement map ($\mathcal{E}^Z_{\Tilde{B} B \to \Tilde{B} Z_B}$) has no effect on the reduced state required by the calculation of the conditional entropy in \eqref{step2}. In fact, since this state is conditioned on the event $\Omega_Z$ corresponding to a $Z$ detection without specifying the outcome, the coarse-grained measurement map ($\mathcal{E}^Z_{B \to \tilde{B} B}$) is enough to fix the state. By using \eqref{rhoZINBE}, we can express the state in \eqref{observation} as follows:
\begin{align}
    \rho_{Z_A  I_A N_A E|\Omega_Z} = \frac{1}{\Pr(Z_B\neq\emptyset|\Gamma_Z)} &\sum_{j=0}^{d-1} \sum_{\mu_i \in\mathcal{S}} \frac{p_{\mu_i}}{d}  \ketbra{j}{j}_{Z_A} \otimes \ketbra{\mu_i}{\mu_i}_{I_A} \otimes \sum_{n=0}^\infty \Pr(n|\mu_i) \ketbra{n}{n}_{N_A} \nonumber\\
    &\otimes  \Pr(Z_B \neq\emptyset|\Gamma_Z,N_A=n,Z_A=j)\, \phi_{E|n,j,\Omega_Z}, \label{sigma-photons}
\end{align}
where we introduced the normalized state $\phi_{E|n,j,\Omega_Z}$ held by Eve in a key generation round, given that Alice sent the symbol $Z_B=j$ encoded in $n$ photons:
\begin{align}
    \phi_{E|n,j,\Omega_Z} := \frac{\Tr_B \left[(\sqrt{Z_\checkmark}\otimes \one_E) U_{BE}\proj{n_{Z_j}}_B \otimes \proj{0}_E U^\dag_{BE}(\sqrt{Z_\checkmark}\otimes \one_E)\right]}{\Pr(Z_B \neq\emptyset|\Gamma_Z,N_A=n,Z_A=j)} \label{phiE}
\end{align}
and the probability:
\begin{align}
    \Pr(Z_B \neq\emptyset|\Gamma_Z,N_A=n,Z_A=j) =\Tr\left[(Z_\checkmark \otimes\one_E) U_{BE} \ketbra{n_{Z_j}}{n_{Z_j}}_B \otimes \ketbra{0}{0}_E U_{BE}^\dag \right] \label{pr(det|j,n)}.
\end{align}

Now, we can view the state in \eqref{sigma-photons}  as a classical-quantum state between the systems $N_A$ and $Z_A I_A E$, by recasting it as follows:
\begin{align}
    \rho_{Z_A I_A N_A E|\Omega_Z} =  \sum_{n=0}^\infty \Pr(n|\Omega_Z) \ketbra{n}{n}_{N_A} \otimes \rho_{Z_A I_A E|n,\Omega_Z}, \label{sigma-photons2}
\end{align}
where we defined the probability of Alice sending exactly $n$ photons, given a key generation round:
\begin{align}
    \Pr(n|\Omega_Z) = \frac{\sum_{\mu_i \in\mathcal{S}}p_{\mu_i} \Pr(n|\mu_i) \sum_{j=0}^{d-1} \Pr(Z_B \neq\emptyset|\Gamma_Z,N_A=n,Z_A=j)/d }{\Pr(Z_B\neq\emptyset|\Gamma_Z)}, \label{pr(n|det)}
\end{align}
and the normalized state:
\begin{align}
     &\rho_{Z_A I_A E|n,\Omega_Z} = \sum_{j=0}^{d-1}\frac{\Pr(Z_B \neq\emptyset|\Gamma_Z,N_A=n,Z_A=j)/d}{\Pr(Z_B\neq\emptyset|\Gamma_Z)\Pr(n|\Omega_Z)} \ketbra{j}{j}_{Z_A}   \otimes\sum_{\mu_i \in\mathcal{S}}p_{\mu_i} \Pr(n|\mu_i) \ketbra{\mu_i}{\mu_i}_{I_A} \otimes \phi_{E|n,j,\Omega_Z}. \label{sigma-nphotons}
\end{align}
Having expressed the state as a classical-quantum state in \eqref{sigma-photons2}, we can decompose its entropy as follows:
\begin{align}
    H(Z_A|I_A N_A  E)_{\rho|\Omega_Z} &= \sum_{n=0}^\infty \Pr(n|\Omega_Z) H(Z_A|I_A E)_{\rho|n,\Omega_Z} \label{step4},
\end{align}
where $H(Z_A|I_A E)_{\rho|n,\Omega_Z}$ is computed on the state \eqref{sigma-nphotons}, which we recast as follows:
\begin{align}
     &\rho_{Z_A I_A E|n,\Omega_Z} = \rho_{I_A|n,\Omega_Z} \otimes \rho_{Z_A E|n,\Omega_Z},\label{sigma-nphotons2}
\end{align}
where we defined the normalized states:
\begin{align}
    \rho_{I_A|n,\Omega_Z} &= \sum_{\mu_i \in\mathcal{S}} \frac{p_{\mu_i} \Pr(n|\mu_i)}{\Pr(n)} \ketbra{\mu_i}{\mu_i}_{I_A} \\
    \rho_{Z_A  E|n,\Omega_Z} &=\sum_{j=0}^{d-1}\frac{\Pr(n)\Pr(Z_B \neq\emptyset|\Gamma_Z,N_A=n,Z_A=j)/d}{\Pr(Z_B\neq\emptyset|\Gamma_Z)\Pr(n|\Omega_Z)} \ketbra{j}{j}_{Z_A} 
     \otimes  \phi_{E|n,j,\Omega_Z} \nonumber\\
     &= \sum_{j=0}^{d-1} \frac{\Pr(Z_B \neq\emptyset|\Gamma_Z,N_A=n,Z_A=j)}{\sum_{l=0}^{d-1} \Pr(Z_B \neq\emptyset|\Gamma_Z,N_A=n,Z_A=l)} \ketbra{j}{j}_{Z_A}\otimes \phi_{E|n,j,\Omega_Z}. 
     \label{sigma-nphotons3}
\end{align}
where we used \eqref{pr(n|det)} and:
\begin{align}
    \Pr(n)=\sum_{\mu_i \in\mathcal{S}}p_{\mu_i} \Pr(n|\mu_i). \label{Pr(n)}
\end{align}
From \eqref{sigma-nphotons2} we observe that that the state of $I_A$ is in tensor product with the state of $Z_AE$, which implies that:
\begin{align}
      H(Z_A|I_A E)_{\rho|n,\Omega_Z} =  H(Z_A|E)_{\rho|n,\Omega_Z}. \label{step4.1}
\end{align}
By combining \eqref{step4} and \eqref{step4.1}, we obtain:
\begin{align}
    H(Z_A|I_A N_A  E)_{\rho|\Omega_Z} &= \sum_{n=0}^\infty \Pr(n|\Omega_Z) H(Z_A| E)_{\rho|n,\Omega_Z} \label{step4.2},
\end{align}
where the entropy on the right-hand side is computed on \eqref{sigma-nphotons3}.

By putting together \eqref{step2} and \eqref{step4.2}, we can lower bound the first entropy in \eqref{DWrate} as follows:
\begin{align}
    H(Z_A|I_A E)_{\rho|\Omega_Z} &\geq \sum_{n=0}^\infty \Pr(n|\Omega_Z) H(Z_A|E)_{\rho|n,\Omega_Z} \nonumber\\
    &=\sum_{n=0}^\infty \frac{Y^Z_{n,(\eta_\downarrow,\eta_\downarrow)} \Pr(n)}{\Pr(Z_B\neq\emptyset|\Gamma_Z)} H(Z_A|E)_{\rho|n,\Omega_Z}, \label{step5}
\end{align}
where we used \eqref{pr(n|det)} and defined the $n$-photon $Z$-basis yield as the detection probability of Bob's key generation measurement when Alice selects the $Z$ basis and sends $n$ photons, as follows:
\begin{align}
    Y^Z_{n,(\eta_\downarrow,\eta_\downarrow)} &:=  \Pr(Z_B \neq\emptyset|\Gamma_Z, N_A=n) =\sum_{j=0}^{d-1} \Pr(Z_B \neq\emptyset|\Gamma_Z,N_A=n,Z_A=j)\frac{1}{d} .\label{yield-KGn}
\end{align}
Note that the yields are not directly observed and are instead estimated with the decoy-state method. Moreover, note that the yield in \eqref{yield-KGn} is well-defined since $1/d=\Pr(Z_A=j)=\Pr(Z_A=j|N_A=n)$, where the last equality is due to the fact that the distribution of the symbols and photons prepared by Alice is factorized: $\Pr(Z_A=j,N_A=n)=\Pr(Z_A=j)\Pr(N_A=n)$.

We can further lower bound the entropy in \eqref{step5} by using the fact that $H(Z_A|E)_{\rho|n,\Omega_Z}\geq 0$ since  $\rho_{Z_A  E|n,\Omega_Z}$ is classical on $Z_A$. We obtain:
\begin{align}
    H(Z_A|I_A E)_{\rho|\Omega_Z} &\geq  \frac{Y^Z_{0,(\eta_\downarrow,\eta_\downarrow)} \Pr(0)}{\Pr(Z_B\neq\emptyset|\Gamma_Z)} H(Z_A|E)_{\rho|0,\Omega_Z} + \frac{Y^Z_{1,(\eta_\downarrow,\eta_\downarrow)} \Pr(1)}{\Pr(Z_B\neq\emptyset|\Gamma_Z)} H(Z_A|E)_{\rho|1,\Omega_Z}, \label{step6}
\end{align}
where we discarded all the terms corresponding to events where Alice sends $n\geq 2$ photons. This is done because our ability to find a non-trivial lower bound on $H(Z_A|E)_{\sigma|n,\Omega_Z}$ for $n\geq 2$ is limited by the fact that we cannot apply the uncertainty relation on such entropies. Indeed, the states sent by Alice in key generation rounds and test rounds can be seen as originating from the same entangled state (which is a precondition for using the uncertainty relation) only at the single-photon level. In addition, in the event that $n\geq 2$, Eve can perform a photon number splitting attack and learn $Z_A$ while remaining undetected, which implies that $H(Z_A|E)_{\sigma|n,\Omega_Z} =0$ for $n\geq 2$.

Now, we observe that the state $\rho_{Z_A  E|0,\Omega_Z}$ in \eqref{sigma-nphotons3} factorizes in the case that Alice sends the vacuum:
\begin{align}
    \rho_{Z_A  E|0,\Omega_Z}=\rho_{Z_A |0,\Omega_Z} \otimes \rho_{E |0,\Omega_Z},
\end{align}
due to the fact that Eve's state \eqref{phiE}, $\phi_{E|0,j,\Omega_Z}$, is independent of $j$. This reflects the fact that Eve is uncorrelated with Alice's outcome $Z_A$ when Alice sends the vacuum. Therefore, we have that:
\begin{align}
    H(Z_A|E)_{\rho|0,\Omega_Z} &= H(Z_A)_{\rho|0,\Omega_Z} \nonumber\\
    &= H\left(\left\lbrace \frac{\Pr(Z_B \neq\emptyset|\Gamma_Z,N_A=0,Z_A=j)}{\sum_{l=0}^{d-1} \Pr(Z_B \neq\emptyset|\Gamma_Z,N_A=0,Z_A=l)}\right\rbrace_{j=0}^{d-1}\right), \label{H(Z-A)}
\end{align}
where the entropy in the second line is the Shannon entropy of the enclosed distribution. Now, from \eqref{pr(det|j,n)} we observe that $\Pr(Z_B \neq\emptyset|\Gamma_Z,N_A=0,Z_A=j)=\Pr(Z_B \neq\emptyset|\Gamma_Z,N_A=0)$, i.e.\ the distribution is independent of the symbol $Z_A$ when Alice sends the vacuum. Thus, the distribution in the Shannon entropy simplifies to a uniform distribution of $d$ outcomes and we obtain:
\begin{align}
    H(Z_A|E)_{\rho|0,\Omega_Z} &= \log_2 d \label{H(Z-A)-final}.
\end{align}
By employing \eqref{H(Z-A)-final} in \eqref{step6}, we obtain:
\begin{align}
    H(Z_A|I_A E)_{\rho|\Omega_Z} &\geq  \frac{Y^Z_{0,(\eta_\downarrow,\eta_\downarrow)} \Pr(0)}{\Pr(Z_B\neq\emptyset|\Gamma_Z)} \log_2 d + \frac{Y^Z_{1,(\eta_\downarrow,\eta_\downarrow)} \Pr(1)}{\Pr(Z_B\neq\emptyset|\Gamma_Z)} H(Z_A|E)_{\rho|1,\Omega_Z}. \label{step7}
\end{align}
\subsubsection{Using uncertainty relations}\label{app:proof3Buncertainty}
We are left to find a lower bound on the entropy $H(Z_A|E)_{\rho|1,\Omega_Z}$ of Alice's outcome $Z_A$ of a key generation round where she sends exactly one photon, conditioned on Eve's quantum side information. The entropy is computed on the state in \eqref{sigma-nphotons3}, which for $n=1$ can be recast as:
\begin{align}
    &\rho_{Z_A  E|1,\Omega_Z} = \sum_{j=0}^{d-1} \frac{1}{d Y^Z_{1,(\eta_\downarrow,\eta_\downarrow)}} \ketbra{j}{j}_{Z_A} \otimes \Tr_{B} \left[(\sqrt{Z_\checkmark} \otimes\one_E) U_{BE} \ketbra{1_{Z_j}}{1_{Z_j}} \otimes \ketbra{0}{0}_E U_{BE}^\dag (\sqrt{Z_\checkmark} \otimes\one_E) \right], \label{sigma-entropy}
\end{align}
where we used \eqref{yield-KGn} and \eqref{phiE}. The sought bound is obtained through the uncertainty relation for entropies with quantum side information, whose statement is reported below.

\begin{lemma}[Entropic uncertainty relation  \cite{entropicUncert,UncertRelSmooth,Furrer:2012um,Furrer:2014ig}]
    Let $\sigma_{ABE}$ be a normalized quantum state and let $\{M^Z_j\}_j$ and $\{M^X_k\}_k$ be two POVMs on $A$. It holds:
    \begin{align}
        H(Z_A |E)_\sigma + H(X_A|B)_\sigma \geq \log_2 \frac{1}{\max_{j,k} \norm{\sqrt{M^Z_j} \sqrt{M^X_k}}_\infty^2} \label{uncert-rel}
    \end{align}
    where the first and second entropy are computed on the classical-quantum states:
    \begin{align}
        \sigma_{Z_A E} &= \sum_j \proj{j}_{Z_A} \otimes \Tr_{AB}\left[(M^Z_j \otimes \one_{BE}) \sigma_{ABE}\right] \label{sigmaZE} \\
        \sigma_{X_A B} &= \sum_k \proj{k}_{X_A} \otimes \Tr_{AE}\left[(M^X_k \otimes \one_{BE}) \sigma_{ABE}\right] \label{sigmaXB},
    \end{align}
    respectively.
\end{lemma}
We choose an appropriate state $\sigma_{ABE}$ and POVMs on $A$ such that the first entropy in \eqref{uncert-rel} coincides with the entropy we need to bound in \eqref{step7}. To this aim, we define the average state sent by Alice in any round, restricted to the single-photon subspace,
\begin{align}
    \tau := \frac{1}{d}\sum_{j=0}^{d-1} \ketbra{1_{Z_j}}{1_{Z_j}} = \frac{1}{d} \sum_{k=0}^{d-1} \ketbra{1_{X_k}}{1_{X_k}},
\end{align}
where the equality between the average state of $Z$-basis rounds and the average state of $X$-basis rounds is due to the assumption \eqref{equal-sum-twobases}. We then define $\tau_{AB}=\proj{\Psi_\tau}$ to be the purification of $\tau$ and define the POVMs $\{M^Z_j\}_j$ and $\{M^X_k\}_k$ acting on the purifying system $A$:
\begin{align}
    M^Z_j &= \frac{1}{d} \left(\tau^{-1/2}\right)^T \left(\proj{1_{Z_j}}\right)^T \left(\tau^{-1/2}\right)^T  \label{Mj}\\
    M^X_k &= \frac{1}{d} \left(\tau^{-1/2}\right)^T \left(\proj{1_{X_k}}\right)^T \left(\tau^{-1/2}\right)^T,  \label{Mk}
\end{align}
where $^T$ is the transpose with respect to the Schmidt basis of $\tau_{AB}$. Then, according to Lemma~14 in \cite{Tomamichel2017}, the reduced states on $B$, after applying the POVMs \eqref{Mj} and \eqref{Mk} on system $A$ of $\tau_{AB}$, are exactly the states Alice prepares in the one-photon subspace (multiplied by their probability):
\begin{align}
    \Tr_A \left[(M^Z_j \otimes\one_B) \tau_{AB}\right] &= \frac{\proj{1_{Z_j}}}{d} \label{Bj}\\
    \Tr_A \left[(M^X_k \otimes\one_B) \tau_{AB}\right] &= \frac{\proj{1_{X_k}}}{d} \label{Bk}.
\end{align}
Moreover, Lemma~14 shows that:
\begin{align}
    \max_{j,k} \norm{\sqrt{M^Z_j} \sqrt{M^X_k}}_\infty^2 &= \max_{0 \leq j,k \leq d-1} \norm{\ketbra{1_{Z_j}}{1_{Z_j}} S^{-1} \ketbra{1_{X_k}}{1_{X_k}}}^2_\infty \nonumber\\
    &=c, \label{compatibility2}
\end{align}
for the above defined POVMs, where $S$ is given in \eqref{equal-sum-twobases}, $\norm{A}_\infty$ is the largest singular value of $A$, and $c$ is given in \eqref{compatibility}.
\begin{remark} \label{rmk:compatibility}
    As discussed in Sec.~\ref{sec:protocol}, the protocol's key rate \eqref{protocol-rate} is maximized when the compatibility coefficient $c$ is minimal: $c=1/d$. Here, we show that a sufficient condition to attain this is that Alice's one-photon states form two mutually unbiased sets:
    \begin{align}
        \abs{\braket{1_{Z_j}|1_{X_{k}}}}^2 =\frac{1}{d} \quad\forall\, j,k. \label{appendix-mutually-unbiased-states}
    \end{align}
    \begin{proof}
    The proof hinges on the fact that when two sets of vectors, $\{\ket{1_{Z_j}}\}_j$ and $\{\ket{1_{X_k}}\}_k$, sum to the same operator $S$ by \eqref{equal-sum-twobases} and are mutually unbiased \eqref{appendix-mutually-unbiased-states}, then they are orthogonal sets.

    To see this, consider the scalar $\braket{1_{Z_j}|S|1_{Z_j}}$. By virtue of $S=\sum_k \proj{1_{X_k}}$ and \eqref{appendix-mutually-unbiased-states}, we get: $\braket{1_{Z_j}|S|1_{Z_j}}=1$. At the same time, by using the other expression for $S$, we get: $\braket{1_{Z_j}|S|1_{Z_j}}=1 + \sum_{j'\neq j} \abs{\braket{1_{Z_j}|1_{Z_{j'}}}}^2 $, which implies that: $\sum_{j'\neq j} \abs{\braket{1_{Z_j}|1_{Z_{j'}}}}^2 = 0$. This implies that the set of vectors $\{\ket{1_{Z_j}}\}_j$ is orthonormal:
    \begin{align}
        \abs{\braket{1_{Z_j}|1_{Z_{j'}}}}^2 = \delta_{j,j'},
    \end{align}
    and similarly for the set $\{\ket{1_{X_k}}\}_k$:
    \begin{align}
        \abs{\braket{1_{X_k}|1_{X_{k'}}}}^2 = \delta_{k,k'}.
    \end{align}
    Thus, we deduce that $S$ is the identity in the subspace defined by $\{\ket{1_{Z_j}}\}_j$ (or $\{\ket{1_{X_k}}\}_k$), which simplifies \eqref{compatibility} as follows:
        \begin{align}
            c &= \max_{0 \leq j,k \leq d-1} \abs{\bra{1_{Z_j}}  \ket{1_{X_k}}}^2 \nonumber\\
            &=\frac{1}{d},
        \end{align}
        where in the second equality we used \eqref{appendix-mutually-unbiased-states}.
    \end{proof}
\end{remark}

Therefore, we define the state $\sigma_{ABE}=\proj{\Psi_\sigma}$, with:
\begin{align}
    \ket{\Psi_\sigma} &= \frac{1}{\sqrt{Y^Z_{1,(\eta_\downarrow,\eta_\downarrow)}}} (\one_{AE} \otimes \sqrt{Z_\checkmark})(\one_A \otimes U_{BE}) \ket{\Psi_\tau}_{AB} \otimes\ket{0}_E,
\end{align}
and apply the uncertainty relation on $\sigma_{ABE}$ with the POVMs \eqref{Mj} and \eqref{Mk}. By computing the states \eqref{sigmaZE} and \eqref{sigmaXB} with these choices and by using \eqref{Bj} and \eqref{Bk}, we get:
\begin{align}
        \sigma_{Z_A E} &= \sum_{j=0}^{d-1} \proj{j}_{Z_A} \otimes \Tr_{AB}\left[(M^Z_j \otimes \one_{BE}) \proj{\Psi_\sigma}\right] \nonumber\\
        &=\rho_{Z_A E|1,\Omega_Z}  \label{sigmaZE2},
\end{align}
where we used \eqref{sigma-entropy}, and
\begin{align}
    \sigma_{X_A B} &= \sum_{k=0}^{d-1} \proj{k}_{X_A} \otimes \Tr_{AE}\left[(M^X_k \otimes \one_{BE}) \proj{\Psi_\sigma}\right]  \nonumber\\
    &=\frac{1}{ Y^Z_{1,(\eta_\downarrow,\eta_\downarrow)} d} \sum_{k=0}^{d-1}  \ketbra{k}{k}_{X_A} \otimes \sqrt{Z_\checkmark}  \sigma_k \sqrt{Z_\checkmark} \label{sigmaXB2},
\end{align}
respectively, where we defined $\sigma_k$ as:
\begin{align}
    \sigma_k &:= \Tr_E \left[ U_{BE} \proj{1_{X_k}}_B \otimes \proj{0}_E U_{BE}^\dag \right] \label{sigmak}.
\end{align}
Note that, due to the assumption \eqref{equal-sum-twobases} and the definition \eqref{yield-KGn} of yields for rounds with $\Gamma_Z$, we have that:
\begin{align}
    \sum_{k=0}^{d-1} \Tr[Z_\checkmark \sigma_k] &= \sum_{j=0}^{d-1} \Pr(Z_B \neq\emptyset|\Gamma_Z,N_A=1,Z_A=j) \nonumber\\
    &= Y^Z_{1,(\eta_\downarrow,\eta_\downarrow)} d,  \label{correct-normalization}
\end{align}
which confirms that the state in \eqref{sigmaXB2} is properly normalized. Thus, the uncertainty relation \eqref{uncert-rel} yields a lower bound on the entropy of interest:
\begin{align}
    H(Z_A|E)_{\rho|1,\Omega_Z} &= H(Z_A|E)_{\sigma} \nonumber\\
    &\geq \log_2 \frac{1}{c} - H(X_A|B)_{\sigma} \label{step9},
\end{align}
where we used \eqref{sigmaZE2}, \eqref{compatibility2} and where the entropy on the right-hand side is computed on the state \eqref{sigmaXB2}. The rest of the proof is devoted to finding an upper bound on $H(X_A|B)_{\sigma}$ in terms of observed statistics.

Since the entropy $H(X_A|B)_{\sigma}$ is defined on a classical-quantum state \eqref{sigmaXB2}, we can apply a measurement map $\mathcal{E}^X_{B \to \tilde{X}_B}$ on its conditioning system which cannot reduce the entropy \cite{NielsenChuang}:
\begin{align}
    H(X_A|B)_{\sigma} \leq H(X_A|\tilde{X}_B)_{\mathcal{E}^X(\sigma)}  \label{step10},
\end{align}
where $\tilde{X}_B$ is now a classical random variable resulting from a fictitious test measurement. Intuitively, a sensible choice for the fictitious measurement on system $B$ could be Bob's test measurement \eqref{Bobtestmeasurement-12}, defined earlier as the reduced POVM of Bob that focuses on the outcomes of the $X$-basis detector, from which Bob can try to guess the symbol $X_A$ sent by Alice in an $X$-basis round. However, the states of system $B$ in \eqref{sigmaXB2} undergo a post-selection corresponding to a detection in Bob's key generation measurement. This post-selection can be described in terms of the first map \eqref{BobKGmeasurement-1} composing Bob's key generation measurement map and of the post-selection map \eqref{postsel-map}, such that the state in \eqref{sigmaXB2} can be recast as:
\begin{align}
    \sigma_{X_A B} = \Tr_{\tilde{B}} \circ \mathcal{P} \circ \mathcal{E}^Z_{B \to \tilde{B}B} \left(\frac{1}{d} \sum_{k=0}^{d-1}  \ketbra{k}{k}_{X_A} \otimes \sigma_k \right). \label{sigmaXB3}
\end{align}
Therefore, the application of Bob's test measurement map \eqref{Bobtestmeasurement-12}, $\mathcal{E}^X_{B \to \Tilde{B} X_B}$, on the state in \eqref{sigmaXB3} would \textit{not} return a classical state whose correlations are observed experimentally. In other words, while the state
\begin{align}
   \sigma'_{X_A X_B \tilde{B}}=  \mathcal{E}^X_{B \to \Tilde{B} X_B} \left(\frac{1}{d} \sum_{k=0}^{d-1}  \ketbra{k}{k}_{X_A} \otimes \sigma_k \right),
\end{align}
could be characterized experimentally with traditional QKD techniques like the decoy-state method, the state
\begin{align}
    \sigma''_{X_A X_B \tilde{B}}=\mathcal{E}^X_{B \to \Tilde{B} X_B} \left(\sigma_{X_AB} \right)
\end{align}
cannot, due to the additional maps $\Tr_{\tilde{B}} \circ \mathcal{P} \circ \mathcal{E}^Z_{B \to \tilde{B}B}$ enacting the post-selection of a detection in Bob's key generation measurement. According to this observation, the measurement map $\mathcal{E}^X_{B \to \tilde{X}_B}$ to be applied in \eqref{step10} should not be chosen to be Bob's test measurement: $\mathcal{E}^X_{B \to \tilde{X}_B}\neq \mathcal{E}^X_{B \to \Tilde{B} X_B}$.

Nevertheless, recall that Bob's test measurement can be decomposed into two CPTP maps \eqref{Bobtestmeasurement-12-decomposed}, where the first map \eqref{Bobtestmeasurement-1}, $\mathcal{E}^X_{B \to \Tilde{B} B}$, is a coarse-grained map that only describes whether there is a detection in the $X$ detector. Then, let us assume for the moment that the detection probability of Bob's key generation and test measurements coincides for every input state, i.e.\ the assumption in \eqref{common-assumption} which is often implicit in standard QKD security proofs. Under the assumption \eqref{common-assumption}, we observe that the following two maps coincide:
\begin{align}
     \Tr_{\tilde{B}} \circ \mathcal{P} \circ \mathcal{E}^Z_{B \to \tilde{B}B} =  \Tr_{\tilde{B}} \circ \mathcal{P} \circ \mathcal{E}^X_{B \to \tilde{B}B},
\end{align}
i.e., post-selecting on a detection in the key generation measurement and in the test measurement has the same effect. This fact suggests a better choice for the fictitious measurement ($\mathcal{E}^X_{B \to \tilde{X}_B}$) to be applied on \eqref{sigmaXB3}: the second CPTP map of the decomposition of Bob's test measurement, i.e.\ the map $\mathcal{E}^X_{ \Tilde{B} B \to \Tilde{B} X_B}$, given in \eqref{Bobtestmeasurement-2}. More precisely, since the state in \eqref{sigmaXB3} is already post-selected on a detection of the test measurement --provided that the assumption \eqref{common-assumption} holds--, we choose $\mathcal{E}^X_{B \to \tilde{X}_B}$ to be the restriction of $\mathcal{E}^X_{ \Tilde{B} B \to \Tilde{B} X_B}$ to the subspace with $\tilde{B}=1$:
\begin{align}
    \mathcal{E}^X_{B \to  \tilde{X}_B} (\cdot) :=  \sum_{\tilde{k}=0}^{d-1}  \Tr\left[ \tilde{X}_{\tilde{k}} \,\cdot\, \right] \ketbra{\tilde{k}}{\tilde{k}}_{\tilde{X}_B}   , \label{Bobtestmeasurement-modified}
\end{align}
with $\tilde{X}_{\tilde{k}}$ defined in \eqref{Xprimek}. As a matter of fact, with the above fictitious measurement map and under the assumption \eqref{common-assumption}, the state $\sigma_{X_AB}$ transforms to:
\begin{align}
    &\mathcal{E}^X_{B \to  \tilde{X}_B} (\sigma_{X_AB}) = \nonumber\\
    &=\mathcal{E}^X_{B \to  \tilde{X}_B} \circ \Tr_{\tilde{B}} \circ \mathcal{P} \circ \mathcal{E}^Z_{B \to \tilde{B}B} \left(\frac{1}{d} \sum_{k=0}^{d-1}  \ketbra{k}{k}_{X_A} \otimes \sigma_k \right) \nonumber\\
    &=\mathcal{E}^X_{B \to  \tilde{X}_B} \circ \Tr_{\tilde{B}} \circ \mathcal{P} \circ \mathcal{E}^X_{B \to \tilde{B}B} \left(\frac{1}{d} \sum_{k=0}^{d-1}  \ketbra{k}{k}_{X_A} \otimes \sigma_k \right) \nonumber\\
    &= \Tr_{\tilde{B}} \circ \mathcal{P} \circ \mathcal{E}^X_{B \to \Tilde{B} X_B} \left(\frac{1}{d} \sum_{k=0}^{d-1}  \ketbra{k}{k}_{X_A} \otimes \sigma_k \right) \nonumber\\
    &= \Tr_{\tilde{B}} \circ \mathcal{P} (\sigma'_{X_A X_B \tilde{B}}), \label{state-with-assumption}
\end{align}
which can be characterized experimentally by traditional QKD methods. In our proof, however, we avoid making the assumption in \eqref{common-assumption} and hence the classical state arising from $\mathcal{E}^X_{B \to  \tilde{X}_B} (\sigma_{X_AB})$, with the fictitious measurement in \eqref{Bobtestmeasurement-modified}, cannot be directly estimated like in most QKD proofs. Nevertheless, we maintain the choice of fictitious measurement map as given in \eqref{Bobtestmeasurement-modified}, such that our proof can reduce to standard QKD proofs in the special case where the assumption \eqref{common-assumption} holds.

With our choice of fictitious test measurement, the entropy on the right-hand side of \eqref{step10} is computed on the normalized state:
\begin{align}
    &\sigma_{X_A \tilde{X}_B}= \mathcal{E}^X_{B \to  \tilde{X}_B} (\sigma_{X_AB}) \nonumber\\
    &= \sum_{k,\tilde{k}=0}^{d-1} \frac{\Tr\left[ \tilde{X}_{\tilde{k}} \sqrt{Z_\checkmark}  \sigma_k \sqrt{Z_\checkmark} \right]}{Y^Z_{1,(\eta_\downarrow,\eta_\downarrow)} d} \ketbra{k}{k}_{X_A} \otimes \ketbra{\tilde{k}}{\tilde{k}}_{\tilde{X}_B} \,, \label{sigma-third}
\end{align}
where we used \eqref{sigmaXB2}. Since \eqref{sigma-third} is a classical-classical state, we can upper bound its conditional Shannon entropy by using Fano's inequality,
\begin{align}
   H(X_A|\tilde{X}_B)_{\mathcal{E}^X(\sigma)}  \leq u(\tilde{e}_{X,1}) , \label{step11}
\end{align}
where $u(x)$ is given in \eqref{u(x)} and where we introduced the phase error rate:
\begin{align}
    \tilde{e}_{X,1} &:= \sum_{k=0}^{d-1} \sum_{\tilde{k}\neq k}
    \frac{\Tr\left[ \tilde{X}_{\tilde{k}} \sqrt{Z_\checkmark}  \sigma_k \sqrt{Z_\checkmark} \right]}{Y^Z_{1,(\eta_\downarrow,\eta_\downarrow)} d}\label{phase-error-rate}.
\end{align}

By combining \eqref{step9}, \eqref{step10}, and \eqref{step11}, we obtain the following lower bound on the remaining entropy in \eqref{step7}:
\begin{align}
    H(Z_A|E)_{\rho|1,\Omega_Z} \geq \log_2 \frac{1}{c} - u(\tilde{e}_{X,1})   \label{step12}.
\end{align}
\subsubsection{Reduction of phase error rate}\label{app:proof3Creduction}
We remark that the derived bound does not yet conclude the proof, since the phase error rate, as defined in \eqref{phase-error-rate}, cannot be directly estimated with the decoy-state method. The reason is that the probability distribution on which the phase error rate is defined is not directly accessible, in general. Indeed, Bob never actually performs the POVM $\{\tilde{X}_{\tilde{k}}\}_{\tilde{k}}$ after post-selecting the state on a detection in the key generation measurement. This complication, however, vanishes in the proofs that assume that the detection probability is independent of the measurement basis, as explained in the following Remark.

\begin{remark} \label{rmk:assumption}
Standard QKD security proofs \cite{finite-key-decoyBB84-security,Tomamichel2017} assume equality between the detection probabilities of Bob's key generation measurement \eqref{Bobs-Z-measurement} and Bob's test measurement \eqref{Bobs-X-measurement} for any input state. That is, they assume the equality:
\begin{align}
    Z_\checkmark = X_\checkmark \label{common-assumption2},
\end{align}
where $Z_\checkmark$ ($X_\checkmark$) is the POVM element corresponding to a detection of the key generation (test) measurement.
Equipped with the condition in \eqref{common-assumption2}, it is immediate to verify that the phase error rate can be directly estimated from the decoy-state method applied on the statistics of Bob's test measurements, thereby completing the security proof.

To see this, we consider the one-photon bit error rate of Bob's test measurement \eqref{Bobs-X-measurement}, which can be defined as follows:
\begin{align}
    e_{X,1} &:= \Pr\left(X_A \neq X_B |T=X,N_A=1,(\eta_\uparrow,\eta_\uparrow),X_B \neq \emptyset \right) \nonumber\\
    &=\sum_{k=0}^{d-1} \sum_{k' \neq k} \Pr(X_A=k,X_B=k'|T=X,N_A=1,(\eta_\uparrow,\eta_\uparrow),X_B \neq \emptyset) \nonumber\\
    &= \sum_{k=0}^{d-1}\sum_{k' \neq k} \frac{\Tr[X_{k'}\sigma_k]}{Y^X_{1,(\eta_\uparrow,\eta_\uparrow)} d}, \label{bit-error-rate}
\end{align}
where we introduced the one-photon $X$-basis yield:
\begin{align}
    Y^X_{1,(\eta_\uparrow,\eta_\uparrow)} &:= \sum_{k=0}^{d-1} \Pr(X_B \neq\emptyset|T=X,N_A=1,X_A=k,(\eta_\uparrow,\eta_\uparrow))\frac{1}{d} \nonumber\\
    &= \sum_{k=0}^{d-1} \frac{1}{d} \Tr\left[(X_\checkmark \otimes \one_E) U_{BE} \ketbra{1_{X_k}}{1_{X_k}}_B \otimes \ketbra{0}{0}_E U_{BE}^\dag \right].\label{yield-test1}
\end{align}
Now, the one-photon $Z$-basis yield \eqref{yield-KGn}, with \eqref{pr(det|j,n)}, can be expressed as:
\begin{align}
    Y^Z_{1,(\eta_\downarrow,\eta_\downarrow)} &=\sum_{j=0}^{d-1} \frac{1}{d} \Tr\left[(Z_\checkmark \otimes\one_E) U_{BE} \ketbra{1_{Z_j}}{1_{Z_j}}_B \otimes \ketbra{0}{0}_E U_{BE}^\dag \right] \nonumber\\
    &= \sum_{k=0}^{d-1} \frac{1}{d} \Tr\left[(X_\checkmark \otimes\one_E) U_{BE} \ketbra{1_{X_k}}{1_{X_k}}_B \otimes \ketbra{0}{0}_E U_{BE}^\dag \right] \nonumber\\
    &= Y^X_{1,(\eta_\uparrow,\eta_\uparrow)}, \label{simplification-remark}
\end{align}
where in the second equality we used the assumption in \eqref{common-assumption2} together with \eqref{equal-sum-twobases}. At the same time, the assumption in \eqref{common-assumption2} allows us to simplify the following expression:
\begin{align}
    \Tr\left[ \tilde{X}_{\tilde{k}} \sqrt{Z_\checkmark}  \sigma_k \sqrt{Z_\checkmark} \right] = \Tr[X_{\tilde{k}} \sigma_k], \label{simplification-remark2}
\end{align}
where we used \eqref{Xprimek}. By employing \eqref{simplification-remark} and \eqref{simplification-remark2} in the expression for the phase error rate \eqref{phase-error-rate}, we obtain:
\begin{align}
    \tilde{e}_{X,1} = \sum_{k=0}^{d-1} \sum_{\tilde{k}\neq k} \frac{ \Tr[X_{\tilde{k}} \sigma_k]}{ Y^X_{1,(\eta_\uparrow,\eta_\uparrow)} d} = e_{X,1}, \label{biterrorrate-testround}
\end{align}
i.e.\ the phase error rate reduces to the bit error rate of Bob's test measurement, which can be estimated with the decoy-state method. This fact would thus conclude the security proof of the protocol, since the phase error rate would be experimentally estimated.
\end{remark}

The absence of the assumption \eqref{common-assumption2} makes the estimation of the phase error rate in \eqref{phase-error-rate} considerably more challenging in our proof, as it cannot be identified anymore with the bit error rate of Bob's test measurement. In the next subsection we show how we can make use of the full statistics collected in the test rounds to provide an accurate estimation of the phase error rate.

\subsection{Phase error rate estimation} \label{sec:phase-error-rate-bound}

In this subsection we develop a method to derive an upper bound on the phase error rate as a function of quantities that can be obtained experimentally with the decoy-state method, without assuming basis-independent detection probabilities, i.e.\ the assumption in \eqref{common-assumption2}. For clarity, we report the definition of the phase error rate from \eqref{phase-error-rate}:
\begin{align}
    \tilde{e}_{X,1} &= \frac{\sum_{k=0}^{d-1} \sum_{k'\neq k} \Tr\left[ \tilde{X}_{k'} \sqrt{Z_\checkmark}  \sigma_k \sqrt{Z_\checkmark} \right]}{Y^Z_{1,(\eta_\downarrow,\eta_\downarrow)} d}\label{phase-error-rate-appendix},
\end{align}
where $Z_\checkmark$ is given in \eqref{Zclick}, $\{\sigma_k\}_k$ is given in \eqref{sigmak}, and $\tilde{X}_{k}$ given in \eqref{Xprimek} and forms a POVM $\{\tilde{X}_{k}\}_{k=0}^{d-1}$.\\

We start with a general overview of the argument, after which we detail all the steps of the derivation.

\subsubsection{Overview of the argument}\label{app:PER0}

Let us consider the subspace with zero or one photon localized in $\mathcal{Z}$, which we recall being the set of modes detected by the $Z$-basis detector. For brevity, we name this subspace as the ($\leq 1$)-subspace and label the average state received by Bob, when Alice sends one photon, as follows:
\begin{align}
    \bar{\sigma} &=  {\textstyle\sum_{k=0}^{d-1}} \frac{\sigma_k}{d} \label{simplifiedproof-bar-sigma} \\
    &= {\textstyle\sum_{k=0}^{d-1}} \Tr_E \left[ U_{BE} \frac{\proj{1_{X_k}}_B}{d} \otimes \proj{0}_E U_{BE}^\dag \right] \nonumber\\
    &={\textstyle\sum_{j=0}^{d-1}} \Tr_E \left[ U_{BE} \frac{\proj{1_{Z_j}}_B}{d} \otimes \proj{0}_E U_{BE}^\dag \right],
\end{align}
where in the last equality we used \eqref{equal-sum-twobases}.

The first step in estimating the phase error rate consists in reducing the calculation of the phase error rate in \eqref{phase-error-rate} to the ($\leq 1$)-subspace. The details of this step are deferred to Appendices~\ref{app:PER1} to \ref{app:PER4}. Indeed, we expect most of the state $\bar{\sigma}$ to lie in the ($\leq 1$)-subspace in an honest implementation of the protocol.

To this aim, we define $\pia$ to be the projector of $n=\alpha$ photons in the set $\mathcal{Z}$ and the identity elsewhere, such that the ($\leq 1$)-subspace is the support of $\Pi^0_{\mathcal{Z}} + \Pi^1_{\mathcal{Z}}$. Then, we can express $Z_\checkmark$ in terms of $\pia$ as follows:
\begin{align}
    Z_\checkmark = {\textstyle\sum_{\alpha=0}^\infty} p_{\checkmark|\alpha} \pia \label{Zclick-expansion},
\end{align}
where $p_{\checkmark|\alpha}$ is the probability that the $Z$-basis detector clicks, given that Bob's input state contains $\alpha$ photons localized in $\mathcal{Z}$. The expression of $p_{\checkmark|\alpha}$ is determined by the dark count probability ($p^Z_d$) and the mode-independent detection efficiency ($\eta_Z$) of the $Z$-basis detector, as well as on the TBS transmittance $\eta_\downarrow$, and is given in \eqref{Pr(oneclick|alpha)} of Appendix~\ref{app:PER2} (in reality, $p_{\checkmark|\alpha}$ does not depend on $\eta_Z$ since the loss of Bob's $Z$-basis detector is factored out of Bob's apparatus and assigned to the quantum channel, see Appendix~\ref{app:PER1}). By employing the expansion \eqref{Zclick-expansion} of $Z_\checkmark$ in the phase error rate \eqref{phase-error-rate}, we can isolate the contribution to the phase error rate given by the fraction of the states $\sigma_k$ living in the ($\leq 1$)-subspace. We remark that we cannot assume the states $\sigma_k$ to be completely confined to that subspace due to potential attacks by Eve, which may, e.g., add photons to Alice's pulses. Therefore, in Appendix~\ref{app:PER3} we show how to employ the detection statistics of the $Z$-basis detector, with the TBS settings $(\eta_\downarrow,\eta_\downarrow)$, $(\eta_\uparrow,\eta_\uparrow)$, and $(\eta_2,\eta_2)$, to derive an upper bound on the weight of $\Bar{\sigma}$ outside the ($\leq 1$)-subspace. Our method draws inspiration from the detector decoy technique \cite{detector-decoy} and leads to the following bound (cf.~Eq.~\eqref{yalpha>1-upperbound} in Appendix~\ref{app:PER3}):
\begin{align}
    {\textstyle\sum_{\alpha=2}^\infty} \Tr[\bar{\sigma} \pia] \leq \overline{w}^{>1}_{\mathcal{Z}} \,,\label{weight-outside}
\end{align}
where $\overline{w}^{>1}_{\mathcal{Z}}$ is reported in \eqref{yalpha>1-upperbound-explicit} in terms of quantities that can be directly estimated with the decoy-state method (the $Z$-basis yields). Then, in Appendix~\ref{app:PER4} we obtain the following upper bound on the phase error rate \eqref{phase-error-rate}, which employs \eqref{Zclick-expansion} and \eqref{weight-outside} to reduce the evaluation of the error rate to the ($\leq 1$)-subspace (cf.~Eq.~\eqref{phase-error-rate-uppbound} in Appendix~\ref{app:PER4}):
\begin{align}
    \tilde{e}_{X,1} Y^Z_{1,(\eta_\downarrow,\eta_\downarrow)} &\leq \frac{1}{d}{\textstyle\sum_{k=0}^{d-1}} \Tr\left[\sigma_k  M^{\leq 1}_{\mathcal{Z}} {\textstyle\sum_{k'\neq k}}\tilde{X}_{k'} M^{\leq 1}_{\mathcal{Z}} \right] \nonumber\\
    &\quad + \overline{\Delta_2} + \overline{w}^{>1}_{\mathcal{Z}} \label{simplifiedproof-phase-error-rate-uppbound},
\end{align}
where $\overline{\Delta_2}$ is a function of $\overline{w}^{>1}_{\mathcal{Z}}$ and $Y^Z_{1,(\eta_\downarrow,\eta_\downarrow)}$ given in \eqref{Delta2bar}, and where the support of the operator:
\begin{align}
    M^{\leq 1}_{\mathcal{Z}}=\sqrt{p_{\checkmark|0}} \Pi^0_{\mathcal{Z}} + \sqrt{p_{\checkmark|1}} \Pi^1_{\mathcal{Z}}  \label{M-operator},
\end{align}
is the ($\leq 1$)-subspace.

In Appendix~\ref{app:PER9}, we simplify the expression in \eqref{simplifiedproof-phase-error-rate-uppbound} by computing explicitly the POVM element $\Tilde{X}_k$ given in \eqref{simplifiedproof-Xprimek}. To this aim, similarly to $\pia$ for the $Z$-basis detector, we introduce the projector $\Pi^\beta_{\mathcal{X}}$ that projects on the subspace with $\beta$ photons in the set of modes $\mathcal{X}$, and acts as the identity elsewhere. Then, the POVM element $X_k$ of Bob's test measurement can be written in terms of, a) projectors similar to $\Pi^\beta_{\mathcal{X}}$, b) the total dark count probability ($p^{\mathcal{X}}_d$) of the $X$-basis detector, c) the TBS transmittance $\eta_\uparrow$, and, d) the ratio between the mode-independent detection efficiency of the $X$-basis and $Z$-basis detector (since we factored out the efficiency of the $Z$-basis detector):
\begin{align}
    \eta_r =\frac{\eta_X}{\eta_Z}.
\end{align}
Note that, by assumption, $\eta_r\leq 1$. The calculation of $\Tilde{X}_k$ in Appendix~\ref{app:PER9} leads to the following operator upper bound (cf.~Eq.~\eqref{Xprimek-appendix5}):
\begin{align}
    {\textstyle\sum_{k' \neq k}}\Tilde{X}_{k'} &\leq \frac{{\textstyle\sum_{k' \neq k}} X_{k'}}{p^{\mathcal{X}}_d + \eta_r \eta_\uparrow (1-p^{\mathcal{X}}_d)} \nonumber\\
    &\quad+  \frac{\eta_r \eta_\uparrow(1-p^{\mathcal{X}}_d)}{p^{\mathcal{X}}_d + \eta_r \eta_\uparrow (1-p^{\mathcal{X}}_d)} \Pi^0_{\mathcal{X}} ,  \label{operator-bound}
\end{align}
which, when used in \eqref{simplifiedproof-phase-error-rate-uppbound}, yields the following upper bound on the phase error rate:
\begin{align}
    \tilde{e}_{X,1} Y^Z_{1,(\eta_\downarrow,\eta_\downarrow)} &\leq \frac{{\textstyle\sum_{k=0}^{d-1}} \Tr\left[\frac{\sigma_k}{d}  M^{\leq 1}_{\mathcal{Z}} {\textstyle\sum_{k'\neq k}} X_{k'} M^{\leq 1}_{\mathcal{Z}} \right]}{p^{\mathcal{X}}_d + \eta_r \eta_\uparrow (1-p^{\mathcal{X}}_d)} \nonumber\\
    &\quad+ \frac{\eta_r \eta_\uparrow(1-p^{\mathcal{X}}_d)}{p^{\mathcal{X}}_d + \eta_r \eta_\uparrow (1-p^{\mathcal{X}}_d)} \Tr\left[\Bar{\sigma}  M^{\leq 1}_{\mathcal{Z}} \Pi^0_{\mathcal{X}} M^{\leq 1}_{\mathcal{Z}} \right] \nonumber\\
    &\quad + \overline{\Delta_2} + \overline{w}^{>1}_{\mathcal{Z}} \label{simplifiedproof-phase-error-rate-uppbound2}.
\end{align}

We now focus on estimating the first term on the right-hand side of \eqref{simplifiedproof-phase-error-rate-uppbound2}. This term is clearly linked to the one-photon bit error rate of Bob's test measurement, \eqref{simplifiedproof-reduced-phase-error-rate}. However, note that we must estimate such error rate when restricting the input state to the ($\leq 1$)-subspace, due to the presence of the operator $M^{\leq 1}_{\mathcal{Z}}$. To this aim, let us introduce the one-photon $X$-basis bit error rates: $e_{X,1,(\eta_i,\eta_l),\checkmark}$ and $e_{X,1,(\eta_i,\eta_l),\emptyset}$, which we simply refer to as the ``test-round bit error rates'' and which correspond to the bit error rate in the $X$-basis when the $Z$-basis detector has a detection or no detection, respectively. They can be expressed as follows (cf.~Eqs.~\eqref{bit-error-rate-click} and \eqref{bit-error-rate-noclick} in Appendix~\ref{app:PER5}):
\begin{align}
    e_{X,1,(\eta_i,\eta_l),\checkmark} &= \frac{{\textstyle\sum_{k=0}^{d-1}} \Tr\left[\frac{\sigma_k}{d}  {\textstyle\sum_{k'\neq k}}X^{(\eta_i,\eta_l)}_{k',\checkmark}  \right]}{Y^{X,\checkmark}_{1,(\eta_i,\eta_l)}} \label{simplifiedproof-bit-error-rate-click} \\
    e_{X,1,(\eta_i,\eta_l),\emptyset} &= \frac{{\textstyle\sum_{k=0}^{d-1}} \Tr\left[\frac{\sigma_k}{d}  {\textstyle\sum_{k'\neq k}}X^{(\eta_i,\eta_l)}_{k',\emptyset}  \right]}{Y^{X,\emptyset}_{1,(\eta_i,\eta_l)}}, \label{simplifiedproof-bit-error-rate-noclick}
\end{align}
where we call $Y^{X,\checkmark}_{1,(\eta_i,\eta_l)}$ and $Y^{X,\emptyset}_{1,(\eta_i,\eta_l)}$ the ``test-round yields'' and where $X^{(\eta_i,\eta_l)}_{k',\checkmark}$ ($X^{(\eta_i,\eta_l)}_{k',\emptyset}$) is Bob's POVM element describing a detection of outcome $k'$ in the $X$-basis detector and a (no) detection in the $Z$-basis detector, given that the TBS is set to $(\eta_i,\eta_l)$. We refer to Appendix~\ref{app:notation} for a precise definition of the test-round yields and bit error rates. Note that, with this formalism, by definition it holds:
\begin{align}
    X_{k'} = X^{(\eta_\uparrow,\eta_\uparrow)}_{k',\checkmark} + X^{(\eta_\uparrow,\eta_\uparrow)}_{k',\emptyset}. \label{Bobs-test-measurement-equality}
\end{align}
The two test-round bit error rates contain more information about the protocol's statistics than the bit error rate of Bob's test measurement alone, \eqref{biterrorrate-testround}, and this will be crucial to accurately estimate the phase error rate.

Similarly to the bound on the phase error rate in \eqref{simplifiedproof-phase-error-rate-uppbound} through the expansion \eqref{Zclick-expansion}, we can use the relation: $\one=\Pi^{\leq 1}_{\mathcal{Z}} + \Pi^{>1}_{\mathcal{Z}}$, with $\Pi^{\leq 1}_{\mathcal{Z}}=\Pi^0_{\mathcal{Z}} + \Pi^1_{\mathcal{Z}}$, to isolate the contribution of the ($\leq 1$)-subspace to the test-round bit error rates \eqref{simplifiedproof-bit-error-rate-click} and \eqref{simplifiedproof-bit-error-rate-noclick}. We do so in Appendix~\ref{app:PER6}, where we obtain in Eqs.~\eqref{bit-error-rate-click-range} and \eqref{bit-error-rate-noclick-range} tight upper and lower bounds on the following quantities:
\begin{align}
    &\left[Y^{X,\checkmark}_{1,(\eta_i,\eta_l)} e_{X,1,(\eta_i,\eta_l),\checkmark}\right]^{\leq 1}_{\mathcal{Z}} = {\textstyle\sum_{k=0}^{d-1}} \Tr\left[\frac{\sigma_k}{d} \Pi^{\leq 1}_{\mathcal{Z}} {\textstyle\sum_{k'\neq k}}X^{(\eta_i,\eta_l)}_{k',\checkmark} \Pi^{\leq 1}_{\mathcal{Z}} \right] \label{bit-error-rate-click-restricted}\\
    &\left[Y^{X,\emptyset}_{1,(\eta_i,\eta_l)} e_{X,1,(\eta_i,\eta_l),\emptyset}\right]^{\leq 1}_{\mathcal{Z}} = {\textstyle\sum_{k=0}^{d-1}} \Tr\left[\frac{\sigma_k}{d} \Pi^{\leq 1}_{\mathcal{Z}} {\textstyle\sum_{k'\neq k}}X^{(\eta_i,\eta_l)}_{k',\emptyset} \Pi^{\leq 1}_{\mathcal{Z}} \right] \label{bit-error-rate-noclick-restricted} ,
\end{align}
where the notation $[\cdot]^{\leq 1}_{\mathcal{Z}}$ indicates that the enclosed quantity is only calculated in the ($\leq 1$)-subspace. The bounds on the above quantities are reported in \eqref{protocol-l-e,click}--\eqref{protocol-u-e,noclick}. Importantly, they can be directly estimated from the protocol's statistics, as they are given in terms of the test-round yields and bit error rates as well as of $\overline{w}^{>1}_{\mathcal{Z}}$.

Unfortunately, the restricted test-round bit error rates in \eqref{bit-error-rate-click-restricted} and \eqref{bit-error-rate-noclick-restricted} cannot be directly employed to quantify the first term on the right-hand side of \eqref{simplifiedproof-phase-error-rate-uppbound2}. This is because the latter is restricted to the ($\leq 1$)-subspace by the operator in \eqref{M-operator}, while the error rates in \eqref{bit-error-rate-click-restricted} and \eqref{bit-error-rate-noclick-restricted} present the projector $\Pi^{\leq 1}_{\mathcal{Z}}$.

This prompts us to expand both the term in \eqref{simplifiedproof-phase-error-rate-uppbound2} and the restricted test-round bit error rates with \eqref{M-operator} and $\Pi^{\leq 1}_{\mathcal{Z}}=\Pi^0_{\mathcal{Z}} + \Pi^1_{\mathcal{Z}}$, respectively, such that each term in the expansion contains either a projection on $\Pi^0_{\mathcal{Z}}$, or a projection on $\Pi^1_{\mathcal{Z}}$, or the off-diagonal blocks. For the restricted test-round bit error rates, we obtain:
\begin{align}
    &\left[Y^{X,\checkmark}_{1,(\eta_i,\eta_l)} e_{X,1,(\eta_i,\eta_l),\checkmark}\right]^{\leq 1}_{\mathcal{Z}} =  \left[Y^{X,\checkmark}_{1,(\eta_i,\eta_l)} e_{X,1,(\eta_i,\eta_l),\checkmark}\right]^{0}_{\mathcal{Z}} +  \left[Y^{X,\checkmark}_{1,(\eta_i,\eta_l)} e_{X,1,(\eta_i,\eta_l),\checkmark}\right]^{0,1}_{\mathcal{Z}} +  \left[Y^{X,\checkmark}_{1,(\eta_i,\eta_l)} e_{X,1,(\eta_i,\eta_l),\checkmark}\right]^{1}_{\mathcal{Z}} , \label{simplifiedproof-bit-error-rate-click-TBS2} \\
    &\left[Y^{X,\emptyset}_{1,(\eta_i,\eta_l)} e_{X,1,(\eta_i,\eta_l),\emptyset}\right]^{\leq 1}_{\mathcal{Z}} =  \left[Y^{X,\emptyset}_{1,(\eta_i,\eta_l)} e_{X,1,(\eta_i,\eta_l),\emptyset}\right]^{0}_{\mathcal{Z}} + \left[Y^{X,\emptyset}_{1,(\eta_i,\eta_l)} e_{X,1,(\eta_i,\eta_l),\emptyset}\right]^{0,1}_{\mathcal{Z}} +  \left[Y^{X,\emptyset}_{1,(\eta_i,\eta_l)} e_{X,1,(\eta_i,\eta_l),\emptyset}\right]^{1}_{\mathcal{Z}} , \label{simplifiedproof-bit-error-rate-noclick-TBS2}
\end{align}
where we defined:
\begin{align}
       &\left[Y^{X,\checkmark}_{1,(\eta_i,\eta_l)} e_{X,1,(\eta_i,\eta_l),\checkmark}\right]^{\alpha}_{\mathcal{Z}} = {\textstyle\sum_{k=0}^{d-1}} \Tr\left[\frac{\sigma_k}{d} \pia {\textstyle\sum_{k'\neq k}}X^{(\eta_i,\eta_l)}_{k',\checkmark} \pia \right] \label{simplifiedproof-bit-error-rate-click-TBS2-Pin}, \\
      &\left[Y^{X,\checkmark}_{1,(\eta_i,\eta_l)} e_{X,1,(\eta_i,\eta_l),\checkmark}\right]^{0,1}_{\mathcal{Z}} = {\textstyle\sum_{k=0}^{d-1}} \Tr\left[\frac{\sigma_k}{d} \left(\Pi^{0}_{\mathcal{Z}} {\textstyle\sum_{k'\neq k}}X^{(\eta_i,\eta_l)}_{k',\checkmark} \Pi^{1}_{\mathcal{Z}} + \mathrm{h.c.} \right)\right] \label{simplifiedproof-bit-error-rate-click-TBS2-Pi01},
\end{align}
for $\alpha=0,1$, and similarly for the terms relative to no click in the $Z$-basis detector. We note that \eqref{simplifiedproof-bit-error-rate-click-TBS2-Pi01} is not null, in general, since the POVM elements $X^{(\eta_i,\eta_l)}_{k',\checkmark}$ are not necessarily block-diagonal in the subspaces defined by $\pia$.

For the first term on the right-hand side of \eqref{simplifiedproof-phase-error-rate-uppbound2}, the expansion through \eqref{M-operator} produces the following expression:
\begin{align}
    {\textstyle\sum_{k=0}^{d-1}} \Tr\left[\frac{\sigma_k}{d}  M^{\leq 1}_{\mathcal{Z}} {\textstyle\sum_{k'\neq k}} X_{k'} M^{\leq 1}_{\mathcal{Z}} \right] &= 
    {\textstyle\sum_{\alpha=0}^1} p_{\checkmark|\alpha} \left(\left[Y^{X,\checkmark}_{1,(\eta_\uparrow,\eta_\uparrow)} e_{X,1,(\eta_\uparrow,\eta_\uparrow),\checkmark}\right]^{\alpha}_{\mathcal{Z}} + \left[Y^{X,\emptyset}_{1,(\eta_\uparrow,\eta_\uparrow)} e_{X,1,(\eta_\uparrow,\eta_\uparrow),\emptyset}\right]^{\alpha}_{\mathcal{Z}}\right) \nonumber\\
    &\quad+ \sqrt{p_{\checkmark|0} p_{\checkmark|1}} \left(\left[Y^{X,\checkmark}_{1,(\eta_\uparrow,\eta_\uparrow)} e_{X,1,(\eta_\uparrow,\eta_\uparrow),\checkmark}\right]^{0,1}_{\mathcal{Z}} + \left[Y^{X,\emptyset}_{1,(\eta_\uparrow,\eta_\uparrow)} e_{X,1,(\eta_\uparrow,\eta_\uparrow),\emptyset}\right]^{0,1}_{\mathcal{Z}}\right) ,  \label{firstterm-expansion}
\end{align}
where we used the definitions \eqref{simplifiedproof-bit-error-rate-click-TBS2-Pin}, \eqref{simplifiedproof-bit-error-rate-click-TBS2-Pi01} and the equality \eqref{Bobs-test-measurement-equality}.

By comparing the expansions of the restricted test-round bit error rates \eqref{simplifiedproof-bit-error-rate-click-TBS2} and \eqref{simplifiedproof-bit-error-rate-noclick-TBS2} with the expansion \eqref{firstterm-expansion} of the term coming from the right-hand side of \eqref{simplifiedproof-phase-error-rate-uppbound2}, we conclude that an accurate estimation of the latter requires individually estimating each of the three terms in the expansions \eqref{simplifiedproof-bit-error-rate-click-TBS2} and \eqref{simplifiedproof-bit-error-rate-noclick-TBS2} of the restricted test-round bit error rates.

The method that we use to estimate each of the three terms on the right-hand side of \eqref{simplifiedproof-bit-error-rate-click-TBS2} and \eqref{simplifiedproof-bit-error-rate-noclick-TBS2} represents one of the novelties of our security proof. The method is based on the observation that when the state (on which the three terms are computed) is projected on the subspaces with zero or one photon in $\mathcal{Z}$, then the unitary action of the TBS on the photons localized in $\mathcal{Z}$ can be factored out and reduces to a simple prefactor.

More specifically, recall that the TBS setting $(\eta_i,\eta_l)$ means that the TBS acts as a beam splitter with transmittance $\eta_i$ ($\eta_l$) for photons localized in (outside of) $\mathcal{Z}$. For illustrative purposes, let us consider the term $\left[Y^{X,\checkmark}_{1,(\eta_i,\eta_l)} e_{X,1,(\eta_i,\eta_l),\checkmark}\right]^{1}_{\mathcal{Z}}$, which can be interpreted as the probability of the following intersection of events (averaged over Alice's random selection of the symbol $X_A=k$): the input state $\sigma_k$ of Bob contains one photon localized in $\mathcal{Z}$, the $Z$-basis detector clicks, and the $X$-basis detector clicks with an erroneous outcome $k'\neq k$ (all with the TBS setting $(\eta_i,\eta_l)$). By following Fig.~\ref{fig:Bobs-setup}, the single photon localized in $\mathcal{Z}$ is either transmitted by the TBS with probability $\eta_i$, in which case the $Z$-basis detector click is caused by a dark count, or the photon is reflected by the TBS with probability $1-\eta_i$, causing a click in the $Z$-basis detector (recall that the efficiency $\eta_Z$ of the $Z$-basis detector is factored out in the quantum channel). Besides, the erroneous outcome in the $X$-basis detector could be caused by the transmitted photon localized in $\mathcal{Z}$, but also by a dark count or by any number of photons not localized in $\mathcal{Z}$. Then, the probability term that we considered can be expressed as a sum of two terms corresponding to the scenarios with the transmitted or reflected photon:
\begin{align}
    \left[Y^{X,\checkmark}_{1,(\eta_i,\eta_l)} e_{X,1,(\eta_i,\eta_l),\checkmark}\right]^{1}_{\mathcal{Z}} &= p^{\mathcal{Z}}_d \eta_i \,\mathbbm{E}^{1t}_{\eta_l} +(1-\eta_i) \,\mathbbm{E}^{1r}_{\eta_l} \label{simplifiedproof-bit-error-rate-click-TBS2-Pi1-2},
\end{align}
where the action of the TBS on the single photon in $\mathcal{Z}$ is made explicit by the prefactors $\eta_i$ and $1-\eta_i$. In \eqref{simplifiedproof-bit-error-rate-click-TBS2-Pi1-2}, the quantities $\mathbbm{E}^{1t}_{\eta_l}$ and $\mathbbm{E}^{1r}_{\eta_l}$ depend on the input states $\sigma_k$ and their interaction with the $X$-basis detector, but also non-trivially depend on the TBS transmittance $\eta_l$ --indeed, $\sigma_k$ can contain an arbitrary number of photons outside of $\mathcal{Z}$. In Appendix~\ref{app:PER7} we provide a rigorous proof of the result in \eqref{simplifiedproof-bit-error-rate-click-TBS2-Pi1-2} and derive similar decompositions for the other terms in \eqref{simplifiedproof-bit-error-rate-click-TBS2} and \eqref{simplifiedproof-bit-error-rate-noclick-TBS2} (cf.~Eqs.~\eqref{bit-error-rate-click-TBS2-Pi0-2} -- \eqref{bit-error-rate-noclick-TBS2-Pi1-2}):
\begin{align}
    \left[Y^{X,\checkmark}_{1,(\eta_i,\eta_l)} e_{X,1,(\eta_i,\eta_l),\checkmark}\right]^{0}_{\mathcal{Z}} &= p^{\mathcal{Z}}_d \,\mathbbm{E}^{0}_{\eta_l}  \label{simplifiedproof-bit-error-rate-click-TBS2-Pi0-2} \\
    \left[Y^{X,\emptyset}_{1,(\eta_i,\eta_l)} e_{X,1,(\eta_i,\eta_l),\emptyset}\right]^{0}_{\mathcal{Z}} &= (1-p^{\mathcal{Z}}_d) \, \mathbbm{E}^{0}_{\eta_l}  \label{simplifiedproof-bit-error-rate-noclick-TBS2-Pi0-2} \\
    \left[Y^{X,\checkmark}_{1,(\eta_i,\eta_l)} e_{X,1,(\eta_i,\eta_l),\checkmark}\right]^{0,1}_{\mathcal{Z}} &= p^{\mathcal{Z}}_d \sqrt{\eta_i} \,\mathbbm{E}^{0,1}_{\eta_l}  \label{simplifiedproof-bit-error-rate-click-TBS2-Pi01-2} \\
    \left[Y^{X,\emptyset}_{1,(\eta_i,\eta_l)} e_{X,1,(\eta_i,\eta_l),\emptyset}\right]^{0,1}_{\mathcal{Z}} &= (1-p^{\mathcal{Z}}_d) \sqrt{\eta_i} \,\mathbbm{E}^{0,1}_{\eta_l}  \label{simplifiedproof-bit-error-rate-noclick-TBS2-Pi01-2} \\
    \left[Y^{X,\emptyset}_{1,(\eta_i,\eta_l)} e_{X,1,(\eta_i,\eta_l),\emptyset}\right]^{1}_{\mathcal{Z}} &= (1-p^{\mathcal{Z}}_d) \eta_i \,\mathbbm{E}^{1t}_{\eta_l}  \label{simplifiedproof-bit-error-rate-noclick-TBS2-Pi1-2} ,
\end{align}
where again the action of the TBS on the photon localized in $\mathcal{Z}$ is singled out by the factors containing $\eta_i$.

The notable fact from \eqref{simplifiedproof-bit-error-rate-click-TBS2-Pi1-2}--\eqref{simplifiedproof-bit-error-rate-noclick-TBS2-Pi1-2} is that $\eta_i$ is factored out of the variables: $\mathbbm{E}^0_{\eta_l}$, $\mathbbm{E}^{0,1}_{\eta_l}$, $\mathbbm{E}^{1r}_{\eta_l}$, and $\mathbbm{E}^{1t}_{\eta_l}$. Therefore, any choice of $\eta_i$ affects the test-round bit error rates only through the coefficients with which the variables $\mathbbm{E}^0_{\eta_l}$, $\mathbbm{E}^{0,1}_{\eta_l}$, $\mathbbm{E}^{1r}_{\eta_l}$, and $\mathbbm{E}^{1t}_{\eta_l}$ appear in \eqref{simplifiedproof-bit-error-rate-click-TBS2-Pi1-2}--\eqref{simplifiedproof-bit-error-rate-noclick-TBS2-Pi1-2}. Thus, by employing the expressions \eqref{simplifiedproof-bit-error-rate-click-TBS2-Pi1-2}--\eqref{simplifiedproof-bit-error-rate-noclick-TBS2-Pi1-2} in the expansions \eqref{simplifiedproof-bit-error-rate-click-TBS2} and \eqref{simplifiedproof-bit-error-rate-noclick-TBS2}, we can derive a linear system of equations (cf.~Eq.~\eqref{bit-error-rate-system}):
\begin{align}
    \left\lbrace \begin{array}{ll}
    \left[Y^{X,\checkmark}_{1,(\eta_i,\eta_l)} e_{X,1,(\eta_i,\eta_l),\checkmark}\right]^{\leq 1}_{\mathcal{Z}} &= p^{\mathcal{Z}}_d \left( \mathbbm{E}^{0}_{\eta_l} +\sqrt{\eta_i}\, \mathbbm{E}^{0,1}_{\eta_l} + \eta_i \,\mathbbm{E}^{1t}_{\eta_l} \right) + (1-\eta_i)\,\mathbbm{E}^{1r}_{\eta_l} \\[2ex]
    \left[Y^{X,\emptyset}_{1,(\eta_i,\eta_l)} e_{X,1,(\eta_i,\eta_l),\emptyset}\right]^{\leq 1}_{\mathcal{Z}} &= (1-p^{\mathcal{Z}}_d) \left( \mathbbm{E}^{0}_{\eta_l} +\sqrt{\eta_i}\, \mathbbm{E}^{0,1}_{\eta_l} + \eta_i \,\mathbbm{E}^{1t}_{\eta_l} \right)
    \end{array} \right. ,\label{simplifiedproof-bit-error-rate-system}
\end{align}
in the variables $\mathbbm{E}^0_{\eta_l}$, $\mathbbm{E}^{0,1}_{\eta_l}$, $\mathbbm{E}^{1r}_{\eta_l}$, and $\mathbbm{E}^{1t}_{\eta_l}$, by simply choosing different values of $\eta_i$. In particular, each equation in the above system corresponds to three different equations depending on the setting $\eta_i\in\{\eta_\uparrow,\eta_\downarrow,\eta_2\}$. Note that the left-hand sides, i.e., the test-round bit error rates restricted to the ($\leq 1$)-subspace \eqref{bit-error-rate-click-restricted} and \eqref{bit-error-rate-noclick-restricted}, can be considered as known quantities since they have been tightly bounded with upper and lower bounds that are given in terms of observed statistics.

In Appendix~\ref{app:PER8}, we solve the linear system in \eqref{simplifiedproof-bit-error-rate-system} and provide the solution for $\mathbbm{E}^0_{\eta_l}$, $\mathbbm{E}^{0,1}_{\eta_l}$, $\mathbbm{E}^{1r}_{\eta_l}$, and $\mathbbm{E}^{1t}_{\eta_l}$ in terms of the test-round bit error rates restricted to the ($\leq 1$)-subspace. Subsequently, we replace the latter with their respective upper or lower bound, in such a way that the resulting expressions are upper bounds on the quantities $\mathbbm{E}^0_{\eta_l}$, $\mathbbm{E}^{0,1}_{\eta_l}$, $\mathbbm{E}^{1r}_{\eta_l}$, and $\mathbbm{E}^{1t}_{\eta_l}$. We indicate such bounds with $\overline{\mathbbm{E}}^0_{\eta_l}$, $\overline{\mathbbm{E}}^{0,1}_{\eta_l}$, $\overline{\mathbbm{E}}^{1r}_{\eta_l}$, and $\overline{\mathbbm{E}}^{1t}_{\eta_l}$ and report them in \eqref{protocol-Eetak-0-upp}--\eqref{protocol-Eetak-1r-upp}. In fact, the variables $\mathbbm{E}^0_{\eta_l}$, $\mathbbm{E}^{0,1}_{\eta_l}$, $\mathbbm{E}^{1r}_{\eta_l}$, and $\mathbbm{E}^{1t}_{\eta_l}$ appear with positive signs in the phase error rate and thus should be replaced by their upper bounds, in case they cannot be computed explicitly.

We can now employ the bounds $\overline{\mathbbm{E}}^0_{\eta_l}$, $\overline{\mathbbm{E}}^{0,1}_{\eta_l}$, $\overline{\mathbbm{E}}^{1r}_{\eta_l}$, and $\overline{\mathbbm{E}}^{1t}_{\eta_l}$ to obtain an upper bound on the first term of the phase error rate bound in \eqref{simplifiedproof-phase-error-rate-uppbound2}. Indeed, by combining \eqref{simplifiedproof-bit-error-rate-click-TBS2-Pi1-2}--\eqref{simplifiedproof-bit-error-rate-noclick-TBS2-Pi1-2} with \eqref{firstterm-expansion}, we obtain the upper bound:
\begin{align}
    &{\textstyle\sum_{k=0}^{d-1}} \Tr\left[\frac{\sigma_k}{d}  M^{\leq 1}_{\mathcal{Z}} {\textstyle\sum_{k'\neq k}} X_{k'} M^{\leq 1}_{\mathcal{Z}} \right] \leq 
    p_{\checkmark|0} \overline{\mathbbm{E}}^0_{\eta_\uparrow} + \sqrt{p_{\checkmark|0} p_{\checkmark|1}} \, \sqrt{\eta_\uparrow} \overline{\mathbbm{E}}^{0,1}_{\eta_\uparrow} + p_{\checkmark|1} \left( \eta_\uparrow \overline{\mathbbm{E}}^{1t}_{\eta_\uparrow} + (1-\eta_\uparrow) \overline{\mathbbm{E}}^{1r}_{\eta_\uparrow}\right)  \label{firstterm-expansion-upp},
\end{align}
where each quantity on the right-hand side is either known or directly expressed in terms of observed statistics. Recall that $p_{\checkmark|\alpha}$ was introduced after Eq.~\eqref{Zclick-expansion}.

The remaining term to be estimated, in order to complete the estimation of the phase error rate, is the second term in the phase error rate bound \eqref{simplifiedproof-phase-error-rate-uppbound2}, which can be interpreted as the weight of the average state received by Bob in the subspace with zero photons in the modes $\mathcal{X}$, restricted to the ($\leq 1$)-subspace. By expanding it through \eqref{M-operator}, we obtain three terms:
\begin{align}
   &\Tr\left[\bar{\sigma}  M^{\leq 1}_{\mathcal{Z}}  \Pi^0_{\mathcal{X}} M^{\leq 1}_{\mathcal{Z}} \right] = p_{\checkmark|0} \Tr\left[\bar{\sigma}  \Pi^{0}_{\mathcal{Z}}  \Pi^0_{\mathcal{X}} \Pi^{0}_{\mathcal{Z}} \right] + \sqrt{p_{\checkmark|0}p_{\checkmark|1}} \Tr\left[\bar{\sigma}  \left(\Pi^{0}_{\mathcal{Z}}  \Pi^0_{\mathcal{X}} \Pi^{1}_{\mathcal{Z}} + \mathrm{h.c.}\right) \right]  + p_{\checkmark|1} \Tr\left[\bar{\sigma}  \Pi^{1}_{\mathcal{Z}}  \Pi^0_{\mathcal{X}} \Pi^{1}_{\mathcal{Z}} \right]  \label{secondterm-expansion},
\end{align}
corresponding to the projections in the subspaces of $\Pi^0_{\mathcal{Z}}$, $\Pi^1_{\mathcal{Z}}$, or the off-diagonal term. We deduce that a correct estimation of the term of interest requires estimating individually the three terms on the right-hand side of \eqref{secondterm-expansion}. To this aim, we need to combine two methods that have already been used in the proof, namely: the detector decoy technique (used to extract the weight of $\bar{\sigma}$ outside the ($\leq 1$)-subspace with \eqref{weight-outside}) and the method based on linear equations (used to estimate the single terms in the expansion \eqref{firstterm-expansion}). For this reason, the accurate estimation of the three terms in \eqref{secondterm-expansion} can be seen as the more sophisticated procedure of the whole proof.

The first step considers the test-round yields $Y^{X,\checkmark}_{1,(\eta_l,\eta_l)}$ and $Y^{X,\emptyset}_{1,(\eta_l,\eta_l)}$, i.e.\ the probability that the $X$-basis detector clicks and the $Z$-basis detector clicks or does not click, respectively (cf. Appendix~\ref{app:notation}). Their sum can be expressed in terms of Bob's POVM elements:
\begin{align}
    Y^{X,\checkmark}_{1,(\eta_l,\eta_l)} + Y^{X,\emptyset}_{1,(\eta_l,\eta_l)} = \Tr\left[\Bar{\sigma} \sum_{k=0}^{d-1} \left(X^{(\eta_l,\eta_l)}_{k,\checkmark} + X^{(\eta_l,\eta_l)}_{k,\emptyset}\right) \right] ,  \label{sum-testround-yields}
\end{align}
where in Appendix~\ref{app:PER5} we show that the following equality holds:
\begin{align}
    {\textstyle\sum_{k=0}^{d-1}} X^{(\eta_l,\eta_l)}_{k,\checkmark} &+ X^{(\eta_l,\eta_l)}_{k,\emptyset} = {\textstyle\sum_{\beta=0}^\infty} \left[1-(1-p^{\mathcal{X}}_d)(1-\eta_r \eta_l)^\beta\right] \Pi^\beta_{\mathcal{X}},
\end{align}
which can be seen as the POVM element corresponding to a detection in the $X$-basis, when the TBS setting is $(\eta_l,\eta_l)$. From the last expression, we can immediately obtain the POVM element describing a no detection in the $X$-basis detector:
\begin{align}
   \one - {\textstyle\sum_{k=0}^{d-1}} \left(X^{(\eta_l,\eta_l)}_{k,\checkmark} + X^{(\eta_l,\eta_l)}_{k,\emptyset}\right) = (1-p^{\mathcal{X}}_d) {\textstyle\sum_{\beta=0}^\infty} (1-\eta_r \eta_l)^\beta \Pi^\beta_{\mathcal{X}} \label{simplifiedproof-Xnodet-etak},
\end{align}
where we used the fact that $\sum_{\beta=0}^\infty \Pi^\beta_{\mathcal{X}}=\one$. Now, we realize that the three terms in \eqref{secondterm-expansion} can be recovered from the right-hand side of \eqref{simplifiedproof-Xnodet-etak} if we project the operator on the subspaces of $\Pi^0_{\mathcal{Z}}$ or $\Pi^1_{\mathcal{Z}}$. By doing so on both sides of the operator equation in \eqref{simplifiedproof-Xnodet-etak}, we obtain:
\begin{align}
    \pia - &\pia \left({\textstyle\sum_{k=0}^{d-1}} X^{(\eta_l,\eta_l)}_{k,\checkmark} + X^{(\eta_l,\eta_l)}_{k,\emptyset}\right)\pia = (1-p^{\mathcal{X}}_d) {\textstyle\sum_{\beta=0}^\infty} (1-\eta_r \eta_l)^\beta \pia\Pi^\beta_{\mathcal{X}} \pia \label{simplifiedproof-Xnodet-etak-projected},
\end{align}
for $\alpha=0,1$. We can now take the expectation value of the operator in \eqref{simplifiedproof-Xnodet-etak-projected} on the state $\bar{\sigma}$ and obtain the following equation, which contains two of the three terms in \eqref{secondterm-expansion}:
\begin{align}
    \Tr[\bar{\sigma} \Pi^\alpha_{\mathcal{Z}}] &-\left(\left[Y^{X,\checkmark}_{1,(\eta_l,\eta_l)}\right]^\alpha_{\mathcal{Z}} + \left[Y^{X,\emptyset}_{1,(\eta_l,\eta_l)}\right]^\alpha_{\mathcal{Z}}\right) = (1-p^{\mathcal{X}}_d) {\textstyle\sum_{\beta=0}^\infty} (1-\eta_r \eta_l)^\beta  \Tr\left[\bar{\sigma} \Pi^\alpha_{\mathcal{Z}} \Pi^\beta_{\mathcal{X}} \Pi^\alpha_{\mathcal{Z}} \right], \label{simplifiedproof-detectordecoy}
\end{align}
where we introduced, similarly to \eqref{simplifiedproof-bit-error-rate-click-TBS2-Pin}, the test-round yields of the states projected on the subspaces of $\Pi^0_{\mathcal{Z}}$ and $\Pi^1_{\mathcal{Z}}$ (for $\alpha=0,1$):
\begin{align}
    \left[Y^{X,\checkmark}_{1,(\eta_i,\eta_l)}\right]^\alpha_{\mathcal{Z}} &= \Tr\left[\Bar{\sigma} \pia {\textstyle\sum_{k=0}^{d-1}} X^{(\eta_i,\eta_l)}_{k,\checkmark} \pia \right] \label{simplifiedproof-yield-Pialpha-click} \\ 
    \left[Y^{X,\emptyset}_{1,(\eta_i,\eta_l)}\right]^\alpha_{\mathcal{Z}} &= \Tr\left[\Bar{\sigma} \pia {\textstyle\sum_{k=0}^{d-1}} X^{(\eta_i,\eta_l)}_{k,\emptyset} \pia \right] \label{simplifiedproof-yield-Pialpha-noclick}.
\end{align}
Now, we show that the left-hand side of \eqref{simplifiedproof-detectordecoy} is known by linking it to observed quantities. In particular, the quantity $\Tr[\bar{\sigma} \Pi^\alpha_{\mathcal{Z}}]$ is nothing more than the weight of the average state received by Bob in the subspace with $\alpha$ photons in $\mathcal{Z}$. In Appendix~\ref{app:PER3}, in a similar manner to the derivation of the bound in \eqref{weight-outside}, we employ the detector decoy technique \cite{detector-decoy} to obtain tight upper ($\overline{w}^\alpha_{\mathcal{Z}}$) and lower ($\underline{w}^\alpha_{\mathcal{Z}}$) bounds on $\Tr[\bar{\sigma} \Pi^\alpha_{\mathcal{Z}}]$, in terms of quantities estimated with the decoy-state method --the $Z$-basis yields. The bounds are reported in \eqref{protocol-y0-upperbound}--\eqref{protocol-y1-lowerbound}.

In parallel, we can again use the relation: $\one=\Pi^{\leq 1}_{\mathcal{Z}} + \Pi^{>1}_{\mathcal{Z}}$, with $\Pi^{\leq 1}_{\mathcal{Z}}=\Pi^0_{\mathcal{Z}} + \Pi^1_{\mathcal{Z}}$, to isolate the contribution of the ($\leq 1$)-subspace to the test-round yields in \eqref{sum-testround-yields}. In particular, in Appendix~\ref{app:PER6} we derive tight upper and lower bounds (summarized in \eqref{protocol-l-Y,click}--\eqref{protocol-u-Y,noclick}) on the left-hand sides of the following equalities:
\begin{align}
    &\left[Y^{X,\checkmark}_{1,(\eta_i,\eta_l)}\right]^{\leq 1}_{\mathcal{Z}} =  \left[Y^{X,\checkmark}_{1,(\eta_i,\eta_l)}\right]^{0}_{\mathcal{Z}} +  \left[Y^{X,\checkmark}_{1,(\eta_i,\eta_l)} \right]^{0,1}_{\mathcal{Z}} +  \left[Y^{X,\checkmark}_{1,(\eta_i,\eta_l)}\right]^{1}_{\mathcal{Z}} \label{simplifiedproof-yield-click-TBS2} \\
    &\left[Y^{X,\emptyset}_{1,(\eta_i,\eta_l)} \right]^{\leq 1}_{\mathcal{Z}} =  \left[Y^{X,\emptyset}_{1,(\eta_i,\eta_l)} \right]^{0}_{\mathcal{Z}} +  \left[Y^{X,\emptyset}_{1,(\eta_i,\eta_l)} \right]^{0,1}_{\mathcal{Z}} +  \left[Y^{X,\emptyset}_{1,(\eta_i,\eta_l)} \right]^{1}_{\mathcal{Z}}  \label{simplifiedproof-yield-noclick-TBS2}
\end{align}
in terms of observed statistics and of $\overline{w}^{>1}_{\mathcal{Z}}$, where we defined:
\begin{align}
       \left[Y^{X,\checkmark}_{1,(\eta_i,\eta_l)}\right]^{0,1}_{\mathcal{Z}} = \Tr\left[\Bar{\sigma} \left(\Pi^{0}_{\mathcal{Z}} {\textstyle\sum_{k=0}^{d-1}} X^{(\eta_i,\eta_l)}_{k,\checkmark} \Pi^{1}_{\mathcal{Z}} + \mathrm{h.c.} \right)\right] \label{simplifiedproof-yield-click-TBS2-Pi01}
\end{align}
and similarly for the case with no $Z$-basis detection.

Now, with the same argument used in \eqref{simplifiedproof-bit-error-rate-click-TBS2-Pi1-2} to single out the effect of the TBS on the single photon localized in $\mathcal{Z}$, we can express the test-round yields projected by $\pia$ in terms of some variables $\mathbbm{Y}^{0}_{\eta_l}$, $\mathbbm{Y}^{0,1}_{\eta_l}$, $\mathbbm{Y}^{1t}_{\eta_l}$, and $\mathbbm{Y}^{1r}_{\eta_l}$, as follows (cf.~Eq.~\eqref{yield-click-TBS2-Pi01-2} and thereafter in Appendix~\ref{app:PER7}):
\begin{align}
    \left[Y^{X,\checkmark}_{1,(\eta_i,\eta_l)} \right]^{0}_{\mathcal{Z}} &= p^{\mathcal{Z}}_d \mathbbm{Y}^{0}_{\eta_l}  \label{simplifiedproof-yield-click-TBS2-Pi0-2} \\
    \left[Y^{X,\emptyset}_{1,(\eta_i,\eta_l)} \right]^{0}_{\mathcal{Z}} &= (1-p^{\mathcal{Z}}_d) \mathbbm{Y}^{0}_{\eta_l}  \label{simplifiedproof-yield-noclick-TBS2-Pi0-2} \\
    \left[Y^{X,\checkmark}_{1,(\eta_i,\eta_l)} \right]^{0,1}_{\mathcal{Z}} &= p^{\mathcal{Z}}_d \sqrt{\eta_i}\, \mathbbm{Y}^{0,1}_{\eta_l} 
 \label{simplifiedproof-yield-click-TBS2-Pi01-2}  \\
    \left[Y^{X,\emptyset}_{1,(\eta_i,\eta_l)} \right]^{0,1}_{\mathcal{Z}} &= (1-p^{\mathcal{Z}}_d)\sqrt{\eta_i}\, \mathbbm{Y}^{0,1}_{\eta_l}   \label{simplifiedproof-yield-noclick-TBS2-Pi01-2} \\
    \left[Y^{X,\checkmark}_{1,(\eta_i,\eta_l)} \right]^{1}_{\mathcal{Z}} &= p^{\mathcal{Z}}_d  \eta_i \,\mathbbm{Y}^{1t}_{\eta_l} + (1-\eta_i)\,\mathbbm{Y}^{1r}_{\eta_l}  \label{simplifiedproof-yield-click-TBS2-Pi1-2}  \\
    \left[Y^{X,\emptyset}_{1,(\eta_i,\eta_l)} \right]^{1}_{\mathcal{Z}} &= (1-p^{\mathcal{Z}}_d)  \eta_i\, \mathbbm{Y}^{1t}_{\eta_l},  \label{simplifiedproof-yield-noclick-TBS2-Pi1-2}
\end{align}
such that \eqref{simplifiedproof-yield-click-TBS2} and \eqref{simplifiedproof-yield-noclick-TBS2} can be interpreted as a linear system in the variables $\mathbbm{Y}^{0}_{\eta_l}$, $\mathbbm{Y}^{0,1}_{\eta_l}$, $\mathbbm{Y}^{1t}_{\eta_l}$, and $\mathbbm{Y}^{1r}_{\eta_l}$ (cf.~Eq.~\eqref{yield-system}):
\begin{align}
    \left\lbrace \begin{array}{ll}
    \left[Y^{X,\checkmark}_{1,(\eta_i,\eta_l)} \right]^{\leq 1}_{\mathcal{Z}} &= p^{\mathcal{Z}}_d \left( \mathbbm{Y}^{0}_{\eta_l} +\sqrt{\eta_i}\, \mathbbm{Y}^{0,1}_{\eta_l} + \eta_i \,\mathbbm{Y}^{1t}_{\eta_l} \right) + (1-\eta_i)\,\mathbbm{Y}^{1r}_{\eta_l} \\[2ex]
    \left[Y^{X,\emptyset}_{1,(\eta_i,\eta_l)} \right]^{\leq 1}_{\mathcal{Z}} &= (1-p^{\mathcal{Z}}_d) \left( \mathbbm{Y}^{0}_{\eta_l} +\sqrt{\eta_i}\, \mathbbm{Y}^{0,1}_{\eta_l} + \eta_i \,\mathbbm{Y}^{1t}_{\eta_l} \right)
    \end{array} \right. , \label{simplifiedproof-yield-system}
\end{align}
where each of the two equations actually corresponds to three equations depending on the choice of $\eta_i \in\{\eta_\uparrow,\eta_\downarrow,\eta_2\}$. In Appendix~\ref{app:PER8} we solve the system in \eqref{simplifiedproof-yield-system} for the variables $\mathbbm{Y}^{0}_{\eta_l}$, $\mathbbm{Y}^{0,1}_{\eta_l}$, $\mathbbm{Y}^{1t}_{\eta_l}$, and $\mathbbm{Y}^{1r}_{\eta_l}$. Given that we previously obtained tight upper and lower bounds on the left-hand sides of the system, this translates to accurate upper and lower bounds on the variables $\mathbbm{Y}^{0}_{\eta_l}$, $\mathbbm{Y}^{0,1}_{\eta_l}$, $\mathbbm{Y}^{1t}_{\eta_l}$, and $\mathbbm{Y}^{1r}_{\eta_l}$, reported in \eqref{protocol-Yetak-upp}--\eqref{protocol-Yetak-1r-low}. 

We have thus argued how to obtain good bounds on the terms in the left-hand side of \eqref{simplifiedproof-detectordecoy}. As a matter of fact, by using \eqref{simplifiedproof-yield-click-TBS2-Pi0-2}--\eqref{simplifiedproof-yield-noclick-TBS2-Pi1-2} in \eqref{simplifiedproof-detectordecoy}, we can express the equation as follows for the case $\alpha=0$:

\begin{align}
    \Tr[\bar{\sigma} \Pi^0_{\mathcal{Z}}] &-\mathbbm{Y}^0_{\eta_l} = (1-p^{\mathcal{X}}_d) \sum_{\beta=0}^\infty (1-\eta_r \eta_l)^\beta  \Tr\left[\bar{\sigma} \Pi^0_{\mathcal{Z}} \Pi^\beta_{\mathcal{X}} \Pi^0_{\mathcal{Z}} \right] \label{simplifiedproof-prob-Xnodet-etak-Pi0-2}
\end{align}
and for $\alpha=1$:
\begin{align}
    \Tr[\bar{\sigma} \Pi^1_{\mathcal{Z}}] &-\left[\eta_l \mathbbm{Y}^{1t}_{\eta_l} + (1-\eta_l)\mathbbm{Y}^{1r}_{\eta_l} \right] = (1-p^{\mathcal{X}}_d) \sum_{\beta=0}^\infty (1-\eta_r \eta_l)^\beta  \Tr\left[\bar{\sigma} \Pi^1_{\mathcal{Z}} \Pi^\beta_{\mathcal{X}} \Pi^1_{\mathcal{Z}} \right] \label{simplifiedproof-prob-Xnodet-etak-Pi1-2},
\end{align}
where $\Tr[\Bar{\sigma}\pia]$ is bounded from above and below in \eqref{protocol-y0-upperbound}--\eqref{protocol-y1-lowerbound}, while $\mathbbm{Y}^{0}_{\eta_l}$, $\mathbbm{Y}^{1t}_{\eta_l}$, and $\mathbbm{Y}^{1r}_{\eta_l}$, are bounded in \eqref{protocol-Yetak-upp}--\eqref{protocol-Yetak-1r-low}. This allows us to consider the left-hand sides in \eqref{simplifiedproof-prob-Xnodet-etak-Pi0-2} and \eqref{simplifiedproof-prob-Xnodet-etak-Pi1-2} as known quantities, thus enabling us to perform the detector decoy technique independently on the two equations, as illustrated in Appendix~\ref{app:PER10}. In this way, we are able to obtain accurate upper bounds on $\Tr\left[\bar{\sigma} \Pi^0_{\mathcal{Z}} \Pi^0_{\mathcal{X}} \Pi^0_{\mathcal{Z}} \right]$ and $\Tr\left[\bar{\sigma} \Pi^1_{\mathcal{Z}} \Pi^0_{\mathcal{X}} \Pi^1_{\mathcal{Z}} \right]$ in terms of observed statistics:
\begin{align}
    \Tr\left[\bar{\sigma} \Pi^0_{\mathcal{Z}} \Pi^0_{\mathcal{X}} \Pi^0_{\mathcal{Z}} \right] \leq \overline{\left[w^0_{\mathcal{X}}\right]^0_{\mathcal{Z}}} 
    \label{secondterm-upp1}
\end{align}
and
\begin{align}
    \Tr\left[\bar{\sigma} \Pi^1_{\mathcal{Z}} \Pi^0_{\mathcal{X}} \Pi^1_{\mathcal{Z}} \right] \leq \overline{\left[w^0_{\mathcal{X}}\right]^1_{\mathcal{Z}}}, \label{secondterm-upp2}
\end{align}
where the explicit expressions of $\overline{\left[w^0_{\mathcal{X}}\right]^0_{\mathcal{Z}}}$ and $\overline{\left[w^0_{\mathcal{X}}\right]^1_{\mathcal{Z}}}$ are given in \eqref{protocol-x0-upperbound000} and \eqref{protocol-x0-upperbound101}.

After obtaining the upper bounds in \eqref{secondterm-upp1} and \eqref{secondterm-upp2}, we now focus on deriving an upper bound on the remaining expectation value in \eqref{secondterm-expansion}, namely $\Tr\left[\Bar{\sigma} (\Pi^0_{\mathcal{Z}} \Pi^0_{\mathcal{X}} \Pi^1_{\mathcal{Z}} + \mathrm{h.c.}) \right]$. To this aim, we define:
\begin{align}
    z:= \Tr\left[ \bar{\sigma} \Pi^0_{\mathcal{Z}} \Pi^0_{\mathcal{X}} \Pi^1_{\mathcal{Z}}\right],
\end{align}
such that the expectation value of interest can be bounded as follows:
\begin{align}
    \Tr\left[\Bar{\sigma} (\Pi^0_{\mathcal{Z}} \Pi^0_{\mathcal{X}} \Pi^1_{\mathcal{Z}} + \mathrm{h.c.}) \right] &= z+ z^* \nonumber\\
    &= 2 \Re (z) \nonumber\\
    &\leq 2 |z| \nonumber\\
    &= 2\abs{\Tr\left[ \bar{\sigma} \Pi^0_{\mathcal{Z}} \Pi^0_{\mathcal{X}} \Pi^1_{\mathcal{Z}}\right]}.
\end{align}
Now, we can use the Cauchy-Schwarz inequality in the following form:
\begin{align}
    \abs{\Tr[A^\dag B]} \leq \sqrt{\Tr[A^\dag A] \Tr[B^\dag B]},
\end{align}
where we choose $A= \sqrt{\bar{\sigma}}$ and $B= \Pi^0_{\mathcal{Z}} \Pi^0_{\mathcal{X}} \Pi^1_{\mathcal{Z}} \sqrt{\Bar{\sigma}}$. We obtain:
\begin{align}
     \Tr\left[\Bar{\sigma} (\Pi^0_{\mathcal{Z}} \Pi^0_{\mathcal{X}} \Pi^1_{\mathcal{Z}} + \mathrm{h.c.}) \right]  &\leq 2 \sqrt{\Tr\left[\bar{\sigma} \Pi^1_{\mathcal{Z}} \Pi^0_{\mathcal{X}} \Pi^0_{\mathcal{Z}} \Pi^0_{\mathcal{X}} \Pi^1_{\mathcal{Z}} \right]} \nonumber\\
    &\leq 2 \sqrt{\Tr\left[\bar{\sigma} \Pi^1_{\mathcal{Z}} \Pi^0_{\mathcal{X}} \Pi^1_{\mathcal{Z}} \right]},
\end{align}
where we used the fact that $ \Pi^0_{\mathcal{Z}} \leq \one$. Finally, we employ the bound already established in \eqref{secondterm-upp2} and obtain the bound on the remaining expectation value:
\begin{align}
    \Tr\left[\Bar{\sigma} (\Pi^0_{\mathcal{Z}} \Pi^0_{\mathcal{X}} \Pi^1_{\mathcal{Z}} + \mathrm{h.c.}) \right] &\leq 2\left(\,\overline{\left[w^0_{\mathcal{X}}\right]^1_{\mathcal{Z}}}\,\right)^{\frac{1}{2}} \label{secondterm-upp3},
\end{align}
where $\overline{\left[w^0_{\mathcal{X}}\right]^1_{\mathcal{Z}}}$ is given in \eqref{protocol-x0-upperbound101}.

By combining \eqref{secondterm-upp1}, \eqref{secondterm-upp2}, and \eqref{secondterm-upp3} with \eqref{secondterm-expansion}, we obtain an upper bound on the second term in \eqref{simplifiedproof-phase-error-rate-uppbound2}:
\begin{align}
   \Tr\left[\bar{\sigma}  M^{\leq 1}_{\mathcal{Z}}  \Pi^0_{\mathcal{X}} M^{\leq 1}_{\mathcal{Z}} \right] &\leq p_{\checkmark|0} \overline{\left[w^0_{\mathcal{X}}\right]^0_{\mathcal{Z}}} + p_{\checkmark|1}  \overline{\left[w^0_{\mathcal{X}}\right]^1_{\mathcal{Z}}} +2\sqrt{p_{\checkmark|0}p_{\checkmark|1}}  \left(\,\overline{\left[w^0_{\mathcal{X}}\right]^1_{\mathcal{Z}}}\,\right)^{\frac{1}{2}} \label{secondterm-expansion-upp},
\end{align}
which is given in terms of observed statistics.

Finally, we use the results in \eqref{firstterm-expansion-upp} and \eqref{secondterm-expansion-upp} in \eqref{simplifiedproof-phase-error-rate-uppbound2}, which provides us with the following upper bound on the phase error rate (cf.~Eq.~\eqref{phase-error-rate-uppbound-final} in Appendix~\ref{app:PER12}):
\begin{align}
    \tilde{e}_{X,1} &\leq \overline{\tilde{e}_{X,1}}, \label{simplifiedproof-phase-error-rate-uppbound3}
\end{align}
where:
\begin{align}
    \overline{\tilde{e}_{X,1}} &= \frac{1}{\underline{Y^Z_{1,(\eta_\downarrow,\eta_\downarrow)}}[p^{\mathcal{X}}_d + \eta_r \eta_\uparrow (1-p^{\mathcal{X}}_d)]}  \left[p_{\checkmark|0} \overline{\mathbbm{E}}^0_{\eta_\uparrow} +  \sqrt{p_{\checkmark|0} p_{\checkmark|1}} \, \sqrt{\eta_\uparrow} \overline{\mathbbm{E}}^{0,1}_{\eta_\uparrow} + p_{\checkmark|1} \left( \eta_\uparrow \overline{\mathbbm{E}}^{1t}_{\eta_\uparrow} + (1-\eta_\uparrow) \overline{\mathbbm{E}}^{1r}_{\eta_\uparrow}\right)\right] \nonumber\\
    &+ \frac{\eta_r \eta_\uparrow(1-p^{\mathcal{X}}_d)}{\underline{Y^Z_{1,(\eta_\downarrow,\eta_\downarrow)}}[p^{\mathcal{X}}_d + \eta_r \eta_\uparrow (1-p^{\mathcal{X}}_d)]}   \left[ p_{\checkmark|0} \overline{\left[w^0_{\mathcal{X}}\right]^0_{\mathcal{Z}}} + 2\sqrt{p_{\checkmark|0}p_{\checkmark|1}} \left(\,\overline{\left[w^0_{\mathcal{X}}\right]^1_{\mathcal{Z}}}\,\right)^{\frac{1}{2}} + p_{\checkmark|1} \overline{\left[w^0_{\mathcal{X}}\right]^1_{\mathcal{Z}}} \right] + \frac{1}{\underline{Y^Z_{1,(\eta_\downarrow,\eta_\downarrow)}}} \left( \overline{\Delta_2} + \overline{w}^{>1}_{\mathcal{Z}}\right) \label{simplifiedproof-phase-error-rate-uppbound-final},
\end{align}
and where we replaced the one-photon $Z$-basis yield with the corresponding lower bound from the decoy-state method. We emphasize that all the quantities appearing in \eqref{simplifiedproof-phase-error-rate-uppbound-final} are either known or obtained with the decoy-state method (cf. Appendix~\ref{app:decoy} for the formulas of the decoy-state method). This concludes the derivation of a computable upper bound on the phase error rate.

\begin{figure}[hbt]
    \centering
    \includegraphics[width=0.7\linewidth,keepaspectratio]{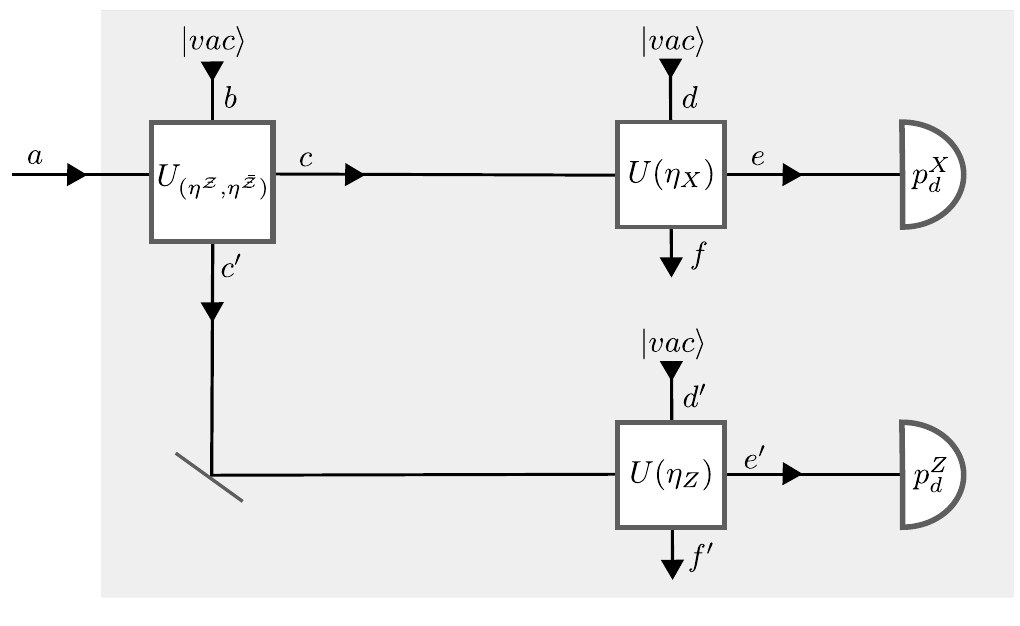}
    \caption{Bob's measurement apparatus. In each round, the incoming signal (mode $a$) enters a tunable beam splitter (TBS) described by the unitary $U_{(\eta_i,\eta_l)}$, where $(\eta_i,\eta_l)$ is the TBS setting for that particular round. The transmitted mode (mode $c$) is detected by an $X$-basis detector with efficiency $\eta_X$ and dark count probability $p_d^X$ for each outcome. This is modeled by a beam splitter ($U(\eta_X)$) followed by a perfect detector with dark count probability $p_d^X$. The reflected mode (mode $c'$) is detected by a $Z$-basis detector with efficiency $\eta_{Z}$ and dark count probability $p_d^{Z_X}$ per outcome, which is similarly modeled. Note that the insertion loss of the TBS is accounted for in the efficiencies of the two detectors.}
    \label{fig:Bobs-setup-appendix}
\end{figure}

\subsubsection{Bob's measurement apparatus}\label{app:PER1}
Let us now detail the missing steps of the argument.
The method that we introduce to estimate the phase error rate  relies on a partial characterization of Bob's measurement apparatus, as described in Sec.~\ref{sec:protocol}. In Fig.~\ref{fig:Bobs-setup-appendix}, we provide a model of Bob's measurement apparatus. Here, the non-unit detection efficiency of the $Z$-basis detector ($X$-basis detector) is modeled by a beam splitter with transmittance $\eta_Z$ ($\eta_X$), described by the unitary $U(\eta_Z)$ ($U(\eta_X)$). The TBS is described by a unitary $U_{(\eta_i,\eta_l)}$ acting on spatial-temporal modes $a^\dag_t$ and $b^\dag_t$ as follows:
\begin{align}
    U_{(\eta_i,\eta_l)} =\left\lbrace\begin{array}{ll}
    U(\eta_i)    &\mbox{if } t\in\mathcal{Z}  \\
     U(\eta_l)    &\mbox{if } t\notin\mathcal{Z}
    \end{array} \right. , \label{TBS-unitary}
\end{align}
where $\mathcal{Z}=\sum_{j=0}^{d-1} \mathcal{Z}_j$ is the union of all the modes that are detected by the $Z$-basis detector. In the case where the $Z$-basis detector is a time-of-arrival measurement, $\mathcal{Z}$ corresponds to the union of all the time bins.  We recall that, in each round, Bob selects one of the following six TBS settings: $(\eta_i,\eta_\uparrow)$ and  $(\eta_i,\eta_\downarrow)$, with $\eta_1=\eta_\uparrow$,  $\eta_3=\eta_\downarrow$ and $\eta_2$ satisfying: $\eta_\downarrow<\eta_2<\eta_\uparrow$. Moreover, we recall that the TBS settings and the detection efficiencies of the two detectors must satisfy \eqref{constraint1} and \eqref{constraint2}.

In our formalism, the unitary $U(\eta)$ maps the input modes ${\rm in}_1^\dag=a^\dag, c^\dag, {c'}^\dag$ and ${\rm in}_2^\dag=b^\dag, d^\dag, {d'}^\dag$ to the output modes ${\rm out}_1^\dag=c^\dag,e^\dag,{e'}^\dag$ and ${\rm out}_2^\dag={c'}^\dag,f^\dag,{f'}^\dag$, respectively, as follows:
\begin{align}
    \begin{pmatrix}
     {\rm out}_1^\dag \\
     {\rm out}_2^\dag 
    \end{pmatrix}
    = U (\eta)
    \begin{pmatrix}
        {\rm in}_1^\dag \\
        {\rm in}_2^\dag 
    \end{pmatrix}, \quad\quad\quad\mbox{with: }\quad
    U (\eta) := 
    \begin{pmatrix}
        \sqrt{\eta} & -\sqrt{1-\eta} \\
        \sqrt{1-\eta} & \sqrt{\eta} 
    \end{pmatrix}. \label{U(eta)}
\end{align}

\begin{figure}[hbt]
    \centering
    \includegraphics[width=0.7\linewidth,keepaspectratio]{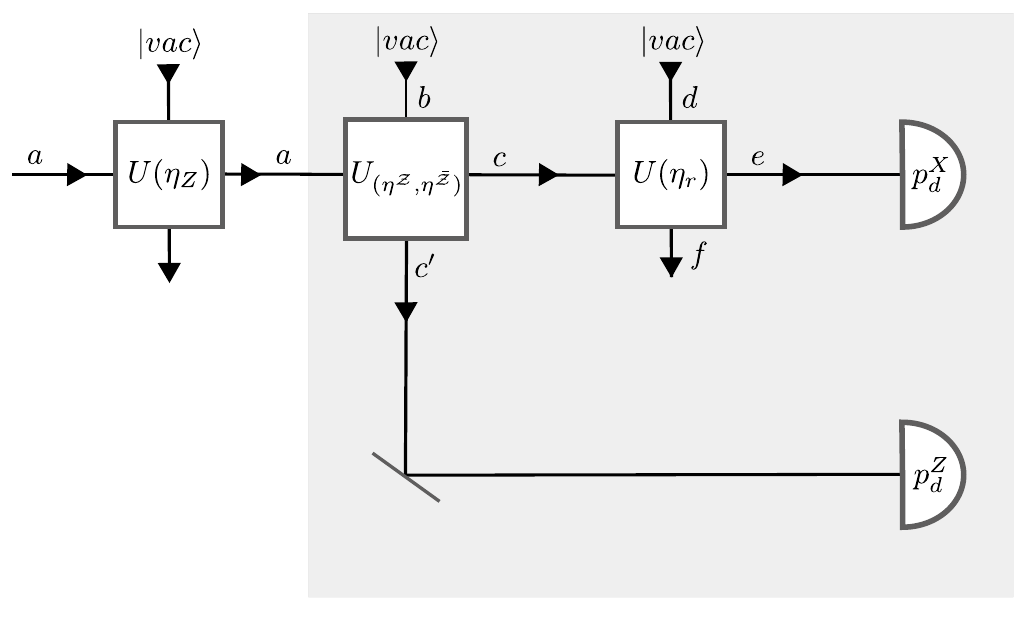}
    \caption{Bob's simplified measurement apparatus. Here, the common loss of the two detectors in Fig.~\ref{fig:Bobs-setup-appendix} -- which amounts to $1-\eta_Z$ -- has been factored outside of Bob's apparatus, leaving an ideal detector in the $Z$ basis and a lossy detector in the $X$ basis with efficiency $\eta_r=\eta_X/\eta_Z$.}
    \label{fig:Bobs-setup-appendix-factored}
\end{figure}

Now, we perform a simplification of Bob's measurement apparatus. The idea is to factor out the common loss in the two detectors and view it as loss occurring in the channel from Alice to Bob, before reaching Bob's measurement apparatus. More specifically, let us define the ratio between the efficiency of the $X$-basis detector and the efficiency of the $Z$-basis detector:
\begin{align}
    \eta_r: = \frac{\eta_X}{\eta_Z} \label{eta-ratio},
\end{align}
where we have that $0<\eta_r<1$ since we assumed in Sec.~\ref{sec:protocol} that the detector used to generate key bits is more efficient. This is a sensible assumption that typically leads to better key rates, however, if not verified, one can always exchange the roles of the two bases. Then, we can re-express the efficiency of the $X$-basis detector as $\eta_X=\eta_Z \eta_r$, which allows us to factor out the transmittance that is shared by the two detectors, namely $\eta_Z$. In particular, we can equivalently describe Bob's apparatus as depicted in Fig.~\ref{fig:Bobs-setup-appendix-factored}, where we introduced a beam splitter with transmittance $\eta_Z$ before Bob's apparatus, which accounts for the common loss of both detectors, and replaced the $Z$-basis ($X$-basis) detector with a lossless detector (lossy detector with efficiency $\eta_r$). This equivalence is well-known in quantum optics, see, e.g.,  \cite{entanglement-ver-detmismatch}. 

Furthermore, we argue that it can only be advantageous to Eve if we give her control over the beam splitter preceding Bob's apparatus ($U(\eta_Z)$) by viewing it as part of the insecure channel between Alice and Bob. Therefore, from now on we ignore the unitary $U(\eta_Z)$ and describe Bob's apparatus as depicted in Fig.~\ref{fig:Bobs-setup-appendix-factored}.\\

\subsubsection{Bob's $Z$-basis detector}\label{app:PER2}
The next step of the proof consists in a reduction of the phase error rate to the subspace with at most one photon in the interval $\mathcal{Z}$, which we call the ``($\leq 1$)-subspace'' for brevity. Intuitively, this is the subspace where the majority of the states in $\{\sigma_k\}_k$ should lie, in the absence of an eavesdropper.

\begin{remark} \label{rmk:no-constraint-sigma}
    We recall that $\sigma_k$ is the state received by Bob, when Alice encodes the test symbol $X_A=k$ on a single photon. However, since we do not restrict Eve's collective attack, we cannot assume that the states $\{\sigma_k\}_k$ are confined to, e.g., the subspace with at most one photon. In fact, Eve could perform attacks where she adds photons to the states sent by Alice. This could allow her to control the click probability of the two bases, if they have asymmetric detection efficiencies. We emphasize that, while previous analytical QKD proofs restrict to the one-photon subspace when dealing with detection efficiency mismatches \cite{detection-efficiency-mismatch-Lo,trushechkin-mismatch1,detection-efficiency-mismatch-Ma}, our proof can accommodate both mode-dependent and mode-independent mismatches in the detection probability without posing any constraint on the states received by Bob.
\end{remark}

Because, as argued in Remark~\ref{rmk:no-constraint-sigma}, we do not constrain the states $\sigma_k$, we are forced to estimate their weight in the ($\leq 1$)-subspace, enabling us to reduce the calculation of the phase error rate to the ($\leq 1$)-subspace.

The estimation of the states' weight in the ($\leq 1$)-subspace is carried out from the statistics of the $Z$-basis detector. Thus, as the first step, we derive an explicit expression for some of the elements of the POVM: $\{(Z_0)_{c'},(Z_1)_{c'},\dots,(Z_{d-1})_{c'},(Z_\emptyset)_{c'}\}$ that describes Bob's $Z$-basis measurement in the reflected mode of the TBS, where the subscript indicates the spatial mode on which the operator acts (cf.~Fig.~\ref{fig:Bobs-setup-appendix-factored}).

Let $\pia$ be the projector of $n=\alpha$ photons in the global detection interval $\mathcal{Z}$, while acting as the identity outside the interval $\mathcal{Z}$. Then, the POVM element corresponding to a detection of one of the $d$ outcomes would be: $(Z_\checkmark)_{c'}=\sum_{j=0}^{d-1} (Z_j)_{c'} =\sum_{\alpha=1}^\infty (\pia)_{c'}$ in the absence of dark counts, i.e.\ the projector in the subspace with at least one photon in $\mathcal{Z}$. Note that this is true since we assumed that multi-click events are always mapped to a single outcome (cf.~Sec.~\ref{sec:protocol}). However, since the detector has a dark count probability $p^Z_d$ per outcome, the POVM element corresponding to a detection becomes:
\begin{align}
    (Z_\checkmark)_{c'} &:= p_d^{\mathcal{Z}} \left[\one - \sum_{\alpha=1}^\infty \pia \right] + \sum_{\alpha=1}^\infty \pia \nonumber\\
    &= p_d^{\mathcal{Z}} \one +(1-p_d^{\mathcal{Z}}) \sum_{\alpha=1}^\infty \pia, \label{Zclick-cprime}
\end{align}
where $p_d^{\mathcal{Z}}$ is defined in \eqref{prob-dk-tot-Z}. Note that if the $Z$ detector implemented a time-of-arrival measurement, the projector $\pia$ would be given by:
\begin{align}
    \pia = \sum_{n=\alpha}^\infty \Pi^\alpha_{\mathcal{Z},n}, \label{pia} 
\end{align}
where $\Pi^\alpha_{\mathcal{Z},n}$ is the projector on the subspace of $n$
photons, with exactly $\alpha$ photons localized in the time interval $\mathcal{Z}$:
\begin{align}
    \Pi^\alpha_{\mathcal{Z},n} = \intop_{\mathcal{Z}} \frac{\text{d}t_1 \dots \text{d}t_\alpha}{\alpha!} \intop_{\overline{\mathcal{Z}}} \frac{\text{d}t_{\alpha+1} \dots \text{d}t_n}{(n-\alpha)!} \proj{1_{t_1}\dots 1_{t_n}}, \label{pia-n}
\end{align}
where $\overline{\mathcal{Z}}$ is the complement of $\mathcal{Z}$ in the time domain and $\ket{1_t}=a_t^\dag \ket{vac}$ represents the unphysical state of one photon at time $t$. We have that $a_t^\dag$ creates a photon at time $t$ as the Fourier transform of a photon with frequency $f$:
\begin{align}
    a^\dag_t = \intop_{-\infty}^\infty \text{d}f e^{2\pi i f t} a^\dag_f.\label{a_t}
\end{align}
Moreover, it holds $[a^\dag_t,a^\dag_{t'}]=0$ and $[a_t,a_{t'}^\dag]=\delta(t-t')\one$, which implies:
\begin{align}
    &\braket{1_{t_1},\dots,1_{t_n}|1_{t'_1},\dots,1_{t'_k}} = \delta_{n,k} \sum_{\pi\in S_n} \delta(t_1 - t'_{\pi(1)}) \dots \delta(t_n - t'_{\pi(n)}) , \label{braket-n-timephotons}
\end{align}
where $S_n$ is the set of all permutations of indexes in $\{1,2,\dots,n\}$. To familiarize with this formalism, which allows $n$ photons to be arbitrarily de-localized in time, we can verify that:
\begin{align}
    \sum_{\alpha=0}^\infty \pia =\one \label{normalization-pia},
\end{align}
holds. The following proof is not essential to the remainder of this Appendix and can be skipped.

\begin{proof}[Proof of Eq.~\eqref{normalization-pia}]
    By using \eqref{pia}, we can rewrite the sum over $\alpha$ as follows:
    \begin{align}
        \sum_{\alpha=0}^\infty \pia &= \sum_{\alpha=0}^\infty \sum_{n=\alpha}^\infty \Pi^\alpha_{\mathcal{Z},n} \nonumber\\
        &= \sum_{n=0}^\infty \sum_{\alpha=0}^n \Pi^\alpha_{\mathcal{Z},n}.
    \end{align}
    By comparing the last expression with:
    \begin{align}
        \one = \sum_{n=0}^\infty \Pi_n,
    \end{align}
    where $\Pi_n$ is the projector on the subspace of $n$ photons:
    \begin{align}
        \Pi_n = \intop_{-\infty}^\infty \frac{\text{d}t_1 \dots \text{d}t_n}{n!} \proj{1_{t_1}\dots 1_{t_n}},
    \end{align}
    we are left to prove that $\Pi_n = \sum_{\alpha=0}^n \Pi^\alpha_{\mathcal{Z},n}$, i.e.\ that:
    \begin{align}
        \intop_{-\infty}^\infty \frac{\text{d}t_1 \dots \text{d}t_n}{n!} \proj{1_{t_1}\dots 1_{t_n}} = \sum_{\alpha=0}^n  \intop_{\mathcal{Z}} \frac{\text{d}t_1 \dots \text{d}t_\alpha}{\alpha!} \intop_{\overline{\mathcal{Z}}} \frac{\text{d}t_{\alpha+1} \dots \text{d}t_n}{(n-\alpha)!} \proj{1_{t_1}\dots 1_{t_n}}. \label{to-prove}
    \end{align}
    To prove \eqref{to-prove}, we independently apply the left-hand side and right-hand side on a generic state and verify that the resulting states coincide. Let $\ket{\Psi}$ be an $n$-photon state defined as:
    \begin{align}
        \ket{\Psi} = \intop_{-\infty}^\infty \text{d}\tau_{1} \dots \text{d}\tau_n f(\tau_1,\dots,\tau_n) \ket{1_{\tau_1},\dots,1_{\tau_n}},
    \end{align}
    where $f(\tau_1,\dots,\tau_n)$ is an arbitrary time distribution function. We have:
    \begin{align}
        \Pi_n \ket{\Psi} &= \intop_{-\infty}^\infty \frac{\text{d}t_1 \dots \text{d}t_n}{n!} \ket{1_{t_1}\dots 1_{t_n}} \sum_{\pi\in S_n} f( t_{\pi(1)}, \dots ,t_{\pi(n)}) \nonumber\\
        &=  \intop_{-\infty}^\infty \text{d}t_1 \dots \text{d}t_n f(t_1 \dots,t_n) \ket{1_{t_1}\dots 1_{t_n}} = \ket{\Psi},
    \end{align}
    where we used \eqref{braket-n-timephotons} in the first equality and renamed $t_{\pi(1)}\to t_1$, \dots, $t_{\pi(n)}\to t_n$ in the second equality. Now we compute:
    \begin{align}
        \sum_{\alpha=0}^n \Pi^\alpha_{\mathcal{Z},n}\ket{\Psi} &= \sum_{\alpha=0}^n  \intop_{\mathcal{Z}} \frac{\text{d}t_1 \dots \text{d}t_\alpha}{\alpha!} \intop_{\overline{\mathcal{Z}}} \frac{\text{d}t_{\alpha+1} \dots \text{d}t_n}{(n-\alpha)!} \ket{1_{t_1}\dots 1_{t_n}} \sum_{\pi\in S_n} f( t_{\pi(1)}, \dots ,t_{\pi(n)}) \nonumber\\
        &= \sum_{\alpha=0}^n \sum_{\pi\in S_n} \intop_{\mathcal{Z}} \frac{\text{d}t_{\pi(1)} \dots \text{d}t_{\pi(\alpha)}}{\alpha!} \intop_{\overline{\mathcal{Z}}} \frac{\text{d}t_{\pi(\alpha+1)} \dots \text{d}t_{\pi(n)}}{(n-\alpha)!} \ket{1_{t_1}\dots 1_{t_n}}  f( t_1, \dots ,t_n),
    \end{align}
    where we renamed $t_{\pi(1)}\to t_1$, \dots, $t_{\pi(n)}\to t_n$ in the second equality. Note that $ \ket{1_{t_1}\dots 1_{t_n}}$ remains unchanged because it is permutationally invariant. Now, we observe that the following equality holds after expanding the left-hand side:
    \begin{align}
        \left(\intop_{\mathcal{Z}}\text{d}t_1 + \intop_{\overline{\mathcal{Z}}}\text{d}t_1\right) \dots \left(\intop_{\mathcal{Z}}\text{d}t_n + \intop_{\overline{\mathcal{Z}}}\text{d}t_n\right) = \sum_{\alpha=0}^n \sum_{\pi\in S_n} \intop_{\mathcal{Z}} \frac{\text{d}t_{\pi(1)} \dots \text{d}t_{\pi(\alpha)}}{\alpha!} \intop_{\overline{\mathcal{Z}}} \frac{\text{d}t_{\pi(\alpha+1)} \dots \text{d}t_{\pi(n)}}{(n-\alpha)!}, \label{to-prove2}
    \end{align}
    where the division by $\alpha!$ and $(n-\alpha)!$ ensures that the permutations where the only terms permuting are within the same integration range are not counted more than once. Then, by combining the previous equality with:
    \begin{align}
        \left(\intop_{\mathcal{Z}}\text{d}t_1 + \intop_{\overline{\mathcal{Z}}}\text{d}t_1\right) \dots \left(\intop_{\mathcal{Z}}\text{d}t_n + \intop_{\overline{\mathcal{Z}}}\text{d}t_n\right) = \intop_{-\infty}^\infty \text{d}t_1 \dots \text{d}t_n,
    \end{align}
    and using this in \eqref{to-prove2}, we obtain:
    \begin{align}
        \sum_{\alpha=0}^n \Pi^\alpha_{\mathcal{Z},n}\ket{\Psi} =  \intop_{-\infty}^\infty \text{d}t_1 \dots \text{d}t_n f(t_1 \dots,t_n) \ket{1_{t_1}\dots 1_{t_n}} = \ket{\Psi},
    \end{align}
    which proves \eqref{to-prove} and thus concludes the proof.
\end{proof}

\begin{remark}
    Note that if Bob's $Z$-basis measurement corresponds to measuring a different degree of freedom than the arrival time of the signal, one can still use the expressions \eqref{Zclick-cprime} and \eqref{pia} for other degrees of freedom, while the time degree of freedom should be replaced by the measured degree of freedom in \eqref{pia-n}.
\end{remark}

We have thus obtained the expression \eqref{Zclick-cprime} for the POVM element representing a detection in the $Z$-basis detector, acting on the reflected mode $c'$. However, the actual measurement statistics are collected from Bob's global measurement on the received signal (mode $a$), which comprises detections in both the $Z$- and $X$-basis detector and is formally introduced in \eqref{Bobs-POVM}. Nevertheless, similarly to \eqref{Bobs-Z-measurement}, from the global POVM \eqref{Bobs-POVM} we can define a POVM element on mode $a$ that corresponds to a detection in the $Z$ detector and ignores the outcome of the $X$-basis detector:
\begin{align}
    Z^{(\eta_l,\eta_l)}_\checkmark := \sum_{j=0}^{d-1} M^{(\eta_l,\eta_l)}_{j,\emptyset} + \sum_{j,k=0}^{d-1} M^{(\eta_l,\eta_l)}_{j,k}  \label{Zclick-a},
\end{align}
where the superscript indicates the TBS setting. Note that $Z^{(\eta_\downarrow,\eta_\downarrow)}_\checkmark=Z_\checkmark$, where $Z_\checkmark$ is the POVM element for a detection in Bob's key generation measurement and it appears in the phase error rate.

In order to obtain an explicit expression for \eqref{Zclick-a}, we first note that the one-photon $Z$-basis yield with TBS setting $(\eta_l,\eta_l)$ is, similarly to \eqref{yield-KGn}, given by:
\begin{align}
    Y^Z_{1,(\eta_l,\eta_l)} &= \sum_{j=0}^{d-1} \frac{1}{d} \Pr(Z_B \neq\emptyset|T=Z,N_A=1,Z_A=j,(\eta_l,\eta_l)) \nonumber\\
    &= \sum_{j=0}^{d-1} \frac{1}{d} \Tr\left[(Z^{(\eta_l,\eta_l)}_\checkmark \otimes\one_E) U_{BE} \ketbra{1_{Z_j}}{1_{Z_j}}_B \otimes \ketbra{0}{0}_E U_{BE}^\dag \right] \nonumber\\
    &=\Tr_a\left[\bar{\sigma} Z^{(\eta_l,\eta_l)}_\checkmark \right] \label{prob-Zclick},
\end{align}
where in the second equality we used a generalization of \eqref{pr(det|j,n)} and in the third equality the fact that \eqref{equal-sum-twobases} holds, together with a short-hand notation for the average state received by Bob:
\begin{align}
    \bar{\sigma} := \frac{1}{d} \sum_{k=0}^{d-1} \sigma_k \label{barsigma},
\end{align}
with $\sigma_k$ given in \eqref{sigmak}. At the same time, by using the scheme in Fig.~\ref{fig:Bobs-setup-appendix-factored} and the definition of the TBS unitary in \eqref{TBS-unitary}, we have that:
\begin{align}
    Y^Z_{1,(\eta_l,\eta_l)} &= \Tr_{cc'}\left[U(\eta_l)\bar{\sigma}_a \otimes\proj{vac}_b U(\eta_l)^\dag \one_c \otimes (Z_\checkmark)_{c'}\right] \nonumber\\
    &= \Tr_{ab}\left[\bar{\sigma}_a \otimes\proj{vac}_b U(\eta_l)^\dag \one_c \otimes (Z_\checkmark)_{c'} U(\eta_l)\right] \nonumber\\
    &= \Tr_{a}\left[\bar{\sigma}_a \bra{vac}_b U(\eta_l)^\dag \one_c \otimes (Z_\checkmark)_{c'} U(\eta_l) \ket{vac}_b\right]  \label{prob-Zclick-alternative}.
\end{align}
By comparing \eqref{prob-Zclick} and \eqref{prob-Zclick-alternative}, we deduce an expression for the operator in \eqref{Zclick-a}:
\begin{align}
    Z^{(\eta_l,\eta_l)}_\checkmark  &= \bra{vac}_b U(\eta_l)^\dag \one_c \otimes (Z_\checkmark)_{c'} U(\eta_l) \ket{vac}_b \nonumber\\
    &= p^{\mathcal{Z}}_d \one + (1-p^{\mathcal{Z}}_d) \bra{vac}_b U(\eta_l)^\dag \one_c \otimes \sum_{\alpha=1}^\infty (\pia)_{c'} U(\eta_l) \ket{vac}_b\label{Zclick-reflectedTBS-simplified-modea},
\end{align}
where in the second equality we used \eqref{Zclick-cprime}. Let us now define the following operator acting on mode $a$:
\begin{align}
    Q := \bra{vac}_b U(\eta_l)^\dag \one_c \otimes \sum_{\alpha=1}^\infty (\pia)_{c'} U(\eta_l) \ket{vac}_b \label{Qa} .
\end{align}
We can explicitly calculate $Q$ by applying the unitary $U(\eta_l)$  according to \eqref{U(eta)}. To do so, one makes $(\pia)_{c'}$ and $\one_c$ explicit according to \eqref{pia}, \eqref{pia-n} and \eqref{normalization-pia}. Then, by linearity one obtains some terms like the following:
\begin{align}
    &\bra{vac}_b U^\dag(\eta_l) c^\dag_{t_1} \dots c^\dag_{t_n} (c')^\dag_{t'_1} \dots (c')^\dag_{t'_m} \proj{vac} c_{t_1} \dots c_{t_n} (c')_{t'_1} \dots (c')_{t'_m} U(\eta_l) \ket{vac}_b \nonumber\\
    &= \eta_l^n (1-\eta_l)^m \proj{1_{t_1},\dots,1_{t_n},1_{t'_1},\dots, 1_{t'_m}}_a, \label{calc}
\end{align}
which corresponds to a pure state on mode $a$. By using \eqref{calc} in \eqref{Qa} and by re-parametrising the resulting sums, one obtains:
\begin{align}
    Q &= \sum_{\alpha=1}^\infty (1-\eta_l^\alpha) (\pia)_a  \label{Qa2}.
\end{align}
By using this expression in \eqref{Zclick-reflectedTBS-simplified-modea}, we find the final expression for the POVM element \eqref{Zclick-a} describing a detection in the $Z$ detector with the TBS setting $(\eta_l,\eta_l)$:
\begin{align}
    Z^{(\eta_l,\eta_l)}_\checkmark &= p^{\mathcal{Z}}_d \one + (1-p^{\mathcal{Z}}_d) \sum_{\alpha=1}^\infty (1-\eta_l^\alpha) \pia \nonumber\\
    &=\sum_{\alpha=0}^\infty \left[1-(1-p^{\mathcal{Z}}_d) \eta_l^\alpha \right] \pia . \label{Zclick-reflectedTBS-simplified-modea2}
\end{align}
This operator has a two-fold purpose. First, by recalling the definition of Bob's key generation measurement in \eqref{Bobs-Z-measurement}, it provides us with the explicit expression for the operator $Z_\checkmark$ that appears in the phase error rate formula \eqref{phase-error-rate-appendix}:
\begin{align}
    Z_\checkmark &= Z^{(\eta_\downarrow,\eta_\downarrow)}_\checkmark \nonumber\\
    &= \sum_{\alpha=0}^\infty p_{\checkmark|\alpha} \pia \label{Zclick-KGmeasurement}.
\end{align}
where we identified $p_{\checkmark|\alpha}$ as the probability of a detection in the $Z$-basis detector, given that the state received by Bob contains $\alpha$ photons localized in the detection interval $\mathcal{Z}$:
\begin{align}
    p_{\checkmark|\alpha} := 1-(1-p^{\mathcal{Z}}_d) \eta_\downarrow^\alpha  \label{Pr(oneclick|alpha)}.
\end{align}
We also specify the above expression for the cases $\alpha=0,1$:
\begin{align}
    p_{\checkmark|0} &= p^{\mathcal{Z}}_d  \label{Pr(oneclick|0)} \\
    p_{\checkmark|1} &= 1-(1-p^{\mathcal{Z}}_d) \eta_\downarrow \label{Pr(oneclick|1)}.
\end{align}
Secondly, the operator in \eqref{Zclick-reflectedTBS-simplified-modea2} allows us to estimate the weight of the states received by Bob in the ($\leq 1$)-subspace, through the detector decoy technique \cite{detector-decoy}.\\

\subsubsection{The detector decoy technique on the $Z$ detector statistics}\label{app:PER3}
The detector decoy technique \cite{detector-decoy} is a powerful tool that allows one to estimate the weight of measured states in various subspaces with a fixed photon number. In our case, for the purpose of reducing the phase error rate calculation to the ($\leq 1$)-subspace, we are interested in estimating the weight of the average state received by Bob in the subspaces defined by $\Pi^0_{\mathcal{Z}}$ and $\Pi^1_{\mathcal{Z}}$. More precisely, let:
\begin{align}
    \Tr\left[\bar{\sigma} \pia\right]\quad, \quad \mbox{for }\alpha=0,1,\dots,\infty, \label{weight-alpha}
\end{align}
be the weight of $\bar{\sigma}$ in the subspace with $\alpha$ photons localized in $\mathcal{Z}$. Then, the goal is to derive upper and lower bounds on $\Tr[\bar{\sigma} \Pi^0_{\mathcal{Z}}]$ and $\Tr[\bar{\sigma} \Pi^1_{\mathcal{Z}}]$.

To this aim, we will use the statistics of the no detection events in the $Z$ detector with TBS setting $(\eta_l,\eta_l)$, which are generated by the following POVM element:
\begin{align}
    Z^{(\eta_l,\eta_l)}_\emptyset &= \one - Z^{(\eta_l,\eta_l)}_\checkmark \nonumber\\
    &=(1-p^{\mathcal{Z}}_d) \sum_{\alpha=0}^\infty \eta_l^\alpha  \pia . \label{Znoclick-reflectedTBS-simplified-modea2}
\end{align}
where we used \eqref{Zclick-reflectedTBS-simplified-modea2}. Therefore, the probability of a no detection in the $Z$-basis detector reads:
\begin{align}
    1-Y^Z_{1,(\eta_l,\eta_l)} &= \Tr\left[\bar{\sigma} Z^{(\eta_l,\eta_l)}_\emptyset\right] \nonumber\\
    &= (1-p^{\mathcal{Z}}_d) \sum_{\alpha=0}^\infty \eta_l^\alpha  \Tr[\bar{\sigma}\pia] . \label{prob-Znoclick}
\end{align}

The detector decoy technique is based on a set of equations of the following form:
\begin{align}
    f_l = \sum_{\alpha=0}^\infty \eta_l^\alpha y_\alpha \label{detector-decoy-Z}
\end{align}
where $f_l$ are known quantities that are experimentally observed, $y_\alpha\in[0,1]$ are the variables that are to be bounded, and $\eta_l$ is a parameter that can be fixed to different values by varying $l$. In our case, the equation \eqref{prob-Znoclick} can be put in the form of \eqref{detector-decoy-Z} by identifying:
\begin{align}
    f_l &= \frac{1-Y^Z_{1,(\eta_l,\eta_l)}}{1-p^{\mathcal{Z}}_d} \\
    y_\alpha &= \Tr[\bar{\sigma}\pia],
\end{align}
from which we deduce the normalization property of the variables $\{y_\alpha\}_\alpha$:
\begin{align}
    {\textstyle\sum_{\alpha=0}^\infty}  y_\alpha &=1 \label{normalization-yalpha} .
\end{align}
We now apply the idea of the detector decoy technique to \eqref{detector-decoy-Z} to derive an upper and a lower bound on $y_0$ and $y_1$. We recall that, among Bob's possible choices for the TBS setting (cf.~Sec.~\ref{sec:protocol}), we have that $\eta_1= \eta_\uparrow$ and $\eta_3= \eta_\downarrow$.

We start by deriving the upper bound on $y_0$. Consider the following system of two inequalities, obtained from \eqref{detector-decoy-Z} and \eqref{normalization-yalpha}:
\begin{align}
    &\left\lbrace \begin{array}{l}
       f_1 \leq y_0 + \eta_1 y_1 + \eta_1^2 (1-y_0 -y_1)  \\
    f_3 \geq y_0 + \eta_3 y_1  
    \end{array}\right. \nonumber\\
    \Rightarrow \quad &\left\lbrace \begin{array}{l}
        y_1 \geq \frac{f_1 -y_0 -\eta_1^2(1-y_0)}{\eta_1 (1-\eta_1)}  \\
     y_0 + \eta_3  \frac{f_1 -y_0 -\eta_1^2(1-y_0)}{\eta_1 (1-\eta_1)} \leq f_3
    \end{array}\right. \nonumber\\
    \Rightarrow \quad &\left\lbrace \begin{array}{l}
        y_1 \geq \frac{f_1 -y_0 -\eta_1^2(1-y_0)}{\eta_1 (1-\eta_1)}  \\
     y_0 \left[1- \eta_3\frac{1+\eta_1}{\eta_1}\right] \leq f_3 - \eta_3 \frac{f_1}{\eta_1 (1-\eta_1)} + \frac{\eta_1 \eta_3}{1-\eta_1}
    \end{array}\right.,  \label{system-y0-upperbound}
\end{align}
where in the second step we used the lower bound on $y_1$ in the second inequality. From the bottom inequality of \eqref{system-y0-upperbound}, we can derive an upper bound on $y_0$, provided that its coefficient is positive. We thus require that:
\begin{align}
    &\eta_1 - \eta_3 (1 + \eta_1) >0 \nonumber\\
    \iff \quad & \eta_\uparrow > \frac{\eta_\downarrow}{1-\eta_\downarrow}, \label{condition-y0-upperbound}
\end{align}
which is indeed satisfied since we assumed \eqref{constraint1}. Then, the upper bound on $y_0$ can be derived from \eqref{system-y0-upperbound} and reads:
\begin{align}
    y_0 \leq \frac{\eta_1 f_3 -\eta_3 f_1 + \eta_1^2 (\eta_3 -f_3) }{(1-\eta_1) \left[\eta_1 -\eta_3 (1 + \eta_1)\right]} . \label{y0-upperbound-2}
\end{align}
Finally, we substitute back the values for $\eta_1$, $\eta_3$, $f_l$ and $y_0$ and obtain the desired upper bound on the weight of the state in the subspace with no photon in $\mathcal{Z}$, in terms of observed quantities:
\begin{align}
          \Tr[\bar{\sigma}\Pi^0_{\mathcal{Z}}] \leq \overline{w}^0_{\mathcal{Z}}  , \label{y0-upperbound}
\end{align}
with:
\begin{align}
    \overline{w}^0_{\mathcal{Z}} := \min\left\lbrace 1,\frac{\overline{\eta_\downarrow Y^Z_{1,(\eta_\uparrow,\eta_\uparrow)} -\eta_\uparrow (1-\eta_\uparrow)Y^Z_{1,(\eta_\downarrow,\eta_\downarrow)}} +\eta_\uparrow (1-\eta_\uparrow) -\eta_\downarrow + \eta_\uparrow^2 \eta_\downarrow (1-p^{\mathcal{Z}}_d)}{(1-p^{\mathcal{Z}}_d)(1-\eta_\uparrow) \left[\eta_\uparrow -\eta_\downarrow (1 + \eta_\uparrow)\right]}\right\rbrace \label{y0-upperbound-explicit},
\end{align}
where we used the fact that $\Pi^0_{\mathcal{Z}}\leq \one$ to impose that $\overline{w}^0_{\mathcal{Z}}\leq 1$ and where we replaced the terms containing the yields $Y^Z_{1,(\eta_l,\eta_l)}$ -- which are unobserved -- with a bespoke upper bound obtained with the decoy-state method (see Appendix~\ref{app:decoy}), such that the resulting expression is still a valid upper bound.

Now we derive the upper bound on $y_1$. To this aim, we consider the following system of inequalities:
\begin{align}
    &\left\lbrace \begin{array}{l}
       f_3 \leq y_0 + \eta_3 y_1 + \eta_3^2 (1-y_0 -y_1)  \\
    f_1 \geq y_0 + \eta_1 y_1  
    \end{array}\right. \nonumber\\
    \Rightarrow \quad &\left\lbrace \begin{array}{l}
        y_0 \geq \frac{f_3 -y_1 \eta_3(1-\eta_3) -\eta_3^2}{1-\eta_3^2}  \\
     \frac{f_3 -y_1 \eta_3(1-\eta_3) -\eta_3^2}{1-\eta_3^2} + \eta_1 y_1 \leq f_1
    \end{array}\right. \nonumber\\
    \Rightarrow \quad &\left\lbrace \begin{array}{l}
       y_0 \geq \frac{f_3 -y_1 \eta_3(1-\eta_3) -\eta_3^2}{1-\eta_3^2}  \\
     y_1 \left[\eta_1 - \frac{\eta_3}{1+\eta_3}\right] \leq f_1 - \frac{f_3}{1-\eta_3^2} + \frac{\eta_3^2}{1-\eta_3^2}
    \end{array}\right.,  \label{system-y1-upperbound}
\end{align}
where we used the lower bound on $y_0$ in the second inequality of the second step. From the bottom inequality of \eqref{system-y1-upperbound}, we can derive an upper bound on $y_1$, provided that its coefficient is positive, i.e.\ provided that:
\begin{align}
    \eta_\uparrow > \frac{\eta_\downarrow}{1+\eta_\downarrow}
\end{align}
holds, which is indeed the case since $\eta_\uparrow>\eta_\downarrow$. Then, from \eqref{system-y1-upperbound} we obtain the following upper bound on $y_1$:
\begin{align}
    y_1 \leq \frac{f_1 -f_3 +\eta_3^2 (1-f_1)}{(1-\eta_3)\left[\eta_1 (1+ \eta_3) - \eta_3\right]}, \label{y1-upperbound-2}
\end{align}
which in the original variables becomes:
\begin{align}
    \Tr[\bar{\sigma}\Pi^1_{\mathcal{Z}}] \leq \overline{w}^1_{\mathcal{Z}}, \label{y1-upperbound}
\end{align}
with:
\begin{align}
     \overline{w}^1_{\mathcal{Z}}:= \min\left\lbrace 1, \frac{\overline{-(1-\eta_\downarrow^2) Y^Z_{1,(\eta_\uparrow,\eta_\uparrow)} +Y^Z_{1,(\eta_\downarrow,\eta_\downarrow)}} -\eta_\downarrow^2 p^{\mathcal{Z}}_d}{(1-p^{\mathcal{Z}}_d)(1-\eta_\downarrow)\left[\eta_\uparrow (1+ \eta_\downarrow) - \eta_\downarrow\right]}\right\rbrace  \label{y1-upperbound-explicit}.
\end{align}

We now turn to the derivation of the lower bounds on $y_0$ and $y_1$. The lower bound on $y_1$ can be obtained by substituting  the upper bound on $y_0$, \eqref{y0-upperbound-2}, into the lower bound on $y_1$ in \eqref{system-y0-upperbound}. By doing so and by simplifying the expression, we obtain:
\begin{align}
    y_1 \geq \frac{f_1 - f_3 -\eta_1^2 (1-f_3)}{(1-\eta_1)\left[\eta_1 -\eta_3 (1 + \eta_1) \right]}, \label{y1-lowerbound}
\end{align}
which in the original variables becomes:
\begin{align}
    \Tr[\bar{\sigma} \Pi^1_{\mathcal{Z}}] \geq \underline{w}^1_{\mathcal{Z}} \quad,\mbox{ with: } \underline{w}^1_{\mathcal{Z}}:=\max\left\lbrace 0, \frac{-\left(\overline{Y^Z_{1,(\eta_\uparrow,\eta_\uparrow)}-(1-\eta_\uparrow^2) Y^Z_{1,(\eta_\downarrow,\eta_\downarrow)}}\right) +\eta_\uparrow^2 p^{\mathcal{Z}}_d}{(1-p^{\mathcal{Z}}_d)(1-\eta_\uparrow)\left[\eta_\uparrow -\eta_\downarrow (1 + \eta_\uparrow) \right]}\right\rbrace. \label{y1-lowerbound-explicit}
\end{align}
In a similar manner, we can substitute the upper bound on $y_1$, \eqref{y1-upperbound-2}, into the lower bound on $y_0$ from \eqref{system-y1-upperbound}. After some simplifications, we obtain:
\begin{align}
    y_0 \geq \frac{\eta_1 f_3 - \eta_3 f_1 -\eta_3^2 (\eta_1 - f_1) }{(1-\eta_3)\left[\eta_1 (1+\eta_3) -\eta_3 \right]}, \label{y0-lowerbound}
\end{align}
which in the original variables becomes:
\begin{align}
    \Tr[\bar{\sigma} \Pi^0_{\mathcal{Z}}] \geq \underline{w}^0_{\mathcal{Z}} \quad,\mbox{ with: } \underline{w}^0_{\mathcal{Z}}:= \max\left\lbrace 0,  \frac{-\left(\overline{-\eta_\downarrow (1-\eta_\downarrow) Y^Z_{1,(\eta_\uparrow,\eta_\uparrow)} + \eta_\uparrow Y^Z_{1,(\eta_\downarrow,\eta_\downarrow)}}\right) +\eta_\uparrow -\eta_\downarrow(1-\eta_\downarrow) - \eta_\downarrow^2 \eta_\uparrow (1-p^{\mathcal{Z}}_d) }{(1-p^{\mathcal{Z}}_d)(1-\eta_\downarrow)\left[\eta_\uparrow (1+\eta_\downarrow) -\eta_\downarrow \right]} \right\rbrace. \label{y0-lowerbound-explicit}
\end{align}

Finally, we derive the upper bound on the weight of the state $\Bar{\sigma}$ in the subspaces with two or more photons localized in $\mathcal{Z}$. To this aim, we consider the following combination from \eqref{detector-decoy-Z}:
\begin{align}
    f_1 -\frac{\eta_1 -\eta_3}{\eta_2 -\eta_3} f_2 + \frac{\eta_1 -\eta_2}{\eta_2 - \eta_3} f_3 &= \sum_{\alpha=0}^\infty y_\alpha \left[\eta_1^\alpha -\frac{\eta_1 -\eta_3}{\eta_2 -\eta_3} \eta_2^\alpha + \frac{\eta_1 -\eta_2}{\eta_2 - \eta_3} \eta_3^\alpha \right] \nonumber\\
    &= y_0 \left[1 -\frac{\eta_1 -\eta_3}{\eta_2 -\eta_3}  + \frac{\eta_1 -\eta_2}{\eta_2 - \eta_3} \right] + y_1 \left[\eta_1 -\frac{\eta_1 -\eta_3}{\eta_2 -\eta_3} \eta_2 + \frac{\eta_1 -\eta_2}{\eta_2 - \eta_3} \eta_3 \right] \nonumber\\
    &\quad + \sum_{\alpha=2}^\infty y_\alpha \left[\eta_1^\alpha -\frac{\eta_1 -\eta_3}{\eta_2 -\eta_3} \eta_2^\alpha + \frac{\eta_1 -\eta_2}{\eta_2 - \eta_3} \eta_3^\alpha \right] \nonumber\\
    &= \sum_{\alpha=2}^\infty y_\alpha \frac{(\eta_2 -\eta_3)\eta_1^\alpha -(\eta_1 -\eta_3)\eta_2^\alpha + (\eta_1 -\eta_2) \eta_3^\alpha}{\eta_2 - \eta_3} 
\end{align}
from which we deduce:
\begin{align}
    (\eta_2 - \eta_3) f_1 -(\eta_1 -\eta_3) f_2 + (\eta_1 -\eta_2)f_3 &=\sum_{\alpha=2}^\infty y_\alpha \left[(\eta_2 -\eta_3)\eta_1^\alpha -(\eta_1 -\eta_3)\eta_2^\alpha + (\eta_1 -\eta_2) \eta_3^\alpha \right] \label{first-combination}.
\end{align}
Now consider the following combination:
\begin{align}
    f_1 (1-\eta_3) -f_3 (1-\eta_1) &= \sum_{\alpha=0}^\infty y_\alpha \left[\eta_1^\alpha (1-\eta_3) - \eta_3^\alpha (1-\eta_1) \right] \nonumber\\
    &=(y_0 + y_1)  (\eta_1 -\eta_3) + \sum_{\alpha=2}^\infty y_\alpha \left[\eta_1^\alpha (1-\eta_3) - \eta_3^\alpha (1-\eta_1) \right] \nonumber\\
    &= (\eta_1 -\eta_3) \left( 1 - \sum_{\alpha=2}^\infty y_\alpha \right) + \sum_{\alpha=2}^\infty y_\alpha  \left[\eta_1^\alpha (1-\eta_3) - \eta_3^\alpha (1-\eta_1) \right] ,
\end{align}
from which we deduce:
\begin{align}
    f_1 (1-\eta_3) -f_3 (1-\eta_1) - (\eta_1 - \eta_3) = \sum_{\alpha=2}^\infty y_\alpha  \left[\eta_1^\alpha (1-\eta_3) - \eta_3^\alpha (1-\eta_1) -(\eta_1 - \eta_3) \right] \label{second-combination}.
\end{align}
We can now combine the two equalities \eqref{first-combination} and \eqref{second-combination} by subtracting the latter from the former. We obtain:
\begin{align}
    (\eta_2 - \eta_3) &f_1 -(\eta_1 -\eta_3) f_2 + (\eta_1 -\eta_2)f_3 - \left[f_1 (1-\eta_3) -f_3 (1-\eta_1) - (\eta_1 - \eta_3)\right]  \nonumber\\
    &= \sum_{\alpha=2}^\infty y_\alpha \left[(\eta_2 -1)\eta_1^\alpha -(\eta_1 -\eta_3)\eta_2^\alpha +(1-\eta_2) \eta_3^\alpha +\eta_1 - \eta_3\right] \nonumber\\
    &= \sum_{\alpha=2}^\infty y_\alpha \left[(\eta_1 - \eta_3)(1-\eta_2^\alpha) - (1-\eta_2)(\eta_1^\alpha -\eta_3^\alpha) \right] \label{third-combination}.
\end{align}
We now study the factors of the variables $y_\alpha$ in the last expression. For brevity, we label them as follows:
\begin{align}
    g(\alpha) := (\eta_1 - \eta_3)(1-\eta_2^\alpha) - (1-\eta_2)(\eta_1^\alpha -\eta_3^\alpha) \label{g(alpha)},
\end{align}
and view $g(\alpha)$ as a continuous function of $\alpha\geq 2$. First, we observe that $g(\alpha)$ is always non-negative:
\begin{align}
    &g(\alpha) \geq 0 \nonumber\\
    \Leftrightarrow\quad &(\eta_1 - \eta_3)(1-\eta_2^\alpha) \geq  (1-\eta_2)(\eta_1^\alpha -\eta_3^\alpha) \nonumber\\
    \Leftrightarrow\quad &(\eta_1 - \eta_3)(1-\eta_2)(1 + \eta_2 + \dots + \eta_2^{\alpha-1}) \geq (1-\eta_2)(\eta_1 -\eta_3)(\eta_1^{\alpha-1} +\eta_1^{\alpha-2} \eta_3 + \dots + \eta_1 \eta_3^{\alpha-2} + \eta_3^{\alpha-1} )  \label{g(alpha)>0},
\end{align}
which is verified since $1\geq \eta_1$ and $\eta_2 \geq \eta_3$. Then, we notice that $g(\alpha)$ is a non-decreasing function of $\alpha$. Indeed, this amounts to showing that:
\begin{align}
    &g(\alpha+1) \geq g(\alpha) \nonumber\\
    \Leftrightarrow\quad & (\eta_1 - \eta_3)(1-\eta_2^{\alpha+1}) - (1-\eta_2)(\eta_1^{\alpha+1} -\eta_3^{\alpha+1}) \geq (\eta_1 - \eta_3)(1-\eta_2^\alpha) - (1-\eta_2)(\eta_1^\alpha -\eta_3^\alpha) \nonumber\\
    \Leftrightarrow\quad &(\eta_1 - \eta_3)(1-\eta_2)\eta_2^\alpha + (1-\eta_2)(1-\eta_1)\eta_1^\alpha -(1-\eta_2)(1-\eta_3)\eta_3^\alpha \geq 0 \nonumber\\
    \Leftrightarrow\quad &(\eta_1 - \eta_3)\eta_2^\alpha + (1-\eta_1)\eta_1^\alpha -(1-\eta_3)\eta_3^\alpha \geq 0,
\end{align}
and a sufficient condition to prove the last inequality is obtained by replacing $\eta_1^\alpha$ with $\eta_2^\alpha$ in the second term (since $\eta_1 >\eta_2$):
\begin{align}
    &(\eta_1 - \eta_3)\eta_2^\alpha + (1-\eta_1)\eta_2^\alpha -(1-\eta_3)\eta_3^\alpha \geq 0 \nonumber\\
    \Leftrightarrow\quad &(1 - \eta_3)\eta_2^\alpha  -(1-\eta_3)\eta_3^\alpha \geq 0,
\end{align}
where the last inequality is true since $\eta_2>\eta_3$. Hence, we showed that $g(\alpha)$ is a non-decreasing function of $\alpha$ and, in particular, it holds that:
\begin{align}
    g(\alpha) \geq g(2) \quad\forall\, \alpha\geq 2.
\end{align}
We now employ the last expression in \eqref{third-combination} and derive the following inequality:
\begin{align}
    (\eta_2 - \eta_3) &f_1 -(\eta_1 -\eta_3) f_2 + (\eta_1 -\eta_2)f_3 - \left[f_1 (1-\eta_3) -f_3 (1-\eta_1) - (\eta_1 - \eta_3)\right]  \nonumber\\
    &\geq g(2) \sum_{\alpha=2}^\infty y_\alpha \nonumber\\
    &= (\eta_1-\eta_3)(1-\eta_2)(1+\eta_2 -\eta_1 -\eta_3) \sum_{\alpha=2}^\infty y_\alpha,
\end{align}
which leads to the following upper bound on the weight of the state $\Bar{\sigma}$ outside of the ($\leq 1$)-subspace:
\begin{align}
    \sum_{\alpha=2}^\infty y_\alpha &\leq \frac{(\eta_2 - \eta_3) f_1 -(\eta_1 -\eta_3) f_2 + (\eta_1 -\eta_2)f_3 - \left[f_1 (1-\eta_3) -f_3 (1-\eta_1) - (\eta_1 - \eta_3)\right]}{(\eta_1-\eta_3)(1-\eta_2)(1+\eta_2 -\eta_1 -\eta_3)} \nonumber\\
    &= \frac{(1-\eta_2 )(f_3- f_1) + (\eta_1 -\eta_3) (1-f_2)}{(\eta_1-\eta_3)(1-\eta_2)(1+\eta_2 -\eta_1 -\eta_3)} \label{third-combination-inequality}.
\end{align}
Then, the upper bound in the original variables reads:
\begin{align}
   \sum_{\alpha=2}^\infty \Tr\left[\bar{\sigma} \pia\right] \leq \overline{w}^{>1}_{\mathcal{Z}}, \label{yalpha>1-upperbound}
\end{align}
with:
\begin{align}
    \overline{w}^{>1}_{\mathcal{Z}}:= \min\left\lbrace 1, \frac{\overline{(\eta_\uparrow - \eta_\downarrow)Y^Z_{1,(\eta_2,\eta_2)} - (1-\eta_2)(Y^Z_{1,(\eta_\downarrow,\eta_\downarrow)} - Y^Z_{1,(\eta_\uparrow,\eta_\uparrow)}) -(\eta_\uparrow -\eta_\downarrow) p^{\mathcal{Z}}_d}}{(1-p^{\mathcal{Z}}_d)(\eta_\uparrow - \eta_\downarrow)(1-\eta_2)(1+\eta_2 -\eta_\uparrow - \eta_\downarrow)}  \right\rbrace, \label{yalpha>1-upperbound-explicit}
\end{align}
where in Appendix~\ref{app:decoy} we derive an upper bound tailored to the specific combination of yields in the numerator.

The bound in \eqref{yalpha>1-upperbound} plays a crucial role in reducing the calculation of the phase error rate to the subspace with at most one photon in $\mathcal{Z}$, as we show in the next part of the proof.\\

\subsubsection{Reduction of the phase error rate to ($\leq 1$)-subspace}\label{app:PER4}
We now employ the expression \eqref{Zclick-KGmeasurement}, derived for the detection POVM element $Z_\checkmark$, in the numerator of the phase error rate \eqref{phase-error-rate-appendix}. By combining this with the upper bound in \eqref{yalpha>1-upperbound}, we reduce the calculation of the phase error rate to the subspace defined by $\Pi^0_{\mathcal{Z}} + \Pi^1_{\mathcal{Z}}$.

To start with, from \eqref{Zclick-KGmeasurement} we compute:
\begin{align}
    \sqrt{Z_\checkmark} &= M^{\leq 1}_{\mathcal{Z}} + M^{> 1}_{\mathcal{Z}},  \label{Zclick-KGmeasurement2}
\end{align}
where we introduced two positive-semidefinite operators:
\begin{align}
    M^{\leq 1}_{\mathcal{Z}} &= \sum_{\alpha=0}^1 \sqrt{p_{\checkmark|\alpha}} \pia  \label{M^leq1_Z}\\
    M^{> 1}_{\mathcal{Z}} &= \sum_{\alpha=2}^\infty \sqrt{p_{\checkmark|\alpha}} \pia.
\end{align}
Using \eqref{Zclick-KGmeasurement2}, the numerator of the phase error rate in \eqref{phase-error-rate-appendix} decomposes into three contributions:
\begin{align}
  \tilde{e}_{X,1} Y^Z_{1,(\eta_\downarrow,\eta_\downarrow)} &= \frac{1}{d} {\textstyle\sum_{k=0}^{d-1}} \Tr\left[\sigma_k  \sqrt{Z_\checkmark} {\textstyle\sum_{k'\neq k}}\tilde{X}_{k'} \sqrt{Z_\checkmark} \right] =\Delta_1 + \Delta_2 +\Delta_3 \label{phase-error-rate-num} ,
\end{align}
where:
\begin{align}
  \Delta_1 &= \frac{1}{d}{\textstyle\sum_{k=0}^{d-1}} \Tr\left[\sigma_k  M^{\leq 1}_{\mathcal{Z}} {\textstyle\sum_{k'\neq k}}\tilde{X}_{k'} M^{\leq 1}_{\mathcal{Z}} \right] \\
  \Delta_2 &= \frac{1}{d}{\textstyle\sum_{k=0}^{d-1}} \Tr\left[\sigma_k  (M^{\leq 1}_{\mathcal{Z}} {\textstyle\sum_{k'\neq k}}\tilde{X}_{k'} M^{> 1}_{\mathcal{Z}} + \mathrm{h.c.}) \right] \\
  \Delta_3 &= \frac{1}{d}{\textstyle\sum_{k=0}^{d-1}} \Tr\left[\sigma_k  M^{> 1}_{\mathcal{Z}} {\textstyle\sum_{k'\neq k}}\tilde{X}_{k'} M^{> 1}_{\mathcal{Z}} \right],
\end{align}
where ``$\mathrm{h.c.}$'' stands for the Hermitian conjugate of the term in the round brackets.

While, for the term $\Delta_1$, the only components of the states $\sigma_k$ that contribute are those localized in the ($\leq 1$)-subspace (due to $M_{\mathcal{Z}}^{\leq 1}$), the other two terms ($\Delta_2$ and $\Delta_3$) receive contributions from infinitely many subspaces with an arbitrary number of photons in the interval $\mathcal{Z}$. As we will show later, the statistics collected for different TBS settings allow us to estimate the contributions to the phase error rate coming from states in the ($\leq 1$)-subspace. Hence, we can precisely estimate the value of $\Delta_1$. Conversely, we upper bound $\Delta_2$ and $\Delta_3$ by virtue of our characterization of the weight of the states $\sigma_k$ in the subspace with more than one photon in $\mathcal{Z}$, embodied by the upper bound \eqref{yalpha>1-upperbound}.

We start by deriving an upper bound on $\Delta_3$. The fact that $\one -{\textstyle\sum_{k'\neq k}}\tilde{X}_{k'} \geq 0$ holds, implies that $M^{> 1}_{\mathcal{Z}} (\one -{\textstyle\sum_{k'\neq k}}\tilde{X}_{k'})M^{> 1}_{\mathcal{Z}} \geq 0$ also holds and thus that:
\begin{align}
    M^{> 1}_{\mathcal{Z}}{\textstyle\sum_{k'\neq k}}\tilde{X}_{k'} M^{> 1}_{\mathcal{Z}} &\leq (M^{> 1}_{\mathcal{Z}})^2 \nonumber\\
    &= \sum_{\alpha=2}^\infty p_{\checkmark|\alpha} \pia \nonumber\\
    &\leq \sum_{\alpha=2}^\infty \pia, \label{e3-bound-comp}
\end{align}
where we used the fact that $\pia$ are orthogonal projectors and that $p_{\checkmark|\alpha}$ are probabilities. Then, \eqref{e3-bound-comp} implies the following bound on $\Delta_3$:
\begin{align}
    \Delta_3 &\leq \sum_{\alpha=2}^\infty \Tr\left[ \bar{\sigma}  \pia \right] \nonumber\\
    &\leq \overline{w}^{>1}_{\mathcal{Z}},  \label{e3-bound}
\end{align}
where we used \eqref{barsigma} and \eqref{yalpha>1-upperbound}.

We now derive an upper bound on the term $\Delta_2$ in \eqref{phase-error-rate-num}, taking inspiration from Appendix~B in \cite{Lutkenhaus-imperfectphaserandom}. First, we recast $\Delta_2$ as follows:
\begin{align}
    \Delta_2 =\frac{1}{d} {\textstyle \sum_{k=0}^{d-1}} \Tr[H_k \Gamma_k],
\end{align}
where for brevity we introduced  the positive operator $\Gamma_k$ and the self-adjoint operator $H_k$:
\begin{align}
    H_k &:= M^{\leq 1}_{\mathcal{Z}} \sigma_k M^{>1}_{\mathcal{Z}}+ \mathrm{h.c.}  \label{Hk} \\
    \Gamma_k &:= {\textstyle\sum_{k'\neq k}}\tilde{X}_{k'}.
\end{align}
Since $H_k$ is self-adjoint, we can decompose it as:
\begin{align}
    H_k = H^+_k - H^-_k \label{Hkdec}
\end{align}
for some positive operators $H^+_k$ and $H^-_k$. Then we can upper bound $\Delta_2$ by:
\begin{align}
    \Delta_2 &=\frac{1}{d} {\textstyle \sum_{k=0}^{d-1}} \left(\Tr[H^+_k \Gamma_k] - \Tr[H^-_k \Gamma_k]\right) \nonumber\\
    &\leq \frac{1}{d} {\textstyle \sum_{k=0}^{d-1}} \Tr[H^+_k \Gamma_k] \nonumber\\
    &\leq \frac{1}{d} {\textstyle \sum_{k=0}^{d-1}} \Tr[H^+_k ], \label{e2-bound1}
\end{align}
where in the first inequality we used the fact that the trace of the product of two positive operators is non-negative, and in the second inequality that $\Gamma_k \leq \one$. Now we use \eqref{barsigma} to define the self-adjoint operator:
\begin{align}
    H:= M^{\leq 1}_{\mathcal{Z}} \bar{\sigma} M^{>1}_{\mathcal{Z}}+ \mathrm{h.c.}  \label{H}
\end{align}
and observe that $H$, through \eqref{Hk} and \eqref{Hkdec}, can be written as the difference of two positive operators:
\begin{align}
    H &=  \frac{1}{d} {\textstyle \sum_{k=0}^{d-1}} H^+_k - \frac{1}{d} {\textstyle \sum_{k=0}^{d-1}} H^-_k \\
    &\equiv H^+ - H^-.
\end{align}
Hence, by definition of the one-norm it holds:
\begin{align}
    \norm{H}_1 = \Tr[H^+] + \Tr[H^-].
\end{align}
Moreover, since $\Tr[H]=0$ by the definition of $H$, we have:
\begin{align}
    \Tr[H^+] = \Tr[H^-] = \frac{1}{2}\norm{H}_1. \label{normHequality}
\end{align}
By using this in \eqref{e2-bound1}, we obtain:
\begin{align}
    \Delta_2 \leq \frac{1}{2}\norm{H}_1 \label{e2-bound2}.
\end{align}
Now, we use the definition of $H$ in \eqref{H} to compute:
\begin{align}
    \Delta_2 &\leq \frac{1}{2} \Tr[\sqrt{H^2}] \nonumber\\
    &= \frac{1}{2}\Tr\left[\sqrt{M^{\leq 1}_{\mathcal{Z}} \bar{\sigma} \left(M^{>1}_{\mathcal{Z}}\right)^2\bar{\sigma} M^{\leq 1}_{\mathcal{Z}}}\right] + \frac{1}{2}\Tr\left[\sqrt{M^{> 1}_{\mathcal{Z}} \bar{\sigma} \left(M^{\leq 1}_{\mathcal{Z}}\right)^2\bar{\sigma} M^{> 1}_{\mathcal{Z}}}\right] \nonumber\\
    &=  \norm{M^{\leq 1}_{\mathcal{Z}} \bar{\sigma} M^{>1}_{\mathcal{Z}}}_1, \label{normHcomp}
\end{align}
where we used the fact that $M^{\leq 1}_{\mathcal{Z}}$ and $M^{>1}_{\mathcal{Z}}$ have orthogonal support and that $\norm{A}_1=\Tr[\sqrt{A^\dag A}]=\Tr[\sqrt{A A^\dag}]$. From Eq.~(3) in \cite{Bhatia-CauchySchwarz}, which we report here for clarity:
\begin{align}
    \norm{A^\dag B}_1 \leq \sqrt{\Tr[AA^\dag] \Tr[BB^\dag]}, \label{BhatiaCS}
\end{align}
we can upper bound the right-hand side in \eqref{normHcomp} with the choices $A^\dag=M^{\leq 1}_{\mathcal{Z}} \sqrt{\bar{\sigma}}$ and $B=\sqrt{\bar{\sigma}} M^{>1}_{\mathcal{Z}}$, as follows:
\begin{align}
    \Delta_2 &\leq  \sqrt{\Tr\left[\bar{\sigma} \left(M^{\leq 1}_{\mathcal{Z}}\right)^2\right] \Tr\left[\bar{\sigma} \left(M^{> 1}_{\mathcal{Z}}\right)^2 \right]} \nonumber\\
    &=\sqrt{\left\lbrace\Tr\left[\bar{\sigma} Z_\checkmark\right] - \Tr\left[\bar{\sigma} \left(M^{> 1}_{\mathcal{Z}}\right)^2 \right] \right\rbrace  \Tr\left[\bar{\sigma} \left(M^{> 1}_{\mathcal{Z}}\right)^2 \right]}, \label{e2-bound3}
\end{align}
where in the last equality we used \eqref{Zclick-KGmeasurement2}. Now we notice that $\Tr\left[\bar{\sigma} Z_\checkmark\right]=Y^Z_{1,(\eta_\downarrow,\eta_\downarrow)}$ by \eqref{prob-Zclick}. Moreover, from the derivation of the bound on $\Delta_3$ we know that:
\begin{align}
    0 \leq \Tr\left[\bar{\sigma} \left(M^{> 1}_{\mathcal{Z}}\right)^2 \right] \leq \overline{w}^{>1}_{\mathcal{Z}}.
\end{align}
Thus, we can upper bound the right-hand side in \eqref{e2-bound3} as follows:
\begin{align}
    \Delta_2 \leq \max_{x \in [0,\overline{w}^{>1}_{\mathcal{Z}}]}  \sqrt{\left(\overline{Y^Z_{1,(\eta_\downarrow,\eta_\downarrow)}} - x \right)  x},
\end{align}
where we replaced the yield with an upper bound derived with the decoy-state method (see Appendix~\ref{app:decoy}). Finally, we solve the optimization in the last expression and obtain:
\begin{align}
    \Delta_2 \leq  \sqrt{\left(\overline{Y^Z_{1,(\eta_\downarrow,\eta_\downarrow)}} - \min\left\lbrace \overline{w}^{>1}_{\mathcal{Z}} , \frac{\overline{Y^Z_{1,(\eta_\downarrow,\eta_\downarrow)}}}{2}\right\rbrace \right)  \min\left\lbrace \overline{w}^{>1}_{\mathcal{Z}} , \frac{\overline{Y^Z_{1,(\eta_\downarrow,\eta_\downarrow)}}}{2}\right\rbrace}. \label{e2-bound-final}
\end{align}
We remark that, in a similar way, one could derive a lower bound on $\Delta_2$. In particular, one can lower bound $\Delta_2$ starting from \eqref{e2-bound1}:
\begin{align}
    \Delta_2 &=\frac{1}{d} {\textstyle \sum_{k=0}^{d-1}} \left(\Tr[H^+_k \Gamma_k] - \Tr[H^-_k \Gamma_k]\right) \nonumber\\
    &\geq -\frac{1}{d} {\textstyle \sum_{k=0}^{d-1}} \Tr[H^-_k \Gamma_k] \nonumber\\
    &\geq -\frac{1}{d} {\textstyle \sum_{k=0}^{d-1}} \Tr[H^-_k ] \nonumber\\
    &= - \Tr[H^-],
\end{align}
and then using the result in \eqref{normHequality} to upper bound $\Tr[H^-]$, thus obtaining:
\begin{align}
    \Delta_2 \geq -  \sqrt{\left(\overline{Y^Z_{1,(\eta_\downarrow,\eta_\downarrow)}} - \min\left\lbrace \overline{w}^{>1}_{\mathcal{Z}} , \frac{\overline{Y^Z_{1,(\eta_\downarrow,\eta_\downarrow)}}}{2}\right\rbrace \right)  \min\left\lbrace \overline{w}^{>1}_{\mathcal{Z}} , \frac{\overline{Y^Z_{1,(\eta_\downarrow,\eta_\downarrow)}}}{2}\right\rbrace}. \label{Delta2-lowerbound}
\end{align}

By employing \eqref{e3-bound} and \eqref{e2-bound-final} in \eqref{phase-error-rate-num}, we arrive at the following upper bound on the numerator of the phase error rate:
\begin{align}
    \tilde{e}_{X,1} Y^Z_{1,(\eta_\downarrow,\eta_\downarrow)} \leq \Delta_1 + \overline{\Delta_2} + \overline{w}^{>1}_{\mathcal{Z}} \label{phase-error-rate-uppbound},
\end{align}
where $\overline{w}^{>1}_{\mathcal{Z}}$ is given in \eqref{yalpha>1-upperbound-explicit}, while $\overline{\Delta_2}$ and $\Delta_1$ are defined as:
\begin{align}
    \overline{\Delta_2} &= \sqrt{\left(\overline{Y^Z_{1,(\eta_\downarrow,\eta_\downarrow)}} - \min\left\lbrace \overline{w}^{>1}_{\mathcal{Z}} , \frac{\overline{Y^Z_{1,(\eta_\downarrow,\eta_\downarrow)}}}{2}\right\rbrace \right)  \min\left\lbrace \overline{w}^{>1}_{\mathcal{Z}} , \frac{\overline{Y^Z_{1,(\eta_\downarrow,\eta_\downarrow)}}}{2}\right\rbrace} \label{Delta2bar}\\
    \Delta_1 &= \frac{1}{d}{\textstyle\sum_{k=0}^{d-1}} \Tr\left[\sigma_k  M^{\leq 1}_{\mathcal{Z}} {\textstyle\sum_{k'\neq k}}\tilde{X}_{k'} M^{\leq 1}_{\mathcal{Z}} \right] \label{e1},
\end{align}
and $M^{\leq 1}_{\mathcal{Z}}$ is given in \eqref{M^leq1_Z}.

As anticipated, the right-hand side of the bound in \eqref{phase-error-rate-uppbound} only depends (through $\Delta_1$) on the statistics generated by states that contain up to one photon in the interval $\mathcal{Z}$. In particular, $\Delta_1$ is linked to the error rate in distinguishing the $X$-basis states sent by Alice with Bob's test measurement, \eqref{Bobs-X-measurement}, and can be estimated with the rich statistics of the test rounds. For this, we now focus on deriving explicit expressions of Bob's POVM elements, \eqref{Bobs-POVM}.\\

\subsubsection{Bob's $X$-basis detector}\label{app:PER5}
The detection statistics that are needed in order to estimate $\Delta_1$ are those generated by the following reduced POVM elements:
\begin{align}
    X^{(\eta_i,\eta_l)}_{k,\checkmark} &:= \sum_{j=0}^{d-1} M^{(\eta_i,\eta_l)}_{j,k} \label{Xkdet-a} \\
    X^{(\eta_i,\eta_l)}_{k,\emptyset} &:= M^{(\eta_i,\eta_l)}_{\emptyset,k}, \label{Xknodet-a}
\end{align}
where we ignore the specific outcome of the $Z$-basis detector and only register whether it clicks or not. Note that Bob's test measurement, as defined in \eqref{Bobs-X-measurement}, is recovered from \eqref{Xkdet-a} and \eqref{Xknodet-a} as follows:
\begin{align}
    X_k = X^{(\eta_\uparrow,\eta_\uparrow)}_{k,\checkmark} +  X^{(\eta_\uparrow,\eta_\uparrow)}_{k,\emptyset}. \label{Bobs-X-measurement-appendix}
\end{align}
Now, our goal is to derive explicit expressions for the POVM elements (acting on mode $a$) given in \eqref{Xkdet-a} and \eqref{Xknodet-a}.

The first step is to describe the ideal $X$-basis detector acting on mode $e$ of Fig.~\ref{fig:Bobs-setup-appendix-factored}. Let $D:=\{0,1,\dots,d-1\}$ be the set of outcomes of the $X$-basis detector on mode $e$ and let $\mathcal{P}(D)$ be the power set of $D$. Recall that we assumed that Bobs maps multi-click events $K \in \mathcal{P}(D)$, i.e.\ events with clicks in the physical modes $\mathcal{X}_l$, for $l\in K$, to a single outcome. This can be formalized by an un-normalized right stochastic matrix $P\in \mathrm{M}_{\mathcal{P}(D)\setminus\emptyset \, \times \, d}$, with elements:
\begin{align}
    \{P\}_{K,k} := \Pr(k|K) \,\,, \label{stochastic-matrix}
\end{align}
where $\Pr(k|K)$ represents the probability that the multi-click event $K$ is mapped to the single outcome $k$. As explained in Sec.~\ref{sec:protocol}, we assume that the following holds:
\begin{align}
    \sum_{k=0}^{d-1} \Pr(k|K) =  1 \quad\forall\, K \in \mathcal{P}(D) \setminus\emptyset, \label{normalization-remapping}
\end{align}
i.e.\ that no event with at least one click is assigned to the no detection event. In other words, every time there is at least one click (event $K\neq \emptyset$), Bob assigns the event to outcome $k$ according to the distribution $\Pr(k|K)$. Then, the POVM element corresponding to a detection of outcome $k$, in the absence of dark counts, would be given by:
\begin{align}
    (X_k)_e &= \sum_{K\in\mathcal{P}(D)\setminus\emptyset} \Pr(k|K) \,\, \Pi^{\geq 1}_K \label{Xk-no-dk}
\end{align}
where $\Pi^{\geq 1}_K$ is the projector on the subspace with at least one photon in mode $\mathcal{X}_l$, for each $l\in K$, and no photons in the other measured modes $\mathcal{X}_m$, for $m\notin K$. Note that the event $K=\emptyset$ is excluded in \eqref{Xk-no-dk} since no detection events are not mapped to measurement outcomes. We now consider that each mode $\mathcal{X}_k$ may click with probability $p^X_d$, given that no photon is in that mode. Thus, we deduce that even the events where some of the clicks are caused by dark counts are mapped according to \eqref{stochastic-matrix}, since Bob has no way to distinguish dark counts from actual detections. Therefore, the POVM element for outcome $k$ in the presence of dark counts becomes:
\begin{align}
    (X_k)_e &= \sum_{K\in\mathcal{P}(D)\setminus\emptyset} \Pr(k|K) \,\, X_K \label{Xk-with-dk}
\end{align}
where $X_K$ is the POVM element corresponding to the multi-click event $K$. Clearly, $X_K$ receives contributions from $\Pi^{\geq 1}_K$, but also from all the projectors $\Pi^{\geq 1}_S$, where $S\in\mathcal{P}(K)$ can be interpreted  as the set of ``genuine'' clicks (i.e.\ clicks caused by photons) in the multi-click event $K$, provided that $|K|-|S|$ dark counts also occur ($|K|$ represents the number of elements in $K$). In conclusion, we have:
\begin{align}
    X_K = \sum_{S \in \mathcal{P}(K)} (p^X_d)^{|K|-|S|}(1-p^X_d)^{d-|K|} \,\,\Pi^{\geq 1}_S. \label{XK-unit-efficiency}
\end{align}

\begin{remark} \label{rmk:reduction-to-Zclick}
    By using \eqref{Xk-with-dk} and \eqref{XK-unit-efficiency} in the computation of the POVM element corresponding to a detection in any outcome, $(X_\checkmark)_e$, one obtains:
    \begin{align}
        (X_\checkmark)_e &= \sum_{k=0}^{d-1} (X_k)_e \\
        &= \left[1-(1-p^X_d)^d\right]  \Pi^{\geq 1}_\emptyset  + \sum_{X\in\mathcal{P}(D)\setminus\emptyset} \Pi^{\geq 1}_X,
    \end{align}
    which coincides with the analogous POVM element in the $Z$ basis, \eqref{Zclick-cprime}, by using \eqref{prob-dk-tot-X} and by identifying: $\Pi^{\geq 1}_\emptyset = \Pi^0_{\mathcal{X}}$ and $\sum_{X\in\mathcal{P}(D)\setminus\emptyset} \Pi^{\geq 1}_X = \sum_{\beta=1}^\infty \Pi^\beta_{\mathcal{X}}$, with $\Pi^\beta_{\mathcal{X}}$ being the analogous for the $X$ basis of $\pia$ for the $Z$ basis, namely the projector on the subspace with $\beta$ photons in the interval $\mathcal{X}$ and any number of photons outside of $\mathcal{X}$. With these substitutions, we obtain:
    \begin{align}
        (X_\checkmark)_e &= p^{\mathcal{X}}_d \one    + (1-p^{\mathcal{X}}_d)  \sum_{\beta=1}^\infty \Pi^\beta_{\mathcal{X}}. \label{Xdet-mode-e}
    \end{align}
\end{remark}

Now, we can compute the POVM element corresponding to outcome $k$ in the $X$-basis detector, acting on the transmitted mode $c$. According to the model in Fig.~\ref{fig:Bobs-setup-appendix-factored} and by following the same approach used to compute \eqref{Zclick-reflectedTBS-simplified-modea} for the $Z$-basis detector, this would be given by:
\begin{align}
    (X_k)_c = \bra{vac}_d U^\dag(\eta_r) (X_k)_e \otimes\one_f U(\eta_r) \ket{vac}_d.  \label{Xk-mode-c2}
\end{align}
Alternatively, a faster way to obtain $(X_k)_c$ is to observe that, when each photon can be lost independently with probability $1-\eta_r$, the POVM element $X_K$ of the multi-click event $K$ receives contributions from any possible configuration of photons arriving at the detector. Thus, the relevant projector to express the operator in \eqref{Xk-mode-c2} becomes:
\begin{align}
    \Pi^{\alpha_0,\alpha_1,\dots,\alpha_{d-1}}_{\mathcal{X}}, \label{proj-a0-a(d-1)}
\end{align}
which is defined as the projector on the subspace with $\alpha_k$ photons in mode $\mathcal{X}_k$, for $k=0,\dots,d-1$. Moreover, we note that $X_K$ will have the same structure as in \eqref{XK-unit-efficiency}, if we interpret $S$ as the set of modes that contain at least one photon after considering the detector's non-unit 
efficiency  (i.e.\ after applying $U(\eta_r)$):
\begin{align}
    X_K = \sum_{S \in \mathcal{P}(K)} (p^X_d)^{|K|-|S|}(1-p^X_d)^{d-|K|} \,\,M^{\geq 1}_S, \label{XK-nonunit-efficiency}
\end{align}
where $M^{\geq 1}_S$ is the POVM element of at least one photon in mode $\mathcal{X}_s$, after loss, for every $s \in S$. Thus, we can easily express $M^{\geq 1}_S$ in terms of the projectors \eqref{proj-a0-a(d-1)} as follows:
\begin{align}
    M^{\geq 1}_S = \sum_{\alpha_0,\alpha_1,\dots,\alpha_{d-1}=0}^\infty \Pi^{\alpha_0,\alpha_1,\dots,\alpha_{d-1}}_{\mathcal{X}} \prod_{s\in S} \left[1-(1-\eta_r)^{\alpha_s}\right] \prod_{r \in S^c} (1-\eta_r)^{\alpha_r}, \label{M_S}
\end{align}
where $S^c = D \setminus S$ is the complement of $S$ in $D$.

In conclusion, the POVM element for outcome $k$ in the $X$-basis detector with efficiency $\eta_r$ and dark count probability $p^X_d$ per mode, is given by:
\begin{align}
    (X_k)_c &= \sum_{K\in\mathcal{P}(D)\setminus\emptyset} \Pr(k|K) \,\, \sum_{S \in \mathcal{P}(K)} (p^X_d)^{|K|-|S|}(1-p^X_d)^{d-|K|} \,\,M^{\geq 1}_S, \label{Xk-mode-c}
\end{align}
with $M^{\geq 1}_S$ given in \eqref{M_S}. Moreover, it will turn out useful to calculate the POVM element corresponding to a detection of any outcome in mode $c$. This is achieved from  \eqref{Xk-mode-c2} as follows:
\begin{align}
    (X_\checkmark)_c &= \sum_{k=0}^{d-1} (X_k)_c \nonumber\\
    &= \bra{vac}_d U^\dag(\eta_r) (X_\checkmark)_e \otimes\one_f U(\eta_r) \ket{vac}_d \nonumber\\
    &= p^{\mathcal{X}}_d \one_c    + (1-p^{\mathcal{X}}_d) \bra{vac}_d U^\dag(\eta_r) \sum_{\beta=1}^\infty (\Pi^\beta_{\mathcal{X}})_e  \otimes\one_f U(\eta_r) \ket{vac}_d 
\end{align}
where we used \eqref{Xdet-mode-e} from Remark~\ref{rmk:reduction-to-Zclick} in the last equality. Now we use the fact that a similar calculation to the one above has already been done for computing \eqref{Qa} and resulted in \eqref{Qa2}. Analogously, we obtain:
\begin{align}
    (X_\checkmark)_c &= p^{\mathcal{X}}_d \one_c    + (1-p^{\mathcal{X}}_d)  \sum_{\beta=1}^\infty \left[1-(1-\eta_r)^\beta\right] (\Pi^\beta_{\mathcal{X}})_c \nonumber\\
    &= \sum_{\beta=0}^\infty \left[1-(1-p^{\mathcal{X}}_d)(1-\eta_r)^\beta\right] (\Pi^\beta_{\mathcal{X}})_c  \label{Xclick-mode-c}.
\end{align}

Now, we have all the ingredients to compute the POVM elements defined in \eqref{Xkdet-a} and \eqref{Xknodet-a}. According to the scheme in Fig.~\ref{fig:Bobs-setup-appendix-factored}, we have:
\begin{align}
    X^{(\eta_i,\eta_l)}_{k,\checkmark} &= \bra{vac}_b U_{(\eta_i,\eta_l)}^\dag (X_k)_c \otimes (Z_\checkmark)_{c'} U_{(\eta_i,\eta_l)} \ket{vac}_b \label{Xkdet-a2} \\
    X^{(\eta_i,\eta_l)}_{k,\emptyset} &= \bra{vac}_b U_{(\eta_i,\eta_l)}^\dag (X_k)_c \otimes (Z_\emptyset)_{c'} U_{(\eta_i,\eta_l)} \ket{vac}_b  , \label{Xknodet-a2}
\end{align}
where $(X_k)_c$ is given in \eqref{Xk-mode-c}, $(Z_\checkmark)_{c'}$ is given in \eqref{Zclick-cprime} and $(Z_\emptyset)_{c'}=\one_{c'} -(Z_\checkmark)_{c'}$.

Now that we derived the POVM elements of interest, in the following we link them to the corresponding one-photon yields and bit error rates, which in turn will be used to bound $\Delta_1$ in \eqref{phase-error-rate-uppbound}.\\

\subsubsection{The one-photon $X$-basis yields and bit error rates}\label{app:PER6}
By applying the decoy-state method on the detection statistics of the test rounds (cf.~Appendix~\ref{app:decoy}), Alice and Bob can estimate the following one-photon $X$-basis yields, which we call the ``test-round yields'':
\begin{align}
    Y^{X,\checkmark}_{1,(\eta_i,\eta_l)} &= {\textstyle\sum_{k=0}^{d-1}} \Tr\left[\frac{\sigma_k}{d}   X^{(\eta_i,\eta_l)}_{\checkmark,\checkmark}  \right] \label{yieldX-click}\\
    Y^{X,\emptyset}_{1,(\eta_i,\eta_l)} &= {\textstyle\sum_{k=0}^{d-1}} \Tr\left[\frac{\sigma_k}{d}   X^{(\eta_i,\eta_l)}_{\checkmark,\emptyset}  \right], \label{yieldX-noclick}
\end{align}
where we defined:
\begin{align}
    X^{(\eta_i,\eta_l)}_{\checkmark,\checkmark} &=\sum_{k=0}^{d-1} X^{(\eta_i,\eta_l)}_{k,\checkmark}  \label{Xdetdet-a}\\
    X^{(\eta_i,\eta_l)}_{\checkmark,\emptyset}&=\sum_{k=0}^{d-1} X^{(\eta_i,\eta_l)}_{k,\emptyset}.
\end{align}
Similarly, by applying the decoy-state method on the $X$-basis gains \eqref{gaintest-Zclick} and \eqref{gaintest-Znoclick} and on the $X$-basis QBERs \eqref{QBER-X-Zclick} and \eqref{QBER-X-Znoclick}, the parties can estimate the following one-photon $X$-basis bit error rates, also called the ``test-round bit error rates'':
\begin{align}
    e_{X,1,(\eta_i,\eta_l),\checkmark} &= \frac{{\textstyle\sum_{k=0}^{d-1}} \Tr\left[\frac{\sigma_k}{d}  {\textstyle\sum_{k'\neq k}}X^{(\eta_i,\eta_l)}_{k',\checkmark}  \right]}{Y^{X,\checkmark}_{1,(\eta_i,\eta_l)}} \label{bit-error-rate-click} \\
    e_{X,1,(\eta_i,\eta_l),\emptyset} &= \frac{{\textstyle\sum_{k=0}^{d-1}} \Tr\left[\frac{\sigma_k}{d}  {\textstyle\sum_{k'\neq k}}X^{(\eta_i,\eta_l)}_{k',\emptyset}  \right]}{Y^{X,\emptyset}_{1,(\eta_i,\eta_l)}}. \label{bit-error-rate-noclick}
\end{align}

We observe that the test-round yields, \eqref{yieldX-click} and \eqref{yieldX-noclick}, and the test-round bit error rates, \eqref{bit-error-rate-click} and \eqref{bit-error-rate-noclick}, do not depend only on the component of $\sigma_k$ in the ($\leq 1$)-subspace. Conversely, this is the only component that matters when computing $\Delta_1$ (cf.~\eqref{e1}). Therefore, similarly to the derivation of the upper bound on the phase error rate \eqref{phase-error-rate-uppbound}, we now derive upper and lower bounds on the test-round yields and test-round bit error rates, such that they only depend on the ($\leq 1$)-subspace component of $\sigma_k$. The derivation of the bounds is identical for each of the four equations: \eqref{yieldX-click}, \eqref{yieldX-noclick}, \eqref{bit-error-rate-click}, and \eqref{bit-error-rate-noclick}; thus, we focus on deriving the bounds for \eqref{bit-error-rate-click}. The first step entails using the fact that $\sum_{\alpha=0}^\infty \pia =\one$, such that the numerator in \eqref{bit-error-rate-click} can be recast as follows:
\begin{align}
    Y^{X,\checkmark}_{1,(\eta_i,\eta_l)} e_{X,1,(\eta_i,\eta_l),\checkmark} &= {\textstyle\sum_{k=0}^{d-1}} \Tr\left[\frac{\sigma_k}{d} ({\textstyle\sum_{\alpha=0}^\infty} \pia ){\textstyle\sum_{k'\neq k}}X^{(\eta_i,\eta_l)}_{k',\checkmark} ({\textstyle\sum_{\alpha=0}^\infty} \pia) \right] \nonumber\\
    &= {\textstyle\sum_{k=0}^{d-1}} \Tr\left[\frac{\sigma_k}{d} \Pi^{\leq 1}_{\mathcal{Z}} {\textstyle\sum_{k'\neq k}}X^{(\eta_i,\eta_l)}_{k',\checkmark} \Pi^{\leq 1}_{\mathcal{Z}} \right] + {\textstyle\sum_{k=0}^{d-1}} \Tr\left[\frac{\sigma_k}{d} \left(\Pi^{\leq 1}_{\mathcal{Z}} {\textstyle\sum_{k'\neq k}}X^{(\eta_i,\eta_l)}_{k',\checkmark} \Pi^{> 1}_{\mathcal{Z}} + \mathrm{h.c.}\right) \right] \nonumber\\
    &\quad+ {\textstyle\sum_{k=0}^{d-1}} \Tr\left[\frac{\sigma_k}{d} \Pi^{> 1}_{\mathcal{Z}} {\textstyle\sum_{k'\neq k}}X^{(\eta_i,\eta_l)}_{k',\checkmark} \Pi^{> 1}_{\mathcal{Z}} \right], \label{bit-error-rate-uppbound}
\end{align}
where we defined the projectors:
\begin{align}
    \Pi^{\leq 1}_{\mathcal{Z}} &:= \Pi^0_{\mathcal{Z}} + \Pi^1_{\mathcal{Z}} \\
    \Pi^{> 1}_{\mathcal{Z}} &:= \sum_{\alpha=2}^\infty \pia.
\end{align}
The three terms in \eqref{bit-error-rate-uppbound} closely resemble the three terms $\Delta_1$, $\Delta_2$ and $\Delta_3$ defined in \eqref{phase-error-rate-num} for the phase error rate. As a matter of fact, one can easily verify that the second and third term in \eqref{bit-error-rate-uppbound} can be bounded in an analogous manner to $\Delta_2$ and $\Delta_3$. In particular, for the third term we obtain:
\begin{align}
    {\textstyle\sum_{k=0}^{d-1}} \Tr\left[\frac{\sigma_k}{d} \Pi^{> 1}_{\mathcal{Z}} {\textstyle\sum_{k'\neq k}}X^{(\eta_i,\eta_l)}_{k',\checkmark} \Pi^{> 1}_{\mathcal{Z}} \right]  &\leq  \overline{w}^{>1}_{\mathcal{Z}} \label{uppbound-third},
\end{align}
where we used \eqref{e3-bound}. For the second term we can follow the derivation of the bound on $\Delta_2$ until \eqref{e2-bound3}, which in this case becomes:
\begin{align}
    {\textstyle\sum_{k=0}^{d-1}} \Tr\left[\frac{\sigma_k}{d} \left(\Pi^{\leq 1}_{\mathcal{Z}} {\textstyle\sum_{k'\neq k}}X^{(\eta_i,\eta_l)}_{k',\checkmark} \Pi^{> 1}_{\mathcal{Z}} + \mathrm{h.c.}\right) \right]&\leq \sqrt{\left\lbrace 1 - \Tr\left[\bar{\sigma} \Pi^{> 1}_{\mathcal{Z}} \right] \right\rbrace  \Tr\left[\bar{\sigma} \Pi^{> 1}_{\mathcal{Z}} \right]} \nonumber\\
    &\leq \sqrt{\overline{w}^{>1}_{\mathcal{Z}}} \label{uppbound-second}
\end{align}
By employing \eqref{uppbound-second} and \eqref{uppbound-third} in \eqref{bit-error-rate-uppbound}, we obtain the following upper bound on the numerator of the test-round bit error rate:
\begin{align}
    Y^{X,\checkmark}_{1,(\eta_i,\eta_l)} e_{X,1,(\eta_i,\eta_l),\checkmark} &\leq  \left[Y^{X,\checkmark}_{1,(\eta_i,\eta_l)} e_{X,1,(\eta_i,\eta_l),\checkmark}\right]^{\leq 1}_{\mathcal{Z}} +   \sqrt{\overline{w}^{>1}_{\mathcal{Z}}} + \overline{w}^{>1}_{\mathcal{Z}} . \label{bit-error-rate-uppbound2}
\end{align}
where $\overline{w}^{>1}_{\mathcal{Z}}$ is given in \eqref{yalpha>1-upperbound-explicit} and where we defined:
\begin{align}
    \left[Y^{X,\checkmark}_{1,(\eta_i,\eta_l)} e_{X,1,(\eta_i,\eta_l),\checkmark}\right]^{\leq 1}_{\mathcal{Z}} := {\textstyle\sum_{k=0}^{d-1}} \Tr\left[\frac{\sigma_k}{d} \Pi^{\leq 1}_{\mathcal{Z}} {\textstyle\sum_{k'\neq k}}X^{(\eta_i,\eta_l)}_{k',\checkmark} \Pi^{\leq 1}_{\mathcal{Z}} \right], \label{bit-error-rate-click-<1subspace}
\end{align}
as the numerator of the bit error rate, when the state is restricted to the ($\leq 1$)-subspace. For deriving a lower bound on the numerator of the test-round bit error rate, we use the fact that:
\begin{align}
    {\textstyle\sum_{k=0}^{d-1}} \Tr\left[\frac{\sigma_k}{d} \Pi^{> 1}_{\mathcal{Z}} {\textstyle\sum_{k'\neq k}}X^{(\eta_i,\eta_l)}_{k',\checkmark} \Pi^{> 1}_{\mathcal{Z}} \right]  &\geq 0 \label{lowbound-third},
\end{align}
since $\Pi^{> 1}_{\mathcal{Z}} {\textstyle\sum_{k'\neq k}}X^{(\eta_i,\eta_l)}_{k',\checkmark} \Pi^{> 1}_{\mathcal{Z}}\geq 0$. For the second term in \eqref{bit-error-rate-uppbound}, we can obtain a lower bound in an analogous way to the derivation of the lower bound on $\Delta_2$ in \eqref{Delta2-lowerbound}. We obtain:
\begin{align}
    {\textstyle\sum_{k=0}^{d-1}} \Tr\left[\frac{\sigma_k}{d} \left(\Pi^{\leq 1}_{\mathcal{Z}} {\textstyle\sum_{k'\neq k}}X^{(\eta_i,\eta_l)}_{k',\checkmark} \Pi^{> 1}_{\mathcal{Z}} + \mathrm{h.c.}\right) \right]
    &\geq - \sqrt{\overline{w}^{>1}_{\mathcal{Z}}} \label{lowbound-second}
\end{align}
By employing \eqref{lowbound-third} and \eqref{lowbound-second} in \eqref{bit-error-rate-uppbound}, we obtain the following lower bound on the numerator of the test-round bit error rate:
\begin{align}
    Y^{X,\checkmark}_{1,(\eta_i,\eta_l)} e_{X,1,(\eta_i,\eta_l),\checkmark} &\geq  \left[Y^{X,\checkmark}_{1,(\eta_i,\eta_l)} e_{X,1,(\eta_i,\eta_l),\checkmark}\right]^{\leq 1}_{\mathcal{Z}} -   \sqrt{\overline{w}^{>1}_{\mathcal{Z}}} . \label{bit-error-rate-lowbound}
\end{align}

Similar bounds to \eqref{bit-error-rate-uppbound2} and \eqref{bit-error-rate-lowbound} can be obtained for the other test-round bit error rate \eqref{bit-error-rate-noclick} and for the test-round yields \eqref{yieldX-click} and \eqref{yieldX-noclick}. Such bounds are important as they allow us to estimate the test-round yields and the (numerator of) the test-round bit error rates, when restricted to the ($\leq 1$)-subspace, with significant accuracy. Indeed, we can derive a narrow interval of existence for the quantity defined in \eqref{bit-error-rate-click-<1subspace}, by combining \eqref{bit-error-rate-uppbound2} and \eqref{bit-error-rate-lowbound}:
\begin{align}
     \underline{\left[Y^{X,\checkmark}_{1,(\eta_i,\eta_l)} e_{X,1,(\eta_i,\eta_l),\checkmark}\right]^{\leq 1}_{\mathcal{Z}}} &\leq \left[Y^{X,\checkmark}_{1,(\eta_i,\eta_l)} e_{X,1,(\eta_i,\eta_l),\checkmark}\right]^{\leq 1}_{\mathcal{Z}} \leq \overline{\left[Y^{X,\checkmark}_{1,(\eta_i,\eta_l)} e_{X,1,(\eta_i,\eta_l),\checkmark}\right]^{\leq 1}_{\mathcal{Z}}} \label{bit-error-rate-click-range} , 
\end{align}
where we defined:
\begin{align}
    \underline{\left[Y^{X,\checkmark}_{1,(\eta_i,\eta_l)} e_{X,1,(\eta_i,\eta_l),\checkmark}\right]^{\leq 1}_{\mathcal{Z}}} &:= \max \left\lbrace 0, \underline{Y^{X,\checkmark}_{1,(\eta_i,\eta_l)} e_{X,1,(\eta_i,\eta_l),\checkmark}} -   \sqrt{\overline{w}^{>1}_{\mathcal{Z}}} - \overline{w}^{>1}_{\mathcal{Z}} \right\rbrace \label{l-e,click} \\
    \overline{\left[Y^{X,\checkmark}_{1,(\eta_i,\eta_l)} e_{X,1,(\eta_i,\eta_l),\checkmark}\right]^{\leq 1}_{\mathcal{Z}}} &:=  \min \left\lbrace 1, \overline{Y^{X,\checkmark}_{1,(\eta_i,\eta_l)} e_{X,1,(\eta_i,\eta_l),\checkmark}} +  \sqrt{\overline{w}^{>1}_{\mathcal{Z}}} \right\rbrace \label{u-e,click},
\end{align}
and used the fact that: $ 0\leq  \Pi^{\leq 1}_{\mathcal{Z}} {\textstyle\sum_{k'\neq k}}X^{(\eta_i,\eta_l)}_{k',\checkmark} \Pi^{\leq 1}_{\mathcal{Z}} \leq \one$. Note that we replaced the product of the test-round yield and bit error rate with the respective upper or lower bound obtained from the decoy-state method (derived in Appendix~\ref{app:decoy}), such that \eqref{l-e,click} and \eqref{u-e,click} are exclusively given in terms of observed quantities.

Analogously, we define the restriction of the test-round bit error rate in \eqref{bit-error-rate-noclick} to the ($\leq 1$)-subspace as follows:
\begin{align}
    \left[Y^{X,\emptyset}_{1,(\eta_i,\eta_l)} e_{X,1,(\eta_i,\eta_l),\emptyset}\right]^{\leq 1}_{\mathcal{Z}}:= {\textstyle\sum_{k=0}^{d-1}} \Tr\left[\frac{\sigma_k}{d} \Pi^{\leq 1}_{\mathcal{Z}} {\textstyle\sum_{k'\neq k}}X^{(\eta_i,\eta_l)}_{k',\emptyset} \Pi^{\leq 1}_{\mathcal{Z}} \right], \label{bit-error-rate-noclick-<1subspace}
\end{align}
and derive the interval of existence:
\begin{align}
     \underline{\left[Y^{X,\emptyset}_{1,(\eta_i,\eta_l)} e_{X,1,(\eta_i,\eta_l),\emptyset}\right]^{\leq 1}_{\mathcal{Z}}} &\leq \left[Y^{X,\emptyset}_{1,(\eta_i,\eta_l)} e_{X,1,(\eta_i,\eta_l),\emptyset}\right]^{\leq 1}_{\mathcal{Z}} \leq \overline{\left[Y^{X,\emptyset}_{1,(\eta_i,\eta_l)} e_{X,1,(\eta_i,\eta_l),\emptyset}\right]^{\leq 1}_{\mathcal{Z}}} \label{bit-error-rate-noclick-range}
\end{align}
with:
\begin{align}
    \underline{\left[Y^{X,\emptyset}_{1,(\eta_i,\eta_l)} e_{X,1,(\eta_i,\eta_l),\emptyset}\right]^{\leq 1}_{\mathcal{Z}}} &:= \max \left\lbrace 0, \underline{Y^{X,\emptyset}_{1,(\eta_i,\eta_l)} e_{X,1,(\eta_i,\eta_l),\emptyset}}  -   \sqrt{\overline{w}^{>1}_{\mathcal{Z}}} - \overline{w}^{>1}_{\mathcal{Z}} \right\rbrace \label{l-e,noclick} \\
    \overline{\left[Y^{X,\emptyset}_{1,(\eta_i,\eta_l)} e_{X,1,(\eta_i,\eta_l),\emptyset}\right]^{\leq 1}_{\mathcal{Z}}} &:=  \min \left\lbrace 1, \overline{Y^{X,\emptyset}_{1,(\eta_i,\eta_l)} e_{X,1,(\eta_i,\eta_l),\emptyset}}  +   \sqrt{\overline{w}^{>1}_{\mathcal{Z}}} \right\rbrace \label{u-e,noclick} .
\end{align}
Similarly, we define the restriction of the test-round yields to the ($\leq 1$)-subspace:
\begin{align}
    \left[Y^{X,\checkmark}_{1,(\eta_i,\eta_l)} \right]^{\leq 1}_{\mathcal{Z}} &:= {\textstyle\sum_{k=0}^{d-1}} \Tr\left[\frac{\sigma_k}{d}  \Pi^{\leq 1}_{\mathcal{Z}} X^{(\eta_i,\eta_l)}_{\checkmark,\checkmark} \Pi^{\leq 1}_{\mathcal{Z}}  \right] \label{yieldX-click-<1subspace}\\
    \left[Y^{X,\emptyset}_{1,(\eta_i,\eta_l)}\right]^{\leq 1}_{\mathcal{Z}} &:= {\textstyle\sum_{k=0}^{d-1}} \Tr\left[\frac{\sigma_k}{d} \Pi^{\leq 1}_{\mathcal{Z}}  X^{(\eta_i,\eta_l)}_{\checkmark,\emptyset} \Pi^{\leq 1}_{\mathcal{Z}} \right]. \label{yieldX-noclick-<1subspace}
\end{align}
Then, we can obtain tight intervals of existence for such quantities:
\begin{align}
     \underline{\left[Y^{X,\checkmark}_{1,(\eta_i,\eta_l)} \right]^{\leq 1}_{\mathcal{Z}}} &\leq  \left[Y^{X,\checkmark}_{1,(\eta_i,\eta_l)} \right]^{\leq 1}_{\mathcal{Z}} \leq \overline{\left[Y^{X,\checkmark}_{1,(\eta_i,\eta_l)} \right]^{\leq 1}_{\mathcal{Z}}}  \label{yield-click-range} \\
     \underline{\left[Y^{X,\emptyset}_{1,(\eta_i,\eta_l)}\right]^{\leq 1}_{\mathcal{Z}}} &\leq  \left[Y^{X,\emptyset}_{1,(\eta_i,\eta_l)}\right]^{\leq 1}_{\mathcal{Z}} \leq \overline{\left[Y^{X,\emptyset}_{1,(\eta_i,\eta_l)}\right]^{\leq 1}_{\mathcal{Z}}} \label{yield-noclick-range},
\end{align}
with
\begin{align}
    \underline{\left[Y^{X,\checkmark}_{1,(\eta_i,\eta_l)} \right]^{\leq 1}_{\mathcal{Z}}} &:= \max \left\lbrace 0, \underline{Y^{X,\checkmark}_{1,(\eta_i,\eta_l)}}  -   \sqrt{\overline{w}^{>1}_{\mathcal{Z}}} - \overline{w}^{>1}_{\mathcal{Z}} \right\rbrace \label{l-Y,click} \\
    \overline{\left[Y^{X,\checkmark}_{1,(\eta_i,\eta_l)} \right]^{\leq 1}_{\mathcal{Z}}} &:= \min \left\lbrace 1, \overline{Y^{X,\checkmark}_{1,(\eta_i,\eta_l)}}  +   \sqrt{\overline{w}^{>1}_{\mathcal{Z}}} \right\rbrace  \label{u-Y,click}\\
    \underline{\left[Y^{X,\emptyset}_{1,(\eta_i,\eta_l)}\right]^{\leq 1}_{\mathcal{Z}}} &:= \max \left\lbrace 0, \underline{Y^{X,\emptyset}_{1,(\eta_i,\eta_l)}}  -   \sqrt{\overline{w}^{>1}_{\mathcal{Z}}} - \overline{w}^{>1}_{\mathcal{Z}} \right\rbrace  \label{l-Y,noclick}\\ 
    \overline{\left[Y^{X,\emptyset}_{1,(\eta_i,\eta_l)}\right]^{\leq 1}_{\mathcal{Z}}} &:= \min \left\lbrace 1, \overline{Y^{X,\emptyset}_{1,(\eta_i,\eta_l)}}  +   \sqrt{\overline{w}^{>1}_{\mathcal{Z}}} \right\rbrace. \label{u-Y,noclick}
\end{align}
Importantly, we expect the intervals in \eqref{bit-error-rate-click-range}, \eqref{bit-error-rate-noclick-range}, \eqref{yield-click-range} and \eqref{yield-noclick-range} to collapse to a single point under nominal conditions, since $\Tr[\Bar{\sigma} \Pi^{>1}_{\mathcal{Z}}]=0$ if there is no eavesdropper adding photons and the bound $\Tr[\Bar{\sigma} \Pi^{>1}_{\mathcal{Z}}]\leq\overline{w}^{>1}_{\mathcal{Z}}$ in \eqref{yalpha>1-upperbound-explicit} is tight under such conditions.

Having obtained narrow intervals for the test-round yields and bit error rates restricted to the ($\leq 1$)-subspace, we are one step closer to estimating $\Delta_1$ as defined in \eqref{e1}. We observe, however, that the definition of $\Delta_1$ restricts the states to the ($\leq 1$)-subspace through the positive-semidefinite operators $M^{\leq 1}_{\mathcal{Z}}$, defined in \eqref{M^leq1_Z} and reported here for clarity:
\begin{align}
    M^{\leq 1}_{\mathcal{Z}} =\sqrt{p_{\checkmark|0}}\,\Pi^0_{\mathcal{Z}} + \sqrt{p_{\checkmark|1}}\,\Pi^1_{\mathcal{Z}},  \label{M^leq1_Z2}
\end{align}
rather than through the projectors $\Pi^{\leq 1}_{\mathcal{Z}}=\Pi^0_{\mathcal{Z}} + \Pi^1_{\mathcal{Z}}$. Therefore, we cannot directly use the bounds derived on the restricted test-round yields and bit error rates to compute $\Delta_1$, since the latter lack the prefactors $\sqrt{p_{\checkmark|0}},\sqrt{p_{\checkmark|1}}$ in their definitions, \eqref{yieldX-click-<1subspace}, \eqref{yieldX-noclick-<1subspace}, \eqref{bit-error-rate-click-<1subspace}, and \eqref{bit-error-rate-noclick-<1subspace}.

Therefore, in order to obtain a precise estimation of $\Delta_1$, we need to expand each quantity restricted to the ($\leq 1$)-subspace, namely \eqref{yieldX-click-<1subspace}, \eqref{yieldX-noclick-<1subspace}, \eqref{bit-error-rate-click-<1subspace}, and \eqref{bit-error-rate-noclick-<1subspace}, through the equality $\Pi^{\leq 1}_{\mathcal{Z}}=\Pi^0_{\mathcal{Z}} + \Pi^1_{\mathcal{Z}}$ and then individually estimate the three resulting terms. The three terms can then be multiplied by the appropriate prefactor in $\Delta_1$ after expanding $\Delta_1$ in a similar way through \eqref{M^leq1_Z2}. We detail this in the following.\\

\subsubsection{Expanding the test-round yields and bit error rates restricted to the ($\leq 1$)-subspace} \label{app:PER7}
Here, we expand the expectation values in \eqref{yieldX-click-<1subspace}, \eqref{yieldX-noclick-<1subspace}, \eqref{bit-error-rate-click-<1subspace}, and \eqref{bit-error-rate-noclick-<1subspace} through $\Pi^{\leq 1}_{\mathcal{Z}}=\Pi^0_{\mathcal{Z}} + \Pi^1_{\mathcal{Z}}$. We show that they result into three independent terms that can be individually estimated thanks to the different statistics collected by tuning the TBS.

We start by expanding the test-round yield defined in \eqref{yieldX-click-<1subspace}. The other test-round yield and bit error rates follow a similar pattern. By using $\Pi^{\leq 1}_{\mathcal{Z}}=\Pi^0_{\mathcal{Z}} + \Pi^1_{\mathcal{Z}}$ and \eqref{barsigma}, we get three terms:
\begin{align}
      \left[Y^{X,\checkmark}_{1,(\eta_i,\eta_l)} \right]^{\leq 1}_{\mathcal{Z}} &=  \left[Y^{X,\checkmark}_{1,(\eta_i,\eta_l)} \right]^{0}_{\mathcal{Z}} +  \left[Y^{X,\checkmark}_{1,(\eta_i,\eta_l)} \right]^{0,1}_{\mathcal{Z}} +  \left[Y^{X,\checkmark}_{1,(\eta_i,\eta_l)} \right]^{1}_{\mathcal{Z}} \label{yield-click-TBS},
\end{align}
where we defined:
\begin{align}
    \left[Y^{X,\checkmark}_{1,(\eta_i,\eta_l)} \right]^{0}_{\mathcal{Z}} &:=   \Tr\left[\bar{\sigma}  \Pi^{0}_{\mathcal{Z}} X^{(\eta_i,\eta_l)}_{\checkmark,\checkmark} \Pi^{0}_{\mathcal{Z}}  \right] \label{yield-click-TBS2-Pi0}\\
    \left[Y^{X,\checkmark}_{1,(\eta_i,\eta_l)} \right]^{0,1}_{\mathcal{Z}} &:=  \Tr\left[\bar{\sigma}  \left(\Pi^{0}_{\mathcal{Z}} X^{(\eta_i,\eta_l)}_{\checkmark,\checkmark} \Pi^{1}_{\mathcal{Z}} + \mathrm{h.c.}\right) \right] \label{yield-click-TBS2-Pi01}\\
    \left[Y^{X,\checkmark}_{1,(\eta_i,\eta_l)} \right]^{1}_{\mathcal{Z}} &:=  \Tr\left[\bar{\sigma}  \Pi^{1}_{\mathcal{Z}} X^{(\eta_i,\eta_l)}_{\checkmark,\checkmark} \Pi^{1}_{\mathcal{Z}}  \right]. \label{yield-click-TBS2-Pi1}
\end{align}
We note that we can easily derive non-trivial intervals of existence for the three terms in \eqref{yield-click-TBS}. In particular, we have (for $\alpha=0,1$):
\begin{align}
     0 \leq \left[Y^{X,\checkmark}_{1,(\eta_i,\eta_l)} \right]^{\alpha}_{\mathcal{Z}}  \leq \overline{w}^\alpha_{\mathcal{Z}}, \label{yield-click-Pialpha-interval}
\end{align}
where we used the fact that $X^{(\eta_i,\eta_l)}_{\checkmark,\checkmark}\leq \one$ and the upper bounds in \eqref{y0-upperbound} and \eqref{y1-upperbound}. By combining \eqref{yield-click-TBS} and \eqref{yield-click-Pialpha-interval}, we obtain an interval for the remaining term:
\begin{align}
    \underline{\left[Y^{X,\checkmark}_{1,(\eta_i,\eta_l)} \right]^{\leq 1}_{\mathcal{Z}}} - (\overline{w}^0_{\mathcal{Z}} + \overline{w}^1_{\mathcal{Z}}) \leq \left[Y^{X,\checkmark}_{1,(\eta_i,\eta_l)} \right]^{0,1}_{\mathcal{Z}} \leq \overline{\left[Y^{X,\checkmark}_{1,(\eta_i,\eta_l)} \right]^{\leq 1}_{\mathcal{Z}}} \label{yield-click-Pi01-interval}.
\end{align}

Now, for each of the terms in \eqref{yield-click-TBS}, we use \eqref{Xdetdet-a} combined with \eqref{Xkdet-a2} and \eqref{Zclick-cprime} to recast the term as follows:
\begin{align}
       \left[Y^{X,\checkmark}_{1,(\eta_i,\eta_l)} \right]^{0}_{\mathcal{Z}} &= \Tr_{ab}\left[\left(\bar{\sigma} \otimes \one_b\right) (\Pi^{0}_{\mathcal{Z}})_a \otimes \proj{vac}_b  U^\dag_{(\eta_i,\eta_l)} (X_\checkmark)_c \otimes (p^{\mathcal{Z}}_d \Pi^0_{\mathcal{Z}} + {\textstyle\sum_{\alpha=1}^\infty} \pia)_{c'} U_{(\eta_i,\eta_l)} (\Pi^{0}_{\mathcal{Z}})_a \otimes\proj{vac}_b \right] \label{yield-click-TBS-Pi0}\\
       \left[Y^{X,\checkmark}_{1,(\eta_i,\eta_l)} \right]^{0,1}_{\mathcal{Z}} &= \Tr_{ab}\left[\left(\bar{\sigma} \otimes \one_b\right) (\Pi^{0}_{\mathcal{Z}})_a \otimes \proj{vac}_b  U^\dag_{(\eta_i,\eta_l)} (X_\checkmark)_c \otimes (p^{\mathcal{Z}}_d \Pi^0_{\mathcal{Z}} + {\textstyle\sum_{\alpha=1}^\infty} \pia)_{c'} U_{(\eta_i,\eta_l)} (\Pi^{1}_{\mathcal{Z}})_a \otimes\proj{vac}_b \right] \nonumber\\
       &+ \Tr_{ab}\left[\left(\bar{\sigma} \otimes \one_b\right) (\Pi^{1}_{\mathcal{Z}})_a \otimes \proj{vac}_b  U^\dag_{(\eta_i,\eta_l)} (X_\checkmark)_c \otimes (p^{\mathcal{Z}}_d \Pi^0_{\mathcal{Z}} + {\textstyle\sum_{\alpha=1}^\infty} \pia)_{c'} U_{(\eta_i,\eta_l)} (\Pi^{0}_{\mathcal{Z}})_a \otimes\proj{vac}_b \right] \label{yield-click-TBS-Pi01} \\
       \left[Y^{X,\checkmark}_{1,(\eta_i,\eta_l)} \right]^{1}_{\mathcal{Z}} &= \Tr_{ab}\left[\left(\bar{\sigma} \otimes \one_b\right) (\Pi^{1}_{\mathcal{Z}})_a \otimes \proj{vac}_b  U^\dag_{(\eta_i,\eta_l)} (X_\checkmark)_c \otimes (p^{\mathcal{Z}}_d \Pi^0_{\mathcal{Z}} + {\textstyle\sum_{\alpha=1}^\infty} \pia)_{c'} U_{(\eta_i,\eta_l)} (\Pi^{1}_{\mathcal{Z}})_a \otimes\proj{vac}_b \right] \label{yield-click-TBS-Pi1}.
\end{align}

Then, we observe that the unitary action of the TBS, as defined in \eqref{TBS-unitary}, preserves the temporal localization of each photon. In other words, if there are $\zeta$ incoming photons localized in $\mathcal{Z}$ among the two incoming modes, then there must be also be $\zeta$ outgoing photons in $\mathcal{Z}$ in the two outgoing modes. Therefore, by recalling that $\one=\sum_\alpha \pia$, we can write that:
\begin{align}
    U_{(\eta_i,\eta_l)} &= \one_c \otimes \one_{c'} U_{(\eta_i,\eta_l)} \one_a \otimes \one_b  \nonumber\\
    &= \sum_{\zeta=0}^\infty \sum_{\beta=0}^\zeta (\Pi^\beta_{\mathcal{Z}})_c \otimes (\Pi^{\zeta-\beta}_{\mathcal{Z}})_{c'}  U_{(\eta_i,\eta_l)} \sum_{\alpha=0}^\zeta (\Pi^\alpha_{\mathcal{Z}})_a \otimes (\Pi^{\zeta-\alpha}_{\mathcal{Z}})_b, \label{TBS-rule}
\end{align}
where the last expression implements the above observation. We now apply the observation in \eqref{TBS-rule} to compute the following quantities appearing in \eqref{yield-click-TBS}:
\begin{align}
    U_{(\eta_i,\eta_l)} (\Pi^{0}_{\mathcal{Z}})_a \otimes\proj{vac}_b  &= (\Pi^0_{\mathcal{Z}})_c \otimes (\Pi^0_{\mathcal{Z}})_{c'} U_{(\eta_i,\eta_l)} (\Pi^0_{\mathcal{Z}})_a \otimes\proj{vac}_b \label{TBS-comp1}\\
    U_{(\eta_i,\eta_l)} (\Pi^{1}_{\mathcal{Z}})_a \otimes\proj{vac}_b  &= \left[(\Pi^0_{\mathcal{Z}})_c \otimes (\Pi^1_{\mathcal{Z}})_{c'} + (\Pi^1_{\mathcal{Z}})_c \otimes (\Pi^0_{\mathcal{Z}})_{c'}\right] U_{(\eta_i,\eta_l)} (\Pi^1_{\mathcal{Z}})_a \otimes\proj{vac}_b . \label{TBS-comp2}
\end{align}
By taking the Hermitian conjugate of the above expressions, we obtain:
\begin{align}
    (\Pi^{0}_{\mathcal{Z}})_a \otimes\proj{vac}_b U^\dag_{(\eta_i,\eta_l)}  &= (\Pi^0_{\mathcal{Z}})_a \otimes\proj{vac}_b U^\dag_{(\eta_i,\eta_l)} (\Pi^0_{\mathcal{Z}})_c \otimes (\Pi^0_{\mathcal{Z}})_{c'} \label{TBS-comp3}\\
    (\Pi^{1}_{\mathcal{Z}})_a \otimes\proj{vac}_b U^\dag_{(\eta_i,\eta_l)}&= (\Pi^1_{\mathcal{Z}})_a \otimes\proj{vac}_b U^\dag_{(\eta_i,\eta_l)} \left[(\Pi^0_{\mathcal{Z}})_c \otimes (\Pi^1_{\mathcal{Z}})_{c'} + (\Pi^1_{\mathcal{Z}})_c \otimes (\Pi^0_{\mathcal{Z}})_{c'}\right] . \label{TBS-comp4}
\end{align}

By employing \eqref{TBS-comp1}--\eqref{TBS-comp4} in \eqref{yield-click-TBS-Pi0}--\eqref{yield-click-TBS-Pi1}, we rewrite the three terms in \eqref{yield-click-TBS} as follows:
\begin{align}
     \left[Y^{X,\checkmark}_{1,(\eta_i,\eta_l)} \right]^{0}_{\mathcal{Z}} &= p^{\mathcal{Z}}_d  \Tr_{ab}\left[\left(\bar{\sigma} \otimes \one_b\right) (\Pi^0_{\mathcal{Z}})_a \otimes \proj{vac}_b  U^\dag_{(\eta_i,\eta_l)} (\Pi^0_{\mathcal{Z}} X_\checkmark \Pi^0_{\mathcal{Z}})_c \otimes (\Pi^0_{\mathcal{Z}})_{c'} U_{(\eta_i,\eta_l)} (\Pi^0_{\mathcal{Z}})_a \otimes\proj{vac}_b \right]  \\
     \left[Y^{X,\checkmark}_{1,(\eta_i,\eta_l)} \right]^{1}_{\mathcal{Z}} &= p^{\mathcal{Z}}_d  \Tr_{ab}\left[\left(\bar{\sigma} \otimes \one_b\right) (\Pi^1_{\mathcal{Z}})_a \otimes \proj{vac}_b  U^\dag_{(\eta_i,\eta_l)} (\Pi^1_{\mathcal{Z}} X_\checkmark \Pi^1_{\mathcal{Z}})_c \otimes (\Pi^0_{\mathcal{Z}})_{c'} U_{(\eta_i,\eta_l)} (\Pi^1_{\mathcal{Z}})_a \otimes\proj{vac}_b \right] \nonumber\\
     &\quad+ \Tr_{ab}\left[\left(\bar{\sigma} \otimes \one_b\right) (\Pi^1_{\mathcal{Z}})_a \otimes \proj{vac}_b  U^\dag_{(\eta_i,\eta_l)} (\Pi^0_{\mathcal{Z}} X_\checkmark \Pi^0_{\mathcal{Z}})_c \otimes (\Pi^1_{\mathcal{Z}})_{c'} U_{(\eta_i,\eta_l)} (\Pi^1_{\mathcal{Z}})_a \otimes\proj{vac}_b \right]  \\
     \left[Y^{X,\checkmark}_{1,(\eta_i,\eta_l)} \right]^{0,1}_{\mathcal{Z}} &= p^{\mathcal{Z}}_d  \Tr_{ab}\left[\left(\bar{\sigma} \otimes \one_b\right) (\Pi^0_{\mathcal{Z}})_a \otimes \proj{vac}_b  U^\dag_{(\eta_i,\eta_l)} (\Pi^0_{\mathcal{Z}} X_\checkmark \Pi^1_{\mathcal{Z}})_c \otimes (\Pi^0_{\mathcal{Z}})_{c'} U_{(\eta_i,\eta_l)} (\Pi^1_{\mathcal{Z}})_a \otimes\proj{vac}_b \right] \nonumber\\
     &\quad+ p^{\mathcal{Z}}_d  \Tr_{ab}\left[\left(\bar{\sigma} \otimes \one_b\right) (\Pi^1_{\mathcal{Z}})_a \otimes \proj{vac}_b  U^\dag_{(\eta_i,\eta_l)} (\Pi^1_{\mathcal{Z}} X_\checkmark \Pi^0_{\mathcal{Z}})_c \otimes (\Pi^0_{\mathcal{Z}})_{c'} U_{(\eta_i,\eta_l)} (\Pi^0_{\mathcal{Z}})_a \otimes\proj{vac}_b \right] .
\end{align}

\begin{remark}
We emphasize that $\left[Y^{X,\checkmark}_{1,(\eta_i,\eta_l)} \right]^{0,1}_{\mathcal{Z}}$ is in general not null.  As a matter of fact, the operator $(X_\checkmark)_c$, representing a detection in the $X$-basis detector and reported in \eqref{Xclick-mode-c}, may not be diagonal in the eigenbasis $\{\pia\}_{\alpha=0}^\infty$ of the operator $(Z_\checkmark)_{c'}$, representing a detection in the $Z$-basis detector and reported in \eqref{Zclick-cprime}.    
\end{remark}

In a similar manner, we expand the other test-round yield in \eqref{yieldX-noclick-<1subspace} as follows:
\begin{align}
    \left[Y^{X,\emptyset}_{1,(\eta_i,\eta_l)} \right]^{\leq 1}_{\mathcal{Z}} &=  \left[Y^{X,\emptyset}_{1,(\eta_i,\eta_l)} \right]^{0}_{\mathcal{Z}} +  \left[Y^{X,\emptyset}_{1,(\eta_i,\eta_l)} \right]^{0,1}_{\mathcal{Z}} +  \left[Y^{X,\emptyset}_{1,(\eta_i,\eta_l)} \right]^{1}_{\mathcal{Z}} \label{yield-noclick-TBS2},
\end{align}
with intervals of existence given by:
\begin{align}
     0 \leq &\left[Y^{X,\emptyset}_{1,(\eta_i,\eta_l)} \right]^{\alpha}_{\mathcal{Z}}  \leq \overline{w}^\alpha_{\mathcal{Z}} \label{yield-noclick-Pialpha-interval} \\
    \underline{\left[Y^{X,\emptyset}_{1,(\eta_i,\eta_l)} \right]^{\leq 1}_{\mathcal{Z}}} - (\overline{w}^0_{\mathcal{Z}} + \overline{w}^1_{\mathcal{Z}}) \leq &\left[Y^{X,\emptyset}_{1,(\eta_i,\eta_l)} \right]^{0,1}_{\mathcal{Z}} \leq \overline{\left[Y^{X,\emptyset}_{1,(\eta_i,\eta_l)} \right]^{\leq 1}_{\mathcal{Z}}} \label{yield-noclick-Pi01-interval},
\end{align}
and where \eqref{Xknodet-a2} and the application of the rules in \eqref{TBS-comp1}--\eqref{TBS-comp4} leads to:
\begin{align}
     \left[Y^{X,\emptyset}_{1,(\eta_i,\eta_l)} \right]^{0}_{\mathcal{Z}} &= \Tr\left[\bar{\sigma}  \Pi^{0}_{\mathcal{Z}} X^{(\eta_i,\eta_l)}_{\checkmark,\emptyset} \Pi^{0}_{\mathcal{Z}}  \right] \nonumber\\
     &=(1-p^{\mathcal{Z}}_d)  \Tr_{ab}\left[\left(\bar{\sigma} \otimes \one_b\right) (\Pi^0_{\mathcal{Z}})_a \otimes \proj{vac}_b  U^\dag_{(\eta_i,\eta_l)} (\Pi^0_{\mathcal{Z}} X_\checkmark \Pi^0_{\mathcal{Z}})_c \otimes (\Pi^0_{\mathcal{Z}})_{c'} U_{(\eta_i,\eta_l)} (\Pi^0_{\mathcal{Z}})_a \otimes\proj{vac}_b \right] \label{yield-noclick-TBS2-Pi0} \\
     \left[Y^{X,\emptyset}_{1,(\eta_i,\eta_l)} \right]^{1}_{\mathcal{Z}} &= \Tr\left[\bar{\sigma}  \Pi^{1}_{\mathcal{Z}} X^{(\eta_i,\eta_l)}_{\checkmark,\emptyset} \Pi^{1}_{\mathcal{Z}}  \right] \nonumber\\
     &=(1-p^{\mathcal{Z}}_d)  \Tr_{ab}\left[\left(\bar{\sigma} \otimes \one_b\right) (\Pi^1_{\mathcal{Z}})_a \otimes \proj{vac}_b  U^\dag_{(\eta_i,\eta_l)} (\Pi^1_{\mathcal{Z}} X_\checkmark \Pi^1_{\mathcal{Z}})_c \otimes (\Pi^0_{\mathcal{Z}})_{c'} U_{(\eta_i,\eta_l)} (\Pi^1_{\mathcal{Z}})_a \otimes\proj{vac}_b \right] \label{yield-noclick-TBS2-Pi1} \\
     \left[Y^{X,\emptyset}_{1,(\eta_i,\eta_l)} \right]^{0,1}_{\mathcal{Z}} &= \Tr\left[\bar{\sigma}  \left( \Pi^{0}_{\mathcal{Z}} X^{(\eta_i,\eta_l)}_{\checkmark,\emptyset} \Pi^{1}_{\mathcal{Z}} + \mathrm{h.c.}\right) \right] \nonumber\\
     &=(1-p^{\mathcal{Z}}_d)  \Tr_{ab}\left[\left(\bar{\sigma} \otimes \one_b\right) (\Pi^0_{\mathcal{Z}})_a \otimes \proj{vac}_b  U^\dag_{(\eta_i,\eta_l)} (\Pi^0_{\mathcal{Z}} X_\checkmark \Pi^1_{\mathcal{Z}})_c \otimes (\Pi^0_{\mathcal{Z}})_{c'} U_{(\eta_i,\eta_l)} (\Pi^1_{\mathcal{Z}})_a \otimes\proj{vac}_b \right] \nonumber\\
     &\quad+ (1-p^{\mathcal{Z}}_d)  \Tr_{ab}\left[\left(\bar{\sigma} \otimes \one_b\right) (\Pi^1_{\mathcal{Z}})_a \otimes \proj{vac}_b  U^\dag_{(\eta_i,\eta_l)} (\Pi^1_{\mathcal{Z}} X_\checkmark \Pi^0_{\mathcal{Z}})_c \otimes (\Pi^0_{\mathcal{Z}})_{c'} U_{(\eta_i,\eta_l)} (\Pi^0_{\mathcal{Z}})_a \otimes\proj{vac}_b \right] \label{yield-noclick-TBS2-Pi01}.
\end{align}

Now, we show that each of the three terms in \eqref{yield-click-TBS} and \eqref{yield-noclick-TBS2} can be estimated by solving a system of linear equations. To see this, we compute each expectation value forming the three terms in \eqref{yield-click-TBS} and \eqref{yield-noclick-TBS2}, bearing in mind that they are repeated in the two expressions. To do so, we use the definition of $U^\dag_{(\eta_i,\eta_l)}$ given in \eqref{TBS-unitary} to replace the creation operators $c^\dag$ and $(c')^\dag$ with the creation operators $a^\dag$ and $b^\dag$. More precisely, we only replace the creation operators of the outer kets and bras in the operators sandwiched by $U^\dag_{(\eta_i,\eta_l)}$ and $U_{(\eta_i,\eta_l)}$. Our computation starts from the following operator, appearing in \eqref{yield-click-TBS2-Pi01} and \eqref{yield-noclick-TBS2-Pi01}, and uses the explicit expression for $\pia$ given in \eqref{pia}:
\begin{align}
    (\Pi^0_{\mathcal{Z}})_a &\otimes \proj{vac}_b U^\dag_{(\eta_i,\eta_l)} (\Pi^0_{\mathcal{Z}} X_\checkmark \Pi^1_{\mathcal{Z}})_c \otimes (\Pi^0_{\mathcal{Z}})_{c'} U_{(\eta_i,\eta_l)} (\Pi^1_{\mathcal{Z}})_a \otimes \proj{vac}_b \nonumber\\
    &= \sum_{n,n'=0}^\infty \sum_{m=1}^\infty 
    \intop_{\overline{\mathcal{Z}}} \frac{\text{d}t_1 \dots \text{d}t_n}{n!}
    \intop_{\mathcal{Z}} \text{d}\tau_1 \intop_{\overline{\mathcal{Z}}} \frac{\text{d}\tau_{2} \dots \text{d}\tau_m}{(m-1)!} \intop_{\overline{\mathcal{Z}}} \frac{\text{d}t'_1 \dots \text{d}t'_{n'}}{n'!} \, \bra{1_{t_1}\dots 1_{t_n}} X_\checkmark \ket{1_{\tau_1}\dots 1_{\tau_m}} \nonumber\\
    &\quad(\Pi^0_{\mathcal{Z}})_a \otimes \proj{vac}_b U^\dag_{(\eta_i,\eta_l)} (\ket{1_{t_1}\dots 1_{t_n}}\bra{1_{\tau_1}\dots 1_{\tau_m}})_c \otimes (\proj{1_{t'_1}\dots 1_{t'_{n'}}})_{c'} U_{(\eta_i,\eta_l)} (\Pi^1_{\mathcal{Z}})_a \otimes \proj{vac}_b \nonumber\\
    &= \sum_{n,n'=0}^\infty \sum_{m=1}^\infty 
    \intop_{\overline{\mathcal{Z}}} \frac{\text{d}t_1 \dots \text{d}t_n}{n!}
    \intop_{\mathcal{Z}} \text{d}\tau_1 \intop_{\overline{\mathcal{Z}}} \frac{\text{d}\tau_{2} \dots \text{d}\tau_m}{(m-1)!} \intop_{\overline{\mathcal{Z}}} \frac{\text{d}t'_1 \dots \text{d}t'_{n'}}{n'!} \, \bra{1_{t_1}\dots 1_{t_n}} X_\checkmark \ket{1_{\tau_1}\dots 1_{\tau_m}}\nonumber\\
    &\quad(\Pi^0_{\mathcal{Z}})_a \otimes \proj{vac}_b U^\dag_{(\eta_i,\eta_l)} c^\dag_{t_1} \dots c^\dag_{t_n} \otimes {c'}^\dag_{t'_1} \dots {c'}^\dag_{t'_{n'}} \proj{vac} {c}_{\tau_1} \dots {c}_{\tau_{m}} \otimes {c'}_{t'_1} \dots {c'}_{t'_{n'}}  U_{(\eta_i,\eta_l)} (\Pi^1_{\mathcal{Z}})_a \otimes \proj{vac}_b \nonumber\\
    &= \sum_{n,n'=0}^\infty \sum_{m=1}^\infty 
    \intop_{\overline{\mathcal{Z}}} \frac{\text{d}t_1 \dots \text{d}t_n}{n!}
    \intop_{\mathcal{Z}} \text{d}\tau_1 \intop_{\overline{\mathcal{Z}}} \frac{\text{d}\tau_{2} \dots \text{d}\tau_m}{(m-1)!} \intop_{\overline{\mathcal{Z}}} \frac{\text{d}t'_1 \dots \text{d}t'_{n'}}{n'!} \, \bra{1_{t_1}\dots 1_{t_n}} X_\checkmark \ket{1_{\tau_1}\dots 1_{\tau_m}}\nonumber\\
    &\quad\sqrt{\eta_i} (1-\eta_l)^{n'} (\sqrt{\eta_l})^{n+m-1}   \ket{1_{t_1} \dots 1_{t_n}, 1_{t'_1} \dots 1_{t'_{n'}}} \bra{1_{\tau_1} \dots 1_{\tau_{m}}, 1_{t'_1} \dots 1_{t'_{n'}}}_a \otimes \proj{vac}_b.
\end{align}
Similarly, for the other operator appearing in \eqref{yield-click-TBS2-Pi01} and \eqref{yield-noclick-TBS2-Pi01} we obtain:
\begin{align}
    (\Pi^1_{\mathcal{Z}})_a &\otimes \proj{vac}_b U^\dag_{(\eta_i,\eta_l)} (\Pi^1_{\mathcal{Z}} X_\checkmark \Pi^0_{\mathcal{Z}})_c \otimes (\Pi^0_{\mathcal{Z}})_{c'} U_{(\eta_i,\eta_l)} (\Pi^0_{\mathcal{Z}})_a \otimes \proj{vac}_b \nonumber\\
    &= \sum_{n,n'=0}^\infty \sum_{m=1}^\infty 
    \intop_{\overline{\mathcal{Z}}} \frac{\text{d}t_1 \dots \text{d}t_n}{n!}
    \intop_{\mathcal{Z}} \text{d}\tau_1 \intop_{\overline{\mathcal{Z}}} \frac{\text{d}\tau_{2} \dots \text{d}\tau_m}{(m-1)!} \intop_{\overline{\mathcal{Z}}} \frac{\text{d}t'_1 \dots \text{d}t'_{n'}}{n'!} \, \bra{1_{\tau_1}\dots 1_{\tau_m}} X_\checkmark \ket{1_{t_1}\dots 1_{t_n}}\nonumber\\
    &\quad\sqrt{\eta_i} (1-\eta_l)^{n'} (\sqrt{\eta_l})^{n+m-1}  \ket{1_{\tau_1}\dots 1_{\tau_m}, 1_{t'_1} \dots 1_{t'_{n'}}} \bra{1_{t_1} \dots 1_{t_n}, 1_{t'_1} \dots 1_{t'_{n'}}}_a \otimes \proj{vac}_b.  
\end{align}
The important point of the last two calculations is the realization that the action of the TBS generates a prefactor  $\sqrt{\eta_i}$, such that $\left[Y^{X,\checkmark}_{1,(\eta_i,\eta_l)} \right]^{0,1}_{\mathcal{Z}}$ (resp. $\left[Y^{X,\emptyset}_{1,(\eta_i,\eta_l)} \right]^{0,1}_{\mathcal{Z}}$) can be rewritten as follows:
\begin{align}
    \left[Y^{X,\checkmark}_{1,(\eta_i,\eta_l)} \right]^{0,1}_{\mathcal{Z}} &= p^{\mathcal{Z}}_d \sqrt{\eta_i}\, \mathbbm{Y}^{0,1}_{\eta_l} 
 \label{yield-click-TBS2-Pi01-2}  \\
    \left[Y^{X,\emptyset}_{1,(\eta_i,\eta_l)} \right]^{0,1}_{\mathcal{Z}} &= (1-p^{\mathcal{Z}}_d)\sqrt{\eta_i}\, \mathbbm{Y}^{0,1}_{\eta_l},   \label{yield-noclick-TBS2-Pi01-2}
\end{align}
for some real number $\mathbbm{Y}^{0,1}_{\eta_l}$ that mathematically corresponds to:
\begin{align}
    \mathbbm{Y}^{0,1}_{\eta_l} &= \sum_{n,n'=0}^\infty \sum_{m=1}^\infty 
    \intop_{\overline{\mathcal{Z}}} \frac{\text{d}t_1 \dots \text{d}t_n}{n!}
    \intop_{\mathcal{Z}} \text{d}\tau_1 \intop_{\overline{\mathcal{Z}}} \frac{\text{d}\tau_{2} \dots \text{d}\tau_m}{(m-1)!} \intop_{\overline{\mathcal{Z}}} \frac{\text{d}t'_1 \dots \text{d}t'_{n'}}{n'!} \, (1-\eta_l)^{n'} (\sqrt{\eta_l})^{n+m-1} \nonumber\\
    &\quad \left(\bra{1_{\tau_1}\dots 1_{\tau_m}} X_\checkmark \ket{1_{t_1}\dots 1_{t_n}} \bra{1_{t_1} \dots 1_{t_n}, 1_{t'_1} \dots 1_{t'_{n'}}} \Bar{\sigma}\ket{1_{\tau_1}\dots 1_{\tau_m}, 1_{t'_1} \dots 1_{t'_{n'}}} + \mathrm{c.c} \right),
\end{align}
but that, we emphasize, cannot be directly observed experimentally --analogously to the test-round yields restricted by a combination of $\Pi^0_{\mathcal{Z}}$ and $\Pi^1_{\mathcal{Z}}$.

Nevertheless, in a similar manner, we compute the other terms in \eqref{yield-click-TBS} and \eqref{yield-noclick-TBS2} and obtain:
\begin{align}
    \left[Y^{X,\checkmark}_{1,(\eta_i,\eta_l)} \right]^{0}_{\mathcal{Z}} &= p^{\mathcal{Z}}_d \mathbbm{Y}^{0}_{\eta_l}  \label{yield-click-TBS2-Pi0-2} \\
    \left[Y^{X,\emptyset}_{1,(\eta_i,\eta_l)} \right]^{0}_{\mathcal{Z}} &= (1-p^{\mathcal{Z}}_d) \mathbbm{Y}^{0}_{\eta_l}  \label{yield-noclick-TBS2-Pi0-2}
\end{align}
for a non-negative real number: 
\begin{align}
    \mathbbm{Y}^{0}_{\eta_l} &= \sum_{n,n'=0}^\infty \sum_{m=0}^\infty 
    \intop_{\overline{\mathcal{Z}}} \frac{\text{d}t_1 \dots \text{d}t_n}{n!}
    \intop_{\overline{\mathcal{Z}}} \frac{\text{d}\tau_1 \dots \text{d}\tau_m}{m!} \intop_{\overline{\mathcal{Z}}} \frac{\text{d}t'_1 \dots \text{d}t'_{n'}}{n'!} \, (1-\eta_l)^{n'} (\sqrt{\eta_l})^{n+m} \nonumber\\
    &\quad \bra{1_{\tau_1}\dots 1_{\tau_m}} X_\checkmark \ket{1_{t_1}\dots 1_{t_n}} \bra{1_{t_1} \dots 1_{t_n}, 1_{t'_1} \dots 1_{t'_{n'}}} \Bar{\sigma}\ket{1_{\tau_1}\dots 1_{\tau_m}, 1_{t'_1} \dots 1_{t'_{n'}}},
\end{align}
and
\begin{align}
    \left[Y^{X,\checkmark}_{1,(\eta_i,\eta_l)} \right]^{1}_{\mathcal{Z}} &= p^{\mathcal{Z}}_d  \eta_i \,\mathbbm{Y}^{1t}_{\eta_l} + (1-\eta_i)\,\mathbbm{Y}^{1r}_{\eta_l}  \label{yield-click-TBS2-Pi1-2}  \\
    \left[Y^{X,\emptyset}_{1,(\eta_i,\eta_l)} \right]^{1}_{\mathcal{Z}} &= (1-p^{\mathcal{Z}}_d)  \eta_i\, \mathbbm{Y}^{1t}_{\eta_l},  \label{yield-noclick-TBS2-Pi1-2}
\end{align}
for non-negative real numbers:
\begin{align}
    \mathbbm{Y}^{1t}_{\eta_l} &= \sum_{n'=0}^\infty \sum_{n,m=1}^\infty 
    \intop_{\mathcal{Z}} \text{d}t_1
    \intop_{\overline{\mathcal{Z}}} \frac{\text{d}t_2 \dots \text{d}t_n}{(n-1)!}
    \intop_{\mathcal{Z}} \text{d}\tau_1 \intop_{\overline{\mathcal{Z}}} \frac{\text{d}\tau_{2} \dots \text{d}\tau_m}{(m-1)!} \intop_{\overline{\mathcal{Z}}} \frac{\text{d}t'_1 \dots \text{d}t'_{n'}}{n'!} \, (1-\eta_l)^{n'} (\sqrt{\eta_l})^{n+m-2} \nonumber\\
    &\quad \bra{1_{\tau_1}\dots 1_{\tau_m}} X_\checkmark \ket{1_{t_1}\dots 1_{t_n}} \bra{1_{t_1} \dots 1_{t_n}, 1_{t'_1} \dots 1_{t'_{n'}}} \Bar{\sigma}\ket{1_{\tau_1}\dots 1_{\tau_m}, 1_{t'_1} \dots 1_{t'_{n'}}} \\
    \mathbbm{Y}^{1r}_{\eta_l} &= \sum_{n,m=0}^\infty \sum_{n'=1}^\infty 
    \intop_{\overline{\mathcal{Z}}} \frac{\text{d}t_1 \dots \text{d}t_n}{n!}
    \intop_{\overline{\mathcal{Z}}} \frac{\text{d}\tau_1 \dots \text{d}\tau_m}{m!}  \intop_{\mathcal{Z}} \text{d}t'_1
    \intop_{\overline{\mathcal{Z}}} \frac{\text{d}t'_2 \dots \text{d}t'_{n'}}{(n'-1)!} \, (1-\eta_l)^{n'-1} (\sqrt{\eta_l})^{n+m} \nonumber\\
    &\quad \bra{1_{\tau_1}\dots 1_{\tau_m}} X_\checkmark \ket{1_{t_1}\dots 1_{t_n}} \bra{1_{t_1} \dots 1_{t_n}, 1_{t'_1} \dots 1_{t'_{n'}}} \Bar{\sigma}\ket{1_{\tau_1}\dots 1_{\tau_m}, 1_{t'_1} \dots 1_{t'_{n'}}}
\end{align}
where the superscript $^t$ ($^r$) stands for the fact that the photon localized in $\mathcal{Z}$ is transmitted (reflected) by the TBS.

As already argued, we cannot directly learn the values of $\mathbbm{Y}^{0}_{\eta_l}$, $\mathbbm{Y}^{0,1}_{\eta_l}$, $\mathbbm{Y}^{1t}_{\eta_l}$, and $\mathbbm{Y}^{1r}_{\eta_l}$ from the experiment statistics. However, we know that the sum of \eqref{yield-click-TBS2-Pi01-2}, \eqref{yield-click-TBS2-Pi0-2} and \eqref{yield-click-TBS2-Pi1-2} (resp. \eqref{yield-noclick-TBS2-Pi01-2}, \eqref{yield-noclick-TBS2-Pi0-2} and \eqref{yield-noclick-TBS2-Pi1-2}) is equal to $\left[Y^{X,\checkmark}_{1,(\eta_i,\eta_l)} \right]^{\leq 1}_{\mathcal{Z}}$ ($\left[Y^{X,\emptyset}_{1,(\eta_i,\eta_l)} \right]^{\leq 1}_{\mathcal{Z}}$), which we managed to tightly bound from above and from below in \eqref{yield-click-range} (resp. \eqref{yield-noclick-range}) in terms of observed statistics. Thus, for the moment, we treat $\left[Y^{X,\checkmark}_{1,(\eta_i,\eta_l)} \right]^{\leq 1}_{\mathcal{Z}}$ and $\left[Y^{X,\emptyset}_{1,(\eta_i,\eta_l)} \right]^{\leq 1}_{\mathcal{Z}}$ as known quantities and write down the following linear system of equations that follows from \eqref{yield-click-TBS} and \eqref{yield-noclick-TBS2}:
\begin{align}
    \left\lbrace \begin{array}{ll}
    \left[Y^{X,\checkmark}_{1,(\eta_i,\eta_l)} \right]^{\leq 1}_{\mathcal{Z}} &= p^{\mathcal{Z}}_d \left( \mathbbm{Y}^{0}_{\eta_l} +\sqrt{\eta_i}\, \mathbbm{Y}^{0,1}_{\eta_l} + \eta_i \,\mathbbm{Y}^{1t}_{\eta_l} \right) + (1-\eta_i)\,\mathbbm{Y}^{1r}_{\eta_l} \\[2ex]
    \left[Y^{X,\emptyset}_{1,(\eta_i,\eta_l)} \right]^{\leq 1}_{\mathcal{Z}} &= (1-p^{\mathcal{Z}}_d) \left( \mathbbm{Y}^{0}_{\eta_l} +\sqrt{\eta_i}\, \mathbbm{Y}^{0,1}_{\eta_l} + \eta_i \,\mathbbm{Y}^{1t}_{\eta_l} \right),
    \end{array} \right. \label{yield-system}
\end{align}
where the left-hand sides are known while the unknowns are the variables $\mathbbm{Y}^{0}_{\eta_l}$, $\mathbbm{Y}^{0,1}_{\eta_l}$, $\mathbbm{Y}^{1t}_{\eta_l}$, and $\mathbbm{Y}^{1r}_{\eta_l}$. Note that the system is composed of pairs of equations like the ones reported, each pair fixed by a choice of $\eta_i$, thus totalling to six equations (recall that Bob can select $\eta_i \in \{\eta_\uparrow,\eta_\downarrow,\eta_2\}$). By solving the system of equations for $\mathbbm{Y}^{0}_{\eta_l}$, $\mathbbm{Y}^{0,1}_{\eta_l}$, $\mathbbm{Y}^{1t}_{\eta_l}$, and $\mathbbm{Y}^{1r}_{\eta_l}$, we can deduce the values of each of the three terms in the expansions \eqref{yield-click-TBS} and \eqref{yield-noclick-TBS2} and use the latter to compute $\Delta_1$.\\

In a similar way to what is done for the test-round yields, we can expand the test-round bit error rates in \eqref{bit-error-rate-click-<1subspace} and \eqref{bit-error-rate-noclick-<1subspace} into three terms:
\begin{align}
    \left[Y^{X,\checkmark}_{1,(\eta_i,\eta_l)} e_{X,1,(\eta_i,\eta_l),\checkmark}\right]^{\leq 1}_{\mathcal{Z}} &=  \left[Y^{X,\checkmark}_{1,(\eta_i,\eta_l)} e_{X,1,(\eta_i,\eta_l),\checkmark}\right]^{0}_{\mathcal{Z}} +  \left[Y^{X,\checkmark}_{1,(\eta_i,\eta_l)} e_{X,1,(\eta_i,\eta_l),\checkmark}\right]^{0,1}_{\mathcal{Z}} +  \left[Y^{X,\checkmark}_{1,(\eta_i,\eta_l)} e_{X,1,(\eta_i,\eta_l),\checkmark}\right]^{1}_{\mathcal{Z}} , \label{bit-error-rate-click-TBS2} \\
    \left[Y^{X,\emptyset}_{1,(\eta_i,\eta_l)} e_{X,1,(\eta_i,\eta_l),\emptyset}\right]^{\leq 1}_{\mathcal{Z}} &=  \left[Y^{X,\emptyset}_{1,(\eta_i,\eta_l)} e_{X,1,(\eta_i,\eta_l),\emptyset}\right]^{0}_{\mathcal{Z}} +  \left[Y^{X,\emptyset}_{1,(\eta_i,\eta_l)} e_{X,1,(\eta_i,\eta_l),\emptyset}\right]^{0,1}_{\mathcal{Z}} +  \left[Y^{X,\emptyset}_{1,(\eta_i,\eta_l)} e_{X,1,(\eta_i,\eta_l),\emptyset}\right]^{1}_{\mathcal{Z}} , \label{bit-error-rate-noclick-TBS2}
\end{align}
where the first term represents the test-round bit error rate restricted by the projector $\Pi^{0}_{\mathcal{Z}}$:
\begin{align}
       \left[Y^{X,\checkmark}_{1,(\eta_i,\eta_l)} e_{X,1,(\eta_i,\eta_l),\checkmark}\right]^{0}_{\mathcal{Z}} &:= {\textstyle\sum_{k=0}^{d-1}} \Tr\left[\frac{\sigma_k}{d} \Pi^{0}_{\mathcal{Z}} {\textstyle\sum_{k'\neq k}}X^{(\eta_i,\eta_l)}_{k',\checkmark} \Pi^{0}_{\mathcal{Z}} \right] \label{bit-error-rate-click-TBS2-Pi0} \\
       \left[Y^{X,\emptyset}_{1,(\eta_i,\eta_l)} e_{X,1,(\eta_i,\eta_l),\emptyset}\right]^{0}_{\mathcal{Z}} &:= {\textstyle\sum_{k=0}^{d-1}} \Tr\left[\frac{\sigma_k}{d} \Pi^{0}_{\mathcal{Z}} {\textstyle\sum_{k'\neq k}}X^{(\eta_i,\eta_l)}_{k',\emptyset} \Pi^{0}_{\mathcal{Z}} \right] \label{bit-error-rate-noclick-TBS2-Pi0},
\end{align}
while the third term is restricted by $\Pi^{1}_{\mathcal{Z}}$:
\begin{align}
    \left[Y^{X,\checkmark}_{1,(\eta_i,\eta_l)} e_{X,1,(\eta_i,\eta_l),\checkmark}\right]^{1}_{\mathcal{Z}} &:= {\textstyle\sum_{k=0}^{d-1}} \Tr\left[\frac{\sigma_k}{d} \Pi^{1}_{\mathcal{Z}} {\textstyle\sum_{k'\neq k}}X^{(\eta_i,\eta_l)}_{k',\checkmark} \Pi^{1}_{\mathcal{Z}} \right] \label{bit-error-rate-click-TBS2-Pi1} \\
       \left[Y^{X,\emptyset}_{1,(\eta_i,\eta_l)} e_{X,1,(\eta_i,\eta_l),\emptyset}\right]^{1}_{\mathcal{Z}} &:= {\textstyle\sum_{k=0}^{d-1}} \Tr\left[\frac{\sigma_k}{d} \Pi^{1}_{\mathcal{Z}} {\textstyle\sum_{k'\neq k}}X^{(\eta_i,\eta_l)}_{k',\emptyset} \Pi^{1}_{\mathcal{Z}} \right] \label{bit-error-rate-noclick-TBS2-Pi1},
\end{align}
and the second term is restricted by a combination of $\Pi^{0}_{\mathcal{Z}}$ and $\Pi^{1}_{\mathcal{Z}}$:
\begin{align}
    \left[Y^{X,\checkmark}_{1,(\eta_i,\eta_l)} e_{X,1,(\eta_i,\eta_l),\checkmark}\right]^{0,1}_{\mathcal{Z}} &:= {\textstyle\sum_{k=0}^{d-1}} \Tr\left[\frac{\sigma_k}{d} \left(\Pi^{0}_{\mathcal{Z}} {\textstyle\sum_{k'\neq k}}X^{(\eta_i,\eta_l)}_{k',\checkmark} \Pi^{1}_{\mathcal{Z}} 
 + \mathrm{h.c.} \right)\right] \label{bit-error-rate-click-TBS2-Pi01} \\
    \left[Y^{X,\emptyset}_{1,(\eta_i,\eta_l)} e_{X,1,(\eta_i,\eta_l),\emptyset}\right]^{0,1}_{\mathcal{Z}} &:= {\textstyle\sum_{k=0}^{d-1}} \Tr\left[\frac{\sigma_k}{d} \left(\Pi^{0}_{\mathcal{Z}} {\textstyle\sum_{k'\neq k}}X^{(\eta_i,\eta_l)}_{k',\emptyset} \Pi^{1}_{\mathcal{Z}} + \mathrm{h.c.} \right)\right] \label{bit-error-rate-noclick-TBS2-Pi01}.
\end{align}
Similarly to \eqref{yield-noclick-Pialpha-interval} and \eqref{yield-noclick-Pi01-interval}, we can derive intervals of existence for each of the above terms:
\begin{align}
     0 \leq &\left[Y^{X,\checkmark}_{1,(\eta_i,\eta_l)} e_{X,1,(\eta_i,\eta_l),\checkmark} \right]^{\alpha}_{\mathcal{Z}}  \leq \overline{w}^\alpha_{\mathcal{Z}} \label{bit-error-rate-click-Pialpha-interval} \\
     0 \leq &\left[Y^{X,\emptyset}_{1,(\eta_i,\eta_l)} e_{X,1,(\eta_i,\eta_l),\emptyset} \right]^{\alpha}_{\mathcal{Z}}  \leq \overline{w}^\alpha_{\mathcal{Z}} \label{bit-error-rate-noclick-Pialpha-interval} \\
     \underline{\left[Y^{X,\checkmark}_{1,(\eta_i,\eta_l)} e_{X,1,(\eta_i,\eta_l),\checkmark} \right]^{\leq 1}_{\mathcal{Z}}} - (\overline{w}^0_{\mathcal{Z}} + \overline{w}^1_{\mathcal{Z}}) \leq &\left[Y^{X,\checkmark}_{1,(\eta_i,\eta_l)} e_{X,1,(\eta_i,\eta_l),\checkmark}\right]^{0,1}_{\mathcal{Z}} \leq \overline{\left[Y^{X,\checkmark}_{1,(\eta_i,\eta_l)} e_{X,1,(\eta_i,\eta_l),\checkmark} \right]^{\leq 1}_{\mathcal{Z}}} \label{bit-error-rate-click-Pi01-interval}\\
    \underline{\left[Y^{X,\emptyset}_{1,(\eta_i,\eta_l)} e_{X,1,(\eta_i,\eta_l),\emptyset} \right]^{\leq 1}_{\mathcal{Z}}} - (\overline{w}^0_{\mathcal{Z}} + \overline{w}^1_{\mathcal{Z}}) \leq &\left[Y^{X,\emptyset}_{1,(\eta_i,\eta_l)} e_{X,1,(\eta_i,\eta_l),\emptyset}\right]^{0,1}_{\mathcal{Z}} \leq \overline{\left[Y^{X,\emptyset}_{1,(\eta_i,\eta_l)} e_{X,1,(\eta_i,\eta_l),\emptyset} \right]^{\leq 1}_{\mathcal{Z}}} \label{bit-error-rate-noclick-Pi01-interval}.
\end{align}
By following analogous calculations to those carried out for the test-round yields, each term in the expansions \eqref{bit-error-rate-click-TBS2} and \eqref{bit-error-rate-noclick-TBS2} can be recast as follows:
\begin{align}
    \left[Y^{X,\checkmark}_{1,(\eta_i,\eta_l)} e_{X,1,(\eta_i,\eta_l),\checkmark}\right]^{0}_{\mathcal{Z}} &= p^{\mathcal{Z}}_d \,\mathbbm{E}^{0}_{\eta_l}  \label{bit-error-rate-click-TBS2-Pi0-2} \\
    \left[Y^{X,\emptyset}_{1,(\eta_i,\eta_l)} e_{X,1,(\eta_i,\eta_l),\emptyset}\right]^{0}_{\mathcal{Z}} &= (1-p^{\mathcal{Z}}_d) \, \mathbbm{E}^{0}_{\eta_l}  \label{bit-error-rate-noclick-TBS2-Pi0-2} \\
    \left[Y^{X,\checkmark}_{1,(\eta_i,\eta_l)} e_{X,1,(\eta_i,\eta_l),\checkmark}\right]^{0,1}_{\mathcal{Z}} &= p^{\mathcal{Z}}_d \sqrt{\eta_i} \,\mathbbm{E}^{0,1}_{\eta_l}  \label{bit-error-rate-click-TBS2-Pi01-2} \\
    \left[Y^{X,\emptyset}_{1,(\eta_i,\eta_l)} e_{X,1,(\eta_i,\eta_l),\emptyset}\right]^{0,1}_{\mathcal{Z}} &= (1-p^{\mathcal{Z}}_d) \sqrt{\eta_i} \,\mathbbm{E}^{0,1}_{\eta_l}  \label{bit-error-rate-noclick-TBS2-Pi01-2} \\
    \left[Y^{X,\checkmark}_{1,(\eta_i,\eta_l)} e_{X,1,(\eta_i,\eta_l),\checkmark}\right]^{1}_{\mathcal{Z}} &= p^{\mathcal{Z}}_d \eta_i \,\mathbbm{E}^{1t}_{\eta_l} +(1-\eta_i) \,\mathbbm{E}^{1r}_{\eta_l} \label{bit-error-rate-click-TBS2-Pi1-2} \\
    \left[Y^{X,\emptyset}_{1,(\eta_i,\eta_l)} e_{X,1,(\eta_i,\eta_l),\emptyset}\right]^{1}_{\mathcal{Z}} &= (1-p^{\mathcal{Z}}_d) \eta_i \,\mathbbm{E}^{1t}_{\eta_l}  \label{bit-error-rate-noclick-TBS2-Pi1-2} ,
\end{align}
for some real numbers $\mathbbm{E}^{0}_{\eta_l}$, $\mathbbm{E}^{0,1}_{\eta_l}$, $\mathbbm{E}^{1t}_{\eta_l}$, and $\mathbbm{E}^{1r}_{\eta_l}$, where again the action of the TBS on the photon localized in $\mathcal{Z}$ is singled out by the factors containing $\eta_i$. Therefore, we can employ \eqref{bit-error-rate-click-TBS2-Pi0-2}--\eqref{bit-error-rate-noclick-TBS2-Pi1-2} in the expansions \eqref{bit-error-rate-click-TBS2} and \eqref{bit-error-rate-noclick-TBS2} to derive the following linear system of equations:
\begin{align}
    \left\lbrace \begin{array}{ll}
    \left[Y^{X,\checkmark}_{1,(\eta_i,\eta_l)} e_{X,1,(\eta_i,\eta_l),\checkmark}\right]^{\leq 1}_{\mathcal{Z}} &= p^{\mathcal{Z}}_d \left( \mathbbm{E}^{0}_{\eta_l} +\sqrt{\eta_i}\, \mathbbm{E}^{0,1}_{\eta_l} + \eta_i \,\mathbbm{E}^{1t}_{\eta_l} \right) + (1-\eta_i)\,\mathbbm{E}^{1r}_{\eta_l} \\[2ex]
    \left[Y^{X,\emptyset}_{1,(\eta_i,\eta_l)} e_{X,1,(\eta_i,\eta_l),\emptyset}\right]^{\leq 1}_{\mathcal{Z}} &= (1-p^{\mathcal{Z}}_d) \left( \mathbbm{E}^{0}_{\eta_l} +\sqrt{\eta_i}\, \mathbbm{E}^{0,1}_{\eta_l} + \eta_i \,\mathbbm{E}^{1t}_{\eta_l} \right).
    \end{array} \right. \label{bit-error-rate-system}
\end{align}
Analogously to the system in \eqref{yield-system} for the test-round yields, the left-hand sides can be treated as known quantities as they are tightly bounded in \eqref{bit-error-rate-click-range} and \eqref{bit-error-rate-noclick-range}. Hence, we can solve the system for $\mathbbm{E}^{0}_{\eta_l}$, $\mathbbm{E}^{0,1}_{\eta_l}$, $\mathbbm{E}^{1t}_{\eta_l}$, and $\mathbbm{E}^{1r}_{\eta_l}$, such that we learn individually each term in the expansions \eqref{bit-error-rate-click-TBS2} and \eqref{bit-error-rate-noclick-TBS2}.\\

\subsubsection{Solving the linear systems}\label{app:PER8}
In this paragraph, we solve the linear systems in \eqref{yield-system} and \eqref{bit-error-rate-system}. Since the two systems are formally identical, we discuss the solution of the one in \eqref{yield-system} and apply the result to the other system.

The linear system in \eqref{yield-system} can be put in the following form, where we selected the second equation three times (for three values of $\eta_i$) and we selected the first equation for $\eta_i=\eta_3$:
\begin{align}
    \begin{pmatrix} \left[Y^{X,\emptyset}_{1,(\eta_1,\eta_l)} \right]^{\leq 1}_{\mathcal{Z}} \\[2ex] \left[Y^{X,\emptyset}_{1,(\eta_2,\eta_l)} \right]^{\leq 1}_{\mathcal{Z}} \\[2ex] \left[Y^{X,\emptyset}_{1,(\eta_3,\eta_l)} \right]^{\leq 1}_{\mathcal{Z}} \\[2ex] \left[Y^{X,\checkmark}_{1,(\eta_3,\eta_l)} \right]^{\leq 1}_{\mathcal{Z}}
    \end{pmatrix} = \Lambda
    \begin{pmatrix} \mathbbm{Y}^{0}_{\eta_l} \\[2ex] \mathbbm{Y}^{0,1}_{\eta_l} \\[2ex] \mathbbm{Y}^{1t}_{\eta_l} \\[2ex] \mathbbm{Y}^{1r}_{\eta_l}
    \end{pmatrix},
\end{align}
with the coefficient matrix given by:
\begin{align}
    \setlength\arraycolsep{1.5ex}
    \Lambda := \begin{pmatrix}
     (1-p^{\mathcal{Z}}_d) &  (1-p^{\mathcal{Z}}_d)\sqrt{\eta_1} & (1-p^{\mathcal{Z}}_d)\eta_1 & 0 \\[2ex]
    (1-p^{\mathcal{Z}}_d) & (1-p^{\mathcal{Z}}_d)\sqrt{\eta_2} & (1-p^{\mathcal{Z}}_d)\eta_2 & 0 \\[2ex]
     (1-p^{\mathcal{Z}}_d) & (1-p^{\mathcal{Z}}_d)\sqrt{\eta_3} & (1-p^{\mathcal{Z}}_d)\eta_3 & 0 \\[2ex]
     p^{\mathcal{Z}}_d & p^{\mathcal{Z}}_d \sqrt{\eta_3} & p^{\mathcal{Z}}_d \eta_3 &  1-\eta_3
    \end{pmatrix},
\end{align}
where the three choices of $\eta_i$ correspond to (cf.~Sec~\ref{sec:protocol}):
\begin{align}
    \eta_1 &= \eta_\uparrow \label{eta1}\\
    \eta_3 &< \eta_2 < \eta_1  \label{eta2}\\
    \eta_3 &= \eta_\downarrow.   \label{eta3}
\end{align}
The system is solved by inverting the coefficient matrix. In the protocol's description (Sec.~\ref{sec:protocol}), we noted that the optimal choice for $\eta_2$ is given by $\eta_2 = \frac14 (\sqrt{\eta_\uparrow} + \sqrt{\eta_\downarrow})^2$, which maximizes the absolute value of the determinant of $\Lambda$ (and corresponds to the mean of the geometric mean and arithmetic mean of $\eta_\uparrow$ and $\eta_\downarrow$). The solution of the system in \eqref{yield-system} reads:
\begin{align}
    \mathbbm{Y}^{0}_{\eta_l} &=\frac{\sqrt{\eta_2 \eta_3}}{\left(\sqrt{\eta_1}-\sqrt{\eta_2}\right)\left(\sqrt{\eta_1}-\sqrt{\eta_3}\right)\left(1-p^{\mathcal{Z}}_d\right)} \left[Y^{X,\emptyset}_{1,(\eta_1,\eta_l)} \right]^{\leq 1}_{\mathcal{Z}} - \frac{\sqrt{\eta_1 \eta_3} }{\left(\sqrt{\eta_1}-\sqrt{\eta_2}\right)
   \left(\sqrt{\eta_2}-\sqrt{\eta_3}\right) \left(1-p^{\mathcal{Z}}_d\right)} \left[Y^{X,\emptyset}_{1,(\eta_2,\eta_l)} \right]^{\leq 1}_{\mathcal{Z}} \nonumber\\
   &\quad+\frac{\sqrt{\eta_1 \eta_2}}{\left(\sqrt{\eta_1}-\sqrt{\eta_3}\right) \left(\sqrt{\eta_2}-\sqrt{\eta_3}\right) 
   \left(1-p^{\mathcal{Z}}_d\right)} \left[Y^{X,\emptyset}_{1,(\eta_3,\eta_l)} \right]^{\leq 1}_{\mathcal{Z}}   \label{Yetak-0}\\
   \mathbbm{Y}^{0,1}_{\eta_l} &= -\frac{\sqrt{\eta_2} +  \sqrt{\eta_3}}{\left(\sqrt{\eta_1}-\sqrt{\eta_2}\right)\left(\sqrt{\eta_1}-\sqrt{\eta_3}\right)\left(1-p^{\mathcal{Z}}_d\right)} \left[Y^{X,\emptyset}_{1,(\eta_1,\eta_l)} \right]^{\leq 1}_{\mathcal{Z}} + \frac{\sqrt{\eta_1} +  \sqrt{\eta_3}}{\left(\sqrt{\eta_1}-\sqrt{\eta_2}\right)\left(\sqrt{\eta_2}-\sqrt{\eta_3}\right)\left(1-p^{\mathcal{Z}}_d\right)} \left[Y^{X,\emptyset}_{1,(\eta_2,\eta_l)} \right]^{\leq 1}_{\mathcal{Z}} \nonumber\\
   &\quad- \frac{\sqrt{\eta_1} +  \sqrt{\eta_2}}{\left(\sqrt{\eta_1}-\sqrt{\eta_3}\right)\left(\sqrt{\eta_2}-\sqrt{\eta_3}\right)\left(1-p^{\mathcal{Z}}_d\right)} \left[Y^{X,\emptyset}_{1,(\eta_3,\eta_l)} \right]^{\leq 1}_{\mathcal{Z}} \label{Yetak-01}\\
   \mathbbm{Y}^{1t}_{\eta_l} &= \frac{1}{\left(\sqrt{\eta_1}-\sqrt{\eta_2}\right)\left(\sqrt{\eta_1}-\sqrt{\eta_3}\right)\left(1-p^{\mathcal{Z}}_d\right)} \left[Y^{X,\emptyset}_{1,(\eta_1,\eta_l)} \right]^{\leq 1}_{\mathcal{Z}} - \frac{1}{\left(\sqrt{\eta_1}-\sqrt{\eta_2}\right)
   \left(\sqrt{\eta_2}-\sqrt{\eta_3}\right) \left(1-p^{\mathcal{Z}}_d\right)} \left[Y^{X,\emptyset}_{1,(\eta_2,\eta_l)} \right]^{\leq 1}_{\mathcal{Z}} \nonumber\\
   &\quad+\frac{1}{\left(\sqrt{\eta_1}-\sqrt{\eta_3}\right) \left(\sqrt{\eta_2}-\sqrt{\eta_3}\right) 
   \left(1-p^{\mathcal{Z}}_d\right)} \left[Y^{X,\emptyset}_{1,(\eta_3,\eta_l)} \right]^{\leq 1}_{\mathcal{Z}}  \label{Yetak-1t} \\
   \mathbbm{Y}^{1r}_{\eta_l} &=\frac{\left[Y^{X,\checkmark}_{1,(\eta_3,\eta_l)} \right]^{\leq 1}_{\mathcal{Z}}}{1-\eta_3} + 
   \frac{\left[Y^{X,\emptyset}_{1,(\eta_3,\eta_l)} \right]^{\leq 1}_{\mathcal{Z}}}{1-\eta_3} \left(1-\frac{1}{1-p^{\mathcal{Z}}_d}\right) . \label{Yetak-1r}
\end{align}

In the exact same fashion, we solve the linear system in \eqref{bit-error-rate-system} and obtain:
\begin{align}
    \mathbbm{E}^{0}_{\eta_l} &=\frac{\sqrt{\eta_2 \eta_3}\left[Y^{X,\emptyset}_{1,(\eta_1,\eta_l)} e_{X,1,(\eta_1,\eta_l),\emptyset}\right]^{\leq 1}_{\mathcal{Z}}}{\left(\sqrt{\eta_1}-\sqrt{\eta_2}\right)\left(\sqrt{\eta_1}-\sqrt{\eta_3}\right)\left(1-p^{\mathcal{Z}}_d\right)}  - \frac{\sqrt{\eta_1 \eta_3} \left[Y^{X,\emptyset}_{1,(\eta_2,\eta_l)} e_{X,1,(\eta_2,\eta_l),\emptyset} \right]^{\leq 1}_{\mathcal{Z}}}{\left(\sqrt{\eta_1}-\sqrt{\eta_2}\right)
   \left(\sqrt{\eta_2}-\sqrt{\eta_3}\right) \left(1-p^{\mathcal{Z}}_d\right)}  \nonumber\\
   &\quad+\frac{\sqrt{\eta_1 \eta_2}}{\left(\sqrt{\eta_1}-\sqrt{\eta_3}\right) \left(\sqrt{\eta_2}-\sqrt{\eta_3}\right) 
   \left(1-p^{\mathcal{Z}}_d\right)} \left[Y^{X,\emptyset}_{1,(\eta_3,\eta_l)} e_{X,1,(\eta_3,\eta_l),\emptyset} \right]^{\leq 1}_{\mathcal{Z}}   \label{Eetak-0}\\
   \mathbbm{E}^{0,1}_{\eta_l} &= -\frac{(\sqrt{\eta_2} +  \sqrt{\eta_3})\left[Y^{X,\emptyset}_{1,(\eta_1,\eta_l)} e_{X,1,(\eta_1,\eta_l),\emptyset} \right]^{\leq 1}_{\mathcal{Z}}}{\left(\sqrt{\eta_1}-\sqrt{\eta_2}\right)\left(\sqrt{\eta_1}-\sqrt{\eta_3}\right)\left(1-p^{\mathcal{Z}}_d\right)}  + \frac{(\sqrt{\eta_1} +  \sqrt{\eta_3}) \left[Y^{X,\emptyset}_{1,(\eta_2,\eta_l)} e_{X,1,(\eta_2,\eta_l),\emptyset} \right]^{\leq 1}_{\mathcal{Z}}}{\left(\sqrt{\eta_1}-\sqrt{\eta_2}\right)\left(\sqrt{\eta_2}-\sqrt{\eta_3}\right)\left(1-p^{\mathcal{Z}}_d\right)}  \nonumber\\
   &\quad- \frac{\sqrt{\eta_1} +  \sqrt{\eta_2}}{\left(\sqrt{\eta_1}-\sqrt{\eta_3}\right)\left(\sqrt{\eta_2}-\sqrt{\eta_3}\right)\left(1-p^{\mathcal{Z}}_d\right)} \left[Y^{X,\emptyset}_{1,(\eta_3,\eta_l)} e_{X,1,(\eta_3,\eta_l),\emptyset}\right]^{\leq 1}_{\mathcal{Z}} \label{Eetak-01}\\
   \mathbbm{E}^{1t}_{\eta_l} &= \frac{\left[Y^{X,\emptyset}_{1,(\eta_1,\eta_l)} e_{X,1,(\eta_1,\eta_l),\emptyset} \right]^{\leq 1}_{\mathcal{Z}}}{\left(\sqrt{\eta_1}-\sqrt{\eta_2}\right)\left(\sqrt{\eta_1}-\sqrt{\eta_3}\right)\left(1-p^{\mathcal{Z}}_d\right)}  - \frac{\left[Y^{X,\emptyset}_{1,(\eta_2,\eta_l)} e_{X,1,(\eta_2,\eta_l),\emptyset} \right]^{\leq 1}_{\mathcal{Z}}}{\left(\sqrt{\eta_1}-\sqrt{\eta_2}\right)
   \left(\sqrt{\eta_2}-\sqrt{\eta_3}\right) \left(1-p^{\mathcal{Z}}_d\right)}  \nonumber\\
   &\quad+\frac{1}{\left(\sqrt{\eta_1}-\sqrt{\eta_3}\right) \left(\sqrt{\eta_2}-\sqrt{\eta_3}\right) 
   \left(1-p^{\mathcal{Z}}_d\right)} \left[Y^{X,\emptyset}_{1,(\eta_3,\eta_l)} e_{X,1,(\eta_3,\eta_l),\emptyset}\right]^{\leq 1}_{\mathcal{Z}}  \label{Eetak-1t} \\
   \mathbbm{E}^{1r}_{\eta_l} &=\frac{\left[Y^{X,\checkmark}_{1,(\eta_3,\eta_l)} e_{X,1,(\eta_3,\eta_l),\checkmark} \right]^{\leq 1}_{\mathcal{Z}}}{1-\eta_3} + 
   \frac{\left[Y^{X,\emptyset}_{1,(\eta_3,\eta_l)} e_{X,1,(\eta_3,\eta_l),\emptyset} \right]^{\leq 1}_{\mathcal{Z}}}{1-\eta_3} \left(1-\frac{1}{1-p^{\mathcal{Z}}_d}\right) . \label{Eetak-1r}
\end{align}

\begin{remark}  \label{rmk:definite-signs}
With the choices made for $\eta_1$, $\eta_2$ and $\eta_3$ in \eqref{eta1}--\eqref{eta3}, one can verify that the sign of each coefficient, of the restricted test-round yields in \eqref{Yetak-0}--\eqref{Yetak-1r} and of the restricted test-round bit error rates in \eqref{Yetak-0}--\eqref{Yetak-1r}, is determined and known.
\end{remark}

We observe that the explicit expressions found for the variables  $\mathbbm{Y}^{0}_{\eta_l}$, $\mathbbm{Y}^{0,1}_{\eta_l}$, $\mathbbm{Y}^{1t}_{\eta_l}$, and $\mathbbm{Y}^{1r}_{\eta_l}$ ($\mathbbm{E}^{0}_{\eta_l}$, $\mathbbm{E}^{0,1}_{\eta_l}$, $\mathbbm{E}^{1t}_{\eta_l}$, and $\mathbbm{E}^{1r}_{\eta_l}$) are given in terms of the test-round yields (test-round bit error rates) restricted to the ($\leq 1$)-subspace, which are not known exactly but are tightly bounded in \eqref{yield-click-range} and \eqref{yield-noclick-range} (\eqref{bit-error-rate-click-range} and \eqref{bit-error-rate-noclick-range}). Then, by virtue of Remark~\ref{rmk:definite-signs}, we can easily replace the restricted test-round yields (test-round bit error rates) with the respective upper or lower bound, depending on whether the yields (bit error rates) contribute positively or negatively to the phase error rate.

In the following, we employ the expressions derived for $\mathbbm{Y}^{0}_{\eta_l}$, $\mathbbm{Y}^{0,1}_{\eta_l}$, $\mathbbm{Y}^{1t}_{\eta_l}$, and $\mathbbm{Y}^{1r}_{\eta_l}$ ($\mathbbm{E}^{0}_{\eta_l}$, $\mathbbm{E}^{0,1}_{\eta_l}$, $\mathbbm{E}^{1t}_{\eta_l}$, and $\mathbbm{E}^{1r}_{\eta_l}$) to obtain the individual values of the three terms in the yields' expansions \eqref{yield-click-TBS} and \eqref{yield-noclick-TBS2} (bit error rates' expansions \eqref{bit-error-rate-click-TBS2} and \eqref{bit-error-rate-noclick-TBS2}). These are then used to derive an upper bound on $\Delta_1$, where it will become clear how to replace the restricted yields (bit error rates) with their respective bounds from \eqref{yield-click-range} and \eqref{yield-noclick-range} (\eqref{bit-error-rate-click-range} and \eqref{bit-error-rate-noclick-range}).\\

\subsubsection{Upper bound on $\Delta_1$ (part one)}\label{app:PER9}
Here, we derive an upper bound on the remaining contribution that is left to be estimated in the phase error rate upper bound, \eqref{phase-error-rate-uppbound}, namely $\Delta_1$ given in \eqref{e1}. Initially, we aim at calculating an upper bound on the operator $\sum_{k'\neq k} \Tilde{X}_{k'}$ appearing in $\Delta_1$, where $\tilde{X}_k$ is the POVM element defined in \eqref{Xprimek} and reported here for clarity:
\begin{align}
    \tilde{X}_k &= (\sqrt{X_\checkmark})^{-1} X_k (\sqrt{X_\checkmark})^{-1}  \oplus \frac{\one^{\perp}_{X_\checkmark}}{d},  \label{Xprimek-appendix}
\end{align}
where both $X_k$ and $X_\checkmark=\sum_k X_k$ are taken from Bob's test measurement, \eqref{Bobs-X-measurement-appendix}. According to the formulas in \eqref{Xkdet-a2} and \eqref{Xknodet-a2}, we can obtain an expression for the POVM element $X_k$ of Bob's test measurement, which corresponds to outcome $X_B=k$ in the $X$-basis detector regardless of what happens in the $Z$-basis detector and for the TBS setting $(\eta_\uparrow,\eta_\uparrow)$, as follows:
\begin{align}
    X_k &= \bra{vac}_b U(\eta_\uparrow)^\dag (X_k)_c \otimes \one_{c'} U(\eta_\uparrow) \ket{vac}_b \label{Xk-Delta1},
\end{align}
where $U(\eta_\uparrow)$ describes the unitary action of the TBS and where $(X_k)_c$ is given in \eqref{Xk-mode-c2}. By employing \eqref{Xk-mode-c2} in \eqref{Xk-Delta1}, we notice that $X_k$ on mode $a$ is just the operator $(X_k)_e$, after letting the incoming mode traverse two lossy elements represented by beam splitters with transmittances $\eta_r$ and $\eta_\uparrow$, respectively. Thus, equivalently, we can compute $X_k$ as resulting from $(X_k)_e$ after one lossy element with a combined transmittance of $\eta_r \eta_\uparrow$, represented by a unitary $U(\eta_r \eta_\uparrow)$ mapping modes $a,b$ to modes $e,f$. Thus, we obtain:
\begin{align}
    X_k = \bra{vac}_b U^\dag(\eta_r \eta_\uparrow) (X_k)_e \otimes\one_f U(\eta_r \eta_\uparrow) \ket{vac}_b.  \label{Xk-Delta1-2}
\end{align}
Then, by analogy with $(X_k)_c$ in \eqref{Xk-mode-c2}, we can use the result in \eqref{Xk-mode-c} to deduce the final form of $X_k$ on mode $a$:
\begin{align}
    X_k &= \sum_{K\in\mathcal{P}(D)\setminus\emptyset} \Pr(k|K) \,\, \sum_{S \in \mathcal{P}(K)} (p^X_d)^{|K|-|S|}(1-p^X_d)^{d-|K|} \,\,(M^{\geq 1}_S)_a, \label{Xk-Delta1-3}
\end{align}
where $(M^{\geq 1}_S)_a$ is defined as:
\begin{align}
   (M^{\geq 1}_S)_a = \sum_{\alpha_0,\alpha_1,\dots,\alpha_{d-1}=0}^\infty \Pi^{\alpha_0,\alpha_1,\dots,\alpha_{d-1}}_{\mathcal{X}} \prod_{s\in S} \left[1-(1-\eta_r \eta_\uparrow)^{\alpha_s}\right] \prod_{r \in S^c} (1-\eta_r \eta_\uparrow)^{\alpha_r}. \label{M_S-Delta1}
\end{align}

In a similar manner, we can obtain the expression for $X_\checkmark$, i.e.\ the POVM element corresponding to a detection in the $X$-basis detector,  regardless of what happens in the $Z$-basis detector and for a TBS setting $(\eta_\uparrow,\eta_\uparrow)$. Indeed, from \eqref{Xk-Delta1-2} we have that:
\begin{align}
    X_\checkmark &= \bra{vac}_b U(\eta_r\eta_\uparrow)^\dag (X_\checkmark)_e \otimes \one_{f} U(\eta_r\eta_\uparrow) \ket{vac}_b \label{Xdet-Delta1},
\end{align}
with $(X_\checkmark)_e$ given in \eqref{Xdet-mode-e}. Then, in analogy with the calculation carried out for $(X_\checkmark)_c$ that led to \eqref{Xclick-mode-c}, from the last expression we obtain the desired expression:
\begin{align}
    X_\checkmark &= \sum_{\beta=0}^\infty \left[1-(1-p^{\mathcal{X}}_d)(1-\eta_r \eta_\uparrow)^\beta\right] (\Pi^\beta_{\mathcal{X}})_a  \label{Xdet-Delta1-2}.
\end{align}
From \eqref{Xdet-Delta1-2}, we can easily obtain the inverse of its square root:
\begin{align}
    (\sqrt{X_\checkmark})^{-1} = \sum_{\beta=0}^\infty \frac{(\Pi^\beta_{\mathcal{X}})_a}{\sqrt{1-(1-p^{\mathcal{X}}_d)(1-\eta_r \eta_\uparrow)^\beta}} \label{sqrt(Xclick)to-1},
\end{align}
where we used the fact that the projectors $\Pi^\beta_{\mathcal{X}}$ are orthogonal. Moreover, since $\sum_{\beta=0}^\infty \Pi^\beta_{\mathcal{X}}=\one$, we deduce that $X_\checkmark$ has support on the whole Hilbert space:
\begin{align}
    \one_{X_\checkmark}=\one \label{Xtilde-calc}.
\end{align}

Now, to calculate \eqref{Xprimek-appendix} we recast the operator  in \eqref{M_S-Delta1} as follows:
\begin{align}
    (M^{\geq 1}_S)_a = \sum_{\beta=0}^\infty \,\,\sum_{\substack{\alpha_0,\alpha_1,\dots,\alpha_{d-1}: \\ \sum_i \alpha_i=\beta}} \Pi^{\alpha_0,\alpha_1,\dots,\alpha_{d-1}}_{\mathcal{X}} \prod_{s\in S} \left[1-(1-\eta_r \eta_\uparrow)^{\alpha_s}\right]  (1-\eta_r \eta_\uparrow)^{\beta - \sum_{s\in S} \alpha_s}, \label{M_S2}
\end{align}
such that, together with \eqref{sqrt(Xclick)to-1}, we get:
\begin{align}
    (\sqrt{X_\checkmark})^{-1}  &(M^{\geq 1}_S)_a (\sqrt{X_\checkmark})^{-1}  = \nonumber\\
    &\sum_{\beta=0}^\infty \frac{1}{1-(1-p^{\mathcal{X}}_d)(1-\eta_r \eta_\uparrow)^\beta} \sum_{\substack{\alpha_0,\alpha_1,\dots,\alpha_{d-1}: \\ \sum_i \alpha_i=\beta}} \Pi^{\alpha_0,\alpha_1,\dots,\alpha_{d-1}}_{\mathcal{X}} \prod_{s\in S} \left[1-(1-\eta_r \eta_\uparrow)^{\alpha_s}\right]  (1-\eta_r \eta_\uparrow)^{\beta - \sum_{s\in S} \alpha_s} \nonumber\\
    &\leq \frac{1}{p^{\mathcal{X}}_d} \, \Pi^0_{\mathcal{X}} \, \delta_{S,\emptyset} +  \frac{1}{1-(1-p^{\mathcal{X}}_d)(1-\eta_r \eta_\uparrow)}  \nonumber\\
    &\hspace{3cm}\sum_{\beta=1}^\infty  \sum_{\substack{\alpha_0,\alpha_1,\dots,\alpha_{d-1}: \\ \sum_i \alpha_i=\beta}} \Pi^{\alpha_0,\alpha_1,\dots,\alpha_{d-1}}_{\mathcal{X}} \prod_{s\in S} \left[1-(1-\eta_r \eta_\uparrow)^{\alpha_s}\right]  (1-\eta_r \eta_\uparrow)^{\beta - \sum_{s\in S} \alpha_s} \nonumber\\
    &= \left( \frac{1}{p^{\mathcal{X}}_d} - \frac{1}{1-(1-p^{\mathcal{X}}_d)(1-\eta_r \eta_\uparrow)}\right) \Pi^0_{\mathcal{X}} \, \delta_{S,\emptyset} + \frac{1}{1-(1-p^{\mathcal{X}}_d)(1-\eta_r \eta_\uparrow)} (M^{\geq 1}_S)_a, \label{sqrt(Xclick)-M_S-sqrt(Xclick)}
\end{align}
where, in the inequality, we pulled out the first term in the sum over $\beta$, noting that it always equals zero except when the set $S$ is empty, and multiplied all the other terms in the sum over $\beta$ with the largest coefficient in the sum. By using the following simplifications:
\begin{align}
     \frac{1}{p^{\mathcal{X}}_d} - \frac{1}{1-(1-p^{\mathcal{X}}_d)(1-\eta_r \eta_\uparrow)} &=  \frac{\eta_r \eta_\uparrow(1-p^{\mathcal{X}}_d)}{p^{\mathcal{X}}_d[p^{\mathcal{X}}_d + \eta_r \eta_\uparrow (1-p^{\mathcal{X}}_d)]}, \\
     \frac{1}{1-(1-p^{\mathcal{X}}_d)(1-\eta_r \eta_\uparrow)} &= \frac{1}{p^{\mathcal{X}}_d + \eta_r \eta_\uparrow (1-p^{\mathcal{X}}_d)},
\end{align}
in \eqref{sqrt(Xclick)-M_S-sqrt(Xclick)}, we obtain:
\begin{align}
    (\sqrt{X_\checkmark})^{-1}  (M^{\geq 1}_S)_a (\sqrt{X_\checkmark})^{-1}  &\leq  \frac{\eta_r \eta_\uparrow(1-p^{\mathcal{X}}_d)}{p^{\mathcal{X}}_d[p^{\mathcal{X}}_d + \eta_r \eta_\uparrow (1-p^{\mathcal{X}}_d)]} \, \Pi^0_{\mathcal{X}} \, \delta_{S,\emptyset} + \frac{1}{p^{\mathcal{X}}_d + \eta_r \eta_\uparrow (1-p^{\mathcal{X}}_d)}   (M^{\geq 1}_S)_a \label{sqrt(Xclick)-M_S-sqrt(Xclick)2}.
\end{align}

By employing the last expression, together with \eqref{Xk-Delta1-3} and \eqref{Xtilde-calc}, in \eqref{Xprimek-appendix}, we can derive an upper bound on the following operator that appears in $\Delta_1$:
\begin{align}
    {\textstyle\sum_{k' \neq k}} \Tilde{X}_{k'} &\leq \frac{1}{p^{\mathcal{X}}_d + \eta_r \eta_\uparrow (1-p^{\mathcal{X}}_d)}  {\textstyle\sum_{k' \neq k}} X_{k'}  \nonumber\\
    &\quad + \frac{\eta_r \eta_\uparrow(1-p^{\mathcal{X}}_d)}{p^{\mathcal{X}}_d[p^{\mathcal{X}}_d + \eta_r \eta_\uparrow (1-p^{\mathcal{X}}_d)]}  \sum_{K\in\mathcal{P}(D)\setminus\emptyset} \left({\textstyle\sum_{k' \neq k}}\Pr(k'|K)\right) \, \sum_{S \in \mathcal{P}(K)} (p^X_d)^{|K|-|S|}(1-p^X_d)^{d-|K|}\, \Pi^0_{\mathcal{X}} \, \delta_{S,\emptyset} .  \label{Xprimek-appendix3}
\end{align}
By using \eqref{normalization-remapping}, which implies that: $\sum_{k' \neq k}\Pr(k'|K) \leq 1$, we can further simplify the last expression as follows:
\begin{align}
    {\textstyle\sum_{k' \neq k}} \Tilde{X}_{k'} \leq \frac{{\textstyle\sum_{k' \neq k}} X_{k'}}{p^{\mathcal{X}}_d + \eta_r \eta_\uparrow (1-p^{\mathcal{X}}_d)}  + \Pi^0_{\mathcal{X}}  \frac{\eta_r \eta_\uparrow(1-p^{\mathcal{X}}_d)}{p^{\mathcal{X}}_d[p^{\mathcal{X}}_d + \eta_r \eta_\uparrow (1-p^{\mathcal{X}}_d)]}  \sum_{K\in\mathcal{P}(D)\setminus\emptyset}  (p^X_d)^{|K|}(1-p^X_d)^{d-|K|} .  \label{Xprimek-appendix4}
\end{align}
Now, consider that the terms in the sum over the sets $K$ only depend on the cardinality of $K$. Thus we can turn the sum into a sum over the allowed cardinalities, by introducing a binomial coefficient that accounts the number of different sets $K$ that can be selected for a given cardinality. This leads to the following chain of equalities:
\begin{align}
    {\textstyle\sum_{k' \neq k}}\Tilde{X}_{k'} &\leq \frac{{\textstyle\sum_{k' \neq k}} X_{k'}}{p^{\mathcal{X}}_d + \eta_r \eta_\uparrow (1-p^{\mathcal{X}}_d)}     + \Pi^0_{\mathcal{X}}  \frac{\eta_r \eta_\uparrow(1-p^{\mathcal{X}}_d)}{p^{\mathcal{X}}_d[p^{\mathcal{X}}_d + \eta_r \eta_\uparrow (1-p^{\mathcal{X}}_d)]}  \sum_{r=1}^d \binom{d}{r}  (p^X_d)^r(1-p^X_d)^{d-r} \nonumber\\
    &= \frac{1}{p^{\mathcal{X}}_d + \eta_r \eta_\uparrow (1-p^{\mathcal{X}}_d)}  {\textstyle\sum_{k' \neq k}} X_{k'}   + \Pi^0_{\mathcal{X}}  \frac{\eta_r \eta_\uparrow(1-p^{\mathcal{X}}_d)}{p^{\mathcal{X}}_d[p^{\mathcal{X}}_d + \eta_r \eta_\uparrow (1-p^{\mathcal{X}}_d)]}  \left[1-(1-p^X_d)^d\right] \nonumber\\
    &=\frac{1}{p^{\mathcal{X}}_d + \eta_r \eta_\uparrow (1-p^{\mathcal{X}}_d)}  {\textstyle\sum_{k' \neq k}} X_{k'}   + \Pi^0_{\mathcal{X}}  \frac{\eta_r \eta_\uparrow(1-p^{\mathcal{X}}_d)}{p^{\mathcal{X}}_d[p^{\mathcal{X}}_d + \eta_r \eta_\uparrow (1-p^{\mathcal{X}}_d)]} p^{\mathcal{X}}_d \nonumber\\
    &=\frac{1}{p^{\mathcal{X}}_d + \eta_r \eta_\uparrow (1-p^{\mathcal{X}}_d)}  {\textstyle\sum_{k' \neq k}} \left( X^{(\eta_\uparrow,\eta_\uparrow)}_{k',\checkmark} + X^{(\eta_\uparrow,\eta_\uparrow)}_{k',\emptyset}\right)   +  \frac{\eta_r \eta_\uparrow(1-p^{\mathcal{X}}_d)}{p^{\mathcal{X}}_d + \eta_r \eta_\uparrow (1-p^{\mathcal{X}}_d)} \Pi^0_{\mathcal{X}} ,  \label{Xprimek-appendix5}
\end{align}
where in the third line we used \eqref{prob-dk-tot-X} and in the fourth line we used the definition of Bob's test measurement from \eqref{Bobs-X-measurement-appendix}.

By employing the operator inequality derived in \eqref{Xprimek-appendix5} in the expression for $\Delta_1$, \eqref{e1}, we obtain the following upper bound on $\Delta_1$:
\begin{align}
    \Delta_1 &\leq \frac{1}{p^{\mathcal{X}}_d + \eta_r \eta_\uparrow (1-p^{\mathcal{X}}_d)}  {\textstyle\sum_{k=0}^{d-1}} \Tr\left[\frac{\sigma_k}{d}  M^{\leq 1}_{\mathcal{Z}} {\textstyle\sum_{k' \neq k}} \left( X^{(\eta_\uparrow,\eta_\uparrow)}_{k',\checkmark} + X^{(\eta_\uparrow,\eta_\uparrow)}_{k',\emptyset}\right)  M^{\leq 1}_{\mathcal{Z}} \right] \nonumber\\
    &\quad+  \frac{\eta_r \eta_\uparrow(1-p^{\mathcal{X}}_d)}{p^{\mathcal{X}}_d + \eta_r \eta_\uparrow (1-p^{\mathcal{X}}_d)}   \Tr\left[\bar{\sigma}  M^{\leq 1}_{\mathcal{Z}}  \Pi^0_{\mathcal{X}} M^{\leq 1}_{\mathcal{Z}} \right]   \label{Delta1-upperbound},
\end{align}
where $M^{\leq 1}_{\mathcal{Z}}$ is given in \eqref{M^leq1_Z} and where we used \eqref{barsigma}. Now, the goal is to link the expectation values on the right-hand side of \eqref{Delta1-upperbound} to quantities that are observed in the experiment. For the expectation values in the first addend, we expand $M^{\leq 1}_{\mathcal{Z}}$ according to \eqref{M^leq1_Z} and employ the definitions in \eqref{bit-error-rate-click-TBS2-Pi0}--\eqref{bit-error-rate-noclick-TBS2-Pi01}, thus obtaining:
\begin{align}
    {\textstyle\sum_{k=0}^{d-1}} &\Tr\left[\frac{\sigma_k}{d}  M^{\leq 1}_{\mathcal{Z}} {\textstyle\sum_{k' \neq k}} \left( X^{(\eta_\uparrow,\eta_\uparrow)}_{k',\checkmark} + X^{(\eta_\uparrow,\eta_\uparrow)}_{k',\emptyset}\right)  M^{\leq 1}_{\mathcal{Z}} \right] = \nonumber\\
    &p_{\checkmark|0} \left(\left[Y^{X,\checkmark}_{1,(\eta_\uparrow,\eta_\uparrow)} e_{X,1,(\eta_\uparrow,\eta_\uparrow),\checkmark}\right]^{0}_{\mathcal{Z}} + \left[Y^{X,\emptyset}_{1,(\eta_\uparrow,\eta_\uparrow)} e_{X,1,(\eta_\uparrow,\eta_\uparrow),\emptyset}\right]^{0}_{\mathcal{Z}}\right) \nonumber\\
    &+ \sqrt{p_{\checkmark|0} p_{\checkmark|1}} \left(\left[Y^{X,\checkmark}_{1,(\eta_\uparrow,\eta_\uparrow)} e_{X,1,(\eta_\uparrow,\eta_\uparrow),\checkmark}\right]^{0,1}_{\mathcal{Z}} + \left[Y^{X,\emptyset}_{1,(\eta_\uparrow,\eta_\uparrow)} e_{X,1,(\eta_\uparrow,\eta_\uparrow),\emptyset}\right]^{0,1}_{\mathcal{Z}}\right) \nonumber\\
    &+ p_{\checkmark|1} \left(\left[Y^{X,\checkmark}_{1,(\eta_\uparrow,\eta_\uparrow)} e_{X,1,(\eta_\uparrow,\eta_\uparrow),\checkmark}\right]^{1}_{\mathcal{Z}} + \left[Y^{X,\emptyset}_{1,(\eta_\uparrow,\eta_\uparrow)} e_{X,1,(\eta_\uparrow,\eta_\uparrow),\emptyset}\right]^{1}_{\mathcal{Z}}\right) \label{first-exp-val-Delta1}. 
\end{align}
Now, we use the results in \eqref{bit-error-rate-click-TBS2-Pi0-2}--\eqref{bit-error-rate-noclick-TBS2-Pi1-2}, obtained for the bit error rates restricted by the projectors $\Pi^0_{\mathcal{Z}}$ and $\Pi^1_{\mathcal{Z}}$, to recast the last expression as follows:
\begin{align}
    {\textstyle\sum_{k=0}^{d-1}} &\Tr\left[\frac{\sigma_k}{d}  M^{\leq 1}_{\mathcal{Z}} {\textstyle\sum_{k' \neq k}} \left( X^{(\eta_\uparrow,\eta_\uparrow)}_{k',\checkmark} + X^{(\eta_\uparrow,\eta_\uparrow)}_{k',\emptyset}\right)  M^{\leq 1}_{\mathcal{Z}} \right] = p_{\checkmark|0} \mathbbm{E}^0_{\eta_\uparrow} +  \sqrt{p_{\checkmark|0} p_{\checkmark|1}} \, \sqrt{\eta_\uparrow} \mathbbm{E}^{0,1}_{\eta_\uparrow} + p_{\checkmark|1} \left( \eta_\uparrow \mathbbm{E}^{1t}_{\eta_\uparrow} + (1-\eta_\uparrow) \mathbbm{E}^{1r}_{\eta_\uparrow}\right) \label{first-exp-val-Delta1-2}, 
\end{align}
where the variables $\mathbbm{E}^{0}_{\eta_\uparrow}$, $\mathbbm{E}^{0,1}_{\eta_\uparrow}$, $\mathbbm{E}^{1t}_{\eta_\uparrow}$, and $\mathbbm{E}^{1r}_{\eta_\uparrow}$ have been obtained in terms of the test-round bit error rates, restricted to the ($\leq 1$)-subspace, in \eqref{Eetak-0}--\eqref{Eetak-1r}. Now, we recall that while the test-round bit error rates restricted to the ($\leq 1$)-subspace are not directly observed, they must belong to tight intervals as derived in \eqref{bit-error-rate-click-range} and \eqref{bit-error-rate-noclick-range}.

Therefore, we now replace the variables $\mathbbm{E}^{0}_{\eta_\uparrow}$, $\mathbbm{E}^{0,1}_{\eta_\uparrow}$, $\mathbbm{E}^{1t}_{\eta_\uparrow}$, and $\mathbbm{E}^{1r}_{\eta_\uparrow}$ in \eqref{first-exp-val-Delta1-2} with the respective upper bounds: $\overline{\mathbbm{E}}^{0}_{\eta_\uparrow}$, $\overline{\mathbbm{E}}^{0,1}_{\eta_\uparrow}$, $\overline{\mathbbm{E}}^{1t}_{\eta_\uparrow}$, and $\overline{\mathbbm{E}}^{1r}_{\eta_\uparrow}$. These bounds are obtained from \eqref{Eetak-0}--\eqref{Eetak-1r} by substituting  each test-round bit error rate, restricted to the ($\leq 1$)-subspace, with the appropriate extremal point from the intervals \eqref{bit-error-rate-click-range} and \eqref{bit-error-rate-noclick-range}, such that the resulting expression is an upper bound on the original expression. By doing so, we obtain an upper bound on \eqref{first-exp-val-Delta1-2} which is entirely given in terms of observed quantities and reads:
\begin{align}
    {\textstyle\sum_{k=0}^{d-1}} &\Tr\left[\frac{\sigma_k}{d}  M^{\leq 1}_{\mathcal{Z}} {\textstyle\sum_{k' \neq k}} \left( X^{(\eta_\uparrow,\eta_\uparrow)}_{k',\checkmark} + X^{(\eta_\uparrow,\eta_\uparrow)}_{k',\emptyset}\right)  M^{\leq 1}_{\mathcal{Z}} \right] \leq p_{\checkmark|0} \overline{\mathbbm{E}}^0_{\eta_\uparrow} +  \sqrt{p_{\checkmark|0} p_{\checkmark|1}} \, \sqrt{\eta_\uparrow} \overline{\mathbbm{E}}^{0,1}_{\eta_\uparrow} + p_{\checkmark|1} \left( \eta_\uparrow \overline{\mathbbm{E}}^{1t}_{\eta_\uparrow} + (1-\eta_\uparrow) \overline{\mathbbm{E}}^{1r}_{\eta_\uparrow}\right) \label{first-exp-val-Delta1-upperbound}, 
\end{align}
where, in the definitions of the following upper bounds, we replaced $\eta_1$ and $\eta_3$ with \eqref{eta1} and \eqref{eta3}, respectively:
\begin{align}
    \overline{\mathbbm{E}}^{0}_{\eta_\uparrow} &:=\min\left\lbrace \overline{w}^0_{\mathcal{Z}}, \frac{\sqrt{\eta_2 \eta_\downarrow}\overline{\left[Y^{X,\emptyset}_{1,(\eta_\uparrow,\eta_\uparrow)} e_{X,1,(\eta_\uparrow,\eta_\uparrow),\emptyset}\right]^{\leq 1}_{\mathcal{Z}}}}{\left(\sqrt{\eta_\uparrow}-\sqrt{\eta_2}\right)\left(\sqrt{\eta_\uparrow}-\sqrt{\eta_\downarrow}\right)\left(1-p^{\mathcal{Z}}_d\right)}  - \frac{\sqrt{\eta_\uparrow \eta_\downarrow} \underline{\left[Y^{X,\emptyset}_{1,(\eta_2,\eta_\uparrow)} e_{X,1,(\eta_2,\eta_\uparrow),\emptyset} \right]^{\leq 1}_{\mathcal{Z}}}}{\left(\sqrt{\eta_\uparrow}-\sqrt{\eta_2}\right)
   \left(\sqrt{\eta_2}-\sqrt{\eta_\downarrow}\right) \left(1-p^{\mathcal{Z}}_d\right)} \right.  \nonumber\\
   &\left.\quad+\frac{\sqrt{\eta_\uparrow \eta_2}}{\left(\sqrt{\eta_\uparrow}-\sqrt{\eta_\downarrow}\right) \left(\sqrt{\eta_2}-\sqrt{\eta_\downarrow}\right) 
   \left(1-p^{\mathcal{Z}}_d\right)} \overline{\left[Y^{X,\emptyset}_{1,(\eta_\downarrow,\eta_\uparrow)} e_{X,1,(\eta_\downarrow,\eta_\uparrow),\emptyset} \right]^{\leq 1}_{\mathcal{Z}}} \right\rbrace   \label{Eetak-0-upp}\\
   \overline{\mathbbm{E}}^{0,1}_{\eta_\uparrow} &:= \min\left\lbrace \frac{\overline{\left[Y^{X,\emptyset}_{1,(\eta_\uparrow,\eta_\uparrow)} e_{X,1,(\eta_\uparrow,\eta_\uparrow),\emptyset} \right]^{\leq 1}_{\mathcal{Z}}}}{(1-p^{\mathcal{Z}}_d)\sqrt{\eta_\uparrow}}, -\frac{(\sqrt{\eta_2} +  \sqrt{\eta_\downarrow})\underline{\left[Y^{X,\emptyset}_{1,(\eta_\uparrow,\eta_\uparrow)} e_{X,1,(\eta_\uparrow,\eta_\uparrow),\emptyset} \right]^{\leq 1}_{\mathcal{Z}}}}{\left(\sqrt{\eta_\uparrow}-\sqrt{\eta_2}\right)\left(\sqrt{\eta_\uparrow}-\sqrt{\eta_\downarrow}\right)\left(1-p^{\mathcal{Z}}_d\right)}  \right. \nonumber\\
   &\left.\quad+ \frac{(\sqrt{\eta_\uparrow} +  \sqrt{\eta_\downarrow}) \overline{\left[Y^{X,\emptyset}_{1,(\eta_2,\eta_\uparrow)} e_{X,1,(\eta_2,\eta_\uparrow),\emptyset} \right]^{\leq 1}_{\mathcal{Z}}}}{\left(\sqrt{\eta_\uparrow}-\sqrt{\eta_2}\right)\left(\sqrt{\eta_2}-\sqrt{\eta_\downarrow}\right)\left(1-p^{\mathcal{Z}}_d\right)} - \frac{(\sqrt{\eta_\uparrow} +  \sqrt{\eta_2}) \underline{\left[Y^{X,\emptyset}_{1,(\eta_\downarrow,\eta_\uparrow)} e_{X,1,(\eta_\downarrow,\eta_\uparrow),\emptyset}\right]^{\leq 1}_{\mathcal{Z}}}}{\left(\sqrt{\eta_\uparrow}-\sqrt{\eta_\downarrow}\right)\left(\sqrt{\eta_2}-\sqrt{\eta_\downarrow}\right)\left(1-p^{\mathcal{Z}}_d\right)}  \right\rbrace \label{Eetak-01-upp}\\
   \overline{\mathbbm{E}}^{1t}_{\eta_\uparrow} &:= \min\left\lbrace \overline{w}^1_{\mathcal{Z}}, \frac{\overline{\left[Y^{X,\emptyset}_{1,(\eta_\uparrow,\eta_\uparrow)} e_{X,1,(\eta_\uparrow,\eta_\uparrow),\emptyset} \right]^{\leq 1}_{\mathcal{Z}}}}{\left(\sqrt{\eta_\uparrow}-\sqrt{\eta_2}\right)\left(\sqrt{\eta_\uparrow}-\sqrt{\eta_\downarrow}\right)\left(1-p^{\mathcal{Z}}_d\right)}  - \frac{\underline{\left[Y^{X,\emptyset}_{1,(\eta_2,\eta_\uparrow)} e_{X,1,(\eta_2,\eta_\uparrow),\emptyset} \right]^{\leq 1}_{\mathcal{Z}}}}{\left(\sqrt{\eta_\uparrow}-\sqrt{\eta_2}\right)
   \left(\sqrt{\eta_2}-\sqrt{\eta_\downarrow}\right) \left(1-p^{\mathcal{Z}}_d\right)} \right.  \nonumber\\
   &\left. \quad+\frac{1}{\left(\sqrt{\eta_\uparrow}-\sqrt{\eta_\downarrow}\right) \left(\sqrt{\eta_2}-\sqrt{\eta_\downarrow}\right) 
   \left(1-p^{\mathcal{Z}}_d\right)} \overline{\left[Y^{X,\emptyset}_{1,(\eta_\downarrow,\eta_\uparrow)} e_{X,1,(\eta_\downarrow,\eta_\uparrow),\emptyset}\right]^{\leq 1}_{\mathcal{Z}}} \right\rbrace \label{Eetak-1t-upp} \\
   \overline{\mathbbm{E}}^{1r}_{\eta_\uparrow} &:= \min\left\lbrace \overline{w}^1_{\mathcal{Z}}, \frac{\overline{\left[Y^{X,\checkmark}_{1,(\eta_\downarrow,\eta_\uparrow)} e_{X,1,(\eta_\downarrow,\eta_\uparrow),\checkmark} \right]^{\leq 1}_{\mathcal{Z}}}}{1-\eta_\downarrow} - 
   \frac{p^{\mathcal{Z}}_d}{(1-\eta_\downarrow)(1-p^{\mathcal{Z}}_d)} \underline{\left[Y^{X,\emptyset}_{1,(\eta_\downarrow,\eta_\uparrow)} e_{X,1,(\eta_\downarrow,\eta_\uparrow),\emptyset} \right]^{\leq 1}_{\mathcal{Z}}} \right\rbrace  . \label{Eetak-1r-upp}
\end{align}
In the last expression, the upper and lower bounds on the test-round bit error rates are reported in \eqref{l-e,click}, \eqref{u-e,click}, \eqref{l-e,noclick}, and \eqref{u-e,noclick}. Moreover, we used the intervals of existence in \eqref{bit-error-rate-click-Pialpha-interval}--\eqref{bit-error-rate-noclick-Pi01-interval} to obtain non-trivial ranges for the upper bounds in \eqref{Eetak-0-upp}--\eqref{Eetak-1r-upp}. For example, \eqref{bit-error-rate-noclick-TBS2-Pi0-2} combined with the interval \eqref{bit-error-rate-noclick-Pialpha-interval} implies the following upper bound:
\begin{align}
    \mathbbm{E}^{0}_{\eta_\uparrow}\leq \overline{w}^0_{\mathcal{Z}}/(1-p^{\mathcal{Z}}_d) \label{bound-E0}.
\end{align}
However, since the upper bound provided by the interval \eqref{bit-error-rate-noclick-Pialpha-interval} is independent of $p^{\mathcal{Z}}_d$ and since $\mathbbm{E}^{0}_{\eta_\uparrow}$ does not implicitly depend on $p^{\mathcal{Z}}_d$, we can set $p^{\mathcal{Z}}_d \to 0$ in the upper bound \eqref{bound-E0}, thereby obtaining the condition in \eqref{Eetak-0-upp}. The same can be argued for $\mathbbm{E}^{1t}_{\eta_\uparrow}$ and $\mathbbm{E}^{1r}_{\eta_\uparrow}$. Conversely, the upper bound on $\mathbbm{E}^{0,1}_{\eta_\uparrow}$ implied by \eqref{bit-error-rate-noclick-Pi01-interval} in general depends on $p^{\mathcal{Z}}_d$ and $\eta_\uparrow$, therefore the two parameters cannot be set to arbitrary values in the denominator of the first term in \eqref{Eetak-01-upp}.

Similarly to \eqref{first-exp-val-Delta1}, we can expand the second expectation value in \eqref{Delta1-upperbound} as follows:
\begin{align}
   \Tr\left[\bar{\sigma}  M^{\leq 1}_{\mathcal{Z}}  \Pi^0_{\mathcal{X}} M^{\leq 1}_{\mathcal{Z}} \right] &= p_{\checkmark|0} \Tr\left[\bar{\sigma}  \Pi^{0}_{\mathcal{Z}}  \Pi^0_{\mathcal{X}} \Pi^{0}_{\mathcal{Z}} \right] + \sqrt{p_{\checkmark|0}p_{\checkmark|1}} \Tr\left[\bar{\sigma}  \left(\Pi^{0}_{\mathcal{Z}}  \Pi^0_{\mathcal{X}} \Pi^{1}_{\mathcal{Z}} + \mathrm{h.c.}\right) \right] + p_{\checkmark|1} \Tr\left[\bar{\sigma}  \Pi^{1}_{\mathcal{Z}}  \Pi^0_{\mathcal{X}} \Pi^{1}_{\mathcal{Z}} \right]  \label{second-exp-val-Delta1}.
\end{align}
The three expectation values on the right-hand side are not directly observed experimentally and can be interpreted as the weights of the average state received by Bob in the subspace with no photons in $\mathcal{X}$ (which is the total set of modes measured by the $X$-basis detector), when restricted by the projectors $\Pi^0_{\mathcal{Z}}$ and $\Pi^1_{\mathcal{Z}}$. In order to estimate such quantities, in the following we link them to the test-round yields restricted to the ($\leq 1$)-subspace \eqref{yieldX-click-<1subspace} and \eqref{yieldX-noclick-<1subspace}, through the detector decoy technique \cite{detector-decoy}.\\

\subsubsection{The detector decoy technique on the $X$ detector statistics}\label{app:PER10}
The statistic collected in the test rounds that is required to estimate the three expectation values in \eqref{second-exp-val-Delta1} is the detection probability in the $X$-basis detector, given the TBS setting $(\eta_l,\eta_l)$ and regardless of the measurement outcome in the $Z$-basis detector. This probability is generated by the following POVM element:
\begin{align}
    X^{(\eta_l,\eta_l)}_\checkmark &:= \sum_{k=0}^{d-1} \left(X^{(\eta_l,\eta_l)}_{k,\checkmark} + X^{(\eta_l,\eta_l)}_{k,\emptyset}\right) \nonumber\\
    &= \bra{vac}_b U^\dag (\eta_l) (X_\checkmark)_c \otimes \one_{c'} U(\eta_l) \ket{vac}_b,
\end{align}
where in the second equality we used \eqref{Xkdet-a2} and \eqref{Xknodet-a2}. Now, by using the same observation made earlier in \eqref{Xk-Delta1-2} to compute an explicit expression for the POVM element $X_k$ of Bob's test measurement, we can view $X^{(\eta_l,\eta_l)}_\checkmark$ as resulting from $(X_\checkmark)_e$ when preceded by two lossy elements with transmittances $\eta_l$ and $\eta_r$, respectively. Therefore, we can recast the last expression as follows:
\begin{align}
    X^{(\eta_l,\eta_l)}_\checkmark &= \bra{vac}_b U^\dag (\eta_r \eta_l) (X_\checkmark)_e \otimes \one_{f} U(\eta_r \eta_l) \ket{vac}_b \nonumber\\
    &= \sum_{\beta=0}^\infty \left[1-(1-p^{\mathcal{X}}_d)(1-\eta_r \eta_l)^\beta\right] \Pi^\beta_{\mathcal{X}} \label{Xdet-etak}
\end{align}
where in the second equality we used the similar result obtained for $X_\checkmark$ in \eqref{Xdet-Delta1-2}. Then, from \eqref{Xdet-etak}, we obtain the POVM element corresponding to a no detection in the $X$-basis detector with TBS setting $(\eta_l,\eta_l)$:
\begin{align}
    X^{(\eta_l,\eta_l)}_\emptyset &:= \one - X^{(\eta_l,\eta_l)}_\checkmark \nonumber\\
    &= (1-p^{\mathcal{X}}_d) \sum_{\beta=0}^\infty (1-\eta_r \eta_l)^\beta  \Pi^\beta_{\mathcal{X}} \label{Xnodet-etak}.
\end{align}
Thus, the probability of a no detection in the $X$-basis detector generated by the average state $\bar{\sigma}$ reads:
\begin{align}
    1-\left(Y^{X,\checkmark}_{1,(\eta_l,\eta_l)} + Y^{X,\emptyset}_{1,(\eta_l,\eta_l)}\right) &= \Tr\left[\bar{\sigma} X^{(\eta_l,\eta_l)}_\emptyset\right] \nonumber\\
    &= (1-p^{\mathcal{X}}_d) \sum_{\beta=0}^\infty (1-\eta_r \eta_l)^\beta  \Tr\left[\bar{\sigma} \Pi^\beta_{\mathcal{X}} \right], \label{prob-Xnodet-etak}
\end{align}
where we used the definition of test-rounds yields in \eqref{yieldX-click} and \eqref{yieldX-noclick}.

Now, from \eqref{prob-Xnodet-etak} we realize that the expectation values required in \eqref{second-exp-val-Delta1} are not exactly related to the test-round yields, but rather to their restrictions in the subspace with zero or one photon in $\mathcal{Z}$. As a matter of fact, by projecting the state in \eqref{prob-Xnodet-etak} through $\Pi^0_{\mathcal{Z}}$ or $\Pi^1_{\mathcal{Z}}$, we obtain the following expressions, where on the right-hand side we recover the expectation values of \eqref{second-exp-val-Delta1}:
\begin{align}
    \Tr[\bar{\sigma} \Pi^0_{\mathcal{Z}}] -\left(\left[Y^{X,\checkmark}_{1,(\eta_l,\eta_l)}\right]^0_{\mathcal{Z}} + \left[Y^{X,\emptyset}_{1,(\eta_l,\eta_l)}\right]^0_{\mathcal{Z}}\right) &= (1-p^{\mathcal{X}}_d) \sum_{\beta=0}^\infty (1-\eta_r \eta_l)^\beta  \Tr\left[\bar{\sigma} \Pi^0_{\mathcal{Z}} \Pi^\beta_{\mathcal{X}} \Pi^0_{\mathcal{Z}} \right] \label{prob-Xnodet-etak-Pi0} \\
    \Tr[\bar{\sigma} \Pi^1_{\mathcal{Z}}] -\left(\left[Y^{X,\checkmark}_{1,(\eta_l,\eta_l)}\right]^1_{\mathcal{Z}} + \left[Y^{X,\emptyset}_{1,(\eta_l,\eta_l)}\right]^1_{\mathcal{Z}}\right) &= (1-p^{\mathcal{X}}_d) \sum_{\beta=0}^\infty (1-\eta_r \eta_l)^\beta  \Tr\left[\bar{\sigma} \Pi^1_{\mathcal{Z}} \Pi^\beta_{\mathcal{X}} \Pi^1_{\mathcal{Z}} \right] \label{prob-Xnodet-etak-Pi1},
\end{align}
where we used the definitions of the restricted test-round yields in \eqref{yield-click-TBS2-Pi0} and \eqref{yield-click-TBS2-Pi1}. Now, we link the yields restricted to the subspace with either zero or one photon in $\mathcal{Z}$ to the variables: $\mathbbm{Y}^0_{\eta_l}$, $\mathbbm{Y}^{1t}_{\eta_l}$ and $\mathbbm{Y}^{1r}_{\eta_l}$, via the relations \eqref{yield-click-TBS2-Pi0-2}, \eqref{yield-noclick-TBS2-Pi0-2}, \eqref{yield-click-TBS2-Pi1-2}, and \eqref{yield-noclick-TBS2-Pi1-2}. We obtain:
\begin{align}
    \Tr[\bar{\sigma} \Pi^0_{\mathcal{Z}}] -\mathbbm{Y}^0_{\eta_l} &= (1-p^{\mathcal{X}}_d) \sum_{\beta=0}^\infty (1-\eta_r \eta_l)^\beta  \Tr\left[\bar{\sigma} \Pi^0_{\mathcal{Z}} \Pi^\beta_{\mathcal{X}} \Pi^0_{\mathcal{Z}} \right] \label{prob-Xnodet-etak-Pi0-2} \\
    \Tr[\bar{\sigma} \Pi^1_{\mathcal{Z}}] -\left[\eta_l \mathbbm{Y}^{1t}_{\eta_l} + (1-\eta_l)\mathbbm{Y}^{1r}_{\eta_l} \right] &= (1-p^{\mathcal{X}}_d) \sum_{\beta=0}^\infty (1-\eta_r \eta_l)^\beta  \Tr\left[\bar{\sigma} \Pi^1_{\mathcal{Z}} \Pi^\beta_{\mathcal{X}} \Pi^1_{\mathcal{Z}} \right] \label{prob-Xnodet-etak-Pi1-2}.
\end{align}
We emphasize that the left-hand sides in the last two equations can be considered, for the moment, known quantities. Indeed, the variables $\mathbbm{Y}^0_{\eta_l}$, $\mathbbm{Y}^{1t}_{\eta_l}$ and $\mathbbm{Y}^{1r}_{\eta_l}$ are expressed in terms of the test-round yields restricted to the ($\leq 1$)-subspace via the solutions \eqref{Yetak-0}--\eqref{Yetak-1r} to the linear system \eqref{yield-system}. In turn, the test-round yields restricted to the ($\leq 1$)-subspace are tightly bounded by observed quantities in \eqref{yield-click-range} and \eqref{yield-noclick-range}. At the same time, the weights of the average state with zero or one photon in $\mathcal{Z}$, namely $\Tr[\bar{\sigma} \Pi^0_{\mathcal{Z}}]$ and $\Tr[\bar{\sigma} \Pi^1_{\mathcal{Z}}]$, have already been bounded in: \eqref{y0-upperbound}, \eqref{y1-upperbound}, \eqref{y1-lowerbound-explicit}, and \eqref{y0-lowerbound-explicit} when we applied the detector decoy technique to the statistics of the $Z$-basis detector.

We can thus employ \eqref{prob-Xnodet-etak-Pi0-2} and \eqref{prob-Xnodet-etak-Pi1-2} to derive an upper bound on the expectation values $\Tr\left[\bar{\sigma} \Pi^0_{\mathcal{Z}} \Pi^0_{\mathcal{X}} \Pi^0_{\mathcal{Z}} \right]$ and $\Tr\left[\bar{\sigma} \Pi^1_{\mathcal{Z}} \Pi^0_{\mathcal{X}} \Pi^1_{\mathcal{Z}} \right]$. To this aim, we observe that the two equations in \eqref{prob-Xnodet-etak-Pi0-2} and \eqref{prob-Xnodet-etak-Pi1-2} are of the form: 
\begin{align}
    f_l= \sum_{\beta=0}^\infty t_k^\beta x_\beta,  \label{detector-decoy}
\end{align}
which allows us to apply the detector decoy technique, where we identified:
\begin{align}
    f_l =\frac{\Tr[\bar{\sigma} \Pi^0_{\mathcal{Z}}] -\mathbbm{Y}^0_{\eta_l}}{1-p^{\mathcal{X}}_d}&\quad\mbox{ or }\quad f_l = \frac{\Tr[\bar{\sigma} \Pi^1_{\mathcal{Z}}] -\left[\eta_l \mathbbm{Y}^{1t}_{\eta_l} + (1-\eta_l)\mathbbm{Y}^{1r}_{\eta_l} \right]}{1-p^{\mathcal{X}}_d} \label{fk}\\
    t_k = 1-\eta_r \eta_l & \label{tk}\\
    x_\beta=\Tr\left[\bar{\sigma} \Pi^0_{\mathcal{Z}} \Pi^\beta_{\mathcal{X}} \Pi^0_{\mathcal{Z}} \right] &\quad\mbox{ or }\quad x_\beta= \Tr\left[\bar{\sigma} \Pi^1_{\mathcal{Z}} \Pi^\beta_{\mathcal{X}} \Pi^1_{\mathcal{Z}} \right]  \label{xbeta},
\end{align}
from which we deduce the following conditions on $x_\beta$:
\begin{align}
    x_\beta &\geq 0 \\
    \sum_{\beta=0}^\infty x_\beta&=C\,,\quad\mbox{with: }C=\Tr[\bar{\sigma} \Pi^0_{\mathcal{Z}}] \quad\mbox{ or }\quad C=\Tr[\bar{\sigma} \Pi^1_{\mathcal{Z}}]. \label{sum-xbeta}
\end{align}

We now apply the detector decoy technique to derive an upper bound on $x_0$, and thereby on two of the expectation values in \eqref{second-exp-val-Delta1}. We recall that $\eta_1=\eta_\uparrow$ and $\eta_3=\eta_\downarrow$ (cf.~Sec.~\ref{sec:protocol}). From \eqref{detector-decoy} and \eqref{sum-xbeta} we obtain the following system of inequalities:
\begin{align}
    &\left\lbrace \begin{array}{l}
       f_3 \leq x_0 + t_3 x_1 + t_3^2 (C-x_0 -x_1)  \\
    f_1 \geq x_0 + t_1 x_1  
    \end{array}\right. \nonumber\\
    \Rightarrow \quad &\left\lbrace \begin{array}{l}
        x_1 \geq \frac{f_3 -x_0(1-t_3^2) -C t_3^2}{t_3 (1-t_3)}  \\
     x_0 + t_1  \frac{f_3 -x_0 (1-t_3^2) -C t_3^2}{t_3 (1-t_3)} \leq f_1
    \end{array}\right. \nonumber\\
    \Rightarrow \quad &\left\lbrace \begin{array}{l}
       x_1 \geq \frac{f_3 -x_0(1-t_3^2) -C t_3^2}{t_3 (1-t_3)} \\
     x_0 \left[1 -\frac{t_1(1+t_3)}{t_3}\right] \leq f_1 - t_1 \frac{f_3 -C t_3^2}{t_3 (1-t_3)}
    \end{array}\right.,  \label{system-x0-upperbound}
\end{align}
where in the second step we used the lower bound on $x_1$ in the second inequality. From the bottom inequality of \eqref{system-x0-upperbound}, we can derive an upper bound on $x_0$, provided that its coefficient is positive. We thus require that:
\begin{align}
    &1- t_1 \frac{1+t_3}{t_3} >0  \\
    \iff \quad & t_3 -t_1 (1+t_3) >0 \\
    \iff \quad & t_3 (1-t_1) >t_1 \\
    \iff \quad & (1- \eta_\downarrow \eta_r)\eta_\uparrow \eta_r > 1-\eta_\uparrow \eta_r \\
    \iff \quad & -\eta^2_r (\eta_\downarrow \eta_\uparrow) + 2 \eta_r \eta_\uparrow-1 >0 \\
    \iff \quad & \eta_r > \frac{1-\sqrt{1-\frac{\eta_\downarrow}{\eta_\uparrow}}}{\eta_\downarrow}, \label{condition-x0-upperbound}
\end{align}
which is indeed satisfied since we assumed \eqref{constraint2}. Then, the upper bound on $x_0$ follows from the bottom inequality in \eqref{system-x0-upperbound} and reads:
\begin{align}
    x_0 &\leq \frac{t_3 f_1-t_1 f_3 + t_3^2 (C t_1 -f_1)}{(1-t_3)[t_3 -t_1 (1+t_3)]}. \label{x0-upperbound}
\end{align}
By substituting the original variables through \eqref{fk}, \eqref{tk} and \eqref{xbeta}, we obtain the desired upper bound on the weight of the state in the subspace with no photons in $\mathcal{X}$ and in $\mathcal{Z}$:
\begin{align}
    \Tr\left[ \Bar{\sigma} \Pi^0_{\mathcal{Z}} \Pi^0_{\mathcal{X}} \Pi^0_{\mathcal{Z}} \right] &\leq \frac{(1-\eta_\uparrow \eta_r) \mathbbm{Y}^0_{\eta_\downarrow} - (1-\eta_\downarrow \eta_r) \mathbbm{Y}^0_{\eta_\uparrow} + \Tr[\Bar{\sigma}\Pi^0_{\mathcal{Z}}] (\eta_\uparrow- \eta_\downarrow) \eta_r +(1-\eta_\downarrow \eta_r)^2 \left\lbrace\mathbbm{Y}^0_{\eta_\uparrow} -\Tr[\Bar{\sigma}\Pi^0_{\mathcal{Z}}][p^{\mathcal{X}}_d + \eta_\uparrow \eta_r (1-p^{\mathcal{X}}_d)]  \right\rbrace}{\eta_\downarrow \eta_r (1-p^{\mathcal{X}}_d)(2 \eta_\uparrow \eta_r -1 -\eta_r^2 \eta_\uparrow \eta_\downarrow)} \nonumber\\
    &=  \frac{(1-\eta_\uparrow \eta_r) \mathbbm{Y}^0_{\eta_\downarrow} - \eta_\downarrow \eta_r (1-\eta_\downarrow \eta_r) \mathbbm{Y}^0_{\eta_\uparrow}  -\Tr[\Bar{\sigma}\Pi^0_{\mathcal{Z}}]p^{\mathcal{X}}_d (1-\eta_\uparrow \eta_r)(1-\eta_\downarrow \eta_r)^2 }{\eta_\downarrow \eta_r (1-p^{\mathcal{X}}_d)(2 \eta_\uparrow \eta_r -1 -\eta_r^2 \eta_\uparrow \eta_\downarrow)} + \frac{\Tr[\Bar{\sigma}\Pi^0_{\mathcal{Z}}]}{1-p^{\mathcal{X}}_d}  \nonumber\\
    &\leq \overline{\left[w^0_{\mathcal{X}}\right]^0_{\mathcal{Z}}} \label{x0-upperbound000} ,
\end{align}
where in the third line we replaced $\Tr[\Bar{\sigma}\Pi^0_{\mathcal{Z}}]$ with its upper or lower bound ($\overline{w}^0_{\mathcal{Z}}$ or $\underline{w}^0_{\mathcal{Z}}$, respectively), such that the resulting expression is still an upper bound of the left-hand side. For the same reason, we defined an upper and a lower bound of the variable $\mathbbm{Y}^0_{\eta_l}$ starting from its expression in \eqref{Yetak-0}:
\begin{align}
    \overline{\mathbbm{Y}}^0_{\eta_l} &:= \min \left\lbrace \overline{w}^0_{\mathcal{Z}}, \frac{\sqrt{\eta_2 \eta_\downarrow}}{\left(\sqrt{\eta_\uparrow}-\sqrt{\eta_2}\right)\left(\sqrt{\eta_\uparrow}-\sqrt{\eta_\downarrow}\right)\left(1-p^{\mathcal{Z}}_d\right)}\overline{\left[Y^{X,\emptyset}_{1,(\eta_\uparrow,\eta_l)}\right]^{\leq 1}_{\mathcal{Z}}}  - \frac{\sqrt{\eta_\uparrow \eta_\downarrow} }{\left(\sqrt{\eta_\uparrow}-\sqrt{\eta_2}\right)
   \left(\sqrt{\eta_2}-\sqrt{\eta_\downarrow}\right) \left(1-p^{\mathcal{Z}}_d\right)} \underline{\left[Y^{X,\emptyset}_{1,(\eta_2,\eta_l)} \right]^{\leq 1}_{\mathcal{Z}}} \right. \nonumber\\
   &\left.\quad+\frac{\sqrt{\eta_\uparrow \eta_2}}{\left(\sqrt{\eta_\uparrow}-\sqrt{\eta_\downarrow}\right) \left(\sqrt{\eta_2}-\sqrt{\eta_\downarrow}\right) 
   \left(1-p^{\mathcal{Z}}_d\right)} \overline{\left[Y^{X,\emptyset}_{1,(\eta_\downarrow,\eta_l)}  \right]^{\leq 1}_{\mathcal{Z}}} \right\rbrace \label{Yetak-upp} \\
   \underline{\mathbbm{Y}}^0_{\eta_l} &:=\max \left\lbrace 0, \frac{\sqrt{\eta_2 \eta_\downarrow}}{\left(\sqrt{\eta_\uparrow}-\sqrt{\eta_2}\right)\left(\sqrt{\eta_\uparrow}-\sqrt{\eta_\downarrow}\right)\left(1-p^{\mathcal{Z}}_d\right)}\underline{\left[Y^{X,\emptyset}_{1,(\eta_\uparrow,\eta_l)}\right]^{\leq 1}_{\mathcal{Z}}}  - \frac{\sqrt{\eta_\uparrow \eta_\downarrow} }{\left(\sqrt{\eta_\uparrow}-\sqrt{\eta_2}\right)
   \left(\sqrt{\eta_2}-\sqrt{\eta_\downarrow}\right) \left(1-p^{\mathcal{Z}}_d\right)} \overline{\left[Y^{X,\emptyset}_{1,(\eta_2,\eta_l)} \right]^{\leq 1}_{\mathcal{Z}}} \right.  \nonumber\\
   &\left.\quad+\frac{\sqrt{\eta_\uparrow \eta_2}}{\left(\sqrt{\eta_\uparrow}-\sqrt{\eta_\downarrow}\right) \left(\sqrt{\eta_2}-\sqrt{\eta_\downarrow}\right) 
   \left(1-p^{\mathcal{Z}}_d\right)} \underline{\left[Y^{X,\emptyset}_{1,(\eta_\downarrow,\eta_l)}  \right]^{\leq 1}_{\mathcal{Z}}} \right\rbrace \label{Yetak-low},
\end{align}
respectively, where the upper and lower bounds on the test-round yields in the above formulas are reported in \eqref{l-Y,click}, \eqref{u-Y,click}, \eqref{l-Y,noclick}, and \eqref{u-Y,noclick}. Note that the non-trivial minimization (resp. maximization) in \eqref{Yetak-upp} (resp. \eqref{Yetak-low}) is obtained from the combination of \eqref{yield-noclick-TBS2-Pi0-2} with \eqref{yield-noclick-Pialpha-interval}, similarly to what has been done for $\overline{\mathbbm{E}}^{0}_{\eta_\uparrow}$ in \eqref{Eetak-0-upp}. In the end, we obtain:
\begin{align}
    \overline{\left[w^0_{\mathcal{X}}\right]^0_{\mathcal{Z}}} &:= \min\left\lbrace \overline{w}^0_{\mathcal{Z}}, \frac{(1-\eta_\uparrow \eta_r) \overline{\mathbbm{Y}}^0_{\eta_\downarrow} - \eta_\downarrow \eta_r (1-\eta_\downarrow \eta_r) \underline{\mathbbm{Y}}^0_{\eta_\uparrow}  -\underline{w}^0_{\mathcal{Z}} p^{\mathcal{X}}_d (1-\eta_\uparrow \eta_r)(1-\eta_\downarrow \eta_r)^2 }{\eta_\downarrow \eta_r (1-p^{\mathcal{X}}_d)(2 \eta_\uparrow \eta_r -1 -\eta_r^2 \eta_\uparrow \eta_\downarrow)} + \frac{\overline{w}^0_{\mathcal{Z}}}{1-p^{\mathcal{X}}_d} \right\rbrace \label{x0-upperbound000-explicit} ,
\end{align}
where $\overline{\mathbbm{Y}}^0_{\eta_l}$ and $\underline{\mathbbm{Y}}^0_{\eta_l}$ are given in \eqref{Yetak-upp} and \eqref{Yetak-low}, while $\overline{w}^0_{\mathcal{Z}}$ and $\underline{w}^0_{\mathcal{Z}}$ are given in \eqref{y0-upperbound-explicit} and \eqref{y0-lowerbound-explicit}.

In a similar manner, by substituting the original variables back into \eqref{x0-upperbound}, we obtain the desired upper bound on the weight of the state in the subspace with no photons in $\mathcal{X}$ and one photon in $\mathcal{Z}$: 
\begin{align}
    \Tr\left[ \Bar{\sigma} \Pi^1_{\mathcal{Z}} \Pi^0_{\mathcal{X}} \Pi^1_{\mathcal{Z}} \right] &\leq \frac{(1-\eta_\uparrow \eta_r)  \left[\eta_\downarrow \mathbbm{Y}^{1t}_{\eta_\downarrow} + (1-\eta_\downarrow)\mathbbm{Y}^{1r}_{\eta_\downarrow} \right] - \eta_\downarrow \eta_r (1-\eta_\downarrow \eta_r) \left[\eta_\uparrow \mathbbm{Y}^{1t}_{\eta_\uparrow} + (1-\eta_\uparrow)\mathbbm{Y}^{1r}_{\eta_\uparrow} \right]}{\eta_\downarrow \eta_r (1-p^{\mathcal{X}}_d)(2 \eta_\uparrow \eta_r -1 -\eta_r^2 \eta_\uparrow \eta_\downarrow)} \nonumber\\
    &\quad- \frac{\Tr[\Bar{\sigma}\Pi^1_{\mathcal{Z}}]p^{\mathcal{X}}_d (1-\eta_\uparrow \eta_r)(1-\eta_\downarrow \eta_r)^2 }{\eta_\downarrow \eta_r (1-p^{\mathcal{X}}_d)(2 \eta_\uparrow \eta_r -1 -\eta_r^2 \eta_\uparrow \eta_\downarrow)} + \frac{\Tr[\Bar{\sigma}\Pi^1_{\mathcal{Z}}]}{1-p^{\mathcal{X}}_d} \nonumber\\
    &\leq \overline{\left[w^0_{\mathcal{X}}\right]^1_{\mathcal{Z}}} \label{x0-upperbound101} ,
\end{align}
where in the second inequality we replaced $\Tr[\Bar{\sigma}\Pi^1_{\mathcal{Z}}]$ with its upper or lower bound ($\overline{w}^1_{\mathcal{Z}}$ or $\underline{w}^1_{\mathcal{Z}}$, respectively) and introduced upper and lower bounds on the variables $\mathbbm{Y}^{1t}_{\eta_l}$ and $\mathbbm{Y}^{1r}_{\eta_l}$, starting from their expressions in \eqref{Yetak-1t} and \eqref{Yetak-1r}. In this way, we obtain:
\begin{align}
    \overline{\left[w^0_{\mathcal{X}}\right]^1_{\mathcal{Z}}} &:= \min\left\lbrace \overline{w}^1_{\mathcal{Z}}, \frac{(1-\eta_\uparrow \eta_r) \left[\eta_\downarrow \overline{\mathbbm{Y}}^{1t}_{\eta_\downarrow} + (1-\eta_\downarrow)\overline{\mathbbm{Y}}^{1r}_{\eta_\downarrow} \right] - \eta_\downarrow \eta_r(1-\eta_\downarrow \eta_r) \left[\eta_\uparrow \underline{\mathbbm{Y}}^{1t}_{\eta_\uparrow} + (1-\eta_\uparrow)\underline{\mathbbm{Y}}^{1r}_{\eta_\uparrow} \right]}{\eta_\downarrow \eta_r (1-p^{\mathcal{X}}_d)(2 \eta_\uparrow \eta_r -1 -\eta_r^2 \eta_\uparrow \eta_\downarrow)} \right. \nonumber\\
    &\left.\quad\quad\quad\quad- \frac{\underline{w}^1_{\mathcal{Z}} p^{\mathcal{X}}_d (1-\eta_\uparrow \eta_r)(1-\eta_\downarrow \eta_r)^2 }{\eta_\downarrow \eta_r (1-p^{\mathcal{X}}_d)(2 \eta_\uparrow \eta_r -1 -\eta_r^2 \eta_\uparrow \eta_\downarrow)} + \frac{\overline{w}^1_{\mathcal{Z}}}{1-p^{\mathcal{X}}_d} \right\rbrace \label{x0-upperbound101-explicit} ,
\end{align}
where $\overline{\mathbbm{Y}}^{1t}_{\eta_l}$, $\overline{\mathbbm{Y}}^{1r}_{\eta_l}$, $\underline{\mathbbm{Y}}^{1t}_{\eta_l}$, and $\underline{\mathbbm{Y}}^{1r}_{\eta_l}$ are defined as follows:
\begin{align}
    \overline{\mathbbm{Y}}^{1t}_{\eta_l} &:= \min\left\lbrace \overline{w}^1_{\mathcal{Z}}, \frac{\overline{\left[Y^{X,\emptyset}_{1,(\eta_\uparrow,\eta_l)} \right]^{\leq 1}_{\mathcal{Z}}}}{\left(\sqrt{\eta_\uparrow}-\sqrt{\eta_2}\right)\left(\sqrt{\eta_\uparrow}-\sqrt{\eta_\downarrow}\right)\left(1-p^{\mathcal{Z}}_d\right)}  - \frac{\underline{\left[Y^{X,\emptyset}_{1,(\eta_2,\eta_l)}  \right]^{\leq 1}_{\mathcal{Z}}} }{\left(\sqrt{\eta_\uparrow}-\sqrt{\eta_2}\right)
   \left(\sqrt{\eta_2}-\sqrt{\eta_\downarrow}\right) \left(1-p^{\mathcal{Z}}_d\right)} \right. \nonumber\\
   &\left.\quad+\frac{\overline{\left[Y^{X,\emptyset}_{1,(\eta_\downarrow,\eta_l)} \right]^{\leq 1}_{\mathcal{Z}}}}{\left(\sqrt{\eta_\uparrow}-\sqrt{\eta_\downarrow}\right) \left(\sqrt{\eta_2}-\sqrt{\eta_\downarrow}\right) 
   \left(1-p^{\mathcal{Z}}_d\right)}  \right\rbrace \label{Yetak-1t-upp} \\
   \overline{\mathbbm{Y}}^{1r}_{\eta_l} &:= \min\left\lbrace \overline{w}^1_{\mathcal{Z}}, \frac{1}{1-\eta_\downarrow} \overline{\left[Y^{X,\checkmark}_{1,(\eta_\downarrow,\eta_l)} \right]^{\leq 1}_{\mathcal{Z}}} - 
   \frac{p^{\mathcal{Z}}_d}{(1-\eta_\downarrow)(1-p^{\mathcal{Z}}_d)} \underline{\left[Y^{X,\emptyset}_{1,(\eta_\downarrow,\eta_l)}  \right]^{\leq 1}_{\mathcal{Z}}} \right\rbrace  \label{Yetak-1r-upp} \\
   \underline{\mathbbm{Y}}^{1t}_{\eta_l} &:= \max\left\lbrace 0, \frac{\underline{\left[Y^{X,\emptyset}_{1,(\eta_\uparrow,\eta_l)} \right]^{\leq 1}_{\mathcal{Z}}}}{\left(\sqrt{\eta_\uparrow}-\sqrt{\eta_2}\right)\left(\sqrt{\eta_\uparrow}-\sqrt{\eta_\downarrow}\right)\left(1-p^{\mathcal{Z}}_d\right)}  - \frac{\overline{\left[Y^{X,\emptyset}_{1,(\eta_2,\eta_l)}  \right]^{\leq 1}_{\mathcal{Z}}}}{\left(\sqrt{\eta_\uparrow}-\sqrt{\eta_2}\right)
   \left(\sqrt{\eta_2}-\sqrt{\eta_\downarrow}\right) \left(1-p^{\mathcal{Z}}_d\right)} \right. \nonumber\\
   &\left.\quad+\frac{\underline{\left[Y^{X,\emptyset}_{1,(\eta_\downarrow,\eta_l)} \right]^{\leq 1}_{\mathcal{Z}}}}{\left(\sqrt{\eta_\uparrow}-\sqrt{\eta_\downarrow}\right) \left(\sqrt{\eta_2}-\sqrt{\eta_\downarrow}\right) 
   \left(1-p^{\mathcal{Z}}_d\right)}  \right\rbrace \label{Yetak-1t-low} \\
   \underline{\mathbbm{Y}}^{1r}_{\eta_l} &:=\max\left\lbrace 0, \frac{1}{1-\eta_\downarrow} \underline{\left[Y^{X,\checkmark}_{1,(\eta_\downarrow,\eta_l)} \right]^{\leq 1}_{\mathcal{Z}}} - 
   \frac{p^{\mathcal{Z}}_d}{(1-\eta_\downarrow)(1-p^{\mathcal{Z}}_d)} \overline{\left[Y^{X,\emptyset}_{1,(\eta_\downarrow,\eta_l)}  \right]^{\leq 1}_{\mathcal{Z}}} \right\rbrace  \label{Yetak-1r-low},
\end{align}
while $\overline{w}^1_{\mathcal{Z}}$ and $\underline{w}^1_{\mathcal{Z}}$ are given in \eqref{y1-upperbound-explicit} and \eqref{y1-lowerbound-explicit}.

After obtaining the upper bounds in \eqref{x0-upperbound000} and \eqref{x0-upperbound101}, we now focus on deriving an upper bound on the remaining expectation value in \eqref{second-exp-val-Delta1}, namely $\Tr\left[\Bar{\sigma} (\Pi^0_{\mathcal{Z}} \Pi^0_{\mathcal{X}} \Pi^1_{\mathcal{Z}} + \mathrm{h.c.}) \right]$. To this aim, we define:
\begin{align}
    z:= \Tr\left[ \bar{\sigma} \Pi^0_{\mathcal{Z}} \Pi^0_{\mathcal{X}} \Pi^1_{\mathcal{Z}}\right],
\end{align}
such that the expectation value of interest can be bounded as follows:
\begin{align}
    \Tr\left[\Bar{\sigma} (\Pi^0_{\mathcal{Z}} \Pi^0_{\mathcal{X}} \Pi^1_{\mathcal{Z}} + \mathrm{h.c.}) \right] = z+ z^* = 2 \Re (z)  \leq 2 |z| = 2\abs{\Tr\left[ \bar{\sigma} \Pi^0_{\mathcal{Z}} \Pi^0_{\mathcal{X}} \Pi^1_{\mathcal{Z}}\right]}.
\end{align}
Now, we can use the Cauchy-Schwarz inequality in the following form:
\begin{align}
    \abs{\Tr[A^\dag B]} \leq \sqrt{\Tr[A^\dag A] \Tr[B^\dag B]},
\end{align}
where we choose $A= \sqrt{\bar{\sigma}}$ and $B= \Pi^0_{\mathcal{Z}} \Pi^0_{\mathcal{X}} \Pi^1_{\mathcal{Z}} \sqrt{\Bar{\sigma}}$. We obtain:
\begin{align}
     \Tr\left[\Bar{\sigma} (\Pi^0_{\mathcal{Z}} \Pi^0_{\mathcal{X}} \Pi^1_{\mathcal{Z}} + \mathrm{h.c.}) \right]  &\leq 2 \sqrt{\Tr\left[\bar{\sigma} \Pi^1_{\mathcal{Z}} \Pi^0_{\mathcal{X}} \Pi^0_{\mathcal{Z}} \Pi^0_{\mathcal{X}} \Pi^1_{\mathcal{Z}} \right]} \\
    &\leq 2 \sqrt{\Tr\left[\bar{\sigma} \Pi^1_{\mathcal{Z}} \Pi^0_{\mathcal{X}} \Pi^1_{\mathcal{Z}} \right]}
\end{align}
where we used the fact that $ \Pi^0_{\mathcal{Z}} \leq \one$. Finally, we employ the bound already established in \eqref{x0-upperbound101} and obtain the bound on the remaining expectation value:
\begin{align}
    \Tr\left[\Bar{\sigma} (\Pi^0_{\mathcal{Z}} \Pi^0_{\mathcal{X}} \Pi^1_{\mathcal{Z}} + \mathrm{h.c.}) \right] &\leq 2\left(\,\overline{\left[w^0_{\mathcal{X}}\right]^1_{\mathcal{Z}}}\,\right)^{\frac{1}{2}} \label{x0-upperbound001+hc}.
\end{align}\\

\subsubsection{Upper bound on $\Delta_1$ (part two)}\label{app:PER11}
We can now conclude the derivation of the upper bound on $\Delta_1$, which appears in the phase error rate upper bound, \eqref{phase-error-rate-uppbound}. In particular, we employ the bounds obtained with the detector decoy technique, namely \eqref{x0-upperbound000}, \eqref{x0-upperbound101}, and \eqref{x0-upperbound001+hc}, in order to obtain an upper bound on \eqref{second-exp-val-Delta1}. By combining this result with \eqref{first-exp-val-Delta1-upperbound} in \eqref{Delta1-upperbound}, we obtain the final expression for the upper bound on $\Delta_1$:
\begin{align}
    \Delta_1 &\leq \frac{1}{p^{\mathcal{X}}_d + \eta_r \eta_\uparrow (1-p^{\mathcal{X}}_d)}  \left[p_{\checkmark|0} \overline{\mathbbm{E}}^0_{\eta_\uparrow} +  \sqrt{p_{\checkmark|0} p_{\checkmark|1}} \, \sqrt{\eta_\uparrow} \overline{\mathbbm{E}}^{0,1}_{\eta_\uparrow} + p_{\checkmark|1} \left( \eta_\uparrow \overline{\mathbbm{E}}^{1t}_{\eta_\uparrow} + (1-\eta_\uparrow) \overline{\mathbbm{E}}^{1r}_{\eta_\uparrow}\right)\right] \nonumber\\
    &\quad+  \frac{\eta_r \eta_\uparrow(1-p^{\mathcal{X}}_d)}{p^{\mathcal{X}}_d + \eta_r \eta_\uparrow (1-p^{\mathcal{X}}_d)}   \left[ p_{\checkmark|0} \overline{\left[w^0_{\mathcal{X}}\right]^0_{\mathcal{Z}}} + 2\sqrt{p_{\checkmark|0}p_{\checkmark|1}} \left(\,\overline{\left[w^0_{\mathcal{X}}\right]^1_{\mathcal{Z}}}\,\right)^{\frac{1}{2}} + p_{\checkmark|1} \overline{\left[w^0_{\mathcal{X}}\right]^1_{\mathcal{Z}}} \right]   \label{Delta1-upperbound-final}.
\end{align}\\

\subsubsection{Upper bound on the phase error rate}\label{app:PER12}
By employing \eqref{Delta1-upperbound-final} in \eqref{phase-error-rate-uppbound}, we obtain the desired upper bound on the phase error rate, given in terms of quantities that are estimated with the decoy-state method:
\begin{align}
    \tilde{e}_{X,1} &\leq \overline{\tilde{e}_{X,1}}, \label{step13}
\end{align}
with:
\begin{align}
    \overline{\tilde{e}_{X,1}} &:= \frac{1}{\underline{Y^Z_{1,(\eta_\downarrow,\eta_\downarrow)}}[p^{\mathcal{X}}_d + \eta_r \eta_\uparrow (1-p^{\mathcal{X}}_d)]}  \left[p_{\checkmark|0} \overline{\mathbbm{E}}^0_{\eta_\uparrow} +  \sqrt{p_{\checkmark|0} p_{\checkmark|1}} \, \sqrt{\eta_\uparrow} \overline{\mathbbm{E}}^{0,1}_{\eta_\uparrow} + p_{\checkmark|1} \left( \eta_\uparrow \overline{\mathbbm{E}}^{1t}_{\eta_\uparrow} + (1-\eta_\uparrow) \overline{\mathbbm{E}}^{1r}_{\eta_\uparrow}\right)\right] \nonumber\\
    &+ \frac{\eta_r \eta_\uparrow(1-p^{\mathcal{X}}_d)}{\underline{Y^Z_{1,(\eta_\downarrow,\eta_\downarrow)}}[p^{\mathcal{X}}_d + \eta_r \eta_\uparrow (1-p^{\mathcal{X}}_d)]}   \left[ p_{\checkmark|0} \overline{\left[w^0_{\mathcal{X}}\right]^0_{\mathcal{Z}}} + 2\sqrt{p_{\checkmark|0}p_{\checkmark|1}} \left(\,\overline{\left[w^0_{\mathcal{X}}\right]^1_{\mathcal{Z}}}\,\right)^{\frac{1}{2}} + p_{\checkmark|1} \overline{\left[w^0_{\mathcal{X}}\right]^1_{\mathcal{Z}}} \right] + \frac{1}{\underline{Y^Z_{1,(\eta_\downarrow,\eta_\downarrow)}}} \left( \overline{\Delta_2} + \overline{w}^{>1}_{\mathcal{Z}}\right) \label{phase-error-rate-uppbound-final},
\end{align}
where $\overline{w}^{>1}_{\mathcal{Z}}$, $\overline{\left[w^0_{\mathcal{X}}\right]^0_{\mathcal{Z}}}$, and $\overline{\left[w^0_{\mathcal{X}}\right]^1_{\mathcal{Z}}}$ are given in \eqref{yalpha>1-upperbound-explicit}, \eqref{x0-upperbound000-explicit}, and \eqref{x0-upperbound101-explicit}, respectively, while $\overline{\mathbbm{E}}^0_{\eta_\uparrow}$, $\overline{\mathbbm{E}}^{0,1}_{\eta_\uparrow}$, $\overline{\mathbbm{E}}^{1t}_{\eta_\uparrow}$, and $\overline{\mathbbm{E}}^{1r}_{\eta_\uparrow}$ are given in \eqref{Eetak-0-upp}--\eqref{Eetak-1r-upp}. Finally,  $p_{\checkmark|0}$, $p_{\checkmark|1}$, $\eta_r$, and $\overline{\Delta_2}$ are given in \eqref{Pr(oneclick|0)}, \eqref{Pr(oneclick|1)}, \eqref{eta-ratio}, and \eqref{Delta2bar}, respectively. The derivation of the bound in \eqref{phase-error-rate-uppbound-final} concludes this subsection.

We observe that all the quantities appearing in \eqref{phase-error-rate-uppbound-final} are either input parameters or are expressed in terms of bounds derived with the decoy-state method. In particular, \eqref{phase-error-rate-uppbound-final} contains bounds on the one-photon $Z$-basis yields ($Y^Z_{1,(\eta_\downarrow,\eta_\downarrow)}$, $Y^Z_{1,(\eta_\uparrow,\eta_\uparrow)}$), but also indirectly contains bounds on the test-round yields ($Y^{X,\emptyset}_{1,(\eta_i,\eta_l)}$, $Y^{X,\checkmark}_{1,(\eta_i,\eta_l)}$) through \eqref{l-Y,click}--\eqref{u-Y,noclick} and on the products between the test-round yields and bit error rates ($Y^{X,\emptyset}_{1,(\eta_i,\eta_\uparrow)} e_{X,1,(\eta_i,\eta_\uparrow),\emptyset}$, $Y^{X,\checkmark}_{1,(\eta_i,\eta_\uparrow)} e_{X,1,(\eta_i,\eta_\uparrow),\checkmark}$) through \eqref{l-e,click}, \eqref{u-e,click}, \eqref{l-e,noclick}, and \eqref{u-e,noclick}. Such bounds on the yields and on the bit error rates are obtained with standard procedures of the decoy-state method, as shown in Appendix~\ref{app:decoy}.

\subsection{Secret key rate}  \label{sec:final-key-rate}

In this subsection we complete the security proof of Protocol~1.

We combine the upper bound obtained on the phase error rate \eqref{step13} with the entropy bound \eqref{step12} to derive the following bound on the entropy of interest:
\begin{align}
    H(Z_A|E)_{\rho|1,\Omega_Z} \geq \log_2 \frac{1}{c} - u(\overline{\tilde{e}_{X,1}})   \label{step14},
\end{align}
where we used the fact that $u(x)$ in \eqref{u(x)} is a monotonically non-decreasing function. We now employ \eqref{step14} in \eqref{step7}, together with \eqref{Pr(n)} and \eqref{Pr(n|mu)}, to derive the following lower bound on the first entropy that appears in the DW rate \eqref{DWrate}:
\begin{align}
    H(Z_A|I_A E)_{\rho|\Omega_Z} &\geq  \sum_{\mu_j \in\mathcal{S}} \frac{p_{\mu_j}}{\Pr(Z_B\neq\emptyset|\Gamma_Z)} \left\lbrace e^{-\mu_j} \underline{Y^Z_{0,(\eta_\downarrow,\eta_\downarrow)}} \log_2 d +  e^{-\mu_j} \mu_j \underline{Y^Z_{1,(\eta_\downarrow,\eta_\downarrow)}} \left[\log_2 \frac{1}{c} - u(\overline{\tilde{e}_{X,1}})\right] \right\rbrace, \label{step15}
\end{align}
where we replaced the yields by their lower bounds obtained with the decoy-state method (cf.~Appendix~\ref{app:decoy}).

Finally, we employ the lower bound in \eqref{step15} and the upper bound in \eqref{Bobs-entropy-bound} in the DW rate \eqref{DWrate}, together with \eqref{prob-OmegaZ}, to obtain the following lower bound on the asymptotic key rate of the protocol:
\begin{align}
    r &\geq p_Z^2 \sum_{j=1}^3 p_{\mu_j} \left\lbrace e^{-\mu_j} \underline{Y^Z_{0,(\eta_\downarrow,\eta_\downarrow)}} \log_2 d +  e^{-\mu_j} \mu_j \underline{Y^Z_{1,(\eta_\downarrow,\eta_\downarrow)}} \left[\log_2 \frac{1}{c} - u(\overline{\tilde{e}_{X,1}})\right] - G^Z_{\mu_j,(\eta_\downarrow,\eta_\downarrow)} u(Q_{Z,\mu_j}) \right\rbrace, \label{rate-bound}
\end{align}
which coincides with the key rate provided in the protocol's description, \eqref{protocol-rate}, thereby concluding the security proof.

\section{decoy-state method}  \label{app:decoy}

In this Appendix, we provide upper and lower bounds on zero-photon and one-photon yields and bit error rates, following the standard procedures of the decoy-state method. Such bounds are required by the phase error rate upper bound, \eqref{protocol-phase-error-rate}, as well as in the final expression of the protocol's key rate \eqref{protocol-rate}. Moreover, we provide bounds on specific linear combinations of one-photon yields, which appear in the expressions \eqref{protocol-yalpha>1}--\eqref{protocol-y1-lowerbound} of Appendix~\ref{app:phase-error-rate-formula}.\\

\subsection{Standard equations of the decoy-state method}

We recall from Sec.~\ref{sec:protocol} that Alice prepares WCPs with three different intensities: $\mathcal{S}=\{\mu_1,\mu_2,\mu_3\}$, with $\mu_1 >\mu_2 + \mu_3$ and $\mu_2>\mu_3 \geq 0$ . The parties use all three intensities both to generate the key and to derive the decoy bounds. The equalities on which the decoy-state method relies on, relate the unobserved yields to the observed gains. For instance, by definition \eqref{gainKG}, the $Z$-basis gain is the probability that Bob has a detection in the $Z$-basis detector, given that Alice sent one of the $Z$-basis states with intensity $\mu_j$ and that Bob chose the TBS setting $(\eta_l,\eta_l)$, and can be expressed as follows:
\begin{align}
    G^Z_{\mu_j,(\eta_l,\eta_l)} &=  \Pr(Z_B \neq \emptyset|T=Z,I_A=\mu_j,(\eta_l,\eta_l)) \nonumber\\
    &= \Tr\left[(Z^{(\eta_l,\eta_l)}_\checkmark \otimes\one_E) U_{BE} {\textstyle\sum_{i=0}^{d-1}} \frac{\rho_{Z_i} (\mu_j)}{d} \otimes \proj{0}_E U^\dag_{BE}\right] , 
\end{align}
where $Z^{(\eta_l,\eta_l)}_\checkmark$ is defined in \eqref{Zclick-a} and $\rho_{Z_i}(\mu_j)$ is given in \eqref{stateKG}. Now, by using \eqref{stateKG} and \eqref{Pr(n|mu)}, we can recast the gain as follows:
\begin{align}
    G^Z_{\mu_j,(\eta_l,\eta_l)} &= \sum_{n=0}^\infty  e^{-\mu_j} \frac{\mu^n_j}{n!} \Tr\left[(Z^{(\eta_l,\eta_l)}_\checkmark \otimes\one_E) U_{BE} {\textstyle\sum_{i=0}^{d-1}} \frac{\proj{n_{Z_i}}}{d} \otimes \proj{0}_E U^\dag_{BE}\right] \nonumber\\
    &= \sum_{n=0}^\infty  e^{-\mu_j} \frac{\mu^n_j}{n!} Y^Z_{n,(\eta_l,\eta_l)} ,  \label{gainKG-decoy}
\end{align}
where in the last equality we defined the $n$-photon $Z$-basis yield ($Y^Z_{n,(\eta_l,\eta_l)}$), by generalizing \eqref{prob-Zclick}.

Similarly, the $X$-basis gain in \eqref{gaintest-Zclick} is defined as the probability that Bob has a detection in the $X$-basis detector and in the $Z$-basis detector, given that Alice sent a state of the $X$-basis with intensity $\mu_j$ and Bob selected the TBS setting $(\eta_i,\eta_l)$. It can be expressed as:
\begin{align}
    G^{X,\checkmark}_{\mu_j,(\eta_i,\eta_l)} &= \Pr(X_B \neq \emptyset,Z_B \neq \emptyset|T=X,I_A=\mu_j,(\eta_i,\eta_l)) \nonumber\\
    &= \Tr\left[(X^{(\eta_i,\eta_l)}_{\checkmark,\checkmark} \otimes\one_E) U_{BE} {\textstyle\sum_{k=0}^{d-1}} \frac{\rho_{X_k} (\mu_j)}{d} \otimes \proj{0}_E U^\dag_{BE}\right]   ,
\end{align}
where $X^{(\eta_i,\eta_l)}_{\checkmark,\checkmark}$ is defined in \eqref{Xdetdet-a} and $\rho_{X_k} (\mu_j)$ is given in \eqref{stateTest}. By using \eqref{stateTest}, we express the gain in terms of the $X$-basis yields:
\begin{align}
    G^{X,\checkmark}_{\mu_j,(\eta_i,\eta_l)}  &=\sum_{n=0}^{\infty} e^{-\mu_j} \frac{\mu^n_j}{n!}  Y^{X,\checkmark}_{n,(\eta_i,\eta_l)}   , \label{gaintest-Zclick-decoy}
\end{align}
where we defined the $n$-photon $X$-basis yields ($Y^{X,\checkmark}_{n,(\eta_i,\eta_l)}$) as a generalization of \eqref{yieldX-click}. Analogously, we express the gain in \eqref{gaintest-Znoclick} as follows:
\begin{align}
    G^{X,\emptyset}_{\mu_j,(\eta_i,\eta_l)}  &=\sum_{n=0}^{\infty} e^{-\mu_j} \frac{\mu^n_j}{n!}  Y^{X,\emptyset}_{n,(\eta_i,\eta_l)}    \label{gaintest-Znoclick-decoy},
\end{align}
where $Y^{X,\emptyset}_{n,(\eta_i,\eta_l)}$ generalizes \eqref{yieldX-noclick} to $n$ photons.

On the same lines, one can relate the $n$-photon $X$-basis bit error rates to the corresponding observed QBERs. In particular, the QBER in \eqref{QBER-X-Zclick} is the probability that Bob's $X$-basis outcome differs from Alice's $X$-basis symbol, given that: Bob had a detection in both detectors, Alice chose intensity $\mu_j$, and Bob chose the TBS setting $(\eta_i,\eta_\uparrow)$. Then, the product of this QBER and the corresponding $X$-basis gain reads: 
\begin{align}
    Q_{X,\mu_j,(\eta_i,\eta_\uparrow),\checkmark} G^{X,\checkmark}_{\mu_j,(\eta_i,\eta_\uparrow)}  &= \Pr(X_B \neq \emptyset, Z_B \neq \emptyset, X_A \neq X_B |T=X, I_A=\mu_j ,(\eta_i,\eta_\uparrow)) \nonumber\\
    &=  {\textstyle\sum_{k=0}^{d-1}\sum_{k'\neq k}} \Tr\left[(X^{(\eta_i,\eta_\uparrow)}_{k',\checkmark} \otimes\one_E) U_{BE} \frac{\rho_{X_k} (\mu_j)}{d} \otimes \proj{0}_E U^\dag_{BE}\right],
\end{align}
where $X^{(\eta_i,\eta_\uparrow)}_{k',\checkmark}$ is defined in \eqref{Xkdet-a}. By using \eqref{stateTest}, we can expand the last expression in terms of the products of the $X$-basis yields and bit error rates, as follows:
\begin{align}
    Q_{X,\mu_j,(\eta_i,\eta_\uparrow),\checkmark} G^{X,\checkmark}_{\mu_j,(\eta_i,\eta_\uparrow)} &=\sum_{n=0}^{\infty} e^{-\mu_j} \frac{\mu^n_j}{n!}   Y^{X,\checkmark}_{n,(\eta_i,\eta_\uparrow)} e_{X,n,(\eta_i,\eta_\uparrow),\checkmark}  \label{QBERGaintest-Zclick-decoy}, 
\end{align}
where we defined the $n$-photon $X$-basis bit error rate ($e_{X,n,(\eta_i,\eta_\uparrow),\checkmark}$) as a generalization of \eqref{bit-error-rate-click}. Analogously, we have:
\begin{align}
    Q_{X,\mu_j,(\eta_i,\eta_\uparrow),\emptyset} G^{X,\emptyset}_{\mu_j,(\eta_i,\eta_\uparrow)} &=\sum_{n=0}^{\infty} e^{-\mu_j} \frac{\mu^n_j}{n!}   Y^{X,\emptyset}_{n,(\eta_i,\eta_\uparrow)} e_{X,n,(\eta_i,\eta_\uparrow),\emptyset}  \label{QBERGaintest-Znoclick-decoy}, 
\end{align}
where $e_{X,n,(\eta_i,\eta_\uparrow),\emptyset}$ generalizes \eqref{bit-error-rate-noclick}.\\

\subsection{Bounds on yields and bit error rates}

We observe that the equations defining the decoy-state method, namely \eqref{gainKG-decoy}, \eqref{gaintest-Zclick-decoy}, \eqref{gaintest-Znoclick-decoy}, \eqref{QBERGaintest-Zclick-decoy}, and \eqref{QBERGaintest-Znoclick-decoy}, share the same structure, which can be exemplified as follows:
\begin{align}
    G_{\mu_j} = \sum_{n=0}^{\infty} e^{-\mu_j} \frac{\mu^n_j}{n!} Y_n, \label{decoy-eq}
\end{align}
where $G_{\mu_j}$ can be replaced by a gain or a product of QBER and gain, while $Y_n$ can be replaced by a yield or a product of yield and bit error rate. Then, starting from \eqref{decoy-eq} and following the seminal paper on the decoy-state method \cite{decoy-bounds}, one obtains the following lower bounds on the vacuum and one-photon yields:
\begin{align}
    \underline{Y_0}  &= \max\left\lbrace 0, \frac{\mu_2 e^{\mu_3} G_{\mu_3} - \mu_3 e^{\mu_2} G_{\mu_2}}{\mu_2 -\mu_3} \right\rbrace  \label{Y0-low} \\
    \underline{Y_1} &= \max\left\lbrace 0, \frac{e^{\mu_2} G_{\mu_2} -  e^{\mu_3} G_{\mu_3} - \frac{\mu_2^2 - \mu_3^2}{\mu_1^2} (e^{\mu_1} G_{\mu_1} -\underline{Y_0})}{(\mu_2 -\mu_3)\left(1-\frac{\mu_2 + \mu_3}{\mu_1}\right)} \right\rbrace. \label{Y1-low}
\end{align}
With similar techniques, one can derive a simple upper bound on the one-photon yield. To derive it, consider the following combination of gains in \eqref{decoy-eq}:
\begin{align}
    e^{\mu_2} G_{\mu_2} - e^{\mu_3} G_{\mu_3} = (\mu_2 - \mu_3) Y_1 + \sum_{n=2}^\infty \frac{\mu_2^n - \mu_3^n}{n!} Y_n .
\end{align}
Now, observe that the sum on the right-hand side is only composed of non-negative terms since $\mu_2 > \mu_3$.
Therefore, we can obtain the following inequality:
\begin{align}
    e^{\mu_2} G_{\mu_2} - e^{\mu_3} G_{\mu_3} \geq (\mu_2 - \mu_3) Y_1,
\end{align}
which provides us with the upper bound on the one-photon yield:
\begin{align}
    Y_1 \leq \overline{Y_1} = \frac{e^{\mu_2} G_{\mu_2} - e^{\mu_3} G_{\mu_3}}{\mu_2 - \mu_3}. \label{Y1-upp-2decoy}
\end{align}
An alternative upper bound on $Y_1$ can be obtained by using the statistics from all the intensity choices of Alice. In particular, we use the following linear combination of three gains,
\begin{align}
    (\mu_2^2-\mu_3^2)e^{\mu_1}G_{\mu_1} - (\mu_1^2-\mu_3^2)e^{\mu_2}G_{\mu_2} + (\mu_1^2-\mu_2^2)e^{\mu_3}G_{\mu_3} &= \sum_{n=0}^\infty \left[(\mu_2^2-\mu_3^2)\frac{\mu_1^n}{n!} - (\mu_1^2-\mu_3^2)\frac{\mu_2^n}{n!} + (\mu_1^2-\mu_2^2)\frac{\mu_3^n}{n!} \right] Y_n \nonumber\\
    &= f(1) Y_1  + \sum_{n=3}^\infty f(n) Y_n, \label{Y1-upp-calc}
\end{align}
where we introduced:
\begin{align}
    f(n):= (\mu_2^2-\mu_3^2)\frac{\mu_1^n}{n!} - (\mu_1^2-\mu_3^2)\frac{\mu_2^n}{n!} + (\mu_1^2-\mu_2^2)\frac{\mu_3^n}{n!} \label{f(n)}.
\end{align}
Now we observe that $f(1)$ is negative:
\begin{align}
 &f(1) < 0 \nonumber\\
 \Leftrightarrow\quad & (\mu_2^2-\mu_3^2) \mu_1 - (\mu_1^2-\mu_3^2)\mu_2 + (\mu_1^2-\mu_2^2)\mu_3 <0 \nonumber\\
 \Leftrightarrow\quad & (\mu_2^2-\mu_3^2) \mu_1 - (\mu_1^2-\mu_2^2)\mu_2 - (\mu_2^2-\mu_3^2)\mu_2 + (\mu_1^2-\mu_2^2)\mu_3 <0 \nonumber\\
 \Leftrightarrow\quad & (\mu_2^2-\mu_3^2)(\mu_1 - \mu_2) - (\mu_1^2-\mu_2^2)(\mu_2 - \mu_3) <0 \nonumber\\
 \Leftrightarrow\quad & (\mu_2-\mu_3)(\mu_1 - \mu_2)\left[\mu_2 + \mu_3 -\mu_1 - \mu_2 \right] <0,
\end{align}
which is true since $\mu_3 < \mu_2 < \mu_1$. At the same time, we can show that $f(n)\geq 0$ for $n \geq 3$, as follows:
\begin{align}
    &f(n) \geq 0 \nonumber\\
    \Leftrightarrow\quad & (\mu_2^2-\mu_3^2)\mu_1^2 \mu_1^{n-2} - (\mu_1^2-\mu_3^2)\mu_2^2 \mu_2^{n-2} + (\mu_1^2-\mu_2^2)\mu_3^2 \mu_3^{n-2} \geq 0 \nonumber\\
    \Leftrightarrow\quad & \mu_1^2 \mu_2^2 (\mu_1^{n-2} - \mu_2^{n-2}) - \mu_1^2 \mu_3^2 (\mu_1^{n-2} - \mu_3^{n-2}) + \mu_2^2 \mu_3^2 (\mu_2^{n-2} - \mu_3^{n-2}) \geq 0 \nonumber\\
    \Leftrightarrow\quad & \mu_1^2 \mu_2^2 (\mu_1^{n-2} - \mu_2^{n-2}) - \mu_1^2 \mu_3^2 (\mu_1^{n-2} - \mu_3^{n-2}) + \mu_2^2 \mu_3^2 \left[(\mu_1^{n-2} - \mu_3^{n-2}) - (\mu_1^{n-2} - \mu_2^{n-2})\right] \geq 0 \nonumber\\
    \Leftrightarrow\quad & \mu_2^2 (\mu_1^{n-2} - \mu_2^{n-2})(\mu_1^2 - \mu_3^2) - \mu_3^2 (\mu_1^{n-2} - \mu_3^{n-2})(\mu_1^2 - \mu_2^2) \geq 0 \nonumber\\
    \Leftrightarrow\quad & (\mu_1 - \mu_2)(\mu_1 - \mu_3)\left[\mu_2^2 (\mu_1+\mu_3)(\mu_1^{n-3} + \mu_1^{n-4}\mu_2 + \dots + \mu_2^{n-3})-\mu_3^2(\mu_1 + \mu_2) (\mu_1^{n-3} + \mu_1^{n-4}\mu_3 + \dots \mu_3^{n-3})\right] \geq 0 \nonumber\\
    \Leftrightarrow\quad & \mu_1 \left[\mu_2^2 (\mu_1^{n-3} + \mu_1^{n-4}\mu_2 + \dots + \mu_2^{n-3}) - \mu_3^2 (\mu_1^{n-3} + \mu_1^{n-4}\mu_3 + \dots \mu_3^{n-3}) \right] \nonumber\\
    & \quad + \mu_2 \mu_3 \left[\mu_2 (\mu_1^{n-3} + \mu_1^{n-4}\mu_2 + \dots + \mu_2^{n-3}) - \mu_3 (\mu_1^{n-3} + \mu_1^{n-4}\mu_3 + \dots \mu_3^{n-3}) \right] \geq 0,
\end{align}
which is verified since $\mu_3 < \mu_2 < \mu_1$. Then, we can derive an upper bound on $Y_1$ by replacing $Y_n$ with $1$ in \eqref{Y1-upp-calc}, which yields:
\begin{align}
    (\mu_2^2-\mu_3^2)e^{\mu_1}G_{\mu_1} - (\mu_1^2-\mu_3^2)e^{\mu_2}G_{\mu_2} + (\mu_1^2-\mu_2^2)e^{\mu_3}G_{\mu_3} &\leq f(1) Y_1  + \sum_{n=3}^\infty f(n) \label{Y1-upp-calc2},
\end{align}
where we can simplify the sum running over $n$ as follows:
\begin{align}
    \sum_{n=3}^\infty f(n) &= (\mu_2^2-\mu_3^2)\left(e^{\mu_1} - 1 - \mu_1 -\frac{\mu_1^2}{2}\right) - (\mu_1^2-\mu_3^2)\left(e^{\mu_2} - 1 - \mu_2 -\frac{\mu_2^2}{2}\right) + (\mu_1^2-\mu_2^2)\left(e^{\mu_3} - 1 - \mu_3 -\frac{\mu_3^2}{2}\right) \nonumber\\
    &=  (\mu_2^2-\mu_3^2)\left(e^{\mu_1} - \mu_1 \right) - (\mu_1^2-\mu_3^2)\left(e^{\mu_2} - \mu_2 \right) + (\mu_1^2-\mu_2^2)\left(e^{\mu_3} - \mu_3 \right) \nonumber\\
    &= -f(1) + (\mu_2^2-\mu_3^2) e^{\mu_1} - (\mu_1^2-\mu_3^2)e^{\mu_2} + (\mu_1^2-\mu_2^2)e^{\mu_3} .
\end{align}
By employing the last expression in \eqref{Y1-upp-calc2}, we obtain the following upper bound on the one-photon yield:
\begin{align}
   \overline{Y_1} = 1 - \frac{(\mu_2^2-\mu_3^2) e^{\mu_1}(1-G_{\mu_1}) - (\mu_1^2-\mu_3^2)e^{\mu_2}(1-G_{\mu_2}) + (\mu_1^2-\mu_2^2)e^{\mu_3}(1-G_{\mu_3})}{(\mu_2^2-\mu_3^2) \mu_1 - (\mu_1^2-\mu_3^2)\mu_2 + (\mu_1^2-\mu_2^2)\mu_3} . \label{Y1-upp-3decoy}
\end{align}
By combining \eqref{Y1-upp-2decoy} and \eqref{Y1-upp-3decoy}, we come to the final upper bound on the one-photon yield:
\begin{align}
    \overline{Y_1} = \min\left\lbrace 1, \frac{e^{\mu_2} G_{\mu_2} - e^{\mu_3} G_{\mu_3}}{\mu_2 - \mu_3} ,1 - \frac{(\mu_2^2-\mu_3^2) e^{\mu_1}(1-G_{\mu_1}) - (\mu_1^2-\mu_3^2)e^{\mu_2}(1-G_{\mu_2}) + (\mu_1^2-\mu_2^2)e^{\mu_3}(1-G_{\mu_3})}{(\mu_2^2-\mu_3^2) \mu_1 - (\mu_1^2-\mu_3^2)\mu_2 + (\mu_1^2-\mu_2^2)\mu_3} \right\rbrace .\label{Y1-upp}
\end{align}

We can now substitute the appropriate yields and bit error rates in the generic bounds in \eqref{Y0-low}, \eqref{Y1-low}, and \eqref{Y1-upp}, to derive the required bounds appearing in the phase error rate upper bound \eqref{phase-error-rate-uppbound-final} and in the key rate expression \eqref{step7}. For instance, by employing the bound \eqref{Y0-low} on the relations \eqref{gainKG-decoy}, \eqref{gaintest-Zclick-decoy}, \eqref{gaintest-Znoclick-decoy}, \eqref{QBERGaintest-Zclick-decoy}, and \eqref{QBERGaintest-Znoclick-decoy} we obtain, respectively,
\begin{align}
    \underline{Y^Z_{0,(\eta_l,\eta_l)}}  &= \max\left\lbrace 0, \frac{\mu_2 e^{\mu_3} G^Z_{\mu_3,(\eta_l,\eta_l)} - \mu_3 e^{\mu_2} G^Z_{\mu_2,(\eta_l,\eta_l)}}{\mu_2 -\mu_3} \right\rbrace  \label{YZ0-low} \\
    \underline{Y^{X,\checkmark}_{0,(\eta_i,\eta_l)}}  &= \max\left\lbrace 0, \frac{\mu_2 e^{\mu_3} G^{X,\checkmark}_{\mu_3,(\eta_i,\eta_l)} - \mu_3 e^{\mu_2} G^{X,\checkmark}_{\mu_2,(\eta_i,\eta_l)}}{\mu_2 -\mu_3} \right\rbrace  \label{YX0-Zclick-low} \\
    \underline{Y^{X,\emptyset}_{0,(\eta_i,\eta_l)}}  &= \max\left\lbrace 0, \frac{\mu_2 e^{\mu_3} G^{X,\emptyset}_{\mu_3,(\eta_i,\eta_l)} - \mu_3 e^{\mu_2} G^{X,\emptyset}_{\mu_2,(\eta_i,\eta_l)}}{\mu_2 -\mu_3} \right\rbrace  \label{YX0-Znoclick-low} \\
    \underline{Y^{X,\checkmark}_{0,(\eta_i,\eta_\uparrow)} e_{X,0,(\eta_i,\eta_\uparrow),\checkmark}}  &= \max\left\lbrace 0, \frac{\mu_2 e^{\mu_3} Q_{X,\mu_3,(\eta_i,\eta_\uparrow),\checkmark} G^{X,\checkmark}_{\mu_3,(\eta_i,\eta_\uparrow)} - \mu_3 e^{\mu_2} Q_{X,\mu_2,(\eta_i,\eta_\uparrow),\checkmark}G^{X,\checkmark}_{\mu_2,(\eta_i,\eta_\uparrow)}}{\mu_2 -\mu_3} \right\rbrace  \label{YeX0-Zclick-low} \\
    \underline{Y^{X,\emptyset}_{0,(\eta_i,\eta_\uparrow)} e_{X,0,(\eta_i,\eta_\uparrow),\emptyset}}  &= \max\left\lbrace 0, \frac{\mu_2 e^{\mu_3} Q_{X,\mu_3,(\eta_i,\eta_\uparrow),\emptyset} G^{X,\emptyset}_{\mu_3,(\eta_i,\eta_\uparrow)} - \mu_3 e^{\mu_2} Q_{X,\mu_2,(\eta_i,\eta_\uparrow),\emptyset}G^{X,\emptyset}_{\mu_2,(\eta_i,\eta_\uparrow)}}{\mu_2 -\mu_3} \right\rbrace  \label{YeX0-Znoclick-low},
\end{align}
for $\eta_i \in \{\eta_\uparrow,\eta_\downarrow,\eta_2\}$ and $\eta_l \in\{\eta_\downarrow,\eta_\uparrow\}$.

Similarly, we employ the bound \eqref{Y1-low} and \eqref{Y1-upp} on \eqref{gainKG-decoy} to obtain:
\begin{align}
    \underline{Y^Z_{1,(\eta_l,\eta_l)}} &= \max\left\lbrace 0, \frac{e^{\mu_2} G^Z_{\mu_2,(\eta_l,\eta_l)} -  e^{\mu_3} G^Z_{\mu_3,(\eta_l,\eta_l)} - \frac{\mu_2^2 - \mu_3^2}{\mu_1^2} \left(e^{\mu_1} G^Z_{\mu_1,(\eta_l,\eta_l)} -\underline{Y^Z_{0,(\eta_l,\eta_l)}}\right)}{(\mu_2 -\mu_3)\left(1-\frac{\mu_2 + \mu_3}{\mu_1}\right)} \right\rbrace \label{YZ1-low} \\
    \overline{Y^Z_{1,(\eta_l,\eta_l)}} &=\min\left\lbrace 1, \frac{e^{\mu_2} G^Z_{\mu_2,(\eta_l,\eta_l)} - e^{\mu_3} G^Z_{\mu_3,(\eta_l,\eta_l)}}{\mu_2 - \mu_3}, \right. \nonumber\\
    &\quad\left. 1 - \frac{(\mu_2^2-\mu_3^2) e^{\mu_1}(1-G^Z_{\mu_1,(\eta_l,\eta_l)}) - (\mu_1^2-\mu_3^2)e^{\mu_2}(1-G^Z_{\mu_2,(\eta_l,\eta_l)}) + (\mu_1^2-\mu_2^2)e^{\mu_3}(1- G^Z_{\mu_3,(\eta_l,\eta_l)})}{(\mu_2^2-\mu_3^2) \mu_1 - (\mu_1^2-\mu_3^2)\mu_2 + (\mu_1^2-\mu_2^2)\mu_3}  \right\rbrace , \label{YZ1-upp}
\end{align}
for $\eta_l \in\{\eta_\downarrow,\eta_\uparrow\}$.

Subsequently, we use the bounds \eqref{Y1-low} and \eqref{Y1-upp} on \eqref{gaintest-Zclick-decoy} and \eqref{gaintest-Znoclick-decoy} to obtain: 
\begin{align}
    \underline{Y^{X,\checkmark}_{1,(\eta_i,\eta_l)}} &= \max\left\lbrace 0, \frac{e^{\mu_2} G^{X,\checkmark}_{\mu_2,(\eta_i,\eta_l)} -  e^{\mu_3} G^{X,\checkmark}_{\mu_3,(\eta_i,\eta_l)} - \frac{\mu_2^2 - \mu_3^2}{\mu_1^2} \left(e^{\mu_1} G^{X,\checkmark}_{\mu_1,(\eta_i,\eta_l)} -\underline{Y^{X,\checkmark}_{0,(\eta_i,\eta_l)}}\right)}{(\mu_2 -\mu_3)\left(1-\frac{\mu_2 + \mu_3}{\mu_1}\right)} \right\rbrace \label{YX1-Zclick-low} \\
    \underline{Y^{X,\emptyset}_{1,(\eta_i,\eta_l)}} &= \max\left\lbrace 0, \frac{e^{\mu_2} G^{X,\emptyset}_{\mu_2,(\eta_i,\eta_l)} -  e^{\mu_3} G^{X,\emptyset}_{\mu_3,(\eta_i,\eta_l)} - \frac{\mu_2^2 - \mu_3^2}{\mu_1^2} \left(e^{\mu_1} G^{X,\emptyset}_{\mu_1,(\eta_i,\eta_l)} -\underline{Y^{X,\emptyset}_{0,(\eta_i,\eta_l)}}\right)}{(\mu_2 -\mu_3)\left(1-\frac{\mu_2 + \mu_3}{\mu_1}\right)} \right\rbrace \label{YX1-Znoclick-low}\\
    \overline{Y^{X,\checkmark}_{1,(\eta_i,\eta_l)}} &= \min\left\lbrace 1, \frac{e^{\mu_2} G^{X,\checkmark}_{\mu_2,(\eta_i,\eta_l)} - e^{\mu_3} G^{X,\checkmark}_{\mu_3,(\eta_i,\eta_l)}}{\mu_2 - \mu_3}, \right.\nonumber\\
    &\quad\left. 1 - \frac{(\mu_2^2-\mu_3^2) e^{\mu_1}(1-G^{X,\checkmark}_{\mu_1,(\eta_i,\eta_l)}) - (\mu_1^2-\mu_3^2)e^{\mu_2}(1-G^{X,\checkmark}_{\mu_2,(\eta_i,\eta_l)}) + (\mu_1^2-\mu_2^2)e^{\mu_3}(1- G^{X,\checkmark}_{\mu_3,(\eta_i,\eta_l)})}{(\mu_2^2-\mu_3^2) \mu_1 - (\mu_1^2-\mu_3^2)\mu_2 + (\mu_1^2-\mu_2^2)\mu_3}  \right\rbrace \label{YX1-Zclick-upp} \\
    \overline{Y^{X,\emptyset}_{1,(\eta_i,\eta_l)}} &= \min\left\lbrace 1, \frac{e^{\mu_2} G^{X,\emptyset}_{\mu_2,(\eta_i,\eta_l)} - e^{\mu_3} G^{X,\emptyset}_{\mu_3,(\eta_i,\eta_l)}}{\mu_2 - \mu_3}, \right.\nonumber\\
    &\quad\left. 1 - \frac{(\mu_2^2-\mu_3^2) e^{\mu_1}(1-G^{X,\emptyset}_{\mu_1,(\eta_i,\eta_l)}) - (\mu_1^2-\mu_3^2)e^{\mu_2}(1-G^{X,\emptyset}_{\mu_2,(\eta_i,\eta_l)}) + (\mu_1^2-\mu_2^2)e^{\mu_3}(1- G^{X,\emptyset}_{\mu_3,(\eta_i,\eta_l)})}{(\mu_2^2-\mu_3^2) \mu_1 - (\mu_1^2-\mu_3^2)\mu_2 + (\mu_1^2-\mu_2^2)\mu_3}\right\rbrace, \label{YX1-Znoclick-upp}
\end{align}
for $\eta_i \in \{\eta_\uparrow,\eta_\downarrow,\eta_2\}$ and $\eta_l \in\{\eta_\downarrow,\eta_\uparrow\}$.

Finally, we use the bounds \eqref{Y1-low} and \eqref{Y1-upp} on \eqref{QBERGaintest-Zclick-decoy} and \eqref{QBERGaintest-Znoclick-decoy} to obtain: 
\begin{align}
    \underline{Y^{X,\checkmark}_{1,(\eta_i,\eta_\uparrow)} e_{X,1,(\eta_i,\eta_\uparrow),\checkmark}} &= \max\left\lbrace 0, \frac{e^{\mu_2} Q_{X,\mu_2,(\eta_i,\eta_\uparrow),\checkmark} G^{X,\checkmark}_{\mu_2,(\eta_i,\eta_\uparrow)} -  e^{\mu_3} Q_{X,\mu_3,(\eta_i,\eta_\uparrow),\checkmark} G^{X,\checkmark}_{\mu_3,(\eta_i,\eta_\uparrow)}}{(\mu_2 -\mu_3)\left(1-\frac{\mu_2 + \mu_3}{\mu_1}\right)} \right. \nonumber\\
    &\quad\quad\quad\left. - \frac{\frac{\mu_2^2 - \mu_3^2}{\mu_1^2} \left(e^{\mu_1} Q_{X,\mu_1,(\eta_i,\eta_\uparrow),\checkmark} G^{X,\checkmark}_{\mu_1,(\eta_i,\eta_\uparrow)} -\underline{Y^{X,\checkmark}_{0,(\eta_i,\eta_\uparrow)} e_{X,0,(\eta_i,\eta_\uparrow),\checkmark}} \right)}{(\mu_2 -\mu_3)\left(1-\frac{\mu_2 + \mu_3}{\mu_1}\right)} \right\rbrace \label{YeX1-Zclick-low} \\
    \underline{Y^{X,\emptyset}_{1,(\eta_i,\eta_\uparrow)} e_{X,1,(\eta_i,\eta_\uparrow),\emptyset}} &= \max\left\lbrace 0, \frac{e^{\mu_2} Q_{X,\mu_2,(\eta_i,\eta_\uparrow),\emptyset} G^{X,\emptyset}_{\mu_2,(\eta_i,\eta_\uparrow)} -  e^{\mu_3} Q_{X,\mu_3,(\eta_i,\eta_\uparrow),\emptyset} G^{X,\emptyset}_{\mu_3,(\eta_i,\eta_\uparrow)}}{(\mu_2 -\mu_3)\left(1-\frac{\mu_2 + \mu_3}{\mu_1}\right)} \right. \nonumber\\
    &\quad\quad\quad\left. - \frac{\frac{\mu_2^2 - \mu_3^2}{\mu_1^2} \left(e^{\mu_1} Q_{X,\mu_1,(\eta_i,\eta_\uparrow),\emptyset} G^{X,\emptyset}_{\mu_1,(\eta_i,\eta_\uparrow)} -\underline{Y^{X,\emptyset}_{0,(\eta_i,\eta_\uparrow)} e_{X,0,(\eta_i,\eta_\uparrow),\emptyset}} \right)}{(\mu_2 -\mu_3)\left(1-\frac{\mu_2 + \mu_3}{\mu_1}\right)} \right\rbrace \label{YeX1-Znoclick-low}\\
    \overline{Y^{X,\checkmark}_{1,(\eta_i,\eta_\uparrow)} e_{X,1,(\eta_i,\eta_\uparrow),\checkmark}} &= \min\left\lbrace 1, \frac{e^{\mu_2} Q_{X,\mu_2,(\eta_i,\eta_\uparrow),\checkmark} G^{X,\checkmark}_{\mu_2,(\eta_i,\eta_\uparrow)} - e^{\mu_3} Q_{X,\mu_3,(\eta_i,\eta_\uparrow),\checkmark} G^{X,\checkmark}_{\mu_3,(\eta_i,\eta_\uparrow)}}{\mu_2 - \mu_3} , \right.\nonumber\\
    &\hspace{-2cm}\left. 1 - \left(\frac{(\mu_2^2-\mu_3^2) e^{\mu_1}(1-Q_{X,\mu_1,(\eta_i,\eta_\uparrow),\checkmark} G^{X,\checkmark}_{\mu_1,(\eta_i,\eta_\uparrow)}) - (\mu_1^2-\mu_3^2)e^{\mu_2}(1-Q_{X,\mu_2,(\eta_i,\eta_\uparrow),\checkmark} G^{X,\checkmark}_{\mu_2,(\eta_i,\eta_\uparrow)})}{(\mu_2^2-\mu_3^2) \mu_1 - (\mu_1^2-\mu_3^2)\mu_2 + (\mu_1^2-\mu_2^2)\mu_3} \right. \right.\nonumber\\
    &\left.\left. + \frac{(\mu_1^2-\mu_2^2)e^{\mu_3}(1- Q_{X,\mu_3,(\eta_i,\eta_\uparrow),\checkmark} G^{X,\checkmark}_{\mu_3,(\eta_i,\eta_\uparrow)})}{(\mu_2^2-\mu_3^2) \mu_1 - (\mu_1^2-\mu_3^2)\mu_2 + (\mu_1^2-\mu_2^2)\mu_3} \right) \right\rbrace, \label{YeX1-Zclick-upp}
\end{align}
and
\begin{align}
    \overline{Y^{X,\emptyset}_{1,(\eta_i,\eta_\uparrow)} e_{X,1,(\eta_i,\eta_\uparrow),\emptyset}} &= \min\left\lbrace 1, \frac{e^{\mu_2} Q_{X,\mu_2,(\eta_i,\eta_\uparrow),\emptyset} G^{X,\emptyset}_{\mu_2,(\eta_i,\eta_\uparrow)} - e^{\mu_3} Q_{X,\mu_3,(\eta_i,\eta_\uparrow),\emptyset} G^{X,\emptyset}_{\mu_3,(\eta_i,\eta_\uparrow)}}{\mu_2 - \mu_3}, \right.\nonumber\\
    &\hspace{-2cm}\left. 1 - \left(\frac{(\mu_2^2-\mu_3^2) e^{\mu_1}(1-Q_{X,\mu_1,(\eta_i,\eta_\uparrow),\emptyset} G^{X,\emptyset}_{\mu_1,(\eta_i,\eta_\uparrow)}) - (\mu_1^2-\mu_3^2)e^{\mu_2}(1-Q_{X,\mu_2,(\eta_i,\eta_\uparrow),\emptyset} G^{X,\emptyset}_{\mu_2,(\eta_i,\eta_\uparrow)})}{(\mu_2^2-\mu_3^2) \mu_1 - (\mu_1^2-\mu_3^2)\mu_2 + (\mu_1^2-\mu_2^2)\mu_3} \right. \right.\nonumber\\
    &\left.\left. + \frac{(\mu_1^2-\mu_2^2)e^{\mu_3}(1- Q_{X,\mu_3,(\eta_i,\eta_\uparrow),\emptyset} G^{X,\emptyset}_{\mu_3,(\eta_i,\eta_\uparrow)})}{(\mu_2^2-\mu_3^2) \mu_1 - (\mu_1^2-\mu_3^2)\mu_2 + (\mu_1^2-\mu_2^2)\mu_3} \right) \right\rbrace, \label{YeX1-Znoclick-upp}
\end{align}
for $\eta_i \in \{\eta_\uparrow,\eta_\downarrow,\eta_2\}$.\\

\subsection{Bounds on linear combinations of yields}

Some of the quantities appearing in the phase error rate bound \eqref{protocol-phase-error-rate}, namely \eqref{protocol-yalpha>1}--\eqref{protocol-y1-lowerbound}, depend on linear combinations of one-photon $Z$-basis yields. In principle, in these expressions we could use the bounds already derived for the single yields. However, in order to obtain a tighter bound, in the following we derive bounds on the whole linear combination of the yields.

To start with, we observe that the linear combinations of yields in \eqref{protocol-y0-upperbound}--\eqref{protocol-y1-lowerbound} can be generically written in the form:
\begin{align}
    c_\uparrow Y^Z_{1,(\eta_\uparrow,\eta_\uparrow)} - c_\downarrow Y^Z_{1,(\eta_\downarrow,\eta_\downarrow)}, \label{linear-comb-yields}
\end{align}
where the coefficients $c_\uparrow$ and $c_\downarrow$ depend on the particular linear combination that is considered. Then, we can obtain a tighter upper bound on \eqref{linear-comb-yields} by considering the following equation:
\begin{align}
    \tilde{G}^Z_{\mu_j} = \sum_{n=0}^{\infty} e^{-\mu_j} \frac{\mu^n_j}{n!} \tilde{Y}^Z_n, \label{decoy-eq-tilde}
\end{align}
where we defined:
\begin{align}
    \tilde{G}^Z_{\mu_j} &:=  c_\uparrow G^Z_{\mu_j,(\eta_\uparrow,\eta_\uparrow)} - c_\downarrow G^Z_{\mu_j,(\eta_\downarrow,\eta_\downarrow)} \label{Gtilde} \\
    \tilde{Y}^Z_n &:= c_\uparrow Y^Z_{n,(\eta_\uparrow,\eta_\uparrow)} - c_\downarrow Y^Z_{n,(\eta_\downarrow,\eta_\downarrow)} \label{Ytilde}.
\end{align}
We observe that \eqref{decoy-eq-tilde} has the same form as the standard equation of the decoy-state method, \eqref{decoy-eq}, with the important difference that the new gains ($\tilde{G}^Z_{\mu_j}$) and new yields ($\tilde{Y}^Z_n$) are not necessarily non-negative. In other words, while $0 \leq Y_n \leq 1$ for every yield, this is not true anymore for $\tilde{Y}^Z_n$. Nevertheless, we can still apply the techniques of standard decoy-state method to the equation \eqref{decoy-eq-tilde}, as far as we can obtain an interval of existence for the new yields $\tilde{Y}^Z_n$.

To this aim, we recall from \eqref{gainKG-decoy} that the $n$-photon $Z$-basis yield can be written as:
\begin{align}
    Y^Z_{n,(\eta_l,\eta_l)} = \Tr\left[\sigma_n Z^{(\eta_l,\eta_l)}_\checkmark \right] \label{n-Z-yield},
\end{align}
with:
\begin{align}
    \sigma_n := \Tr_E \left[ U_{BE} {\textstyle\sum_{i=0}^{d-1}} \frac{\proj{n_{Z_i}}}{d} \otimes \proj{0}_E U^\dag_{BE}\right].
\end{align}
Now, by using \eqref{Zclick-reflectedTBS-simplified-modea2} for $Z^{(\eta_l,\eta_l)}_\checkmark$ in \eqref{n-Z-yield}, we can express the new yield \eqref{Ytilde} as follows:
\begin{align}
    \tilde{Y}^Z_n  &=\sum_{\alpha=0}^\infty \Tr\left[\sigma_n \pia\right] \left\lbrace c_\uparrow\left[1-(1-p^{\mathcal{Z}}_d) \eta_\uparrow^\alpha \right] -c_\downarrow \left[1-(1-p^{\mathcal{Z}}_d) \eta_\downarrow^\alpha \right] \right\rbrace  . \label{Ytilde2}
\end{align}
From the last expression, we can immediately derive an interval of existence for $\tilde{Y}^Z_n$:
\begin{align}
   \underline{\tilde{Y}^Z} \leq  \tilde{Y}^Z_n \leq \overline{\tilde{Y}^Z}, \label{Ytilde-interval} 
\end{align}
where we defined:
\begin{align}
    \underline{\tilde{Y}^Z} &= \min_{\alpha \geq 0} \left\lbrace c_\uparrow\left[1-(1-p^{\mathcal{Z}}_d) \eta_\uparrow^\alpha \right] -c_\downarrow \left[1-(1-p^{\mathcal{Z}}_d) \eta_\downarrow^\alpha \right] \right\rbrace \label{Ytilde-min} \\
    \overline{\tilde{Y}^Z} &= \max_{\alpha \geq 0} \left\lbrace c_\uparrow\left[1-(1-p^{\mathcal{Z}}_d) \eta_\uparrow^\alpha \right] -c_\downarrow \left[1-(1-p^{\mathcal{Z}}_d) \eta_\downarrow^\alpha \right] \right\rbrace \label{Ytilde-max}.
\end{align}
Clearly, the interval of existence of $ \tilde{Y}^Z_n$ is independent of $n$ (as for the standard yields) but depends on the particular linear combination of yields, i.e.\ on $c_\uparrow$ and $c_\downarrow$. It is useful to derive a necessary and sufficient condition for the terms in \eqref{Ytilde-min} and \eqref{Ytilde-max} to be increasing with $\alpha$:
\begin{align}
    & c_\uparrow\left[1-(1-p^{\mathcal{Z}}_d) \eta_\uparrow^\alpha \right] -c_\downarrow \left[1-(1-p^{\mathcal{Z}}_d) \eta_\downarrow^\alpha \right] \leq c_\uparrow\left[1-(1-p^{\mathcal{Z}}_d) \eta_\uparrow^{\alpha+1} \right] -c_\downarrow \left[1-(1-p^{\mathcal{Z}}_d) \eta_\downarrow^{\alpha+1} \right] \nonumber\\
    \iff \quad & c_\downarrow \eta^\alpha_\downarrow -c_\uparrow \eta^\alpha_\uparrow \leq c_\downarrow \eta^{\alpha+1}_\downarrow -c_\uparrow \eta^{\alpha+1}_\uparrow \nonumber\\
    \iff \quad & c_\downarrow (1-\eta_\downarrow) \eta^\alpha_\downarrow  \leq c_\uparrow (1-\eta_\uparrow) \eta^\alpha_\uparrow , \label{cond-increasing-alpha}
\end{align}
which can then be applied to the specific cases of $c_\uparrow$ and $c_\downarrow$.

We can now derive an upper bound on \eqref{linear-comb-yields} by following the same procedure that leads to \eqref{Y1-upp-2decoy}. In particular, we consider the following combination of new gains in \eqref{decoy-eq-tilde}:
\begin{align}
    e^{\mu_2} \tilde{G}^Z_{\mu_2} - e^{\mu_3} \tilde{G}^Z_{\mu_3} &= (\mu_2 - \mu_3) \tilde{Y}^Z_1 + \sum_{n=2}^\infty \frac{\mu_2^n - \mu_3^n}{n!} \tilde{Y}^Z_n \nonumber\\
    &\geq (\mu_2 - \mu_3) \tilde{Y}^Z_1 + \underline{\tilde{Y}^Z} \sum_{n=2}^\infty \frac{\mu_2^n - \mu_3^n}{n!} \nonumber\\
    &= (\mu_2 - \mu_3) \tilde{Y}^Z_1 + \underline{\tilde{Y}^Z} \left(e^{\mu_2}-e^{\mu_3} + \mu_3 -\mu_2 \right),  \label{Y1-2decoy-calc}
\end{align}
which leads to the following upper bound on $\tilde{Y}^Z_1$:
\begin{align}
    \tilde{Y}^Z_1= c_\uparrow Y^Z_{1,(\eta_\uparrow,\eta_\uparrow)} - c_\downarrow Y^Z_{1,(\eta_\downarrow,\eta_\downarrow)} \leq \overline{c_\uparrow Y^Z_{1,(\eta_\uparrow,\eta_\uparrow)} - c_\downarrow Y^Z_{1,(\eta_\downarrow,\eta_\downarrow)}},
\end{align}
with:
\begin{align}
    \overline{c_\uparrow Y^Z_{1,(\eta_\uparrow,\eta_\uparrow)} - c_\downarrow Y^Z_{1,(\eta_\downarrow,\eta_\downarrow)}} := \min &\left\lbrace \overline{\tilde{Y}^Z}, \frac{c_\uparrow (e^{\mu_2} G^Z_{\mu_2,(\eta_\uparrow,\eta_\uparrow)} - e^{\mu_3} G^Z_{\mu_3,(\eta_\uparrow,\eta_\uparrow)}) - c_\downarrow (e^{\mu_2} G^Z_{\mu_2,(\eta_\downarrow,\eta_\downarrow)} - e^{\mu_3} G^Z_{\mu_3,(\eta_\downarrow,\eta_\downarrow)})}{\mu_2-\mu_3} \right. \nonumber\\
    &\left.\quad -\frac{ \underline{\tilde{Y}^Z} \left(e^{\mu_2}-e^{\mu_3} + \mu_3 -\mu_2 \right)}{\mu_2 -\mu_3} \right\rbrace. \label{Y1tilde-upp}
\end{align}
This upper bound can be applied to each expression in \eqref{protocol-y0-upperbound}--\eqref{protocol-y1-lowerbound} by replacing $c_\uparrow$ and $c_\downarrow$ with the appropriate coefficients from \eqref{protocol-yalpha>1}--\eqref{protocol-y1-lowerbound} and by employing the explicit expressions for $\underline{\tilde{Y}^Z}$ and $\overline{\tilde{Y}^Z}$. The latter are obtained from \eqref{Ytilde-min} and \eqref{Ytilde-max} for specific choices of $c_\uparrow$ and $c_\downarrow$:
\begin{itemize}
    \item for $c_\uparrow=\eta_\downarrow$ and $c_\downarrow=\eta_\uparrow(1-\eta_\uparrow)$:\\
    In this case the condition in \eqref{cond-increasing-alpha} becomes $(1-\eta_\downarrow)\eta_\downarrow^{\alpha-1} \leq \eta_\uparrow^{\alpha-1}$, which is true if $\alpha \geq 1$. Moreover, since the following inequality:
    \begin{align}
        &\left\lbrace c_\uparrow\left[1-(1-p^{\mathcal{Z}}_d) \eta_\uparrow^\alpha \right] -c_\downarrow \left[1-(1-p^{\mathcal{Z}}_d) \eta_\downarrow^\alpha \right] \right\rbrace_{\alpha=0} \geq \left\lbrace c_\uparrow\left[1-(1-p^{\mathcal{Z}}_d) \eta_\uparrow^\alpha \right] -c_\downarrow \left[1-(1-p^{\mathcal{Z}}_d) \eta_\downarrow^\alpha \right] \right\rbrace_{\alpha=1} \nonumber\\
        \iff \quad & \eta_\uparrow \geq \frac{\eta_\downarrow}{1-\eta_\downarrow}
    \end{align}
    is true due to \eqref{constraint1}, we have that \eqref{Ytilde-min} and \eqref{Ytilde-max} are given by:
    \begin{align}
    \underline{\tilde{Y}^Z} &= \left\lbrace c_\uparrow\left[1-(1-p^{\mathcal{Z}}_d) \eta_\uparrow^\alpha \right] -c_\downarrow \left[1-(1-p^{\mathcal{Z}}_d) \eta_\downarrow^\alpha \right] \right\rbrace_{\alpha=1} = \eta_\downarrow - \eta_\uparrow(1-\eta_\uparrow) - (1-p^{\mathcal{Z}}_d) \eta_\downarrow \eta^2_\uparrow  \label{Ytilde-min-w0upp} \\
    \overline{\tilde{Y}^Z} &= \max \left[\left\lbrace c_\uparrow\left[1-(1-p^{\mathcal{Z}}_d) \eta_\uparrow^\alpha \right] -c_\downarrow \left[1-(1-p^{\mathcal{Z}}_d) \eta_\downarrow^\alpha \right] \right\rbrace_{\alpha=0},\left\lbrace c_\uparrow\left[1-(1-p^{\mathcal{Z}}_d) \eta_\uparrow^\alpha \right] -c_\downarrow \left[1-(1-p^{\mathcal{Z}}_d) \eta_\downarrow^\alpha \right] \right\rbrace_{\alpha \to \infty}\right] \nonumber\\
    &= \max\left\lbrace p^{\mathcal{Z}}_d \left[\eta_\downarrow - \eta_\uparrow(1-\eta_\uparrow)\right],\eta_\downarrow - \eta_\uparrow(1-\eta_\uparrow)\right\rbrace \label{Ytilde-max-w0upp}.
    \end{align}
    
    \item for $c_\uparrow=-\eta_\downarrow (1-\eta_\downarrow)$ and $c_\downarrow=-\eta_\uparrow$:\\
    In this case the condition in \eqref{cond-increasing-alpha} becomes $\eta_\downarrow^{\alpha-1} \geq (1-\eta_\uparrow) \eta_\uparrow^{\alpha-1}$, which is verified for $\alpha=0,1$ but it is not clear for what $\alpha$ it does not hold. To find good estimations of $\underline{\tilde{Y}^Z}$ and $\overline{\tilde{Y}^Z}$ in this case, we define $f(\alpha)$ to be the argument of the optimizations in \eqref{Ytilde-min} and \eqref{Ytilde-max}:
    \begin{align}
        f(\alpha) := \eta_\uparrow - \eta_\downarrow (1-\eta_\downarrow) + (1-p^{\mathcal{Z}}_d)\eta_\uparrow \eta_\downarrow \left[(1-\eta_\downarrow) \eta_\uparrow^{\alpha-1} - \eta_\downarrow^{\alpha-1} \right] ,
    \end{align}
    and find the condition for which its first derivative is non-negative: $f'(\alpha) \geq 0$. We obtain:
    \begin{align}
        \alpha \leq 1 + \frac{\ln\left[(1-\eta_\downarrow)\frac{\ln\eta_\uparrow}{\ln\eta_\downarrow} \right]}{\ln \eta_\downarrow - \ln \eta_\uparrow},
    \end{align}
    which implies $\alpha<1$. Thus, we conclude that $f(\alpha)$ is minimal at the two extremes of $\alpha$'s range: $f(0)$ and $f(\infty)$, and since $\eta_\uparrow > \eta_\downarrow (1-\eta_\downarrow)$, we get:
    \begin{align}
    \underline{\tilde{Y}^Z} &= \left\lbrace c_\uparrow\left[1-(1-p^{\mathcal{Z}}_d) \eta_\uparrow^\alpha \right] -c_\downarrow \left[1-(1-p^{\mathcal{Z}}_d) \eta_\downarrow^\alpha \right] \right\rbrace_{\alpha=0} = p^{\mathcal{Z}}_d \left[\eta_\uparrow - \eta_\downarrow (1-\eta_\downarrow)\right]  \label{Ytilde-min-w0low} \\
    \overline{\tilde{Y}^Z} &= \eta_\uparrow \label{Ytilde-max-w0low},
    \end{align}
    while for $\overline{\tilde{Y}^Z}$ we used the definition \eqref{Ytilde} and the fact that $0 \leq Y^Z_n \leq 1$.

    \item for $c_\uparrow= - (1-\eta^2_\downarrow)$ and $c_\downarrow=-1$:\\
     In this case the condition in \eqref{cond-increasing-alpha} becomes $\eta_\downarrow^{\alpha} \geq (1+\eta_\downarrow)(1-\eta_\uparrow) \eta_\uparrow^{\alpha}$, which is verified for $\alpha=0$. Similarly to before, we define $f(\alpha)$ to be the argument of the optimizations in \eqref{Ytilde-min} and \eqref{Ytilde-max}:
    \begin{align}
        f(\alpha) := 1- (1-\eta^2_\downarrow) + (1-p^{\mathcal{Z}}_d) \left[(1-\eta^2_\downarrow) \eta_\uparrow^{\alpha} - \eta_\downarrow^{\alpha} \right] ,
    \end{align}
    and find the condition for which its first derivative is non-negative: $f'(\alpha) \geq 0$. We obtain:
    \begin{align}
        \alpha \leq  \frac{\ln\left[(1-\eta^2_\downarrow)\frac{\ln\eta_\uparrow}{\ln\eta_\downarrow} \right]}{\ln \eta_\downarrow - \ln \eta_\uparrow},
    \end{align}
    which implies that $f(\alpha)$ stops increasing for a value $\alpha>0$. Thus, $f(\alpha)$ has its global minimum in either $\alpha=0$ or $\alpha=\infty$. We obtain:
    \begin{align}
    \underline{\tilde{Y}^Z} &= p^{\mathcal{Z}}_d \left[1-  (1-\eta^2_\downarrow)\right]  \label{Ytilde-min-w1upp} \\
    \overline{\tilde{Y}^Z} &= 1 \label{Ytilde-max-w1upp},
    \end{align}
    where for $\overline{\tilde{Y}^Z}$ we again used the definition \eqref{Ytilde} and the fact that $0 \leq Y^Z_n \leq 1$.

    \item for $c_\uparrow= 1$ and $c_\downarrow=1-\eta^2_\uparrow$:\\
     In this case the condition in \eqref{cond-increasing-alpha} becomes $(1+\eta_\uparrow)(1-\eta_\downarrow) \eta_\downarrow^{\alpha} \leq \eta_\uparrow^{\alpha}$, which is not verified for $\alpha=0$ due to \eqref{constraint1}. Therefore, given that $f(\alpha)$ is the argument of the optimizations in \eqref{Ytilde-min} and \eqref{Ytilde-max}:
    \begin{align}
        f(\alpha) := 1- (1-\eta^2_\uparrow) + (1-p^{\mathcal{Z}}_d) \left[(1-\eta^2_\uparrow) \eta_\downarrow^{\alpha} - \eta_\uparrow^{\alpha} \right] ,
    \end{align}
    we found that $f(0)> f(1)$. Now we find the condition for which its first derivative is negative: $f'(\alpha) < 0$. We obtain:
    \begin{align}
        \alpha <  \frac{\ln\left[\frac{\ln\eta_\uparrow}{(1-\eta^2_\uparrow) \ln\eta_\downarrow} \right]}{\ln \eta_\downarrow - \ln \eta_\uparrow},
    \end{align}
    and since we know that $f(0)> f(1)$, it must be that $f(\alpha)$ stops decreasing at some point greater than zero. Then, we conclude that $f(\alpha)$ is maximal at the extremes and we obtain:
    \begin{align}
    \underline{\tilde{Y}^Z} &= -(1-\eta_\uparrow^2)  \label{Ytilde-min-w1low} \\
    \overline{\tilde{Y}^Z} &= \eta_\uparrow^2 \label{Ytilde-max-w1low},
    \end{align}
    where for $\underline{\tilde{Y}^Z}$ we used the definition \eqref{Ytilde} and the fact that $0 \leq Y^Z_n \leq 1$.
\end{itemize}

Finally, we derive an upper bound on the linear combination of yields that appears in \eqref{protocol-yalpha>1}, namely:
\begin{align}
    (\eta_\uparrow - \eta_\downarrow)Y^Z_{1,(\eta_2,\eta_2)} - (1-\eta_2)(Y^Z_{1,(\eta_\downarrow,\eta_\downarrow)} - Y^Z_{1,(\eta_\uparrow,\eta_\uparrow)}) -(\eta_\uparrow -\eta_\downarrow) p^{\mathcal{Z}}_d . \label{linear-comb-yields2}
\end{align}
This can again be done by applying the decoy-state method on the formula in \eqref{decoy-eq-tilde}, where this time we define the gain and the yield to be:
\begin{align}
    \tilde{G}^Z_{\mu_j} &:=  (\eta_\uparrow - \eta_\downarrow) G^Z_{\mu_j,(\eta_2,\eta_2)} - (1-\eta_2)\left( G^Z_{\mu_j,(\eta_\downarrow,\eta_\downarrow)} - G^Z_{\mu_j,(\eta_\uparrow,\eta_\uparrow)}\right) -(\eta_\uparrow -\eta_\downarrow) p^{\mathcal{Z}}_d \label{Gtilde-weight} \\
    \tilde{Y}^Z_n &:= (\eta_\uparrow - \eta_\downarrow)Y^Z_{n,(\eta_2,\eta_2)} - (1-\eta_2)\left( Y^Z_{n,(\eta_\downarrow,\eta_\downarrow)} - Y^Z_{n,(\eta_\uparrow,\eta_\uparrow)}\right) -(\eta_\uparrow -\eta_\downarrow) p^{\mathcal{Z}}_d \label{Ytilde-weight}.
\end{align}
Our goal is to derive an upper bound on $\tilde{Y}^Z_1$. Similarly to \eqref{Ytilde2}, we can recast the newly defined yield in \eqref{Ytilde-weight} as follows:
\begin{align}
    \tilde{Y}^Z_n  &=\sum_{\alpha=0}^\infty \Tr\left[\sigma_n \pia\right] \gamma_\alpha  , \label{Ytilde2-weight}
\end{align}
where we introduced:
\begin{align}
    \gamma_\alpha := (1-p^{\mathcal{Z}}_d) \left[(\eta_\uparrow - \eta_\downarrow)(1-\eta_2^\alpha) - (1-\eta_2)(\eta_\uparrow^\alpha - \eta_\downarrow^\alpha)\right] \label{gamma-alpha}.
\end{align}
To derive an interval of existence for $\tilde{Y}^Z_n$, i.e.\ $\underline{\tilde{Y}^Z} \leq  \tilde{Y}^Z_n \leq \overline{\tilde{Y}^Z}$, we notice that $\gamma_0=\gamma_1=0$ and that $\gamma_\alpha \geq 0$ for every $\alpha\geq 2$ (this last result can be directly inferred from the study of $g(\alpha)$ in \eqref{g(alpha)>0}). Then, we have
\begin{align}
    \underline{\tilde{Y}^Z} =\gamma_0 = 0 \label{lower-Ytilde-weight}.
\end{align}
For the upper bound, we observe that $\gamma_\alpha \leq \gamma_{\alpha+1}$ for $\alpha \geq 2$. Indeed:
\begin{align}
    &\gamma_\alpha \leq \gamma_{\alpha+1} \nonumber\\
    \Leftrightarrow\quad &(\eta_\uparrow - \eta_\downarrow)(-\eta_2^\alpha) - (1-\eta_2)(\eta_\uparrow^\alpha - \eta_\downarrow^\alpha) \leq (\eta_\uparrow - \eta_\downarrow)(-\eta_2^{\alpha+1}) - (1-\eta_2)(\eta_\uparrow^{\alpha+1} - \eta_\downarrow^{\alpha+1}) \nonumber\\
    \Leftrightarrow\quad &- (1-\eta_2)(\eta_\uparrow^\alpha - \eta_\downarrow^\alpha) \leq (\eta_\uparrow - \eta_\downarrow)\eta_2^\alpha (1-\eta_2) - (1-\eta_2)(\eta_\uparrow^{\alpha+1} - \eta_\downarrow^{\alpha+1}) \nonumber\\
    \Leftrightarrow\quad & 0 \leq (\eta_\uparrow^\alpha - \eta_\downarrow^\alpha) + (\eta_\uparrow - \eta_\downarrow)\eta_2^\alpha  - (\eta_\uparrow^{\alpha+1} - \eta_\downarrow^{\alpha+1}) \nonumber\\
    \Leftrightarrow\quad & 0 \leq (\eta_\uparrow - \eta_\downarrow)(\eta_2^\alpha + \eta_\uparrow^{\alpha-1} + \eta_\uparrow^{\alpha-2}\eta_\downarrow + \dots + \eta_\uparrow \eta_\downarrow^{\alpha-2} + \eta_\downarrow^{\alpha-1}) \nonumber\\
    &\quad\quad - (\eta_\uparrow - \eta_\downarrow)(\eta_\uparrow^{\alpha} + \eta_\uparrow^{\alpha-1}\eta_\downarrow + \dots + \eta_\uparrow \eta_\downarrow^{\alpha-1} + \eta_\downarrow^{\alpha}) \nonumber\\
    \Leftrightarrow\quad & \eta_2^\alpha + \eta_\uparrow^{\alpha-1} + \eta_\uparrow^{\alpha-2}\eta_\downarrow + \dots + \eta_\uparrow \eta_\downarrow^{\alpha-2} + \eta_\downarrow^{\alpha-1} \geq \eta_\uparrow^{\alpha} + \eta_\uparrow^{\alpha-1}\eta_\downarrow + \dots + \eta_\uparrow \eta_\downarrow^{\alpha-1} + \eta_\downarrow^{\alpha} \nonumber\\
    \Leftrightarrow\quad & \eta_\uparrow^{\alpha-1} (1-\eta_\uparrow) + \eta_\uparrow^{\alpha-2}\eta_\downarrow (1-\eta_\uparrow) + \dots + \eta_\uparrow \eta_\downarrow^{\alpha-2} (1-\eta_\uparrow) + \eta_\downarrow^{\alpha-1} (1-\eta_\uparrow)  + (\eta_2^\alpha - \eta_\downarrow^\alpha) \geq 0 
\end{align}
which is true since $\eta_2>\eta_\downarrow$ and $1\geq \eta_\uparrow$. Then, we have that:
\begin{align}
    \overline{\tilde{Y}^Z} = \gamma_{\infty} = (1-p^{\mathcal{Z}}_d)(\eta_\uparrow - \eta_\downarrow).
\end{align}
We can now derive an upper bound on $\tilde{Y}^Z_1$ in the same way used to derive an upper bound on the one-photon yield with two or three decoy intensities, \eqref{Y1-upp}. We start by obtaining the bound with two decoy intensities. Consider the decoy-state formula in \eqref{decoy-eq-tilde} and consider the combination of gains that leads to \eqref{Y1-2decoy-calc}. We obtain the following upper bound on $\tilde{Y}^Z_1$:
\begin{align}
    \tilde{Y}^Z_1  \leq C(\mu_2,\mu_3),
\end{align}
where we defined:
\begin{align}
   &C(\mu_2,\mu_3) := \frac{ e^{\mu_2} \tilde{G}^Z_{\mu_2} - e^{\mu_3} \tilde{G}^Z_{\mu_3}-\underline{\tilde{Y}^Z} \left(e^{\mu_2}-e^{\mu_3} + \mu_3 -\mu_2 \right)}{\mu_2 - \mu_3} .
\end{align}
By substituting $\underline{\tilde{Y}^Z}=0$ and $\Tilde{G}_{\mu_j}$ with \eqref{Gtilde-weight} in the last expression, we obtain:
\begin{align}
   C(\mu_2,\mu_3) &= e^{\mu_2}\frac{ (\eta_\uparrow - \eta_\downarrow) \left(G^Z_{\mu_2,(\eta_2,\eta_2)}-p^{\mathcal{Z}}_d \right) - (1-\eta_2)\left( G^Z_{\mu_2,(\eta_\downarrow,\eta_\downarrow)} - G^Z_{\mu_2,(\eta_\uparrow,\eta_\uparrow)}\right)}{\mu_2-\mu_3} \nonumber\\
   &\quad- e^{\mu_3} \frac{ (\eta_\uparrow - \eta_\downarrow) \left(G^Z_{\mu_3,(\eta_2,\eta_2)}-p^{\mathcal{Z}}_d \right) - (1-\eta_2)\left( G^Z_{\mu_3,(\eta_\downarrow,\eta_\downarrow)} - G^Z_{\mu_3,(\eta_\uparrow,\eta_\uparrow)}\right)}{\mu_2 - \mu_3} . \label{Y1tilde-upp1}
\end{align}

For the case of three decoy intensities, we consider the same combination of gains as in \eqref{Y1-upp-calc}. Then, in the resulting expression, we can replace $\tilde{Y}^Z_n$ with $\overline{\tilde{Y}^Z}$ (instead of $1$) and follow the same steps leading to \eqref{Y1-upp-3decoy}. By doing so, we obtain:
\begin{align}
   &C(\mu_1,\mu_2,\mu_3) := \overline{\tilde{Y}^Z} - \frac{(\mu_2^2-\mu_3^2) e^{\mu_1}(\overline{\tilde{Y}^Z} - \tilde{G}^Z_{\mu_1}) - (\mu_1^2-\mu_3^2)e^{\mu_2}(\overline{\tilde{Y}^Z}-\tilde{G}^Z_{\mu_2}) + (\mu_1^2-\mu_2^2)e^{\mu_3}(\overline{\tilde{Y}^Z}-\tilde{G}^Z_{\mu_3})}{(\mu_2^2-\mu_3^2) \mu_1 - (\mu_1^2-\mu_3^2)\mu_2 + (\mu_1^2-\mu_2^2)\mu_3}  .
\end{align}
By substituting $\overline{\tilde{Y}^Z}= (1-p^{\mathcal{Z}}_d)(\eta_\uparrow - \eta_\downarrow)$ and $\Tilde{G}_{\mu_j}$ with \eqref{Gtilde-weight} in the last expression, we obtain:
\begin{align}
   &C(\mu_1,\mu_2,\mu_3)=  (1-p^{\mathcal{Z}}_d)(\eta_\uparrow - \eta_\downarrow) - \frac{F}{(\mu_2^2-\mu_3^2) \mu_1 - (\mu_1^2-\mu_3^2)\mu_2 + (\mu_1^2-\mu_2^2)\mu_3}  , \label{Y1tilde-upp2}
\end{align}
where we defined:
\begin{align}
     F:=&(\mu_2^2-\mu_3^2) e^{\mu_1}\left[(\eta_\uparrow - \eta_\downarrow)(1-G^Z_{\mu_1,(\eta_2,\eta_2)}) + (1-\eta_2)(G^Z_{\mu_1,(\eta_\downarrow,\eta_\downarrow)} - G^Z_{\mu_1,(\eta_\uparrow,\eta_\uparrow)})\right]  \nonumber\\
     &- (\mu_1^2-\mu_3^2)e^{\mu_2}\left[(\eta_\uparrow - \eta_\downarrow)(1-G^Z_{\mu_2,(\eta_2,\eta_2)}) + (1-\eta_2)(G^Z_{\mu_2,(\eta_\downarrow,\eta_\downarrow)} - G^Z_{\mu_2,(\eta_\uparrow,\eta_\uparrow)})\right]  \nonumber\\
     & + (\mu_1^2-\mu_2^2)e^{\mu_3}\left[(\eta_\uparrow - \eta_\downarrow)(1-G^Z_{\mu_3,(\eta_2,\eta_2)}) + (1-\eta_2)(G^Z_{\mu_3,(\eta_\downarrow,\eta_\downarrow)} - G^Z_{\mu_3,(\eta_\uparrow,\eta_\uparrow)})\right].
\end{align}
Finally, we combine the two upper bounds derived in \eqref{Y1tilde-upp1} and \eqref{Y1tilde-upp2} to obtain the final expression (analogous to \eqref{Y1-upp}) for our upper bound on the linear combination of yields in \eqref{linear-comb-yields2}:
\begin{align}
    &\overline{(\eta_\uparrow - \eta_\downarrow)Y^Z_{1,(\eta_2,\eta_2)} - (1-\eta_2)(Y^Z_{1,(\eta_\downarrow,\eta_\downarrow)} - Y^Z_{1,(\eta_\uparrow,\eta_\uparrow)}) -(\eta_\uparrow -\eta_\downarrow) p^{\mathcal{Z}}_d}:=\min\left\lbrace(1-p^{\mathcal{Z}}_d)(\eta_\uparrow - \eta_\downarrow), C(\mu_2,\mu_3), C(\mu_1,\mu_2,\mu_3) \right\rbrace.
\end{align}

\section{Simulations formulas} \label{app:simulations}
In this appendix we report the formulas used to generate the simulations of Sec.~\ref{sec:simulations}.

\subsection{High-dimensional time-bin QKD} \label{app:simulations-honest}

According to the channel model employed for simulating the time-bin QKD protocol (Protocol~1) in Figs.~\ref{fig:keyrate_maxloss} and \ref{fig:keyrate_eta}, in the absence of eavesdroppers, the $Z$-basis gain reads:
\begin{align}
    G^Z_{\mu_j,(\eta_l,\eta_l)} = 1- (1-p^{\mathcal{Z}}_d) e^{-\mu_j \eta\eta_Z  (1-\eta_l)}, \label{gainZhonest}
\end{align}
where $\eta$ is the transmittance of the quantum channel. For the $X$-basis gains, we recall that we assumed that the signal arriving at Bob is entirely contained in the detection window $\mathcal{Z}$. Hence, the gains are not affected by the transmittance of the TBS outside of the interval  $\mathcal{Z}$. Thus, we get:
\begin{align}
     G^{X,\checkmark}_{\mu_j,(\eta_i,\eta_l)} &= G^Z_{\mu_j,(\eta_i,\eta_i)} \left[1-(1-p^{\mathcal{X}}_d)e^{-\mu_j \eta\eta_X \eta_i}\right] \\
     G^{X,\emptyset}_{\mu_j,(\eta_i,\eta_l)} &= \left(1-G^Z_{\mu_j,(\eta_i,\eta_i)}\right) \left[1-(1-p^{\mathcal{X}}_d)e^{-\mu_j \eta\eta_X \eta_i}\right]. \label{gainXhonest}
\end{align}

Instead of modeling noise sources in the quantum channel and in the detectors, we assume that each click caused by a photon detection of the $Z$ ($X$) basis detector is affected by a channel-intrinsic QBER of $q_Z$ ($q_X$), while the clicks caused by dark counts are random and hence contribute with an error rate of $1-1/d$. Thus, we obtain the following QBER of the key generation rounds:
\begin{align}
    Q_{Z,\mu_j} &= \frac{q_Z (1- e^{-\mu_j \eta\eta_Z  (1-\eta_\downarrow)}) + \left(1-\frac{1}{d}\right) p^{\mathcal{Z}}_d e^{-\mu_j \eta\eta_Z  (1-\eta_\downarrow)}}{G^Z_{\mu_j,(\eta_\downarrow,\eta_\downarrow)}}. \label{QZsim}
\end{align}
Similarly, the QBERs of the test rounds are given by:
\begin{align}
    Q_{X,\mu_j,(\eta_i,\eta_\uparrow),\checkmark} &= \frac{q_X (1-e^{-\mu_j \eta\eta_X \eta_i}) + \left(1-\frac{1}{d}\right)p^{\mathcal{X}}_d e^{-\mu_j \eta\eta_X \eta_i}}{1-(1-p^{\mathcal{X}}_d)e^{-\mu_j \eta\eta_X \eta_i}} \label{QXsim} \\
    Q_{X,\mu_j,(\eta_i,\eta_\uparrow),\emptyset} &= Q_{X,\mu_j,(\eta_i,\eta_\uparrow),\checkmark}, \label{QXsim2}
\end{align}
where we observe that in our error model the test round QBERs are independent of whether the $Z$-basis detector clicks or does not click.

The quantities appearing in the decoy-BB84 key rate \eqref{protocol-rate-withassumption} are obtained with a standard application of the decoy-state method (cf.~Appendix~\ref{app:decoy}). In particular, we have:
\begin{align}
    \overline{e_{X,1}} &= \min\left\lbrace 1, \frac{e^{\mu_2} Q_{X,\mu_2} G^X_{\mu_2} - e^{\mu_3} Q_{X,\mu_3} G^X_{\mu_3}}{\underline{Y^X_{1}} (\mu_2 - \mu_3)} \right\rbrace  \label{bit-error-rate-decoy} ,
\end{align}
for the bit error rate upper bound, while the lower bounds on the one-photon yields $\underline{Y^X_{1}}$ and $\underline{Y^Z_{1}}$ are reported in \eqref{Y1-low}. The lower bound $\underline{Y^Z_{0}}$ is given by \eqref{Y0-low}. Note that these formulas need to be evaluated with the appropriate $Z$ or $X$ basis gains. Indeed, in Sec.~\ref{sec:simulations} we argued that even in the honest implementation of the protocol there is an asymmetry in the gains of the two bases due to asymmetric detectors. We derive the gains appearing in \eqref{protocol-rate-withassumption} as follows:
\begin{align}
    G^Z_{\mu_j} &= G^Z_{\mu_j,(\eta_\downarrow,\eta_\downarrow)} \\
    G^X_{\mu_j} &= G^{X,\checkmark}_{\mu_j,(\eta_\uparrow,\eta_\uparrow)} + G^{X,\emptyset}_{\mu_j,(\eta_\uparrow,\eta_\uparrow)},
\end{align}
where the gains on the right-hand side are given in \eqref{gainZhonest}--\eqref{gainXhonest}. This is due to the fact that the TBS is not needed in the decoy-BB84 proof, hence its only use in that protocol is to select the measurement basis, with setting $(\eta_\downarrow,\eta_\downarrow)$ that selects the $Z$ basis and setting $(\eta_\uparrow,\eta_\uparrow)$ that selects the $X$ basis. For the QBERs appearing in \eqref{protocol-rate-withassumption}, we have:
\begin{align}
    Q_{Z,\mu_j} &= \frac{q_Z (1- e^{-\mu_j \eta\eta_Z  (1-\eta_\downarrow)}) + \left(1-\frac{1}{d}\right) p^{\mathcal{Z}}_d e^{-\mu_j \eta\eta_Z  (1-\eta_\downarrow)}}{G^Z_{\mu_j}} \\
    Q_{X,\mu_j} &= \frac{q_X (1-e^{-\mu_j \eta\eta_X \eta_\uparrow}) + \left(1-\frac{1}{d}\right)p^{\mathcal{X}}_d e^{-\mu_j \eta\eta_X \eta_\uparrow}}{G^X_{\mu_j}}.
\end{align}

\subsection{Attack-induced efficiency mismatch} \label{app:simulations-adv}

In order to study the attack presented in Sec~\ref{sec:adversarial-implementation}, we assume that Alice can deterministically send one-photon pulses in each round. This allows us to simplify our proof by removing the decoy-state method, which is no longer needed since in this setting one-photon yields and error rates are directly observable quantities.

Here we report the new expressions for the one-photon yields and error rates that enter our key rate \eqref{protocol-rate-nodecoy} and the BB84 key rate \eqref{BB84-rate}, when Alice sends one-photon pulses.

\subsubsection{Protocol statistics without Eve's attack} \label{app:honest-implementation}

First we present the formulas for the case where there is no attack by Eve and Alice and Bob are linked by a channel with transmittance $\eta$ and intrinsic QBERs $q_Z$ and $q_X$. The one-photon yields are:
\begin{align}
     Y^{Z, \textrm{hon.}}_{1,(\eta_l,\eta_l)} &= \eta (1-\eta_l ) \eta_Z + \left[ 1-\eta (1-\eta_l)\eta_Z \right] p^{\mathcal{Z}}_d \label{yieldZ-nodecoy-honest} \\
     Y^{X,\textrm{hon.}}_{1,(\eta_l,\eta_l)} &= \eta \eta_l \eta_X + \left[ 1-\eta \eta_l \eta_X \right] p^{\mathcal{X}}_d \label{yieldX-nodecoy-honest} \\
     Y^{X,\checkmark,\textrm{hon.}}_{1,(\eta_i,\eta_l)} &= \eta \eta_i \eta_X p^{\mathcal{Z}}_d + \eta (1-\eta_i) \eta_Z  p^{\mathcal{X}}_d + p^{\mathcal{Z}}_d p^{\mathcal{X}}_d \left[1-\eta + \eta \eta_i (1-\eta_X) + \eta (1-\eta_i) (1-\eta_Z) \right] \label{yieldXclick-nodecoy-honest} \\
     Y^{X,\emptyset,\textrm{hon.}}_{1,(\eta_i,\eta_l)} &=(1-p^{\mathcal{Z}}_d) \left[ \eta \eta_i \eta_X + \eta \eta_i (1-\eta_X) p^{\mathcal{X}}_d + \eta (1-\eta_i) (1-\eta_Z) p^{\mathcal{X}}_d  + (1-\eta)p^{\mathcal{X}}_d \right]. \label{yieldXnoclick-nodecoy-honest}
\end{align}
The one-photon error rates satisfy the following equalities:
\begin{align}
     Y^{Z,\textrm{hon.}}_{1,(\eta_l,\eta_l)} e^{\textrm{hon.}}_{Z,1,(\eta_l,\eta_l)} &= q_Z \eta (1-\eta_l ) \eta_Z + \left(1-\frac{1}{d}\right)\left[ 1-\eta (1-\eta_l)\eta_Z \right] p^{\mathcal{Z}}_d \label{yieldZerror-nodecoy-honest} \\
     Y^{X,\textrm{hon.}}_{1,(\eta_l,\eta_l)} e^{\textrm{hon.}}_{X,1,(\eta_l,\eta_l)} &= q_X \eta \eta_l \eta_X + \left(1-\frac{1}{d}\right) \left[ 1-\eta \eta_l \eta_X \right] p^{\mathcal{X}}_d \label{yieldXerror-nodecoy-honest} \\
     Y^{X,\checkmark,\textrm{hon.}}_{1,(\eta_i,\eta_l)} e^{\textrm{hon.}}_{X,1,(\eta_i,\eta_l),\checkmark} &= q_X \eta \eta_i \eta_X p^{\mathcal{Z}}_d + \left(1-\frac{1}{d}\right)\left(Y^{X,\checkmark,\textrm{hon.}}_{1,(\eta_i,\eta_l)} -\eta \eta_i \eta_X p^{\mathcal{Z}}_d \right)  \label{yieldXclickerror-nodecoy-honest} \\
     Y^{X,\emptyset,\textrm{hon.}}_{1,(\eta_i,\eta_l)} e^{\textrm{hon.}}_{X,1,(\eta_i,\eta_l),\emptyset} &=q_X \eta \eta_i \eta_X (1-p^{\mathcal{Z}}_d) + \left(1-\frac{1}{d}\right)\left(Y^{X,\emptyset,\textrm{hon.}}_{1,(\eta_i,\eta_l)}-\eta \eta_i \eta_X (1-p^{\mathcal{Z}}_d) \right), \label{yieldXnoclickerror-nodecoy-honest}
\end{align}
where the yields appearing on the right-hand side are given in \eqref{yieldXclick-nodecoy-honest} and \eqref{yieldXnoclick-nodecoy-honest}.

By employing the formulas \eqref{yieldZ-nodecoy-honest}--\eqref{yieldXnoclickerror-nodecoy-honest} to replace to corresponding bounds in the phase error rate upper bound \eqref{protocol-phase-error-rate}, we obtain significant simplifications compared to the expressions in Appendix~\ref{app:phase-error-rate-formula}, which are obtained when Alice prepares phase-randomized coherent pulses for the decoy-state method. In particular, thanks to the knowledge of the true value of the one-photon yields, the estimation of the weight of Bob's received state in the ($\leq 1$)-subspace becomes tight. Namely, the upper bound in \eqref{protocol-yalpha>1} becomes null ($\overline{w}^{>1}_{\mathcal{Z}} =0$) once we replace the yields with \eqref{yieldZ-nodecoy-honest}. This fact greatly improves the tightness of our key rate bound, so much that it coincides with the asymptotic key rate of the BB84 protocol without decoys, as shown in Fig.~\ref{fig:keyrate_attack} of the main text, for $w_Z=0$.

\subsubsection{Protocol statistics with Eve's attack} \label{app:adver-implementation}

Eve intercepts and measures Alice's pulses in the $Z$ ($X$) basis with probability $w_Z$ ($w_X$). Eve can perfectly distinguish Alice's symbols when guessing the right basis. She then prepares a tailored time (frequency) pulse, corresponding to the $Z$-basis ($X$-basis) outcome she observed, and sends it through a pure-loss channel with transmittance $\xi_Z$ ($\xi_X$). Her goal is to make her attack undetectable in the standard BB84 protocol. In other words, she aims at introducing little noise in the events where Alice, Bob and Eve choose the same basis, while decreasing the detection probability at Bob for the events where she chooses the wrong basis, thereby keeping the QBERs low.

Before describing Eve's attack, we review Bob's measurement apparatus for the case of one-photon signals. Bob's measurement apparatus comprises the TBS with setting $(\eta_i,\eta_l)$, followed by a time-of-arrival measurement in the reflected port with efficiency $\eta_Z$. The measurement returns outcome $j'$ if the photon is detected in the time interval centered in $t_{j'}$ with width $\Delta_j$ for each bin. The total detection window in the $Z$ basis is $\Delta_t=d \Delta_j$. The photon transmitted by the TBS undergoes a dispersive medium modeled by the unitary:
\begin{align}
    U_{\rm dis} = \int_{-\infty}^\infty \text{d}f \, e^{i 2\pi^2 \Phi_2 f^2} \ketbra{f}{f},
\end{align}
with $\Phi_2$ the group delay dispersion (GDD) coefficient of the dispersive medium, followed by a time-of-arrival measurement with efficiency $\eta_X$, where outcome $k'$ corresponds to a detection in the time bin centered at $\tilde{t}_{k'}$ with width $\Delta_k$ for each bin. The total detection window in the $X$ basis is $\Delta_f=d \Delta_k$.

In order to remain undetected, when intercepting in the $Z$ basis, Eve prepares the following one-photon Gaussian time pulse corresponding to outcome $j$:
\begin{align}
    \ket{E_Z(j)} := \int_{-\infty}^\infty \text{d}t \, \frac{e^{-\frac{(t-t_j)^2}{2s^2}}}{\sqrt{\sqrt{\pi}s}} \ket{t}, \label{EveattackZ}
\end{align}
where $s$ is a free parameter that Eve can freely tune. Similarly, Eve prepares the following one-photon Gaussian-modulated frequency pulse when intercepting in the $X$ basis, corresponding to outcome $k$:

\begin{align}
    \ket{E_X(k)} := \int_{-\infty}^\infty \text{d}f \, \frac{e^{-\frac{f^2}{2\sigma^2}}}{\sqrt{\sqrt{\pi}\sigma}} e^{i(2\pi \tilde{t}_k f -2\pi^2 \Phi_2 f^2)}\ket{f}, \label{EveattackX}
\end{align}
where $\sigma$ is a free parameter. In \eqref{EveattackZ} and \eqref{EveattackX}, $a^\dag_f \ket{vac}=\ket{f}$ ($a^\dag_t \ket{vac}=\ket{t}$) represents an unphysical state of a single photon at frequency $f$ (time $t$). It holds $\ket{t}=\int_{-\infty}^\infty \text{d}f e^{2i\pi ft} \ket{f}$. 

We now compute the probabilities of Bob obtaining certain outcomes, given that Eve performed the attack in the $Z$ basis, \eqref{EveattackZ}, or $X$ basis, \eqref{EveattackX}.

The probability that Bob obtains outcome $j'$ in the $Z$ basis, given that Eve sent the pulse corresponding to outcome $j$ in \eqref{EveattackZ}, is given by:
\begin{align}
    \Pr(Z_B=j'|Z_E=j) &= (1-\eta_i)\eta_Z \int_{t_{j'}-\Delta_j/2}^{t_{j'}+\Delta_j/2} \text{d}t \frac{e^{-\frac{(t-t_j)^2}{s^2}}}{\sqrt{\pi}s} \nonumber\\
    &= \frac{(1-\eta_i)\eta_Z}{2} \left[\Erf\left(\frac{t_{j'}-t_j+\Delta_j/2}{s}\right) + \Erf\left(\frac{-t_{j'}+t_j+\Delta_j/2}{s}\right)\right],  \label{Prj'|j}
\end{align}
with $\Erf(x)=(2/\sqrt{\pi})\int_0^x \text{d}t e^{-t^2}$ the error function. We observe that Eve can tune the width of her Gaussian pulse, $s$, in order to introduce little noise in the $Z$-basis outcomes. This is achieved by choosing $s \ll \Delta_j$, so that $\Pr(Z=j'|Z=j) \propto \delta_{j'j}$. Note that the probability in \eqref{Prj'|j} is computed modulo the transmittance $\xi_Z$ of Eve's channel. The probability that Bob obtains outcome $k'$ in the $X$ basis, given that Eve sent the pulse corresponding to outcome $j$ in \eqref{EveattackZ}, is given by:
\begin{align}
    \Pr(X_B=k'|Z_E=j) = &\eta_X \int_{\tilde{t}_{k'}-\Delta_k/2}^{\tilde{t}_{k'}+\Delta_k/2} \text{d}t \frac{e^{-\frac{(t-t_j)^2}{s^2+(\Phi_2/s)^2}}}{4\sqrt{\pi}\sqrt{s^2 + (\Phi_2/s)^2}} \nonumber\\
    &\Bigg\lvert 2\sqrt{\eta_l} + (\sqrt{\eta_i} -\sqrt{\eta_l})\left[\Erf\left(\frac{(1+i)s^2(2t+\Delta_t) + (1-i)(2t_j+\Delta_t)\Phi_2}{4s\sqrt{\Phi_2}\sqrt{s^2 - i \Phi_2}}\right)  \right. \nonumber\\
    &\left. - \Erf\left(\frac{(1+i)s^2(2t-\Delta_t) + (1-i)(2t_j-\Delta_t)\Phi_2}{4s\sqrt{\Phi_2}\sqrt{s^2 - i \Phi_2}}\right) \right] \Bigg\rvert^2,  \label{Prk'|j}
\end{align}
and is also computed modulo the transmittance of Eve's channel, $\xi_X$. We observe that Eve can tune $s$ such that $\Phi_2/s \gg \Delta_f$. In this way, she can reduce the detection probability at Bob when they choose opposite bases. In summary, when intercepting in the $Z$ basis, Eve prepares pulses given by \eqref{EveattackZ}, with $s$ satisfying both $s \ll \Delta_j$ and  $\Phi_2/s \gg \Delta_f$, so that Bob observes little noise in the $Z$ basis detections and a reduced detection probability in the $X$ basis.

The probability that Bob obtains outcome $j'$ in the $Z$ basis, given that Eve sent the pulse corresponding to outcome $k$ in \eqref{EveattackX}, is given by:
\begin{align}
    \Pr(Z_B=j'|X_E=k) &= \frac{(1-\eta_i)\eta_Z}{2} \left[\Erf\left(\frac{t_{j'}-\tilde{t}_k+\Delta_j/2}{\sqrt{2}\sigma'}\right) + \Erf\left(\frac{-t_{j'}+\tilde{t}_k+\Delta_j/2}{\sqrt{2}\sigma'}\right)\right],  \label{Prj'|k}
\end{align}
where $\sigma':=\sigma\abs{1/(2\sigma^2)+2i\pi^2 \Phi_2}/(\sqrt{2}\pi)$. Note that Eve can increase the value of $\sigma$ such that $\sigma'\gg \Delta_t$, thereby reducing the detection probability in the $Z$ basis. The probability that Bob obtains outcome $k'$ in the $X$ basis, given that Eve sent the pulse corresponding to outcome $k$ in \eqref{EveattackX}, is given by:
\begin{align}
    \Pr(X_B=k'|X_E=k) = &\eta_X \int_{\tilde{t}_{k'}-\Delta_k/2}^{\tilde{t}_{k'}+\Delta_k/2} \text{d}t \frac{\sqrt{\pi}\sigma}{2}e^{-4\pi^2 \sigma^2 (t-\tilde{t}_k)^2} \nonumber\\
    &\Bigg\lvert 2\sqrt{\eta_l} + (\sqrt{\eta_i} -\sqrt{\eta_l})\left[\Erf\left(\frac{2t+\Delta_t+8i\pi^2 \sigma^2 \Phi_2 (t-\tilde{t}_k)}{2\sqrt{2\Phi_2}\sqrt{4\pi^2 \sigma^2\Phi_2 -i}}\right)  \right. \nonumber\\
    &\left. - \Erf\left(\frac{2t-\Delta_t+8i\pi^2 \sigma^2 \Phi_2 (t-\tilde{t}_k)}{2\sqrt{2\Phi_2}\sqrt{4\pi^2 \sigma^2\Phi_2 -i}}\right) \right]\Bigg\rvert^2. \label{Prk'|k}
\end{align}
Similarly, Eve can increase $\sigma$  to satisfy $(4\pi^2 \sigma^2)^{-1/2} \ll \Delta_k$, such that the error introduced in the $X$-basis outcomes is negligible: $\Pr(X_B=k'|X_B=k) \propto \delta_{kk'}$.

Given the attacks described in \eqref{EveattackZ} and \eqref{EveattackX}, the resulting one-photon yields read:
\begin{align}
     Y^Z_{1,(\eta_l,\eta_l)} &= (1-w_X-w_Z)Y^{Z,\textrm{hon.}}_{1,(\eta_l,\eta_l)} + w_Z\left[p^{\mathcal{Z}}_d +   \xi_Z \Pr(Z \checkmark|Z)(1-p^{\mathcal{Z}}_d)\right] + w_X\left[p^{\mathcal{Z}}_d +   \xi_X \Pr(Z \checkmark|X)(1-p^{\mathcal{Z}}_d)\right] \label{yieldZ-nodecoy-dishonest} \\
     Y^X_{1,(\eta_l,\eta_l)} &= (1-w_X-w_Z)Y^{X,\textrm{hon.}}_{1,(\eta_l,\eta_l)} + w_Z\left[p^{\mathcal{X}}_d +  \xi_Z \Pr(X \checkmark|Z)(1-p^{\mathcal{X}}_d)\right] + w_X\left[p^{\mathcal{X}}_d +  \xi_X \Pr(X \checkmark|X)(1-p^{\mathcal{X}}_d)\right]\label{yieldX-nodecoy-dishonest} \\
     Y^{X,\checkmark}_{1,(\eta_i,\eta_l)} &=  (1-w_X-w_Z)Y^{X,\checkmark,\textrm{hon.}}_{1,(\eta_i,\eta_l)} \nonumber\\
     &+ w_Z \left[ \xi_Z \Pr(X \checkmark|Z) p^{\mathcal{Z}}_d +   \xi_Z \Pr(Z \checkmark|Z) p^{\mathcal{X}}_d + \left(1- \xi_Z \Pr(X \checkmark|Z) -  \xi_Z \Pr(Z\checkmark|Z)\right)p^{\mathcal{Z}}_d p^{\mathcal{X}}_d\right] \nonumber\\
     &+ w_X \left[  \xi_X \Pr(X \checkmark|X) p^{\mathcal{Z}}_d +   \xi_X \Pr(Z \checkmark|X) p^{\mathcal{X}}_d + \left(1- \xi_X \Pr(X \checkmark|X) -  \xi_X \Pr(Z\checkmark|X)\right)p^{\mathcal{Z}}_d p^{\mathcal{X}}_d\right]\label{yieldXclick-nodecoy-dishonest} \\
     Y^{X,\emptyset}_{1,(\eta_i,\eta_l)} &= (1-w_X-w_Z)Y^{X,\emptyset,\textrm{hon.}}_{1,(\eta_i,\eta_l)} + w_Z \left[   \xi_Z \Pr(X \checkmark|Z) (1-p^{\mathcal{Z}}_d) + \left(1-  \xi_Z \Pr(X \checkmark|Z) -  \xi_Z \Pr(Z\checkmark|Z)\right)p^{\mathcal{X}}_d (1-p^{\mathcal{Z}}_d) \right] \nonumber\\
     &+ w_X \left[   \xi_X \Pr(X \checkmark|X) (1-p^{\mathcal{Z}}_d) + \left(1-  \xi_X \Pr(X \checkmark|X) -  \xi_X \Pr(Z\checkmark|X)\right)p^{\mathcal{X}}_d (1-p^{\mathcal{Z}}_d) \right],
     \label{yieldXnoclick-nodecoy-dishonest}
\end{align}
where $\Pr(Z\checkmark|Z)$ is the probability that Bob obtains a detection in the $Z$ basis, given that Eve intercepted in the $Z$ basis. In particular, we have:
\begin{align}
    \Pr(Z\checkmark|Z)&= \frac{1}{d} \sum_{j,j'=0}^{d-1} \Pr(Z_B=j'|Z_E=j),
\end{align}
where we accounted for the fact that Alice sends uniformly random symbols and that Eve can perfectly distinguish them. Similarly, we have:
\begin{align}
    \Pr(X\checkmark|Z)&= \frac{1}{d} \sum_{k',j=0}^{d-1} \Pr(X_B=k'|Z_E=j),
\end{align}
where we accounted for the fact that Eve obtains a uniformly random outcome when measuring Alice's pulses in the opposite basis. Analogously, we have:
\begin{align}
    \Pr(Z\checkmark|X)&= \frac{1}{d} \sum_{k,j'=0}^{d-1} \Pr(Z_B=j'|X_E=k) \\
    \Pr(X\checkmark|X)&= \frac{1}{d} \sum_{k',k=0}^{d-1} \Pr(X_B=k'|X_E=k).
\end{align}

Finally, the one-photon error rates in the adversarial implementation satisfy the following equalities:
\begin{align}
     Y^{Z}_{1,(\eta_l,\eta_l)} e_{Z,1,(\eta_l,\eta_l)} &= (1-w_Z -w_X)Y^{Z,\textrm{hon.}}_{1,(\eta_l,\eta_l)} e^{\textrm{hon.}}_{Z,1,(\eta_l,\eta_l)} + w_Z \left[\xi_Z \Pr(Z \textrm{err}|Z) + \left(1-\xi_Z \Pr(Z \checkmark|Z)\right)p^{\mathcal{Z}}_d \left(1-\frac{1}{d}\right)\right] \nonumber\\
     &+ w_X \left[\xi_X \Pr(Z \textrm{err}|X) + \left(1-\xi_X \Pr(Z \checkmark|X)\right)p^{\mathcal{Z}}_d \left(1-\frac{1}{d}\right)\right] \label{yieldZerror-nodecoy-dishonest} \\
     Y^{X}_{1,(\eta_l,\eta_l)} e_{X,1,(\eta_l,\eta_l)} &= (1-w_Z -w_X) Y^{X,\textrm{hon.}}_{1,(\eta_l,\eta_l)} e^{\textrm{hon.}}_{X,1,(\eta_l,\eta_l)}  + w_Z \left[\xi_Z \Pr(X \textrm{err}|Z) + \left(1-\xi_Z \Pr(X \checkmark|Z)\right)p^{\mathcal{X}}_d \left(1-\frac{1}{d}\right)\right] \nonumber\\
     &+ w_X \left[\xi_X \Pr(X \textrm{err}|X) + \left(1-\xi_X \Pr(X \checkmark|X)\right)p^{\mathcal{X}}_d \left(1-\frac{1}{d}\right)\right] \label{yieldXerror-nodecoy-dishonest} \\
     Y^{X,\checkmark}_{1,(\eta_i,\eta_l)} e_{X,1,(\eta_i,\eta_l),\checkmark} &= (1-w_Z -w_X)Y^{X,\checkmark,\textrm{hon.}}_{1,(\eta_i,\eta_l)} e^{\textrm{hon.}}_{X,1,(\eta_i,\eta_l),\checkmark} + w_Z \left[\xi_Z \Pr(X \textrm{err}|Z)p^{\mathcal{Z}}_d + \xi_Z \Pr(Z \checkmark|Z)p^{\mathcal{X}}_d \left(1-\frac{1}{d}\right) \right. \nonumber\\
     &\left.+ (1-\xi_Z \Pr(X\checkmark|Z)-\xi_Z \Pr(Z\checkmark|Z))p^{\mathcal{X}}_d p^{\mathcal{Z}}_d \left(1-\frac{1}{d}\right) \right] \nonumber\\
     &+ w_X \left[\xi_X \Pr(X \textrm{err}|X)p^{\mathcal{Z}}_d + \xi_X \Pr(Z \checkmark|X)p^{\mathcal{X}}_d \left(1-\frac{1}{d}\right) \right. \nonumber\\
     &\left.+ (1-\xi_X \Pr(X\checkmark|X)-\xi_X \Pr(Z\checkmark|X))p^{\mathcal{X}}_d p^{\mathcal{Z}}_d \left(1-\frac{1}{d}\right) \right]\label{yieldXclickerror-nodecoy-dishonest}
\end{align}
\begin{align}
     Y^{X,\emptyset}_{1,(\eta_i,\eta_l)} e_{X,1,(\eta_i,\eta_l),\emptyset} &=(1-w_Z -w_X)Y^{X,\emptyset,\textrm{hon.}}_{1,(\eta_i,\eta_l)} e^{\textrm{hon.}}_{X,1,(\eta_i,\eta_l),\emptyset} + w_Z \left[\xi_Z \Pr(X \textrm{err}|Z) (1-p^{\mathcal{Z}}_d)  \right. \nonumber\\
     &\left.+ (1-\xi_Z \Pr(X\checkmark|Z)-\xi_Z \Pr(Z\checkmark|Z))p^{\mathcal{X}}_d (1-p^{\mathcal{Z}}_d) \left(1-\frac{1}{d}\right) \right] \nonumber\\
     &+ w_X \left[\xi_X \Pr(X \textrm{err}|X)(1-p^{\mathcal{Z}}_d) + (1-\xi_X \Pr(X\checkmark|X)-\xi_X \Pr(Z\checkmark|X))p^{\mathcal{X}}_d (1-p^{\mathcal{Z}}_d) \left(1-\frac{1}{d}\right) \right] \label{yieldXnoclickerror-nodecoy-dishonest},
\end{align}
where $\Pr(Z \textrm{err}|Z)$ is the probability that Bob obtains a click in the $Z$ basis and the outcome is different from Alice's, given that Eve attacked in the same basis:
\begin{align}
    \Pr(Z\textrm{err}|Z)&= \frac{1}{d} \sum_{j=0}^{d-1} \sum_{j' \neq j} \Pr(Z_B=j'|Z_E=j),
\end{align}
where again we assumed that Eve can perfectly distinguish Alice's symbols with no errors when she chooses the same basis. When Eve chooses the opposite basis, the probability that Bob gets an error is given by:
\begin{align}
    \Pr(Z\textrm{err}|X)&= \frac{1}{d} \sum_{j,k=0}^{d-1} \sum_{j' \neq j} \Pr(Z_B=j'|X_E=k) \Pr(X_E=k|Z_A=j) \nonumber\\
    &= \frac{1}{d^2} \sum_{j,k=0}^{d-1} \sum_{j' \neq j} \Pr(Z_B=j'|X_E=k),
\end{align}
where we used the fact that Eve's outcome is uniformly random when measuring in the opposite basis than Alice's. In the same vein, we have:
\begin{align}
    \Pr(X\textrm{err}|X)&= \frac{1}{d} \sum_{k=0}^{d-1} \sum_{k' \neq k} \Pr(X_B=k'|X_E=k) \\
    \Pr(X\textrm{err}|Z)&=\frac{1}{d^2} \sum_{j,k=0}^{d-1} \sum_{k' \neq k} \Pr(X_B=k'|Z_E=j).
\end{align}

\subsubsection{Upper bound on secret key rate} \label{app:upperbound-keyrate}
Here we derive the upper bound on the asymptotic secret key rate achievable by Alice and Bob under Eve's attack. The upper bound is plotted in Fig.~\ref{fig:keyrate_attack} together with the BB84 key rate. The fact that the latter surpasses the former indicates that the BB84 protocol is insecure and returns overly optimistic key rates, as discussed in Sec.~\ref{sec:discussion}.

The upper bound is obtained by considering the raw key bits per pulse shared by Alice and Bob, $r_{\rm raw}$, and by subtracting the amount of raw key bits per pulse known to Eve, $r_{\rm Eve}$. For $r_{\rm raw}$, we have:
\begin{align}
    r_{\rm raw} = p_Z^2  Y^Z_{1,(\eta_\downarrow,\eta_\downarrow)} \log_2 d,
\end{align}
since the key bits are only generated from the events where both Alice and Bob choose the $Z$ basis and Bob gets a detection. The factor $\log_2 d$ accounts for the fact that each symbol contains $\log_2 d$ bits of information. For $r_{\rm Eve}$, we recall that Eve knows perfectly the shared raw key bits in the rounds where Alice, Bob and Eve all choose the $Z$ basis and Bob gets a detection in the $Z$ basis. Moreover, Eve learns the bits exchanged by Alice and Bob for error correction. This amounts to the following rate of raw key bits known to Eve:
\begin{align}
    r_{\rm Eve} = p_Z^2  w_Z\left[p^{\mathcal{Z}}_d +   \xi_Z \Pr(Z \checkmark|Z)(1-p^{\mathcal{Z}}_d)\right] \log_2 d + p_Z^2  Y^Z_{1,(\eta_\downarrow,\eta_\downarrow)} u(e_{Z,1,(\eta_\downarrow,\eta_\downarrow)}).
\end{align}
Hence, the upper bound on the secret key rate is given by:
\begin{align}
    r_{\rm raw} - r_{\rm Eve} =  p_Z^2  Y^Z_{1,(\eta_\downarrow,\eta_\downarrow)} \left[\log_2 d - u(e_{Z,1,(\eta_\downarrow,\eta_\downarrow)}) \right] - p_Z^2  w_Z\left[p^{\mathcal{Z}}_d +   \xi_Z \Pr(Z \checkmark|Z)(1-p^{\mathcal{Z}}_d)\right] \log_2 d. \label{keyrate-upperbound}
\end{align}

\subsection{Secret key rate vs (mode-independent) detection efficiency mismatch} \label{app:simulations-weightestimation}

In this section, we study the performance of the key rate from our proof, Eq.~\eqref{protocol-rate}, for varying asymmetries in the mode-independent detection efficiency of the two bases, i.e., as a function of $\eta_r=\eta_X /\eta_Z$. This study stems from the observation made in Sec.~\ref{sec:discussion} that our key rate does not match the decoy-BB84 key rate in case of an honest implementation of the protocol and that the difference may be due to the asymmetric detection efficiency of the two bases.

\begin{figure}
    \centering
    \includegraphics[width=0.8\columnwidth]{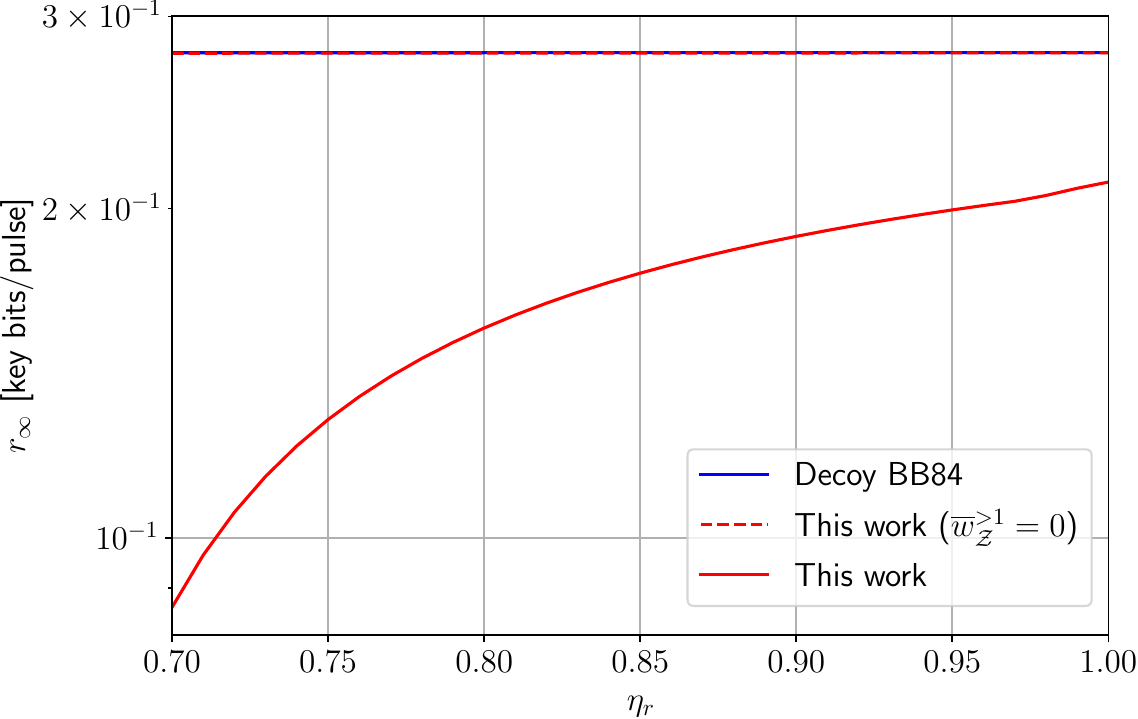}
    \caption{The secret key rate in Eq.~\eqref{protocol-rate} in case of the honest implementation of Sec.~\ref{sec:honest-implementation}, as a function of the detection efficiency mismatch ($\eta_r$). Note that we fix $\eta_Z=0.9\cdot 10^{-1/10}$ and vary $\eta_X$ to achieve different values of $\eta_r$. We also fix the intrinsic QBERs to $q_Z = q_X = 0.02$. Red, solid: The secure key rate using the bound on $\overline{w}^{>1}_{\mathcal{Z}}$ reported in \eqref{protocol-yalpha>1}. Red, dashed: the key rate when we impose $\overline{w}^{>1}_{\mathcal{Z}}=0$. Blue, solid: The decoy-BB84 key rate in \eqref{protocol-rate-withassumption}. We observe that the red dashed line perfectly overlaps the blue solid line, indicating that our proof would become tight and match the decoy-BB84 rate, when a tight estimation of $\overline{w}^{>1}_{\mathcal{Z}}$ is available.}
    \label{fig:keyrate_etar}
\end{figure}

In Fig.~\ref{fig:keyrate_etar}, we plot the key rate \eqref{protocol-rate} as a function of $\eta_r$ and compare it with the decoy-BB84 rate \eqref{protocol-rate-withassumption} and with the same key rate \eqref{protocol-rate} obtained by artificially setting the weight of the state received by Bob outside the $(\leq 1)$-subspace to zero: $\overline{w}^{>1}_{\mathcal{Z}} = 0$. This would correspond to the scenario with perfect knowledge of the parameter $\overline{w}^{>1}_{\mathcal{Z}}$, which is indeed null in the considered honest implementation. The other parameters are set as in Fig.~\ref{fig:keyrate_eta} for the honest implementation of the protocol, where $\eta_Z=0.9 \cdot 10^{-1/10}$ and the channel loss is fixed to $\eta=10^{-1/10}$. 

From Fig.~\ref{fig:keyrate_etar}, we observe that in the case of perfect knowledge of the parameter $\overline{w}^{>1}_{\mathcal{Z}}$, the detection efficiency mismatch does not influence the key rate \eqref{protocol-rate} and the latter matches the decoy-BB84 key rate perfectly. Conversely, when using our bound, Eq.~\eqref{protocol-yalpha>1}, on the parameter $\overline{w}^{>1}_{\mathcal{Z}}$, the key rate of our proof presents a gap to the decoy-BB84 rate that increases as $\eta_r$ decreases. Therefore, we deduce that the non-tight estimation of $\overline{w}^{>1}_{\mathcal{Z}}$ is the main cause of loss of performance, becoming more problematic as the asymmetry in the detection efficiencies increases.

\end{document}